\documentclass[a4paper]{article}
\pdfoutput=1 
\usepackage{jheppub} 
\usepackage[T1]{fontenc}
\usepackage[italian,english]{babel}
\usepackage{hyperref}
\usepackage{ifpdf}
\usepackage{subfigure}
\usepackage{amssymb}
\usepackage{amsfonts}
\usepackage{epsf}
\usepackage{rotating}
\usepackage{graphicx}
\usepackage{amsmath}
\usepackage{fancyhdr}
\usepackage{lineno}
\usepackage{babel}
\usepackage{cancel}
\usepackage{graphics}
\usepackage{pstricks}
\usepackage{color}
\usepackage{diagbox}
\usepackage{multirow}
\usepackage{nicefrac}
\usepackage{bm}
\usepackage{slashed}
\usepackage{float}
\usepackage{array}
\usepackage{boldline}
\usepackage{mathrsfs}
\usepackage{xcolor}

\usepackage{hyperref}
\hypersetup{
	colorlinks=true,
	linkcolor=[rgb]{0.0,0.32,0.89},
	filecolor=magenta,      
	urlcolor=[rgb]{0.0,0.42,0.1},
	citecolor=[rgb]{0.64,0.0,0.0},
}
\newcommand{\gsim}{\raisebox{-0.13cm}{~\shortstack{$>$ \\[-0.07cm]
			$\sim$}}~}

\def\etmiss{\not\!\!{E_T}}
\def\ptmiss{\not\!\!{p_T}}

\definecolor{Gr}{rgb}{0.0, 0.66, 0.15}

\usepackage[numbers,sort&compress]{natbib}
\usepackage{tikz}
\usepackage{tikz-feynman}
\tikzfeynmanset{compat=1.0.0}

\newcommand*\xbar[1]{%
	\hbox{%
		\vbox{%
			\hrule height 0.65pt 
			\kern0.4ex
			\hbox{%
				\kern-0.05em
				\ensuremath{#1}%
				\kern0.0em
			}%
		}%
	}%
}

\allowdisplaybreaks

\title{\boldmath Displaced Higgs production in Type-III Seesaw at the LHC/FCC, MATHUSLA and Muon collider}
\preprint{IITH-PH-0001/21}
\author[a]{Chandrima Sen}
\author[b]{Priyotosh Bandyopadhyay,}
\author[c]{Saunak Dutta,}
\author[d]{Aleesha KT}

\affiliation[a,b,c]{Indian Institute of Technology Hyderabad, Kandi,  Sangareddy-502285, Telengana, India}
\affiliation[c]{2SGTB Khalsa College, University of Delhi, New Delhi, India.}
\affiliation[d]{Department of Physical Sciences, Indian Institute of Science Education and Research Kolkata, Mohanpur, 741246, India}
\emailAdd{ph19resch11014@iith.ac.in }
\emailAdd{bpriyo@phy.iith.ac.in}
\emailAdd{saunak100@gmail.com}
\emailAdd{aleeshakt949@gmail.com }

\abstract{In this article, we explore the possibility of displaced Higgs production from the  decays of the  heavy fermions in  the  Type-III seesaw extension of the Standard Model  at the LHC/FCC and the muon collider. The displaced heavy fermions and the Higgs boson can be traced back by  measuring the displaced charged tracks of the charged leptons along with the $b$-jets. A very small Yukawa coupling  can lead to two successive displaced decays which makes the phenomenology even more interesting. The prospects of the transverse and longitudinal displaced decay lengths are extensively studied in the context of the boost at the LHC/FCC. Due to the parton distribution function, the longitudinal boosts leads to larger displacement compared to the transverse one, which can reach MATHUSLA and beyond. The longitudinal measurements are indeed possible by the  visible part of the finalstate, which captures the complete information about the longitudinal momenta. The comparative studies  are made at the LHC/FCC with  the centre of mass energies of 14, 27 and 100 TeV, respectively. A futuristic study of the muon collider where the collision happen in the centre of mass frame is analysed for centre of mass energies of 3.5, 14 and 30 TeV. Contrary to LHC/FCC, here the transverse momentum diverges, however, the maximum reach in both the direction are identical due to the constant total momentum in each collision. The reach of the Yukawa couplings and fermion masses are appraised for both the colliders. FCC at 100 TeV can probe a mass of 4.25 TeV and a lowest Yukawa coupling of  $\mathcal{O}(5 \times 10^{-11})$.}

\begin{document}
	
\maketitle
\flushbottom

\section{Introduction}
Neutrinos are massless in Standard Model (SM), but observation of   neutrino oscillation  data \cite{nuOscl1,nuOscl2} needs neutrinos to be massive but very tiny.  Explanation of tiny  neutrino mass can be elucidated via Seesaw mechanisms, and one of such existing mechanism is Type-III \cite{Foot:1988aq,Bajc:2006ia,Franceschini:2008pz,Bajc:2007zf,Ma:1998dn}, where the new beyond Standard Model (BSM) fermions are in spin one representation of $SU(2)$. There  are numerous phenomenological studies at the LHC and other colliders on Type-III seesaw probing different aspects of the BSM scenarios \cite{Arhrib:2009mz,Bandyopadhyay:2009xa,Bandyopadhyay:2011aa,Eboli:2011ia,Cai:2017mow, Goswami:2017jqs,delAguila:2008hw,delAguila:2008cj,Bandyopadhyay:2020mnp,Agostinho:2017biv,Ibanez:2009du, Das:2020uer,Ashanujjaman:2021jhi,Das:2020gnt,Abada:2008ea,Escribano:2021css} along with the generic displaced decay phenomenology  \cite{Jana:2020qzn}. The Type-III seesaw model has been studied considering electroweak vacuum stability \cite{Goswami:2018jar}, and limits on heavy fermion generation have  been drawn for inverse Type-III scenario from the perturbativity of SU(2) gauge coupling \cite{Bandyopadhyay:2020djh}.
Recent searches of  Type-III fermions at the LHC has put a lower limit on the mass of Type-III fermions as $\sim 740$ GeV \cite{CMS:2017ybg,ATLAS:2020wop,CMS:2019lwf} at $2\sigma$ level, which  leaves the possibility of such heavy fermions around TeV along with the indirect bounds \cite{Biggio:2019eeo}. However, such bounds are with the assumption of capturing  the prompt leptons only and the lower  mass  bound can reduce for the  displaced  lepton as some of them will be missed by the detectors.

The explanation of atmospheric neutrino mass required relatively small neutrino Yukawa couplings $\mathcal{O}(10^{-7})$ for  $\mathcal{O}(100)$ GeV heavy neutrino mass scale. For solar neutrinos mass scale, the couplings will be one order of magnitude less. As we focus on the  displaced decay signatures of these Type-III fermions, we  consider  one generation of  Yukawa couplings small, whereas  the other two remain relatively  large satisfying the  light neutrino masses mixing \cite{Ibarra:2003up,Ibarra:2005qi}. The tiny neutrino Yukawa coupling can lead to the displaced production of  the Higgs boson, which can be reconstructed  via its dominant decay mode $b\bar{b}$. The displaced signature is clean from any SM background, and it can be used to constrain the Yukawa couplings. Such a scenario is  explored in the context of supersymmetry  while considering the decays of the scalar partners of the heavy neutrinos \cite{Bandyopadhyay:2010wp}.

In this article, at first, we explore the displaced Higgs production at the LHC/FCC with the centre of mass energies of 14, 27 and 100 TeV in the detectors CMS, ATLAS \cite{Cepeda:2019klc} and MATHUSLA \cite{MATHUSLA:2019qpy, Chou:2016lxi,Curtin:2018mvb,Coccaro:2016lnz}. 
MATHUSLA is a proposed detector for studying Long Lived Particles (LLP) produced by the High-Luminosity LHC (HL-LHC) from displaced charged tracks, with a modified range of 68-168 m in the longitudinal direction and 60-85 m in the transverse direction from the CMS/ATLAS interaction point (IP)\cite{Alpigiani:2020iam}, allowing for even smaller effective coupling in constraining or discovering new physics. Earlier the propose detector was  with a $100-300$ m in the longitudinal and  $100-125$ m in the transverse direction and in  this study we present our results for  both proposed lengths. However, the muon collider can open up new frontiers with energies of  3.5, 14 and 30 TeV as we propose similar detectors for long lived particles. 

In particular, we devote our analysis to include the boost effect in the transverse and the longitudinal directions. At  the LHC, colliding parton momenta are unknown and   governed by the parton distribution functions. This causes a boosted system mostly in the longitudinal directions. If the BSM particle decays into a complete visible mode, we can measure the momenta and the exact displaced decay length of such particles.  MATHUSLA detector will be only on one side of the LHC detectors, thus asymmetric in the $z$-axis. While the transverse direction is symmetric, the amount of boost is negligible compared to the longitudinal one. Additionally, if there is only one missing particle in the finalstate  the momentum can be reconstructed as the total transverse momentum is zero at any given point. Thus the momentum of the actual BSM particle can be constructed, enabling us to estimate the boosted  decay  length of  that   BSM particle. A comparative study for transverse and longitudinal decay lengths are given for CMS, ATLAS and MATHUSLA for the centre of mass energies of 14, 27 and 100 TeV at the LHC/FCC and model independent limits  on the parameter space based on these collider searches are derived. 

Recently muon collider \cite{Palmer:1996gs,Ankenbrandt:1999cta,Delahaye:2019omf,Bartosik:2019dzq,Bartosik:2020xwr} is receiving lots of attention for studying various BSM  scenarios due to the precision measurements,  centre of mass frame, no initial state QCD radiation, etc \cite{AlAli:2021let,Costantini:2020stv,Buttazzo:2018qqp,Huang:2021nkl,Huang:2021biu,Asadi:2021gah,Capdevilla:2020qel,Long:2020wfp,Han:2020pif,Han:2020uak,Capdevilla:2021rwo,Han:2021udl,Bandyopadhyay:2021pld,Liu:2021jyc}. The long lived particles  can also be explored at the muon collider, where we proposed such displaced vertex measurement facility, which can be instrumental in exploring some of these BSM scenarios. As opposed to LHC/FCC, in a muon collider, the total momentum is constant for each collision since it happens in the centre of mass frame. One interesting fact is that the transverse momentum for the Type-III fermion diverges perpendicular to the beam axis, i.e. pseudorapidity goes to zero. This results in slight enhancement of the number of events for higher transverse decay length compared to longitudinal ones. The highest reach of the decay length is constrained by the maximum momentum, which in this case is the net momentum, and thus for both scenarios, maximum decay length will be identical. The analysis has been carried out for centre of mass energies of 3.5, 14 and  30 TeV with 1000, 10000 and 30000 fb$^{-1}$ and limits on mass and coupling are drawn.

This article is organised as follows. In \autoref{model}, we describe the model and the decay modes. The allowed parameter space is discussed in \autoref{bpdis}, and we choose our benchmark points for the displaced vertex. \autoref{setup} goes over the collider setups. The simulations  at the LHC/FCC, including the kinematics, displaced vertex and results are given in \autoref{LHCF}. Similarly, these are detailed in \autoref{muonc} for the muon collider. In \autoref{reach}, we summaries the coupling versus mass regions that can be investigated, and in \autoref{concl}, we present our conclusions.  

\section{Type III Seesaw: The Model}\label{model}

The new $SU(2)$ triplet fermions ($N$)  with hypercharge ($Y$)  zero can be added to the SM Lagrangian with the addition of the following terms as  given in \autoref{lag}

\begin{equation}\label{lag}
\mathcal{L}_N = \rm Tr(\bar{N}\,\cancel{D}\,N)~-~\frac{1}{4}M_{N}\,\rm Tr \left[\bar{N}N\right]~-~Y_{N} \left(\tilde{\phi}^\dagger\bar{N}L~+~\bar{L}N\tilde{\phi}\right),
\end{equation}
where $\phi=\begin{pmatrix} G^\pm\\ \frac{1}{\sqrt{2}}(h + v +i G^0)\end{pmatrix} $ is the Higgs doublet in Standard Model and, $L=\begin{pmatrix}
\nu_{\ell} \\ \ell \end{pmatrix}$ is the left handed lepton doublet. Here $G^\pm$ and $G^0$ are charged and neutral Goldstone bosons, $h$ is the physical Higgs boson and $v$ is the vacuum expectation value (vev) which generates the mass for the SM particles. In \autoref{lag}, we use  $\tilde{\phi} = i\sigma^2\phi^*$ and the $N_i$ are represented by  \autoref{sigmai}. $N_i$  has one pair of charged fermion ($N^\pm_i$) and one neutral component ($N^0_i$) for each generation $i$ ($i=1,2,3$) respectively.

\begin{equation}\label{sigmai}
N_i = \begin{pmatrix}
N^0_i && \sqrt{2} N^+ _i\\
\sqrt{2}N^-_i && -N^0_i
\end{pmatrix}  \qquad (i = 1,2,3)
\end{equation}

For the simplicity of the collider study, we consider  the $N_i$ in diagonal basis along with the  $Y_{N _i}$. As the Higgs boson gets vev, i.e. $<\phi>= \frac{v}{\sqrt{2}}$, the light neutrino mass  is generated, which is scaled by the Type-III fermion  mass as  given below
\begin{eqnarray}
m^{\nu}_{ij}= \frac{Y_{N_i} Y_{N_j} v^2}{2M_N}. 
\end{eqnarray}

\begin{figure*}[h]
	\begin{center}
		\hspace*{-0.5cm}
		\mbox{\subfigure[]{\includegraphics[width=0.40\linewidth,angle=-0]{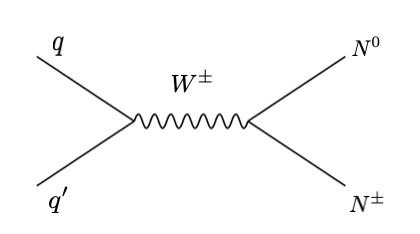}\label{}}
			\subfigure[]{\includegraphics[width=0.40\linewidth,angle=-0]{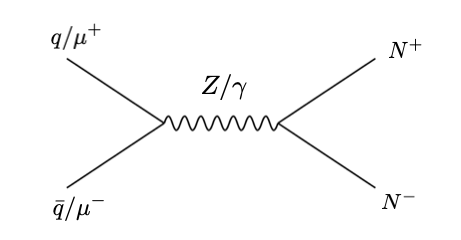}\label{}}}			
		\hspace*{-0.5cm}
		\mbox{\subfigure[]{\includegraphics[width=0.35\linewidth,angle=-0]{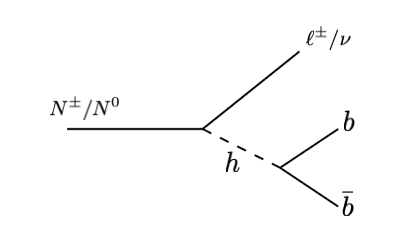}\label{}}
			\subfigure[]{\includegraphics[width=0.35\linewidth,angle=-0]{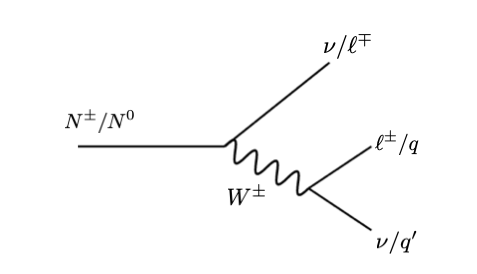}\label{}}
			\subfigure[]{\includegraphics[width=0.35\linewidth,angle=-0]{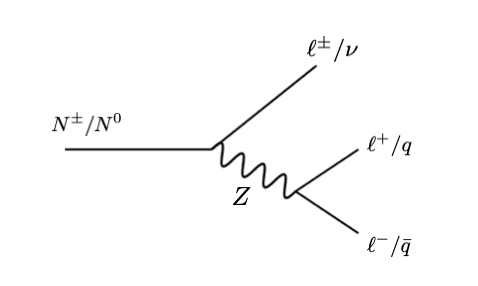}\label{}}}
		\caption{Feynman diagrams of the production and decay of Type-III fermions $N^0$ and $N^{\pm}$. (a) shows the associate production at the LHC/FCC and (b) depicts the pair production of $N^\pm$ at the LHC/FCC and the muon collider. The decay of $N^0$ and $N^{\pm}$ to Higgs boson and SM gauge bosons are shown in (c), (d) and (e). }\label{Fy_diag}
	\end{center}
\end{figure*}

At the LHC, associated $N^0 N^\pm$ can be  produced via $W^\pm$ boson exchange as well as the pair of $N^\pm$ via photon and $Z$ boson which can be seen from \autoref{Fy_diag}(a), (b). Unlike at the LHC, in the muon collider, $N^0 N^\pm$  cannot be produced.

The hard scattering  cross-sections can be calculated as 
\begin{eqnarray}
\frac{d\sigma}{d\hat{s}^2} = \sum_q \int dx \int dy\, f_q(x)f_{\bar{q}}(y) \frac{d\hat{\sigma}}{d\hat{s}^2},
\end{eqnarray} 
where we need to use the parton distribution functions $f_q(x)$ at the LHC/FCC. For this article, we have used NNPDF23$\_$lo$\_$as$\_$0130$\_$qed \cite{NNPDF:2014otw}.  The partonic differential distribution of the cross-section can be written as 
\begin{eqnarray}
\frac{d\hat{\sigma}}{d\cos \theta}&=& \frac{V_L^2+V_R^2}{64\pi}\,\beta\,N_c\left(4M_N^2+\hat{s}-\beta^2\hat{s}\cos^2\theta\right),
\end{eqnarray}
where $N_c=3$ for quarks (=1 for leptons), $\beta = \sqrt{1-\frac{4M_N^2}{\hat{s}}}$ \cite{Cirelli:2005uq} is the velocity of Type-III fermions ($0\leq \beta \leq 1$) and 
\begin{eqnarray}
V_A&=& \frac{Q_fe^2}{\hat{s}}+ \frac{g_A^fg_2^2}{\hat{s}-M_Z^2} ~~~~~~~\text{for } q\,\bar{q}/\ell^+\ell^-\to N^+N^-\\
V_A&=& \frac{g_2^2}{\hat{s}-M_W^2}\frac{\delta_{AL}}{\sqrt{2}} \hspace*{1.5cm}\text{for } q\,\bar{q}'\to N^0N^{\pm}.
\end{eqnarray}
Here the $Z$ coupling for the fermion $f$ is given by $g_A^f=T_3-s_W^2\,Q_f$, where $Q_f$ is the electric charge factor corresponding to quarks or leptons and $A$ corresponds to the chirality $\{L,R\}$.

Due to the vev of the Higgs bosons, such heavy fermions decays to SM particles like gauge bosons, Higgs boson and fermions.  \autoref{Fy_diag} shows the production channels of $N^\pm$ and $N^0$, which further decays to SM particles, where we only focus on $h \to b\bar{b}$ for our study. 
The two-body decay widths for $N^\pm$ and $N^0$ are proportional to $Y^2_N$ \cite{Franceschini:2008pz} and are given by 

\small
\begin{eqnarray}
\Gamma(N^0\to \nu_{\ell} h) &=& \frac{1}{8}\frac{Y^2_N M_N}{8\pi}\left(1-\frac{M_h^2}{M_N^2}\right)^2,\\
\Gamma(N^0\to \nu_{\ell}Z)  &=& \frac{1}{8}\frac{Y^2_N M_N}{8\pi}\left(1-\frac{M_Z^2}{M_N^2}\right)^2 \left(1+2\frac{M_Z^2}{M_N^2}\right),\\
%
\Gamma(N^0\to W^{\pm} \ell^{\mp}) &=& \frac{1}{4}\frac{Y^2_N M_N}{8\pi}\left(1-\frac{M_W^2}{M_N^2}\right)^2 \left(1+2\frac{M_W^2}{M_N^2}\right)
\end{eqnarray}
and
\begin{eqnarray}
\Gamma(N^{\pm}\to \ell^{\pm}h) &=& \frac{1}{4}\frac{Y^2_N M_N}{8\pi}\left(1-\frac{M_h^2}{M_N^2}\right)^2,\\
\Gamma(N^{\pm}\to \ell^{\pm}Z) &=& \frac{1}{4}\frac{Y^2_N M_N}{8\pi}\left(1-\frac{M_Z^2}{M_N^2}\right)^2 \left(1+2\frac{M_Z^2}{M_N^2}\right), \\
\Gamma(N^{\pm}\to \nu_{\ell} W^{\pm}) &=& \frac{1}{2}\frac{Y^2_N M_N}{8\pi}\left(1-\frac{M_W^2}{M_N^2}\right)^2 \left(1+2\frac{M_W^2}{M_N^2}\right),
\end{eqnarray}

\normalsize

respectively, where we drop the generation index $i$ for simplicity.

There are another decay modes possible considering the loop generated mass of $N^\pm$ and $N^0$, which gives a mass splitting of $\Delta M =\mathcal{O}(166)$ MeV \cite{Cirelli:2005uq}. It opens up the following decays with the widths given by
\begin{eqnarray}
\Gamma(N^{\pm}\to N^0 \pi^{\pm}) &=& \frac{2G_F^2V_{ud}^2\Delta M^3 f_{\pi}^2}{\pi} \sqrt{1-\frac{m_{\pi}^2}{\Delta M^2}} ,\\
\Gamma(N^{\pm}\to N^0 e^{\pm} \nu_e) &=& \frac{2G_F^2\Delta M^5}{15\pi^3}  ,\\
\Gamma(N^{\pm}\to N^0 \mu^{\pm} \nu_{\mu}) &=& 0.12\, \Gamma(N^{\pm}\to N^0 e^{\pm} \nu_e)   ,
\end{eqnarray}
where $f_{\pi}= 130.5\,$MeV is the pion form factor, $G_F$ is the Fermi constant and $V_{ud}$ is the CKM mixing matrix.  
\begin{figure*}[hbt]
	\begin{center}
		\hspace*{-0.5cm}
		\mbox{\subfigure[]{\includegraphics[width=0.35\linewidth,angle=-0]{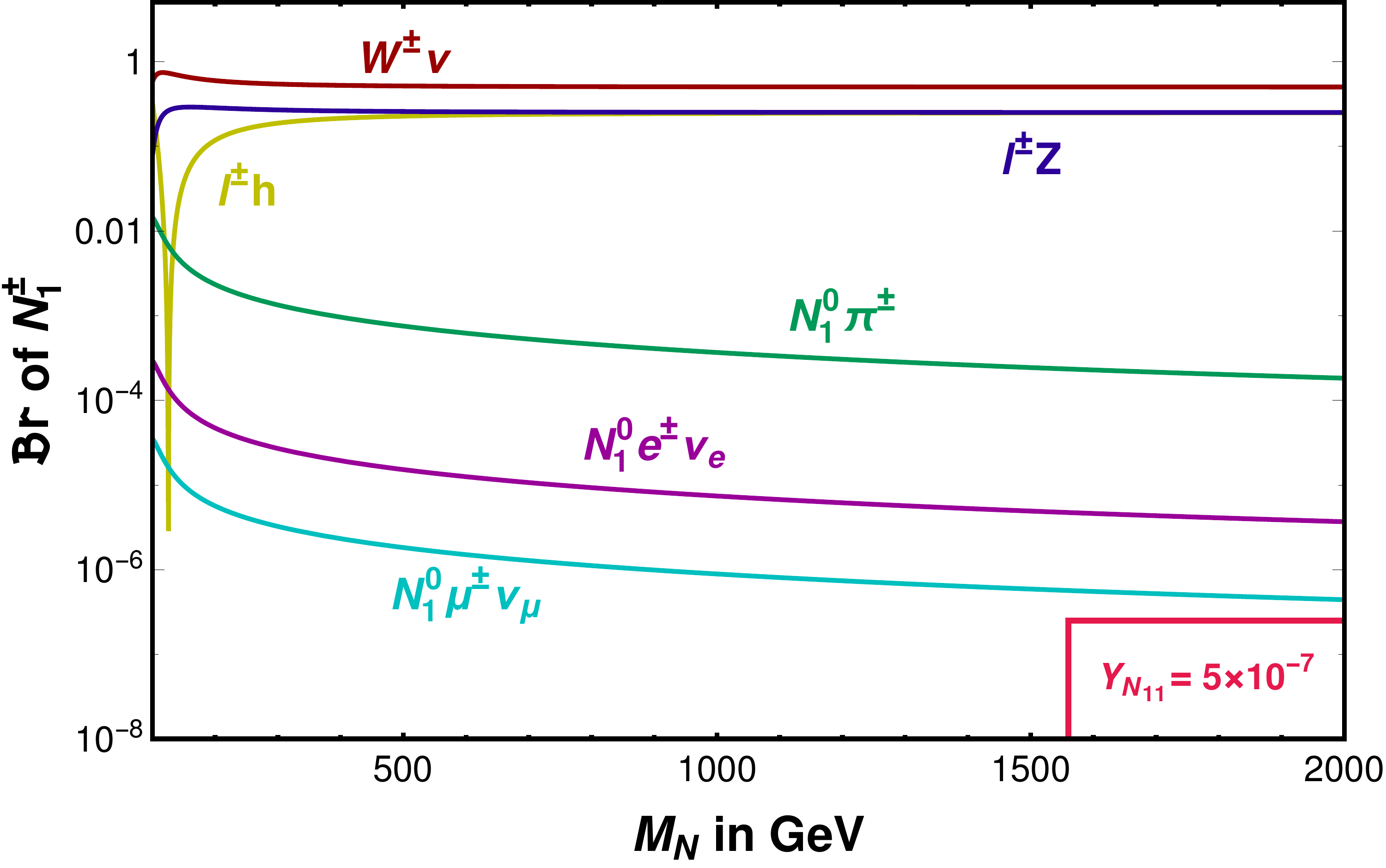}\label{}}\quad
			\subfigure[]{\includegraphics[width=0.35\linewidth,angle=-0]{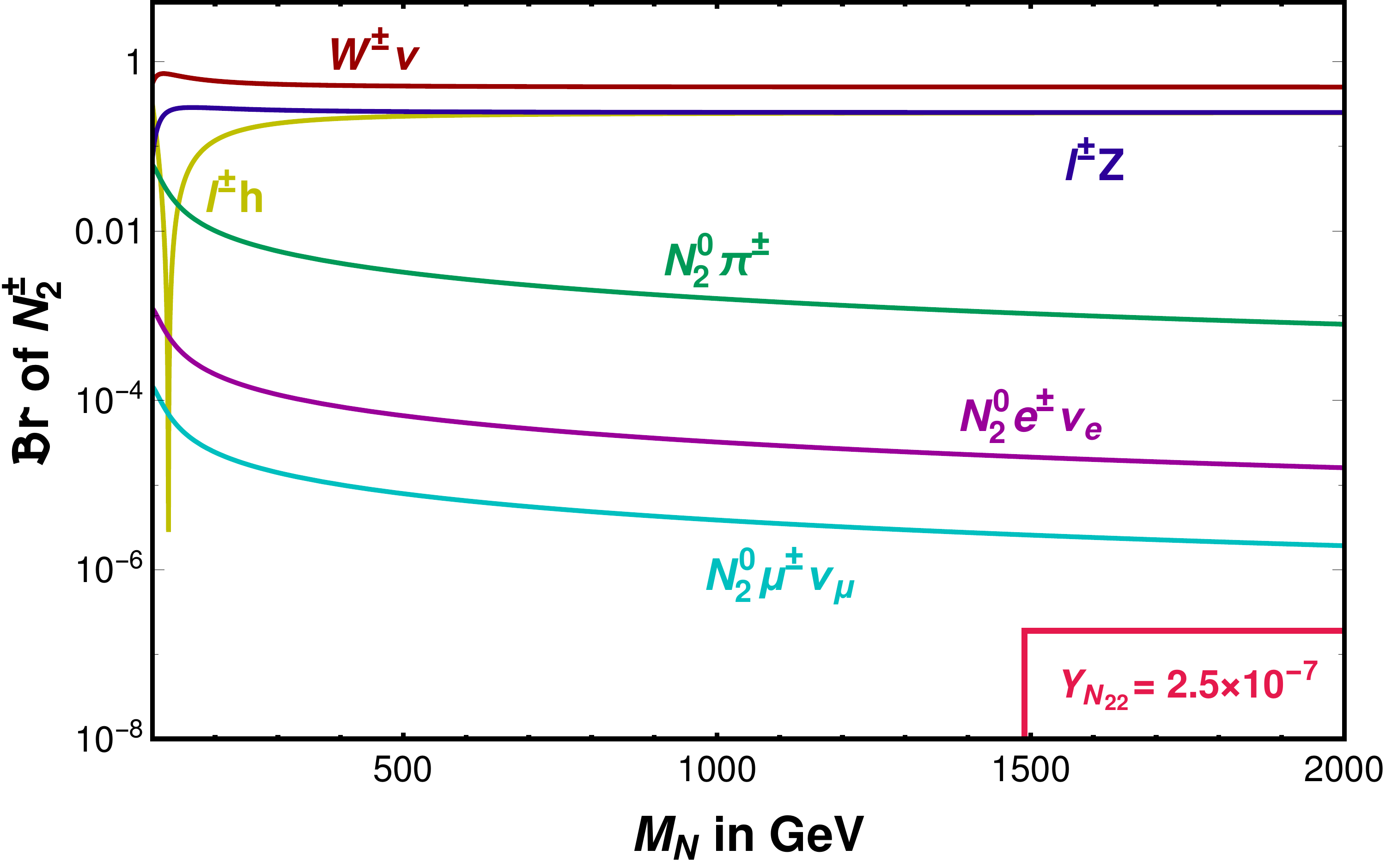}\label{}}\quad
			\subfigure[]{\includegraphics[width=0.35\linewidth,angle=-0]{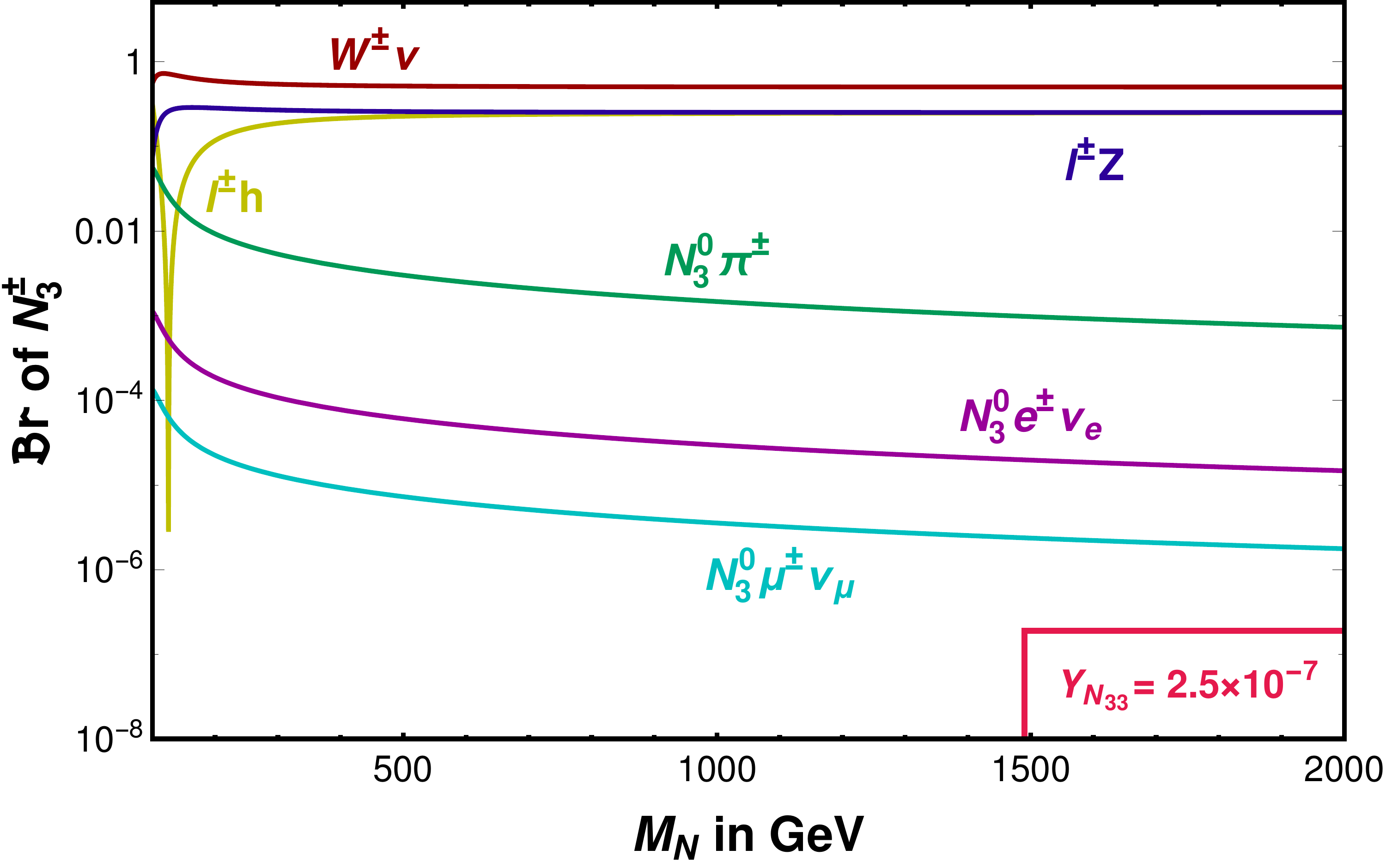}\label{}}}
		\hspace*{-0.5cm}	
		\mbox{\subfigure[]{\includegraphics[width=0.35\linewidth,angle=-0]{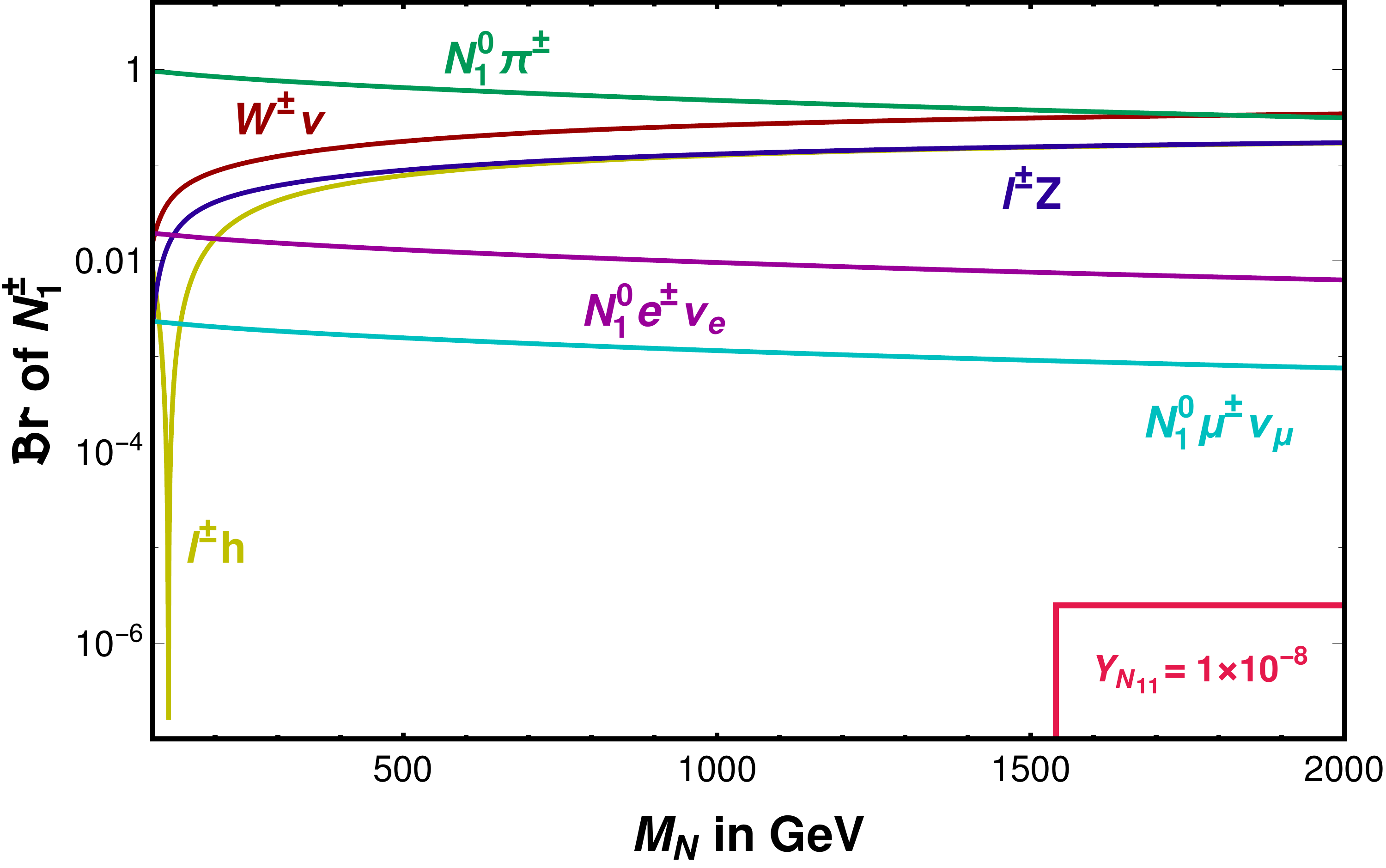}\label{}}\quad
			\subfigure[]{\includegraphics[width=0.35\linewidth,angle=-0]{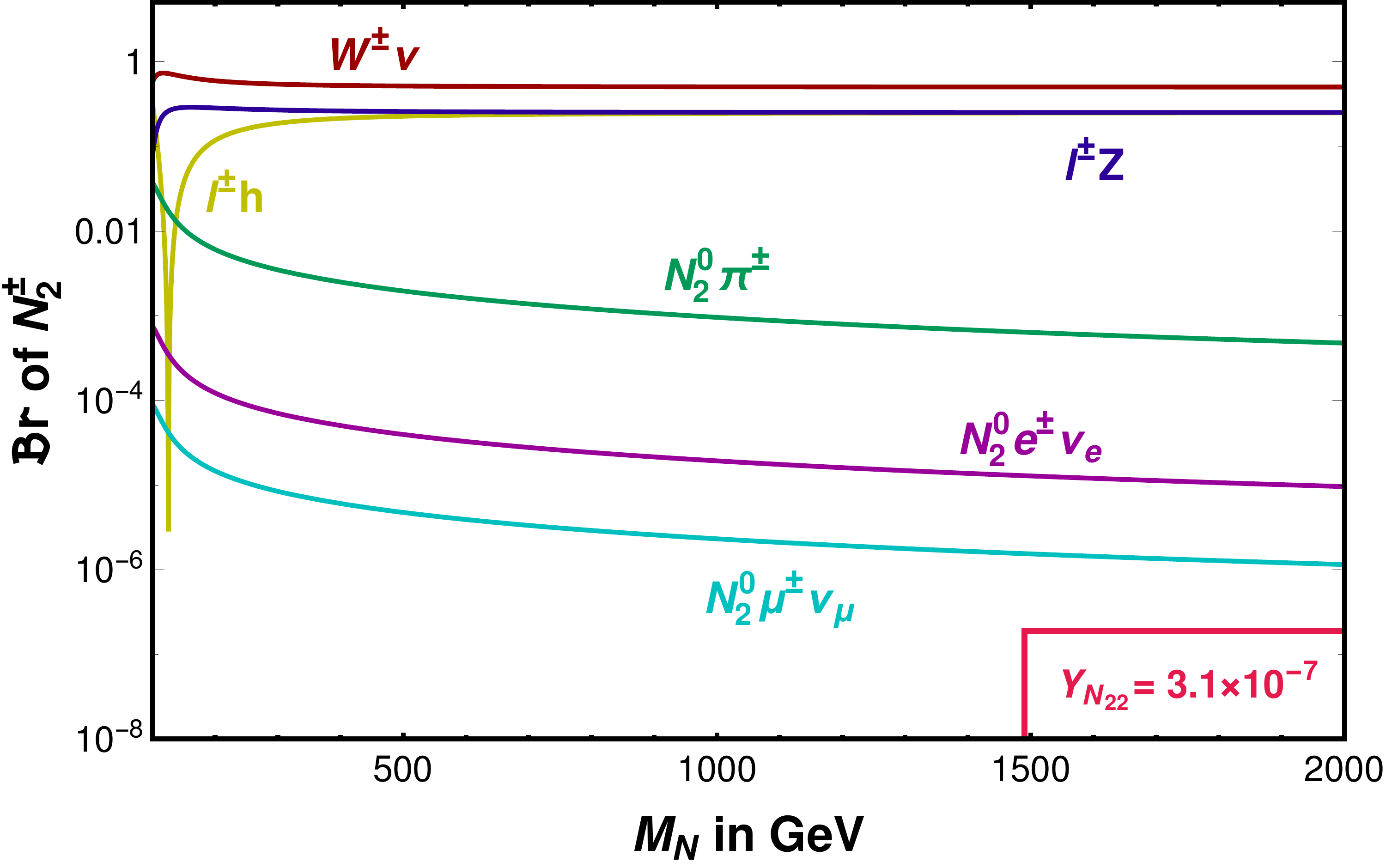}\label{}}\quad
			\subfigure[]{\includegraphics[width=0.35\linewidth,angle=-0]{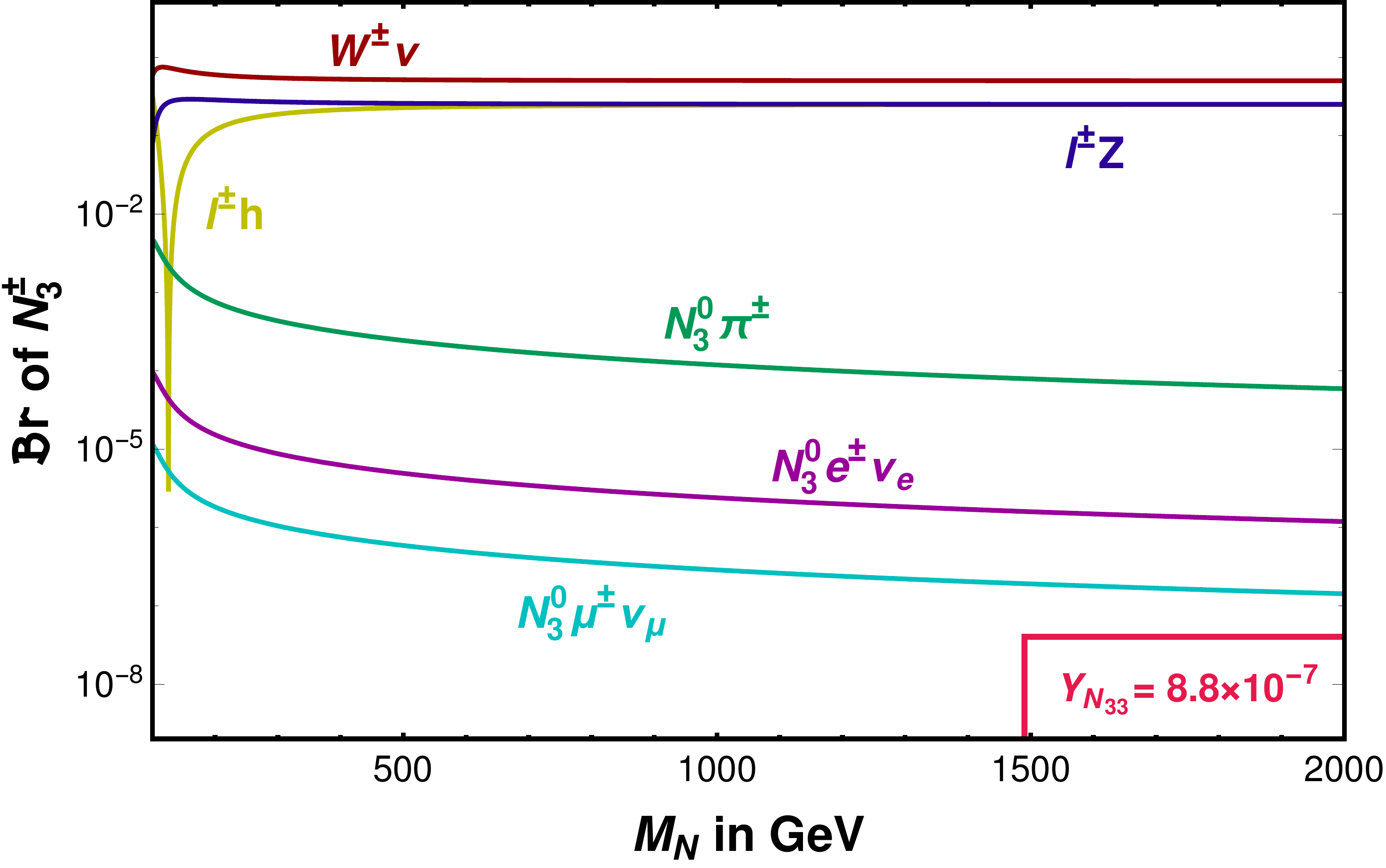}\label{}}}
		\hspace*{-0.5cm}	
		\mbox{\subfigure[]{\includegraphics[width=0.35\linewidth,angle=-0]{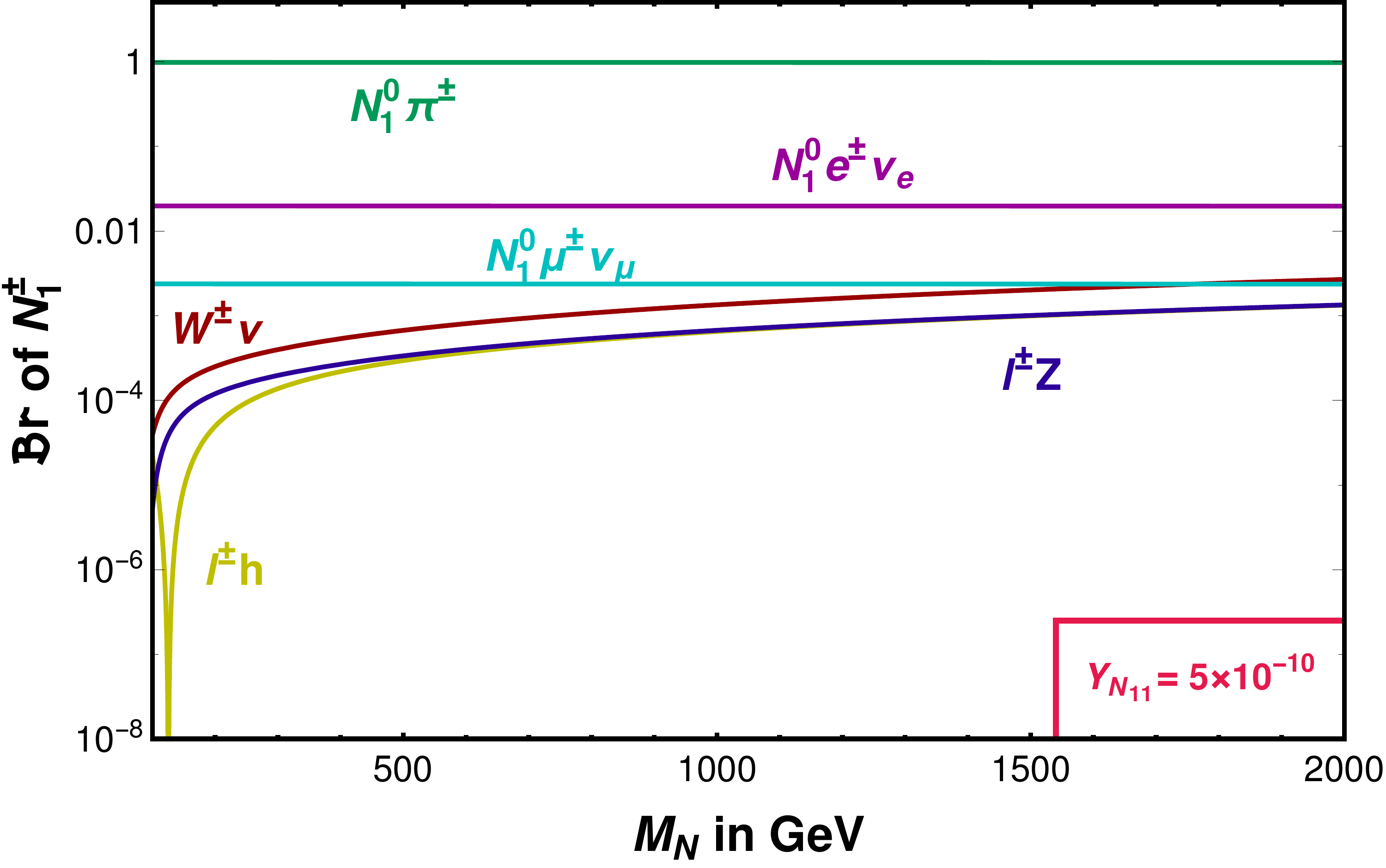}\label{}}\quad
			\subfigure[]{\includegraphics[width=0.35\linewidth,angle=-0]{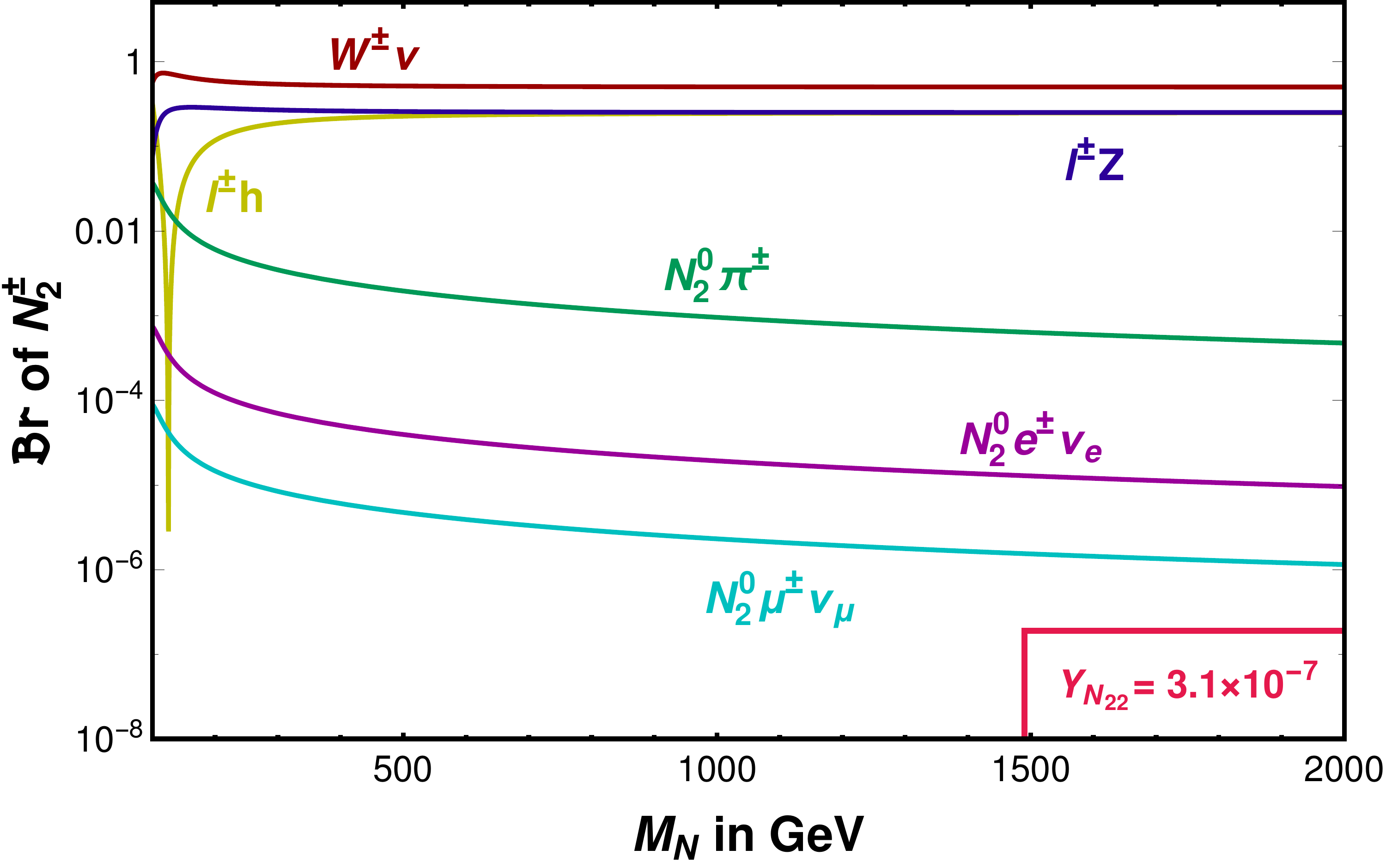}\label{}}\quad
			\subfigure[]{\includegraphics[width=0.35\linewidth,angle=-0]{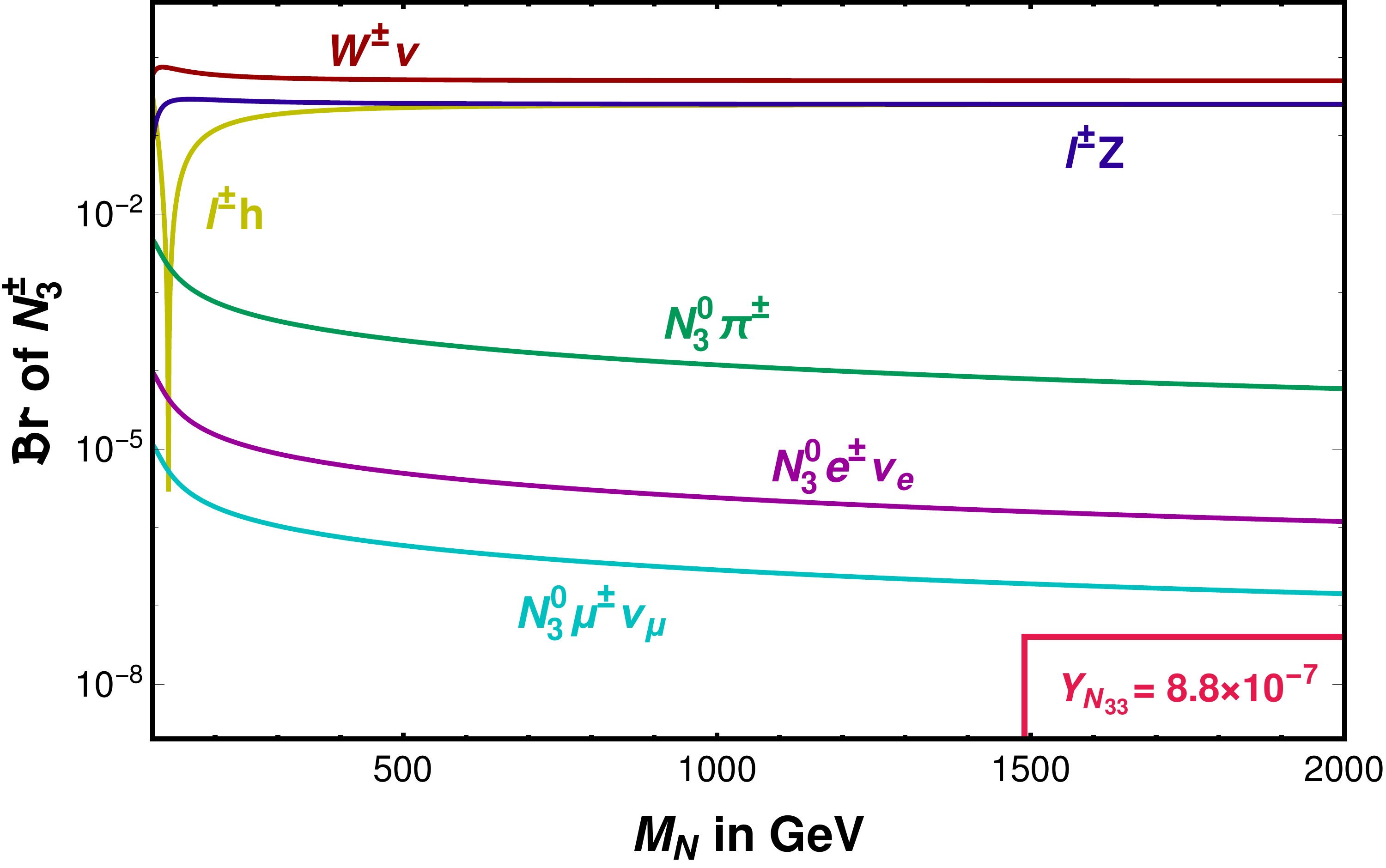}\label{}}}		
		\caption{Charged $SU(2)_L$ triplet heavy fermion ($N^\pm$) branching ratios to different decay channels as function of the triplet mass. Column wise we depict the branching ratios of $N_1^{\pm}$, $N_2^{\pm}$ and $N_3^{\pm}$, respectively for the different choices of Yukawa couplings satisfying light neutrino masses and UPMNS mixing matrix. }\label{branching}
	\end{center}
\end{figure*}

\begin{figure*}[h]
	\begin{center}
		\includegraphics[width=0.55\linewidth,angle=-0]{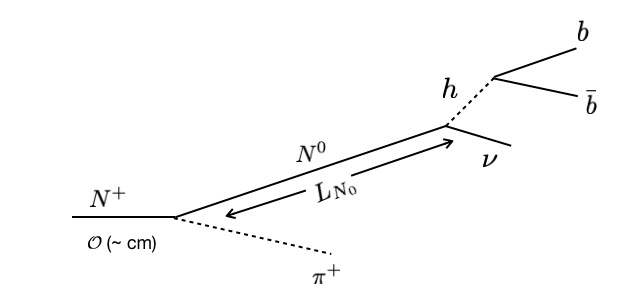}\label{}
		\caption{Schematic diagram of $N^{\pm}\to N^0 \pi^{\pm} $ and $N^0 \to h \nu $  decays.  }\label{Npmdcy}
	\end{center}
\end{figure*}


Considering one generation Type-III fermion $Y_N=Y_{N_{11}} =5\times 10^{-7}$ and $M_N=1$ TeV,  $\mathcal{B}(N^{\pm}\to N^0 \pi^{\pm}) \sim 0.1\%$ and has negligible effect on our collider study for this choice of Yukawa. Such behaviour can be visible from \autoref{branching}(a) with respect to $M_N$. However, for  even lower choices of $Y_N$ i.e. for $10^{-8}$ (\autoref{branching}(d)), $ 5\times 10^{-10}$ (\autoref{branching}(g)), the decay of $N^{\pm}=N^{\pm}_1\to N^0 \pi^{\pm}$ becomes dominant with branching ratios of  47.3\% and 97.5\%, respectively for $M_N=1$ TeV. So for these lower $Y_N$ values, we see a  displacement of $N^\pm \to \pi^\pm,\, N^0$, which is of the order of cm before the $N^0$ decay gives another recoil. The total displaced decay length will thus depend on both the decays of $N^\pm$ and $N^0$ as can be seen from the schematic diagram of \autoref{Npmdcy}.

Considering the general structure of the currents (charged and neutral) both diagonal and off-diagonal Yukawa couplings can play important roles for both one and three generation cases \cite{Li:2009mw}. Nonetheless, if we observe \autoref{Npmdcy}, $N^\pm \to N^0 \pi^\pm$ decay is controlled by the $SU(2)$ gauge coupling  $g_2$, thus there is no generation mixing in the first vertex. However, the second vertex, which causes $N^0 \to h \nu$ decay, is governed by the Yukawa coupling $Y_N$. For our analysis here, we have considered one generation of  the Type-III fermions and the diagonal Yukawa coupling $Y_N$, which results in the decays of $N^0 \to h \nu_e$.  Even if we consider the off-diagonal Yukawa couplings, which are always lower than the diagonal ones while satisfying the UPMNS mixing matrix \cite{Casas:2001sr}, the decays of $N^0  \to h \nu_\mu, h \nu_\tau$  with corresponding branching fractions mediated by $Y_{N_{12}, N_{13}}$ will contribute in Higgs plus missing energy finalstate. Thus for our analysis the choice of diagonal Yukawa will be sufficient. It can be noted that for TeV scale Type-III fermions, the constraints coming from charged lepton flavour violating processes  $\ell_i \to \ell_j \gamma $ are less \cite{Abada:2008ea} compared to those coming from UPMNS mixing matrix.

One can consider three generations of  TeV scale Type-III fermions to see the overall contribution to the desired finalstate. However, while satisfying the light neutrino masses and UPMNS mixing matrix for the benchmark choices of $Y_{N_{11}}=5\times 10^{-7}, 10^{-8}, 5 \times 10^{-10} $, give rise to $Y_{N_{22}}= 2.5\times 10^{-7}, 3.1 \times 10^{-7}, 3.1 \times 10^{-7}$ and $Y_{N_{22}}= 2.5\times 10^{-7}, 8.8 \times 10^{-7}, 8.8 \times 10^{-7}$, respectively.  We present the corresponding decay branching fractions of $N^\pm_{2}$ in \autoref{branching}(b),(e),(h)  (second column)and $N^\pm_{3}$ in \autoref{branching}(c),(f),(i) (third column),   respectively. We see that for the second and third generation Type-III fermions, $N^\pm \to N^0 \pi^\pm$ decay is less dominant and thus misses the first displacement. The second recoil, $N^0 \to h \nu$ also tends to have prompt decay due to relatively large Yukawa couplings. 
It is evident from these set of numbers as well as from the detailed simulation  discussed in \autoref{results_pp}
and \autoref{muon_m} that the second and third generations fail to contribute for the finalstates in MATHUSLA range, similar to the first generation with $Y_{N_{11}}=5\times 10^{-7}$. Thus for simplicity, we present our analysis with one generation considering the diagonal Yukawa couplings. 

\section{Benchmark points and Displaced vertex}\label{bpdis}

\begin{figure*}[h]
	\begin{center}
		\hspace*{-0.5cm}
		\mbox{\subfigure[]{\includegraphics[width=0.52\linewidth,angle=-0]{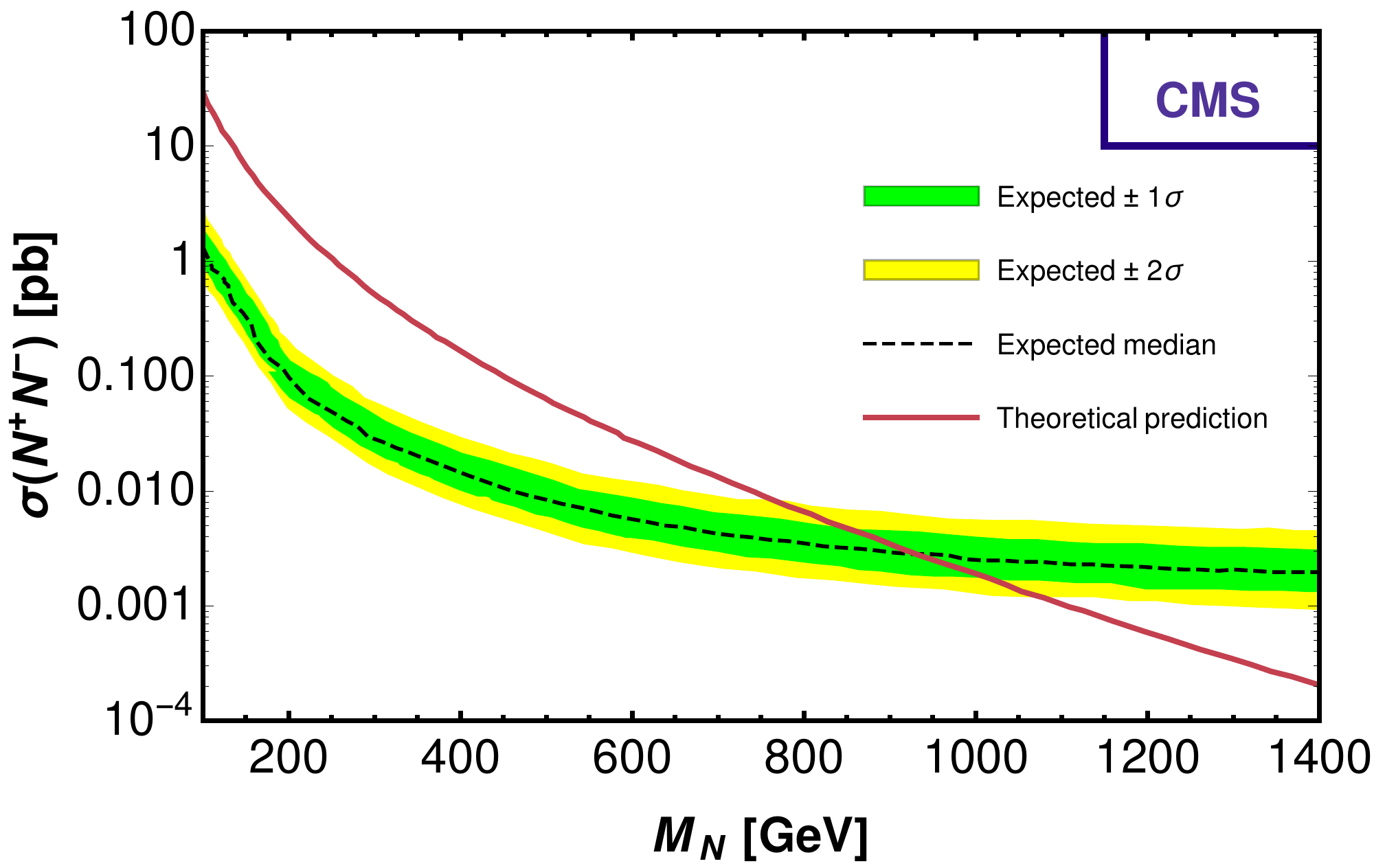}\label{}}\quad\quad
		\subfigure[]{\includegraphics[width=0.5\linewidth,angle=-0]{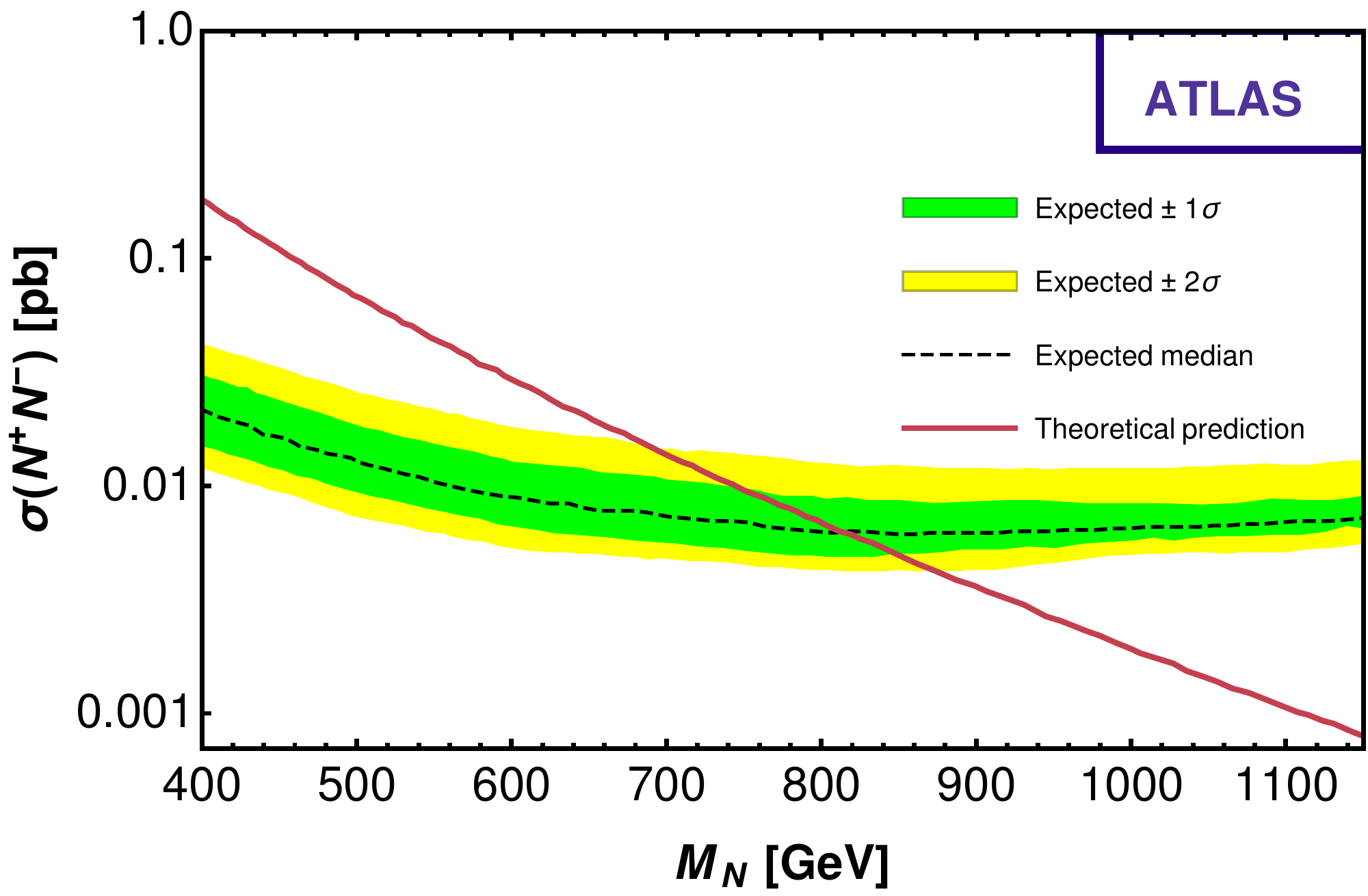}\label{}}}		
		\caption{(a) CMS \cite{CMS:2019lwf} and (b) ATLAS \cite{ATLAS:2020wop} bounds on the Type-III fermions when only one generation is light. The black dashed lines are the expected medians whereas the green and yellow bands present the $1\sigma$ and $2\sigma$ regions respectively. For both the cases the red line denotes the theoretical prediction for the pair production of Type-III fermions. } \label{bound}
	\end{center}
\end{figure*}

The recent collider  searches at CMS \cite{CMS:2017ybg, CMS:2019lwf} and ATLAS \cite{ATLAS:2020wop} have put a lower  bound of  740 GeV and 680 GeV, respectively \autoref{bound}  at $2\sigma$ level, on the Type-III fermion mass with one generation.
In this article we consider only one generation of Type-III fermion as light, whereas the other two can be heavy in explaining the neutrino masses and mixing. However, for a conservative choice, we choose  $M_N=1000,\,  1500$ GeV for BP1, BP2 for the collider study as we list them in \autoref{bps}. For  relatively larger Yukawa coupling i.e., $ \gsim 5\times 10^{-7}$, the branching ratio in $N^\pm \to N^0 \pi^\pm $ is negligible i.e. $\lesssim 0.1\%$. However, for much lower choices of Yukawa  couplings this mode can be dominating as we describe various decay branching ratios in  \autoref{bps} for the benchmark points.
For the neutral heavy fermions $N^0$,  the dominant decay modes  $W^\pm \ell^\mp , \, Z  \nu$ and $h \nu$ are with the branching ratio 2:1:1.  The SM like Higgs boson around $125.5$  GeV, mostly   decays to $b\bar{b}$.

\begin{table}[!htb]
	\begin{center}
		\hspace*{-1.4cm}
		\renewcommand{\arraystretch}{1.3}
		\begin{tabular}{|c|c|c|c||c|c|c|}\hline 
			\multirow{2}{*}{Modes} &  \multicolumn{3}{c||}{BP1} & \multicolumn{3}{c|}{BP2} \\
		      & \multicolumn{3}{c||}{$M_N = 1000\,\rm{GeV}$} & \multicolumn{3}{c|}{$M_N = 1500\,\rm{GeV}$} \\ \hline
		      &  $Y_N=5\times 10^{-7}$ & $Y_N= 10^{-8}$ &  $Y_N=5\times 10^{-10}$ & $Y_N=5\times 10^{-7}$ & $Y_N= 10^{-8}$ & $Y_N= 5\times 10^{-10}$ \\ \hline \hline
			
			$N^\pm \to h l^\pm$  &  24.4 & 12.6 & $\leqslant 0.1$ & 24.7 & 15.2 & 0.1 \\ \hline
			$N^\pm \to Z l^\pm$  &  25.2 & 13.0 & $\leqslant 0.1$ & 25.1 & 15.5 & 0.1 \\ \hline
			$N^\pm \to W^\pm \overset{\textbf{\fontsize{1.5pt}{1.5pt}\selectfont(--)}}{\nu}$ & 50.4 & 26.0 & 0.1 & 50.2 & 30.9 & 0.2 \\ \hline 
			$N^\pm \to N^0 \pi^\pm$ &  $\leqslant 0.1$ & 47.3 & 97.5 & $\leqslant 0.1$ & 37.5 & 97.4\\ \hline
			$N^\pm \to N^0 e^\pm \nu_e$ & $\leqslant 0.1$  & 0.9 & 2.0 & $\leqslant 0.1$ & 0.7 & 2.0\\ \hline
			$N^\pm \to N^0 \mu^\pm \nu_{\mu}$ & $\leqslant 0.1$  & 0.1 & 0.2 & $\leqslant 0.1$ & 0.1 & 0.2\\ \hline
		\end{tabular}
		\caption{Masses of the Type-III fermions ($M_N$) and their branching ratios (in \%) for the different decay modes of $N^{\pm}$. Here $N^{\pm}$ decay branching fractions are for $Y_N=5\times 10^{-7},\,\, 1\times 10^{-8} \,\, \rm and \,\, 5\times 10^{-10}$.}\label{bps}
	\end{center}
\end{table}

\begin{figure}[h]
	\begin{center}
		\mbox{\subfigure[]{\includegraphics[width=0.45\linewidth,angle=-0]{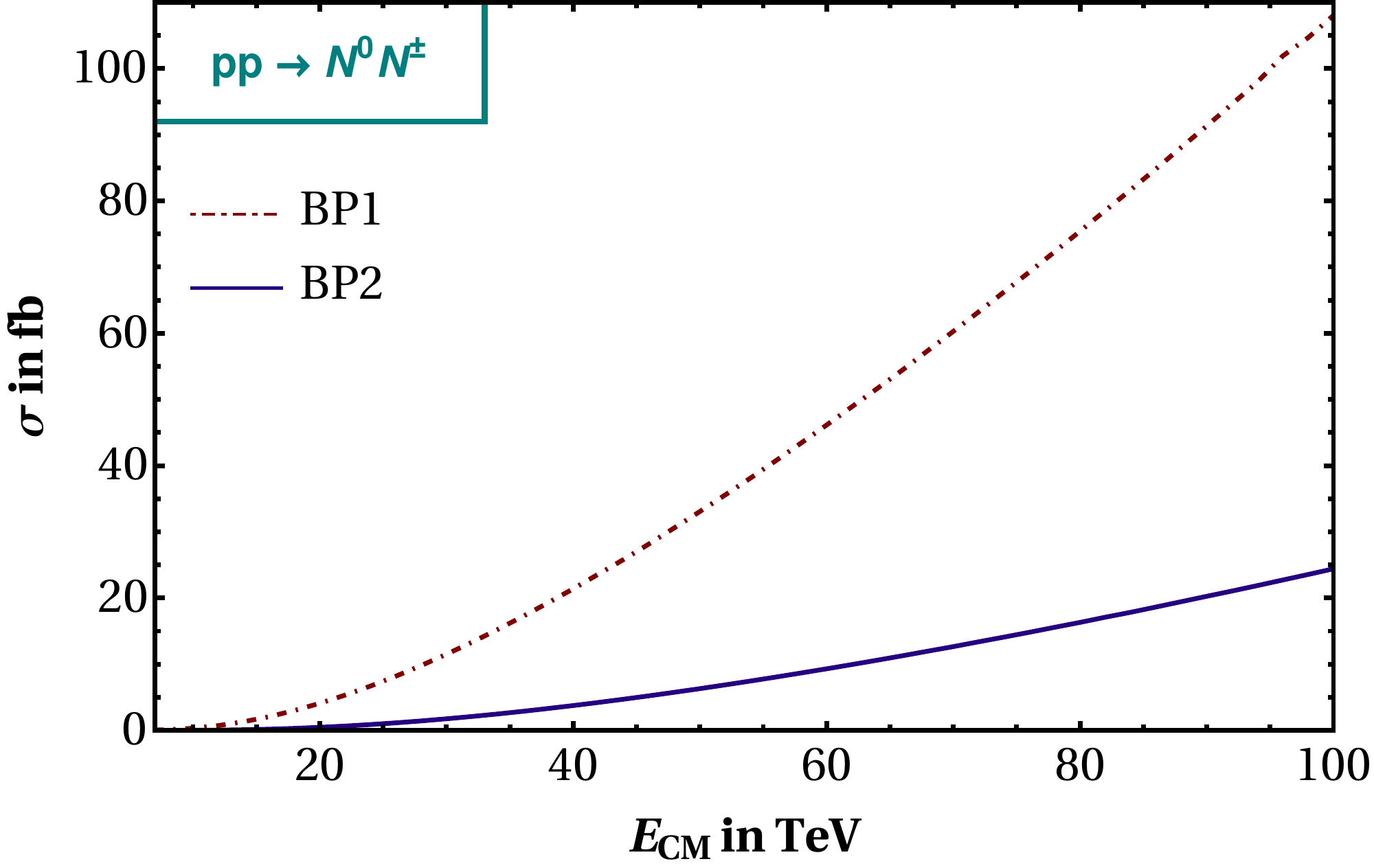}\label{}}\quad \quad
		\subfigure[]{\includegraphics[width=0.45\linewidth,angle=-0]{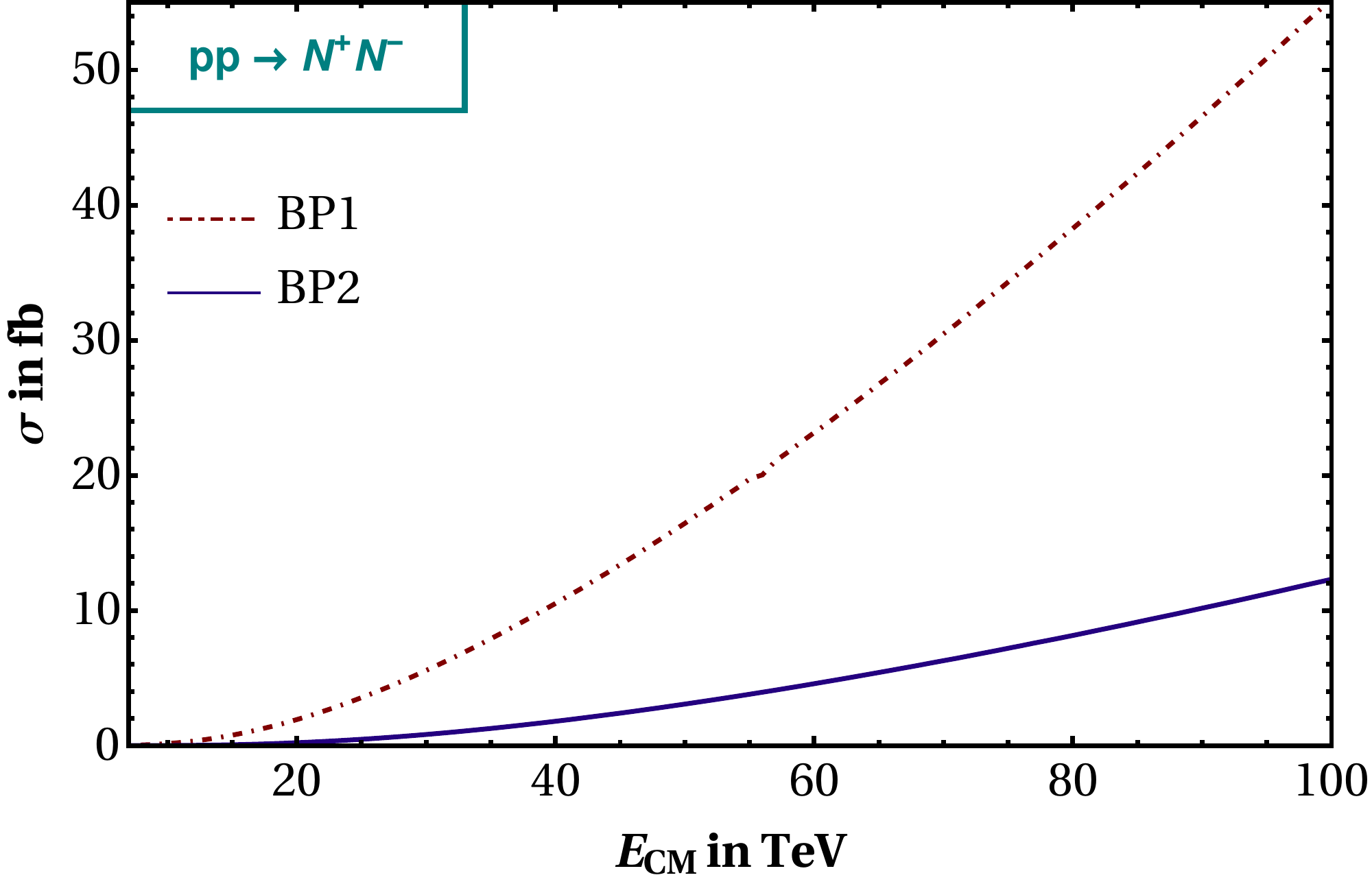}\label{}}}		
		\caption{Cross-section (in fb) as a function of centre of mass energy ($E_{\rm CM}$) for the benchmark points at the LHC for the processes $p\,p\to N^0N^{\pm}$ (a) and $p\,p\to  N^+N^-$ (b). }\label{ppcrosecfig}
	\end{center}
\end{figure}

The production cross-sections in  both LHC, FCC with next-to-leading order (NLO) correction \cite{Ruiz:2015zca} for the centre of mass energies of 14, 27 and 100 TeV are given in \autoref{ppcrosec}, where {\tt NNPDF23$\_$lo$\_$as$\_$0130$\_$qed} \cite{NNPDF:2014otw} is used as the parton distribution function. It is interesting to see the enhancement of the cross-sections as we increase the centre of mass energy at the LHC/FCC from \autoref{ppcrosecfig}. Due to the parton distribution function we always find the on-shell production even at higher energies, contrary to the muon collider. Unlike LHC/FCC, in muon collider, we cannot produce $N^\pm\, N^0$ and we rely only on the pair production of $N^\pm N^\mp$ as we present the cross-sections in \autoref{mucrosec} with the centre  of mass energies of  3.5, 14 and 30 TeV for the benchmark points. Due to $Y=0$ nature, the pair production of  $T_3=0$ component of $SU(2)$, i.e. $N^0$ is not possible in both of the colliders. However, via $N^{\pm}$ decay, the pair production of $N^0$ is possible for lower Yukawa couplings as $N^{\pm}\to N^0 \pi^{\pm}$ decay mode gets dominant. 

\begin{table}[!h]
	\begin{center}
		\renewcommand{\arraystretch}{1.3}
		\begin{tabular}{|c|c|c|c|c|c|c|}\hline 
			& \multicolumn{6}{c|}{Cross-sections (in fb) with the $E_{CM}$ } \\ \cline{2-7}
			Benchmark  & \multicolumn{2}{c|}{ 14\,TeV} &  \multicolumn{2}{c|}{27\,TeV} &  \multicolumn{2}{c|}{100\,TeV}  \\
			\cline{2-7}
			points	&     $ N^0 N^\pm$ & $N^+ N^-$ &  $N^0 N^{\pm}$ & $N^+N^-$ & $N^0 N^\pm$ & $N^+N^-$\\ \hline
			BP1 & 1.5  & 0.64 & 10.0 & 4.6 & 108.0 & 55.2 \\ \hline
			BP2 & 0.11  & 0.05 & 1.5 & 0.7 & 24.4 & 12.3 \\ \hline 
		\end{tabular}
		\caption{Production cross-sections (in fb) with NLO correction of the processes $p\,p \to N^0 N^{\pm}$ and $p\,p \to  N^+ N^-$, for the benchmark points at the LHC for 14\,TeV, 27\,TeV and 100\,TeV centre of mass energies.}\label{ppcrosec}
	\end{center}
\end{table}

The chosen benchmark points can give rise to displaced decays with the rest mass decay lengths for $N^0$ ranging from  mm to hundreds of meters for neutrino Yukawa couplings $Y_N \sim 5\times 10^{-7}- 5\times 10^{-10}$ as shown in \autoref{dcyLT}. However, the decay length for $N^\pm$ can differ from that of  the $N^0$ depending on  dominance of additional mode of  $N^\pm \to N^0 \pi^\pm$. For  $Y_N \sim 5\times 10^{-7}$, the decay length is similar to that of $N^0$.  However, for lower $Y_N$, i.e. $10^{-8}-10^{-10}$,  $N^\pm \to N^0 \pi^\pm$ is the pre-potent and the  displaced decay length in this can go up to cm. 

\begin{table}	
	\begin{center}
		\renewcommand{\arraystretch}{1.4}
		\begin{tabular}{ |c| c|c|c| }
			\hline
			
			{\multirow{2}{*}{\diagbox[width=3.25cm]{$M_N$}{$Y_N$}}}&
			\multicolumn{3}{c|}{Rest Mass Decay Length}\\
			\cline{2-4}
			&\multicolumn{1}{c|}{$5\times 10^{-7}$}&\multicolumn{1}{c|}{$1\times10^{-8}$}&\multicolumn{1}{c|}{$5\times 10^{-10}$}\\ 
			\cline{2-4}
			\hline	
			$1.0\,\rm{TeV}$ & $0.02\,\rm{mm}$ & $5.06\,\rm{cm}$ & $20.26\,\rm{m}$ \\
			\hline
			$1.5\,\rm{TeV}$ & $0.01\,\rm{mm}$ & $3.36\,\rm{cm}$ & $13.42\,\rm{m}$ \\
			\hline 
			
		\end{tabular}
		\caption{Rest mass decay length of $N^0$  for the chosen benchmark points. }  \label{dcyLT}
	\end{center}	
\end{table}


Such rest mass decay lengths will be further affected due to the probability distribution of the decay and also owing to the boost  of  the decaying particle i.e. $N^{\pm(0)}$ . In particular, we focus on the effect of the latter, where LHC/FCC and muon collider behave differently. For this purpose, we perform a detailed simulation by  {\tt PYTHIA8} \cite{Sjostrand:2014zea}, which takes care of the boost effect and  the  decay  distributions, etc.
We see in the following sections that in the case of LHC, longitudinal boost plays a major role  in enhancing the decay lengths and at the  muon collider, the transverse momenta diverges affecting the corresponding  decay lengths.  The collider setup is described  below before presenting the relevant kinematical distributions and analysis.

\section{Setup for collider simulation}\label{setup}
The model has been implemented in {\tt SARAH 4.13.0}\cite{Staub:2013tta} to generate the model
files for {\tt CalcHEP$\_$3.8.7}\cite{Belyaev:2012qa}. The cross-sections at tree-level and event generation are performed in {\tt CalcHEP} with the parton distribution function {\tt NNPDF23$\_$lo$\_$as$\_$0130$\_$qed}\cite{NNPDF:2014otw}. 
The events are then analysed in {\tt PYTHIA8} \cite{Sjostrand:2014zea} with initial, final state radiations and subsequent hadronisation. Such hadrons are then fed to {\tt Fastjet$\_$3.2.3} \cite{Staub:2013tta} for jet formation with the following specifications. 

\begin{itemize}
	\item Calorimeter coverage: $|\eta| < 2.5$.
	\item Jet clustering is done by {\tt ANTI-KT}  algorithm with jet Radius parameter, R = 0.5.
	\item Minimum transverse momentum of each jet: $p_{T,min}^{jet} = 20.0$ GeV; jets are $p_{T}$-ordered.
	\item No hard leptons are inside the jets. 
	\item Minimum transverse momentum cut for each detected lepton: $p_{T,min}^{lep} = 20.0$\,GeV.
	\item Detected leptons are hadronically clean, which implies hadronic activity within a cone of $\Delta R < 0.3$ around each lepton is less than $15\%$ of the leptonic transverse momentum, \textit{i.e.} $ p_{T}^{\mathrm{had}}< 0.15 p_{T}^{\text{lep}}$ within the cone.
	\item Leptons are distinctly registered from the jets produced with an isolation cut $\Delta R_{lj} > 0.4$.
\end{itemize}
We reconstruct the Higgs bosons via $b\bar{b}$ mode, where the $b$-jets are tagged via the  secondary vertex reconstruction with $b$-jet tagging efficiency of maximum 85\% \cite{CMS:2012jki,CMS:2017wtu}.

\section{Simulations  at the LHC/FCC}\label{LHCF}
In this section, we describe all the crucial kinematical distributions leading to  displaced Higgs reconstructions. The effect of  heavy fermion mass, Yukawa and the centre of mass energy of the collider are explored in detail while projecting the  number of events  at the LHC/FCC  and MATHUSLA. 

\subsection{Kinematical distributions}\label{kindis}
\begin{figure}[hbt]
	\begin{center}
		\hspace*{-0.5cm}
		\mbox{\subfigure[] {\includegraphics[width=0.45\linewidth,angle=-0]{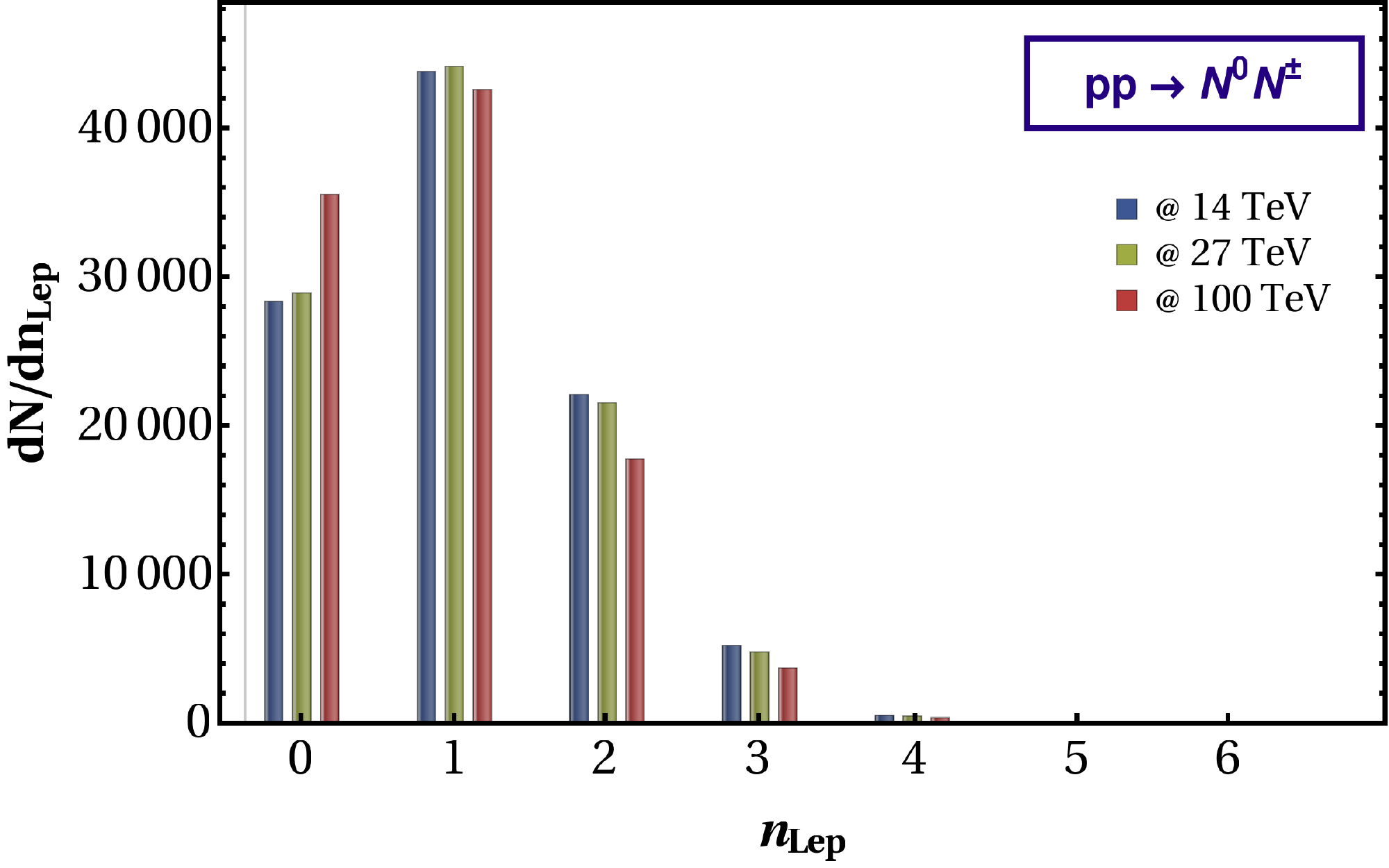}\quad\label{}}\quad\quad
		\subfigure[] {\includegraphics[width=0.45\linewidth,angle=-0]{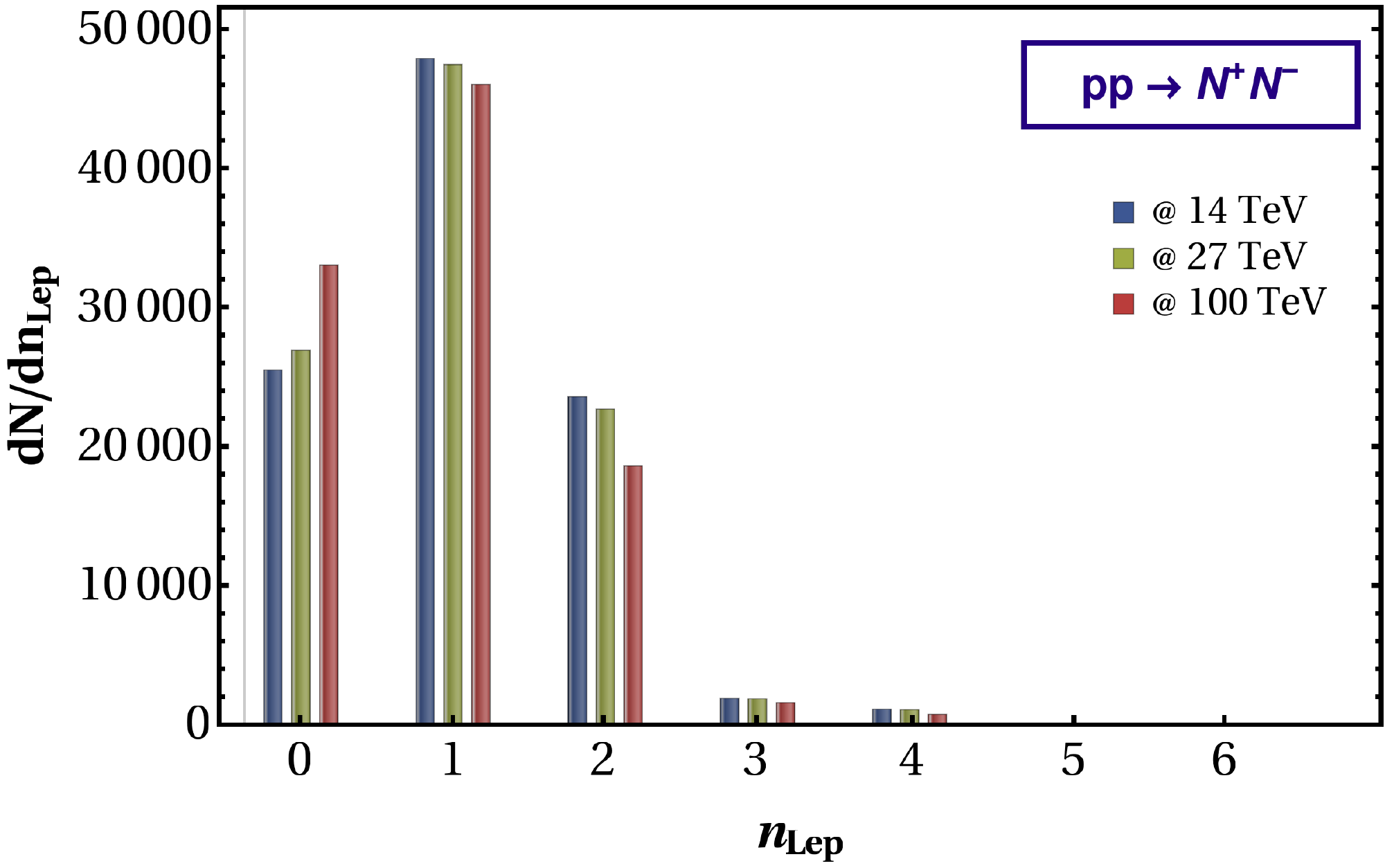}\label{}}}
		\caption{Multiplicity ($n_{\rm Lep}$) distributions of the charged leptons for the process $p\,p\to N^0\,N^{\pm}$ (a) and $p\,p\to N^+\,N^-$ (b) at the three different centre of mass energies of 14\,TeV, 27\,TeV and 100\,TeV for $M_N=1\,\rm{TeV}$ (BP1) and $Y_N=5\times 10^{-7}$. }\label{pp_lep_mul}
	\end{center}
\end{figure}

We describe the lepton ($\mu, e$) multiplicity distributions  at the LHC/FCC in \autoref{pp_lep_mul} with three  different centre of mass energies for the first benchmark point (BP1). In \autoref{pp_lep_mul}(a) and (b), the multiplicity for the processes $N^0\,N^{\pm}$ and $N^+\,N^-$ are depicted, respectively. After imposing the isolation criteria, although the distribution peaks around one, the lepton multiplicity reaches till four. 

\begin{figure}[hbt]
	\begin{center}
		\hspace*{-1.0cm}
		\mbox{\subfigure[]{\includegraphics[width=0.37\linewidth,angle=-0]{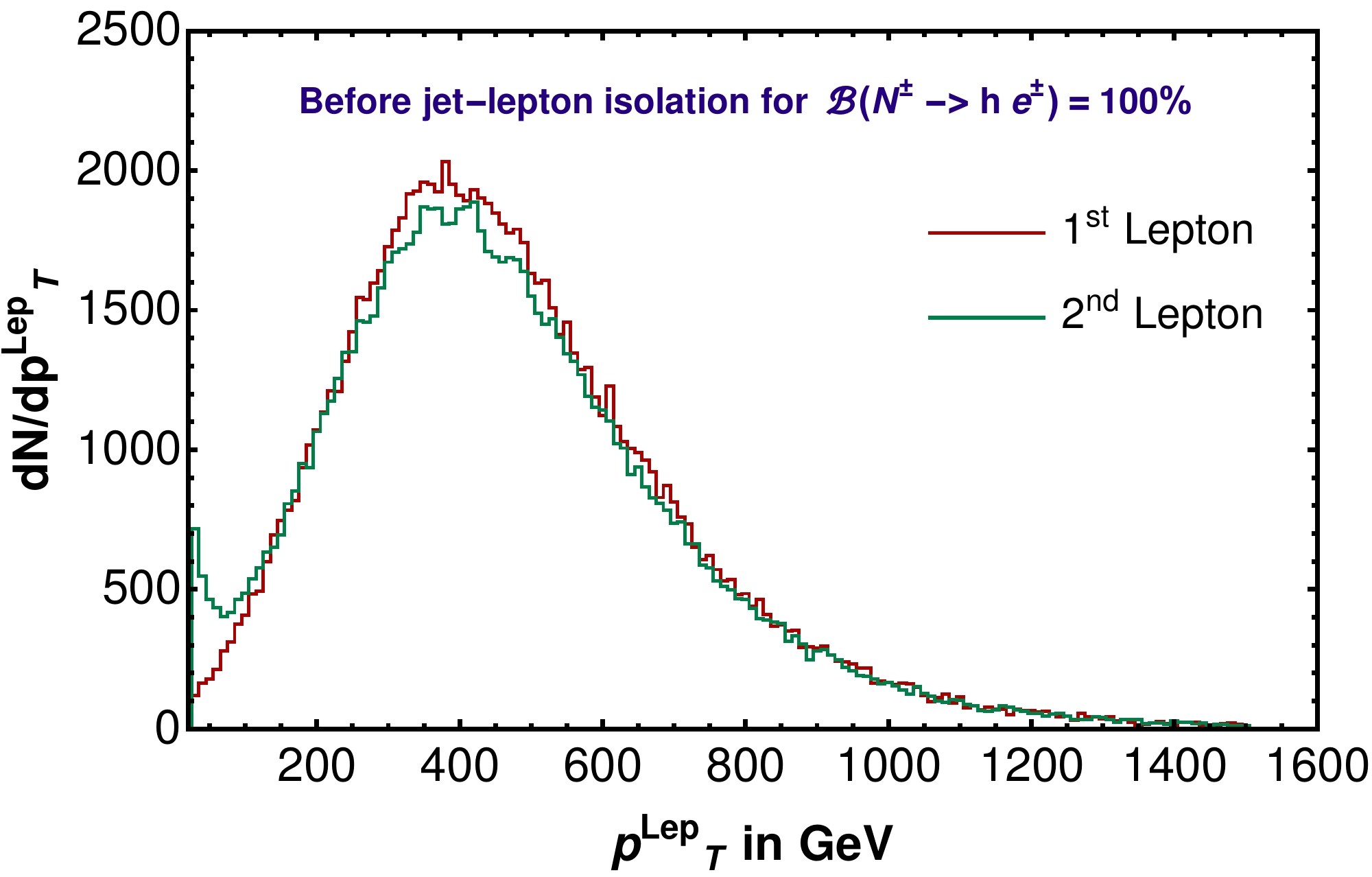}\label{}}\quad
			\subfigure[]{\includegraphics[width=0.37\linewidth,angle=-0]{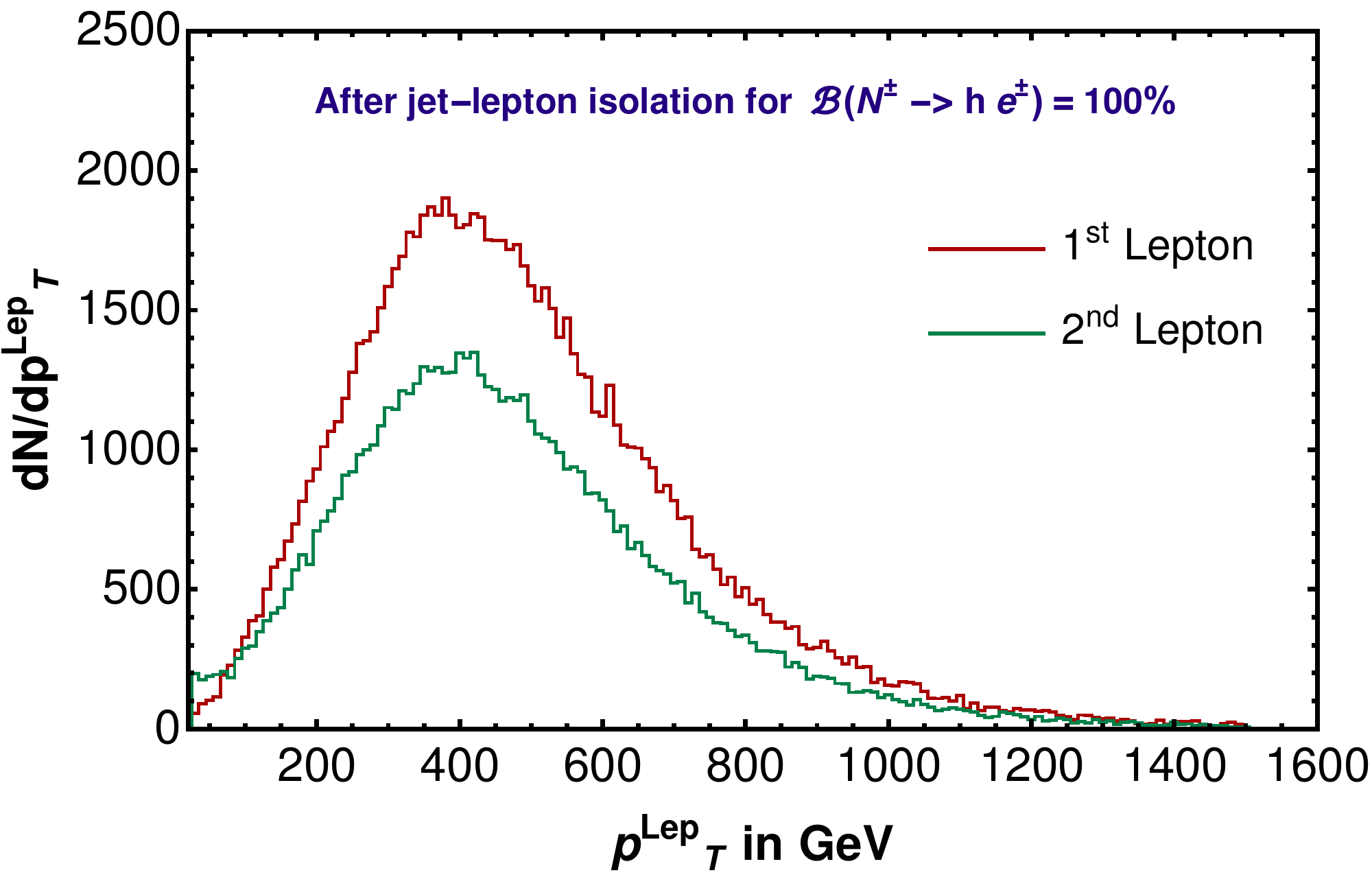}\label{}}\quad
			\subfigure[]{\includegraphics[width=0.37\linewidth,angle=-0]{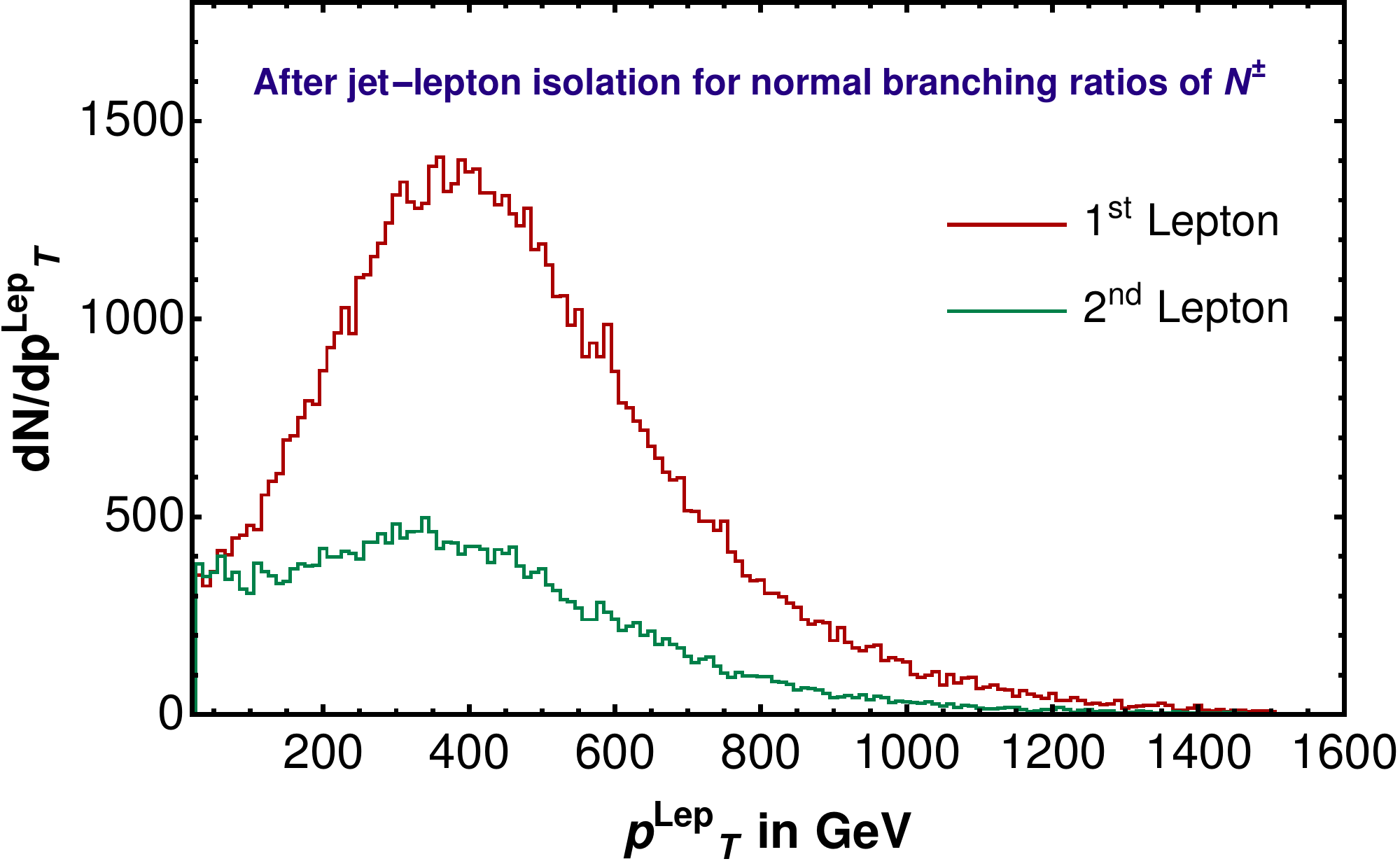}\label{}}}		
		\caption{The transverse momentum ($p_T$) distributions of charged leptons ($1^{st}$ and $2^{nd}$) at the LHC for the process $p\,p\to N^+\,N^-$ with $M_N=1\,\rm{TeV}$, $Y_N=5\times 10^{-7}$ and center of mass energy 14\,TeV. (a) and (b) represent the distributions of two leptons for pre and post jet-lepton isolation, respectively for $\mathcal{B}(N^{\pm}\to h\,e^{\pm})=100\%$. (c) describes the scenario for the isolated leptons obeying the branching ratio of BP1. }\label{lepPT_pp}
	\end{center}
\end{figure}


\autoref{lepPT_pp} demonstrates  the lepton $p_T$ distributions ($p^{\rm Lep}_T$)  for BP1. \autoref{lepPT_pp}(a) corresponds to the production mode of $p\,p\to N^+\,N^-$, where   $N^{\pm}\to h\,e^{\pm}$ is kept at 100\% and jet-lepton isolation is not demanded. Indeed, both the leptons (red and green curves) coming from $N^\pm$ decays have identical $p_T$ distributions. The effect of jet-lepton  isolation can be observed in \autoref{lepPT_pp}(b). Though the distributions are identical, the isolation cuts are responsible for the relatively low yield of the second lepton. Finally, we have the distribution in \autoref{lepPT_pp}(c), where the branching fractions of the decay mode are kept as in BP1 for $Y_N = 5\times 10^{-7}$. In this case, the second leptons can be from the gauge boson decays, thus lower in $p_T$ (green curve).

\begin{figure}[h]
	\begin{center}
		\hspace*{-0.5cm}
		\mbox{\subfigure[] {\includegraphics[width=0.44\linewidth,angle=-0]{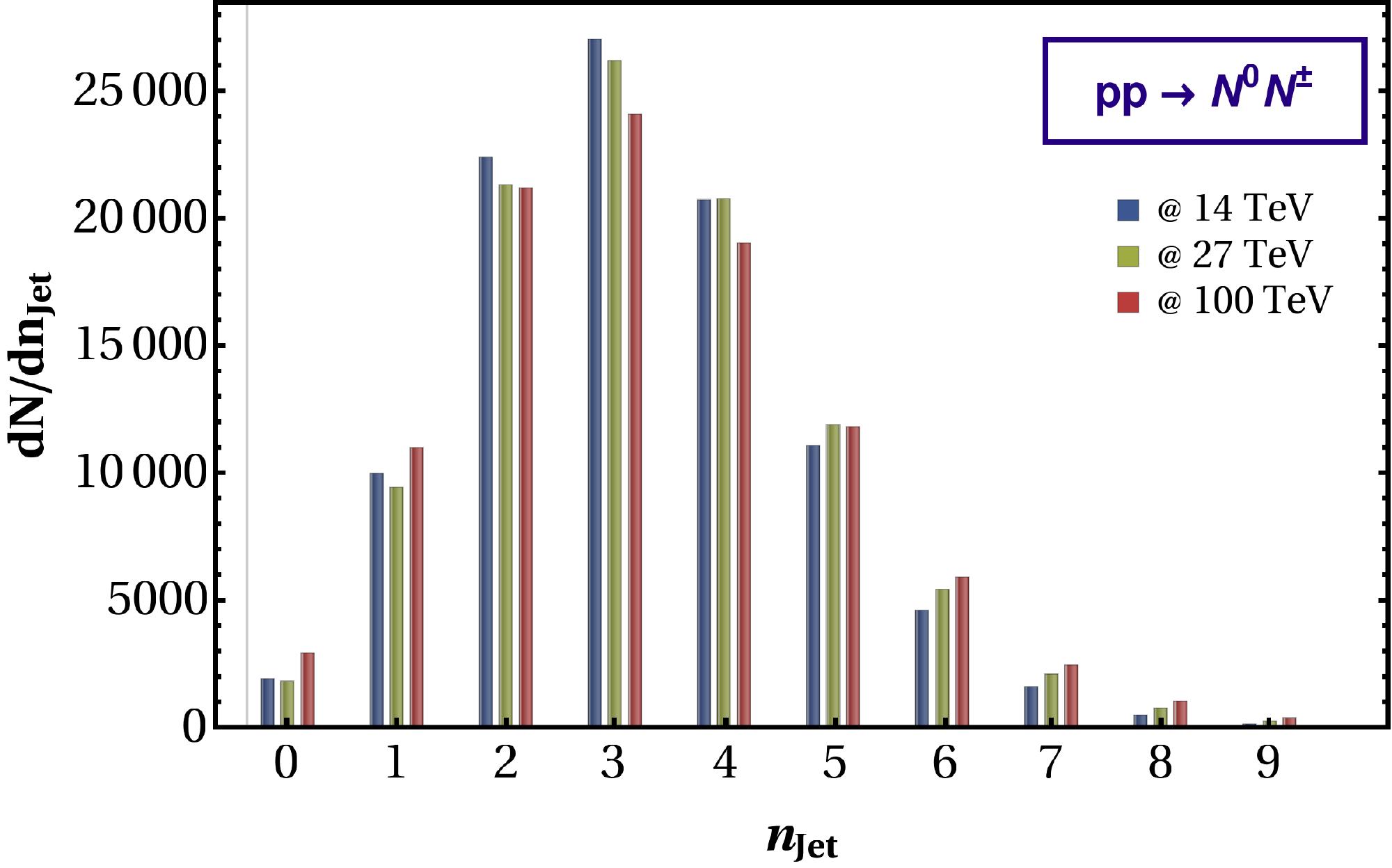}\label{}}\quad \quad
			\subfigure[] {\includegraphics[width=0.44\linewidth,angle=-0]{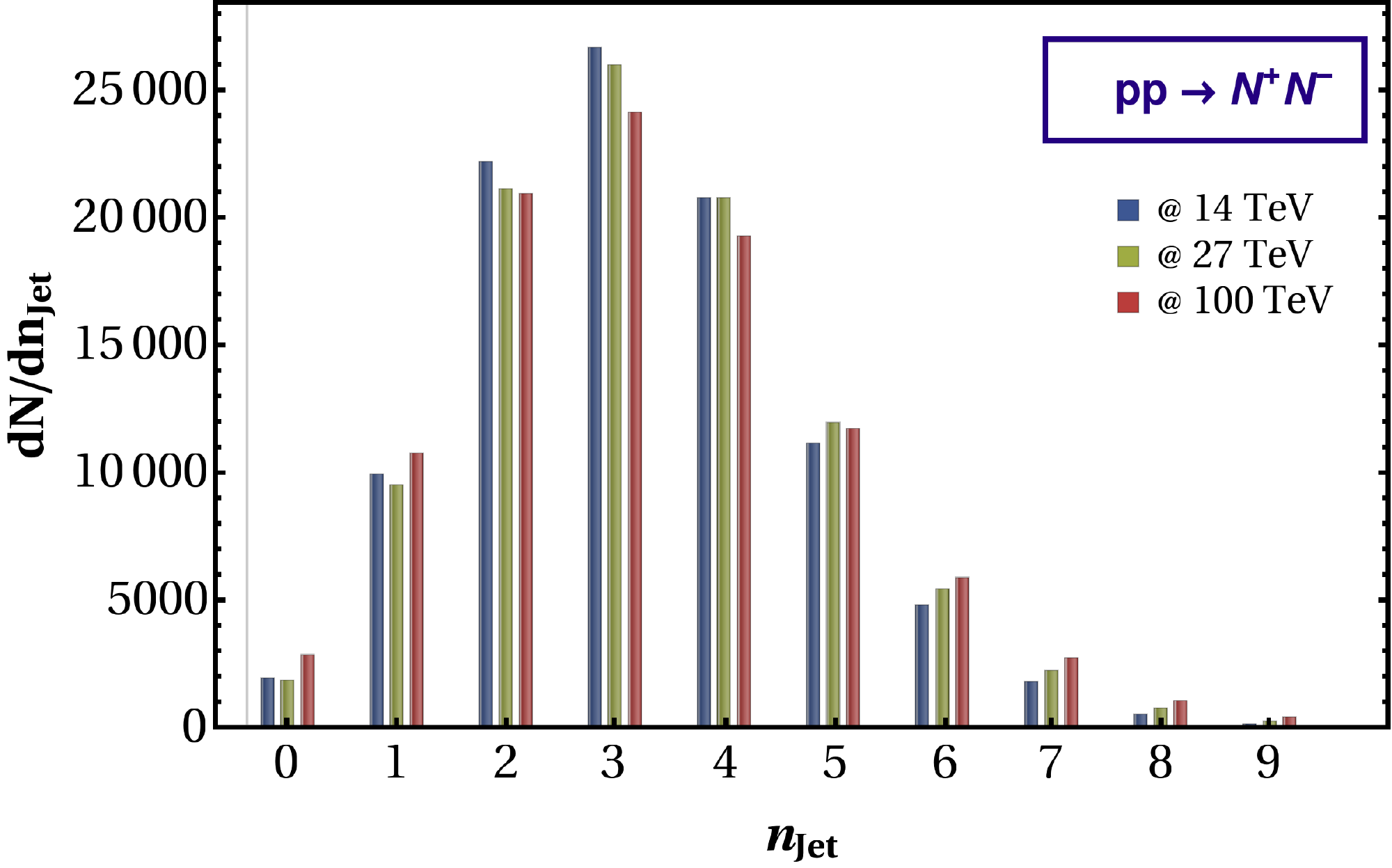}\label{}}}
		\caption{Jet multiplicity ($n_{\rm Jet}$)  distributions for the process $p\,p\to N^0\,N^{\pm}$ (a) and $p\,p\to N^+\,N^-$ (b) at $E_{CM} = $ 14\,TeV, 27\,TeV and 100\,TeV for $Y_N=5\times 10^{-7}$ and $M_N=1\,\rm{TeV}$ (BP1).}\label{pp_jet_mul}
	\end{center}
\end{figure}

\autoref{pp_jet_mul} illustrates the jet multiplicity  distributions for BP1 for the centre of mass energies of 14\,TeV, 27\,TeV and 100\,TeV. \autoref{pp_jet_mul}(a) and (b) represents  the process of  $p\,p\to N^0\,N^{\pm}$  and $p\,p\to N^+\,N^-$, respectively.  The  distributions  in blue, olive green and red corresponds to centre of mass energies of 14, 27 and 100  TeV, respectively. In both cases, the multiplicity peak around three and ISR/FSR jets increase as we go for the higher centre of mass energy.

\begin{figure}[hbt]
	\begin{center}
		\hspace*{-0.5cm}
		\mbox{\subfigure[]{\includegraphics[width=0.44\linewidth,angle=-0]{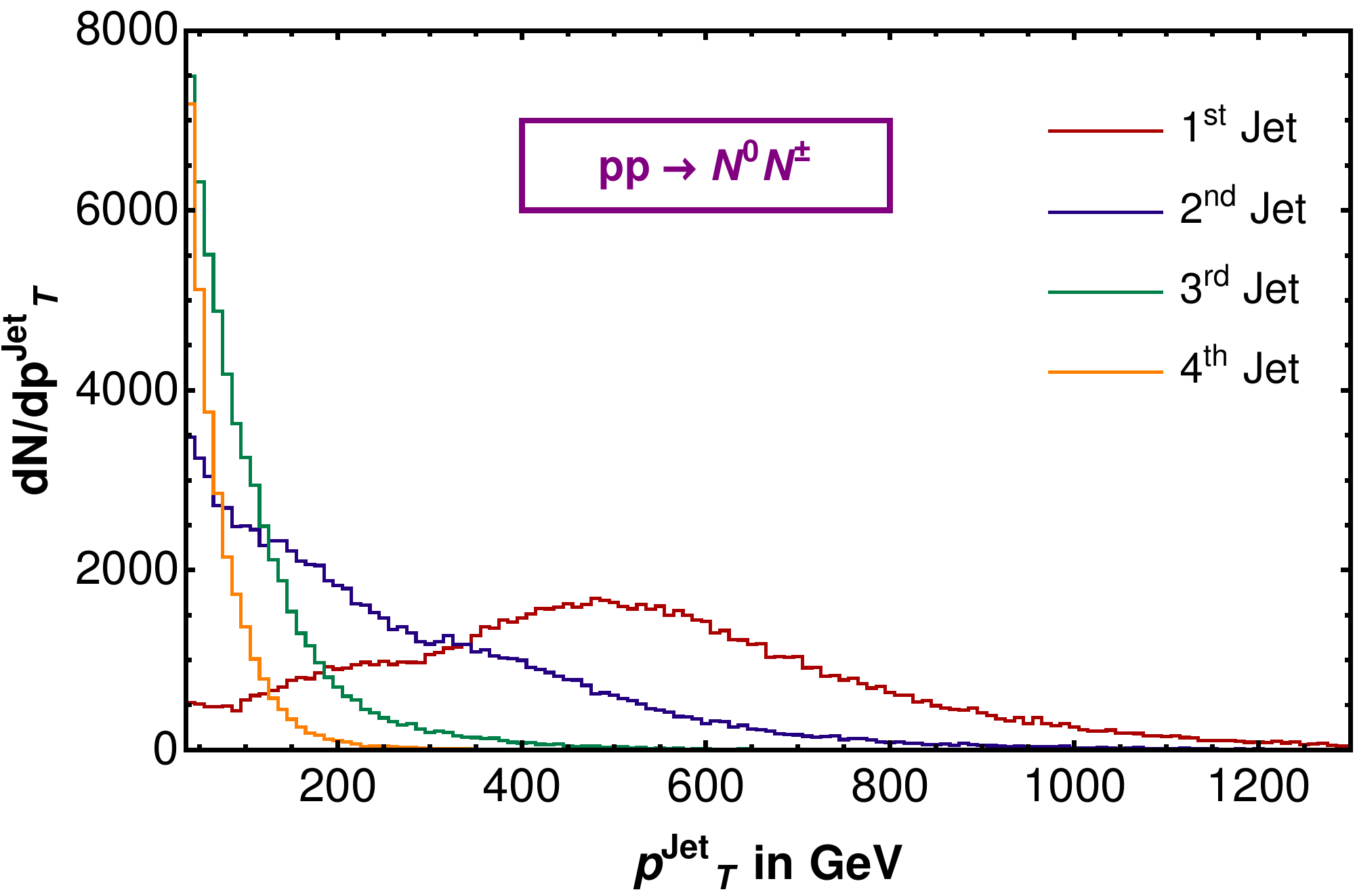}\label{}}\quad \quad
		\subfigure[]{\includegraphics[width=0.44\linewidth,angle=-0]{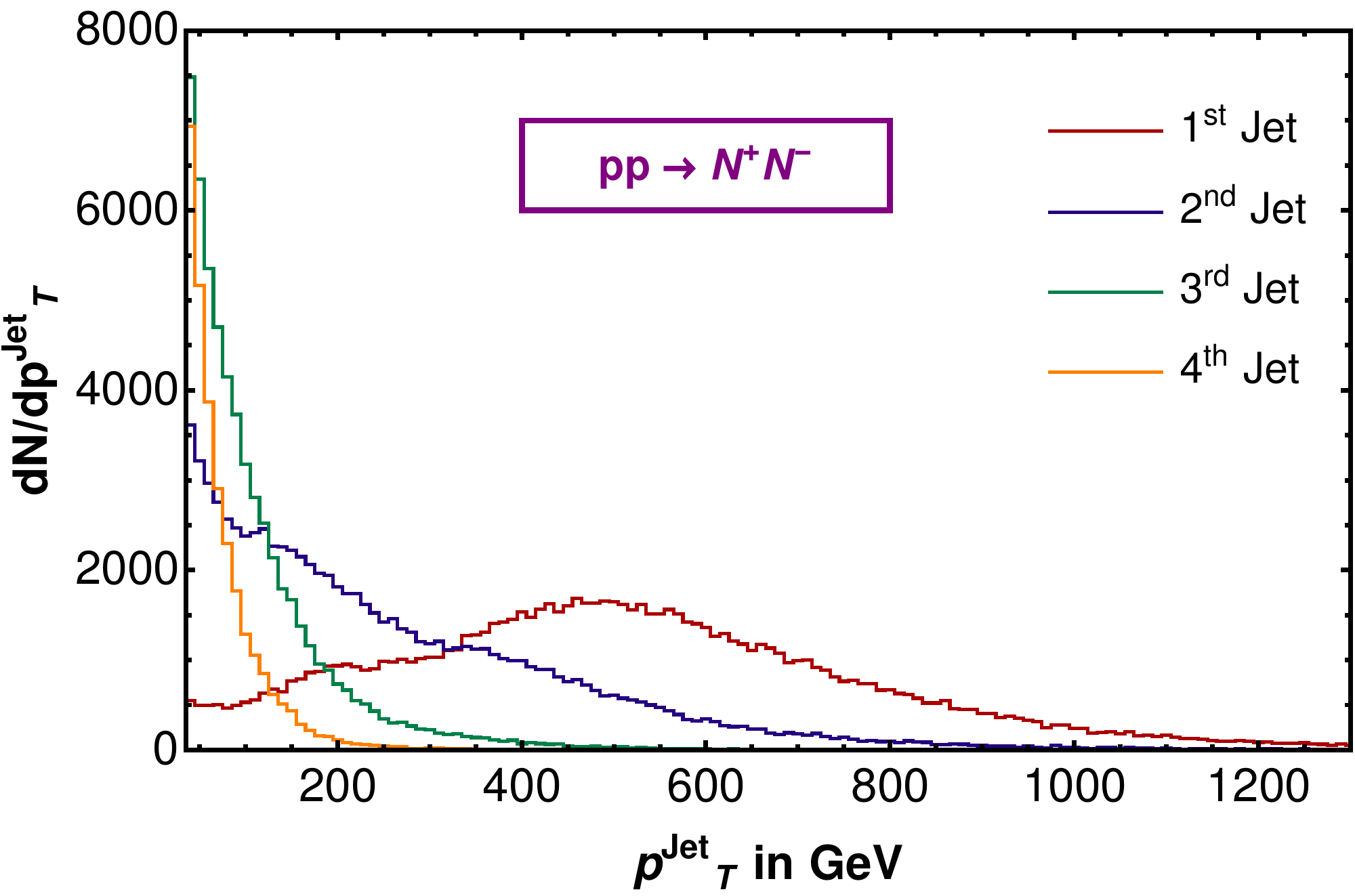}\label{}}}		
		\caption{The transverse momentum ($p_T$) distributions of the first four $p_T$ ordered jets at the LHC for the process $p\,p\to N^0\,N^{\pm}$ (a) and $p\,p\to N^+\,N^-$ (b) with $M_N=1\,\rm{TeV}$ (BP1), $Y_N=5\times 10^{-7}$ and center of mass energy of 14\,TeV.} \label{jetPT_pp}
	\end{center}
\end{figure}


In \autoref{jetPT_pp}, we present the jet $p_T$ distributions ($p^{\rm Jet}_T$) of the first four $p_T$ ordered jets for BP1 at the centre of  mass energy of 14 TeV. The leading jet (in red) peaks around $\frac{M_N}{2}$ as  the two jets  coming from the gauge bosons or Higgs boson become collinear due to large boost effect and combine as single jet. The second jet (in blue) either comes from the gauge  boson decay or the other BSM fermion and is relatively harder. The third and fourth jets (in green and  orange) which come from the decay of gauge bosons are of lower $p_T$.  

\begin{figure*}[hbt]
	\begin{center}
		\hspace*{-0.5cm}
		\mbox{\subfigure[] {\includegraphics[width=0.45\linewidth,angle=-0]{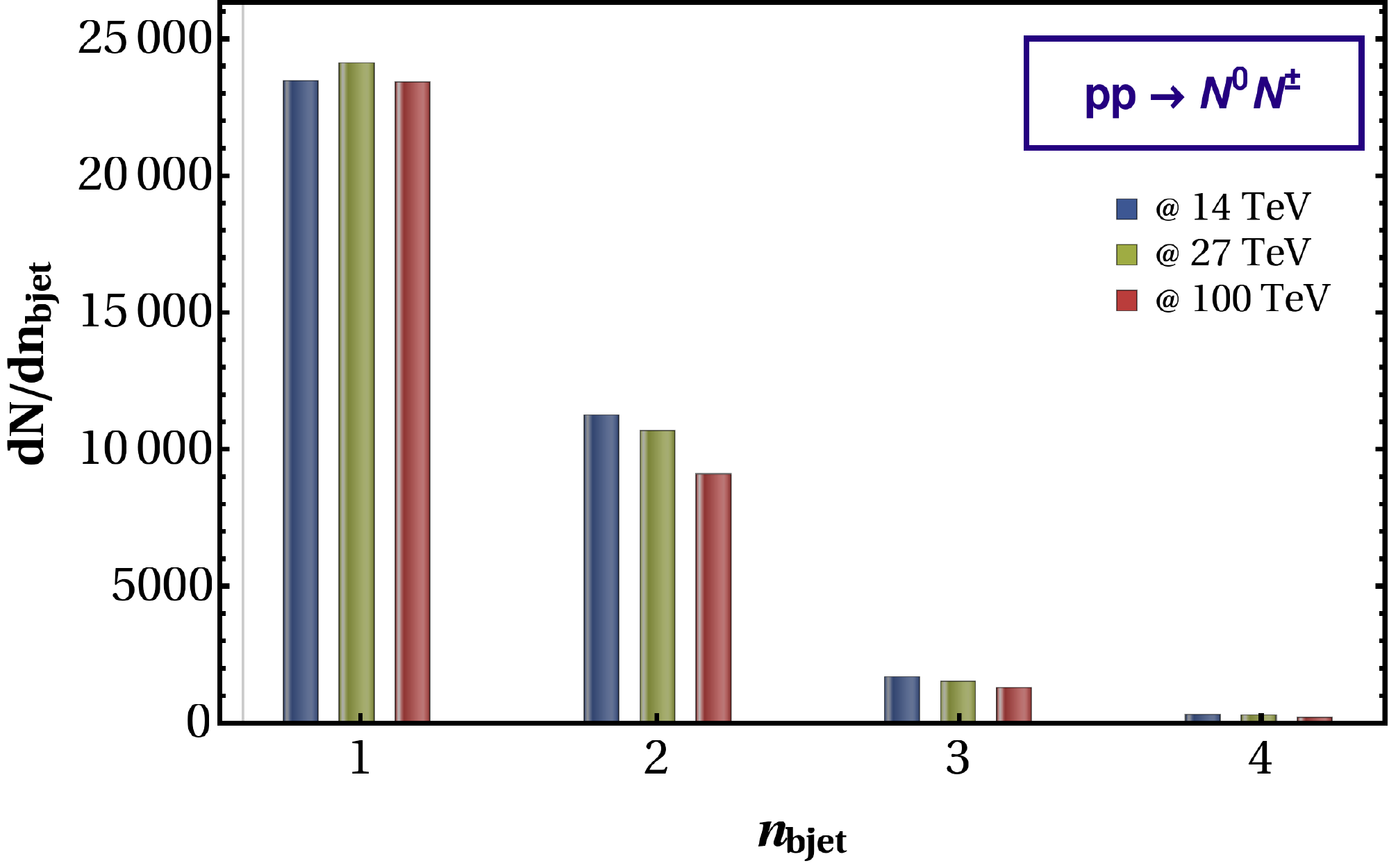}\label{}}\quad \quad
		\subfigure[] {\includegraphics[width=0.45\linewidth,angle=-0]{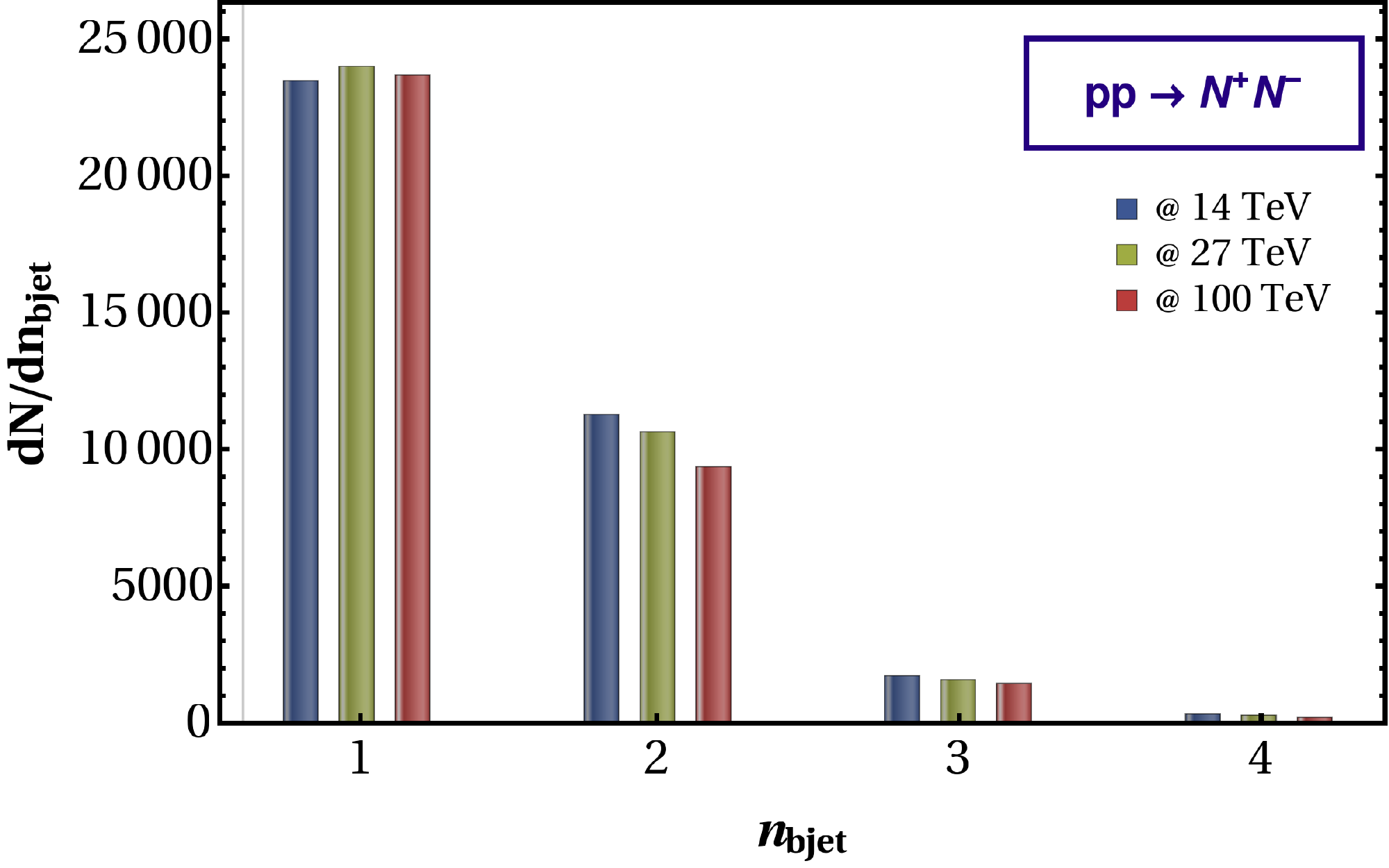}\label{}}}
		\caption{b-jet multiplicity ($n_{\rm bjet}$) distributions for the process (a) $p\,p\to N^0\,N^{\pm}$ and (b)$p\,p\to N^+\,N^-$ at $E_{CM} = $ 14\,TeV, 27\,TeV and 100\,TeV for $M_N=1\,\rm{TeV}$ (BP1) and $Y_N=5\times 10^{-7}$.}\label{pp_bjet_mul}
	\end{center}
\end{figure*}

\autoref{pp_bjet_mul} (a) and (b) describes the multiplicity distributions for the $b$-jets coming from the Higgs boson decays for the production modes of $p\,p\to N^0\,N^{\pm}$  and $p\,p\to N^+\,N^-$, respectively at three different centre of mass energies, 14\,TeV, 27\,TeV and 100\,TeV with $M_N=1\,\rm{TeV}$ (BP1) and $Y_N=5\times 10^{-7}$. The Higgs boson and the $Z$ boson coming from $N^{\pm(0)}$ are the two sources of such  $b$-jets. The unwanted background $b$-jet coming from $Z$ decay can be eliminated via the reconstruction of di-$b$-jet invariant mass distributions. The $b$-jets coming from the Higgs bosons or $Z$ boson decays can be traced back by reconstructing their invariant mass distributions. The $b$-jets are tagged via the secondary vertex reconstitution of the $b$-hadrons associated with the jets. We follow the $b$-jet tagging efficiency of CMS with a maximum of 85\% \cite{CMS:2012jki,CMS:2017wtu}. The multiplicity of $b$-jets can go up to four  as they come from both the Higgs bosons decays.

At the LHC, we do not know the centre of mass frame of the collision due to unknown longitudinal boost governed by the parton distribution functions which dictates the momentum sharing of the colliding partons. The sharing of the colliding energies  of the  quarks in this case also varies with centre of mass energies. If we evaluate the heavy fermion decays in complete  visible mode i.e. $N^\pm \to h \ell^\pm$, the reconstruction of the heavy fermion momentum is possible and allows us to measure  both  $p_T$ and $p_z$ of  $N^\pm$.  This further enables us to determine the longitudinal decay length, in addition to the transverse decay length. On the contrary, it is not possible to reconstruct the total momentum of the neutrino coming from $N^0 \to h \nu$ decay due to the lack of  knowledge of the longitudinal boost of the partonic system. However, as the net transverse momentum is zero, for $pp \to N^0 N^\pm \to h\, h\, \ell^{\pm}\, \nu $, where only a single neutrino contributes towards missing momentum, we can easily calculate the transverse missing momentum as neutrino $p_T$. Thus, for  $pp \to N^0 N^\pm $, only transverse momentum and consequently the transverse decay length can be calculated, whereas, for $pp \to N^+ N^- $, we can compute the displaced decay length in both transverse and longitudinal direction. Such conclusion is however, bound to change as we move towards lower Yukawa coupling i.e., $Y_N \leq 10^{-8}$ where, $N^\pm \to \pi^\pm N^0$ becomes dominant and multiple neutrinos coming from pion and other decays can smear the invariant mass distributions.

\begin{figure}[h]
	\begin{center}		
		\hspace*{-0.5cm}
		\mbox{\subfigure[]{\includegraphics[width=0.47\linewidth,angle=-0]{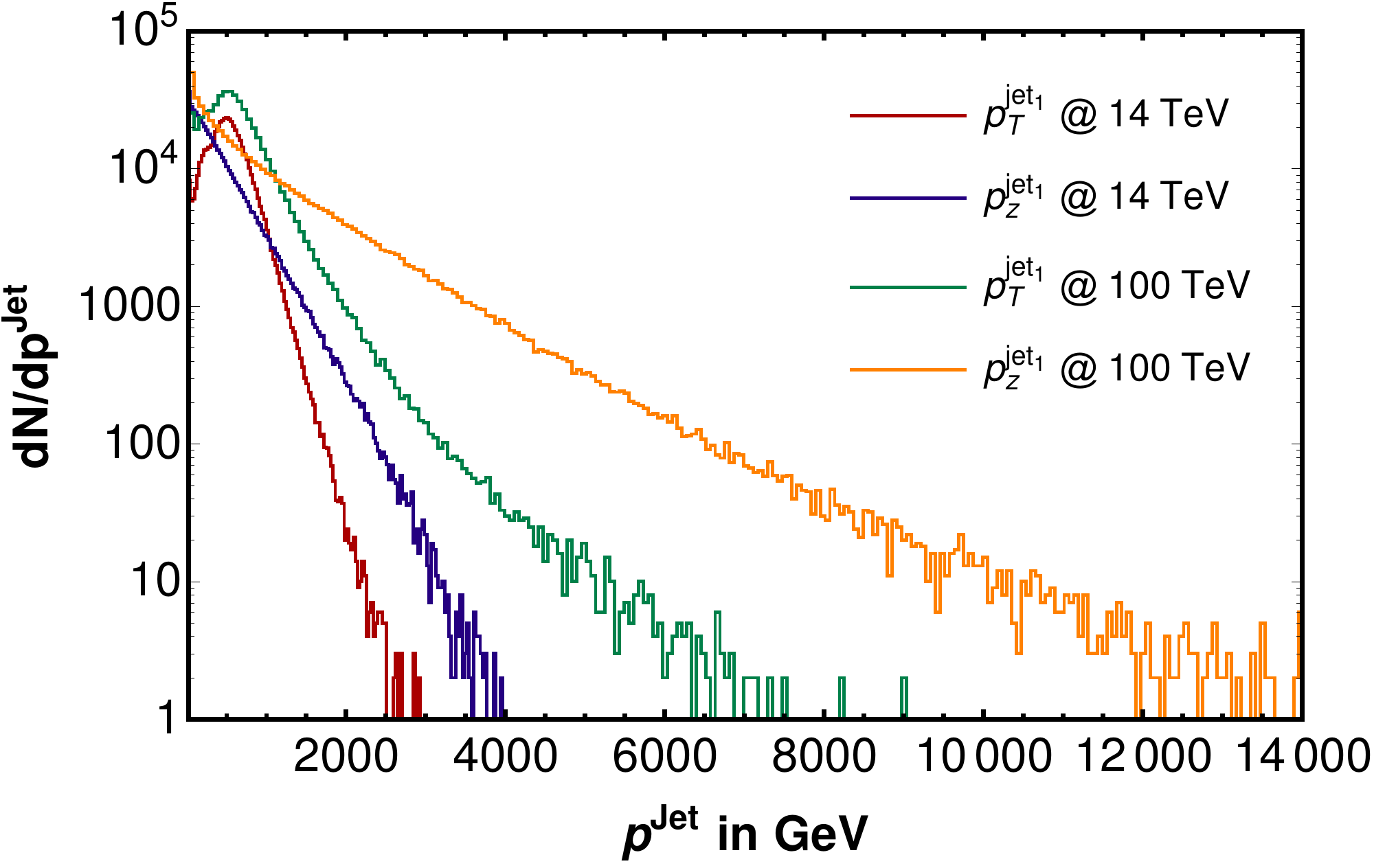}\label{}}\quad \quad
		\subfigure[]{\includegraphics[width=0.47\linewidth,angle=-0]{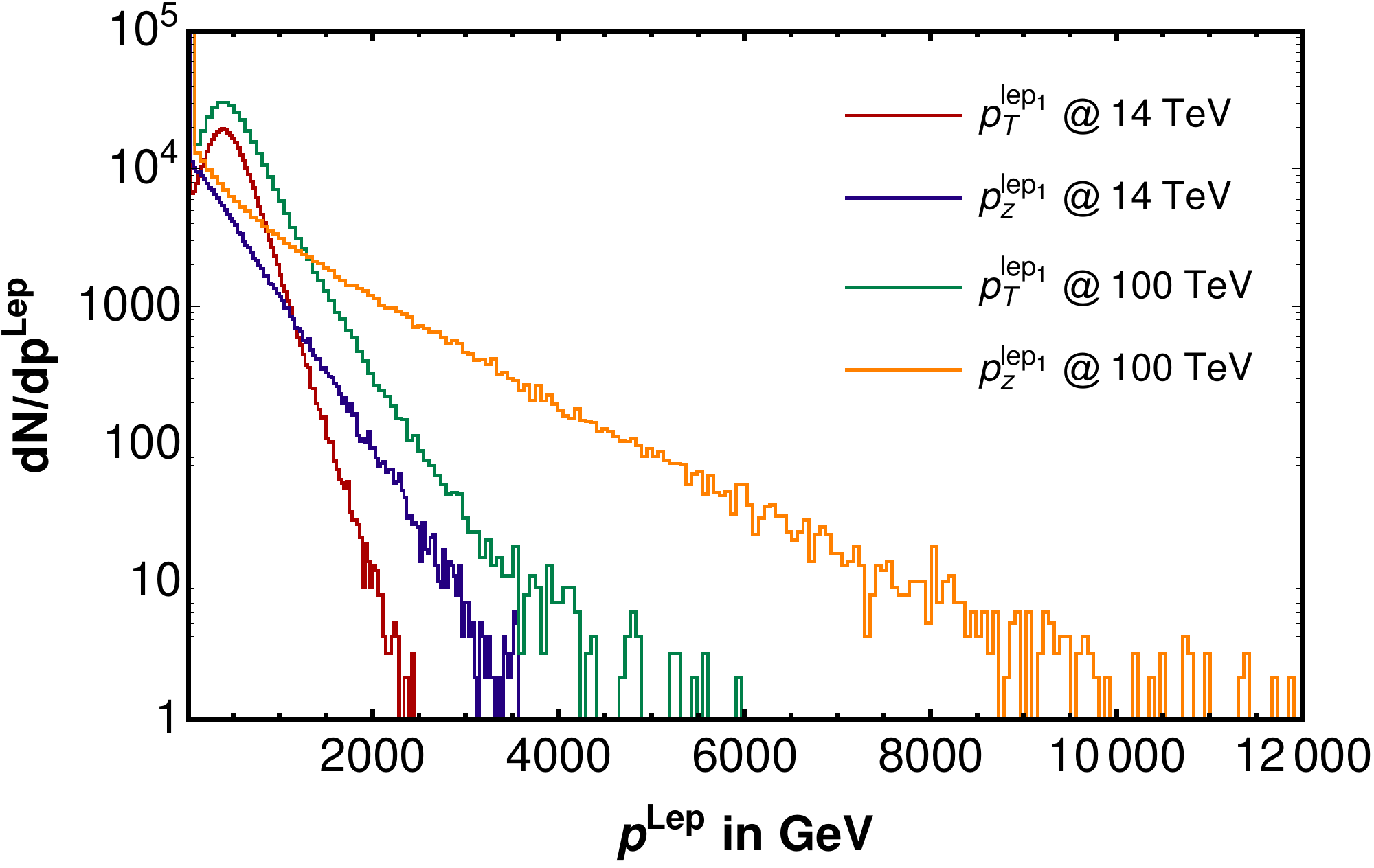}\label{}}}	
		\caption{The transverse and longitudinal momenta distributions of the leading jet (a) and the first lepton (b) for the process $p\,p\to N^+\,N^{-}$,  with $M_N=1\,\rm{TeV}$ (BP1), $Y_N=5\times 10^{-7}$ and center of mass energies of 14 TeV and 100 TeV at the LHC. }  \label{pp_jet_p}
	\end{center}
\end{figure}


Another intriguing feature that we observe is that the enhanced displaced decay length in the longitudinal direction due to the large longitudinal boost compared to the transverse one. This can be inferred from  \autoref{pp_jet_p} which shows the comparison  between transverse and longitudinal momentum of the jets and leptons at 14 TeV and 100 TeV for BP1 with $Y_N=5\times 10^{-7}$ . \autoref{pp_jet_p}(a) depicts the comparison between $p^j_T$ and $|p^j_z|$ at the centre of mass energies of 14 TeV and 100 TeV. In case of 14 TeV, the longitudinal momentum of the jet goes till 3.8 TeV as compared to 2.5 TeV of $p^j_T$ and the cross-over of the two distributions happen around 1.3 TeV. Similarly, at 100 TeV, $p^j_T$ reaches up to 7.5 TeV, while $p^j_z$ reaches till 14 TeV with a cross-over around 1.5 TeV. Certainly, the extra boost in the longitudinal direction can further push the displaced decay compared to the transverse one.  In \autoref{pzpT} we see the number of events in percentage where $P_z^{\rm{jet}}> P_c^{\rm{jet}}$ and  $P_c^{\rm{jet}}$ is the cross-over point of $p_z$ and $p_T$ as can be seen from \autoref{pp_jet_p}(a). It is evident that as we increase the centre of mass energy, the  longitudinal boost increases more and thus events for which $P_z^{\rm{jet}}$ is greater than $P_c^{\rm{jet}}$  can be around 30\% for 100\,TeV centre of mass energy. Such effect can be seen in the corresponding displaced decay lengths as we discuss them next. 

\begin{table}[hbt]	
	\begin{center}
		\renewcommand{\arraystretch}{1.4}
		\begin{tabular}{|c|c|c|}
			\hline
			
			{\multirow{2}{*}{Centre of mass}}&
			\multicolumn{2}{c|}{$\%$ of events for $P_z^{\rm{jet}}> P_c^{\rm{jet}}$}\\
			\cline{2-3}
			energy&\multicolumn{1}{c|}{BP1}&\multicolumn{1}{c|}{BP2}\\ 
			\cline{2-3}
			\hline	
			\hline
			$14\,\rm{TeV}$  & $8.4\%$ & $3.9\%$ \\
			\hline
			$27\rm{TeV}$  & $18.9\%$ & $11.9\%$ \\
			\hline
			$100\,\rm{TeV}$ & $31.3\%$ & $29.6\%$ \\
			\hline			
		\end{tabular}
		\caption{The percentage of events for $P_z^{\rm{jet}}> P_c^{\rm{jet}}$ distribution for the benchmark points at centre of mass of energies of 14, 27 and 100 TeV, where $P_c^{\rm{jet}}$ is the point of cross-over for $P_z^{\rm{jet}}$ and $P_T^{\rm{jet}}$ distributions as shown in \autoref{pp_jet_p}.} \label{pzpT}
	\end{center}	
\end{table}


Similarly \autoref{pp_jet_p}(b) depicts comparison between the charged lepton momentum distributions for BP1 at the centre of mass energies of 14 and 100 TeV. We can see similar effect as the jet momentum distributions in \autoref{pp_jet_p}(a). The longitudinal momentum is boosted as compared to the transverse one. In the  following subsection, we examine both the transverse and the longitudinal displaced decays  along with their reaches  at CMS, ATLAS and MATHUSLA. 

\subsection{Displaced decays at the CMS, ATLAS and MATHUSLA}

\begin{figure*}[h]
	\begin{center}
		\centering
		\mbox{\subfigure{\includegraphics[width=0.6\linewidth,angle=-0]{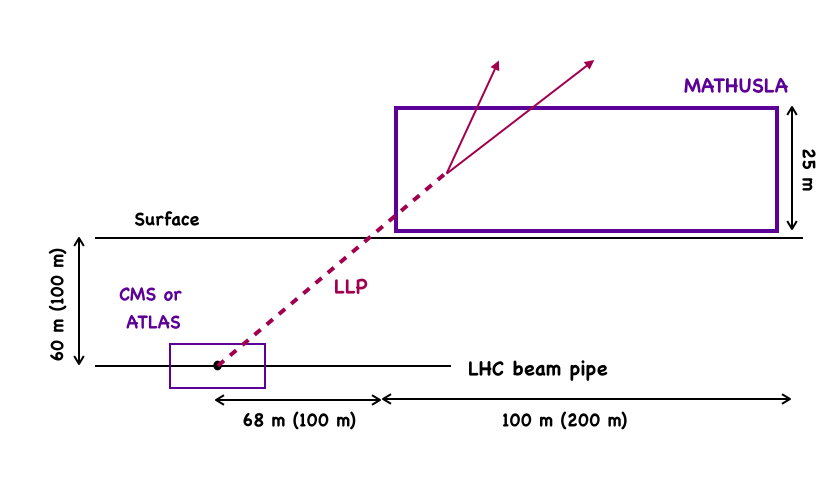}\label{}}}		
		\caption{Schematic diagram of the MATHUSLA detector.  In transverse direction the range of MATHUSLA is taken as $60-85$\,m ($100-125$\,m), whereas in the longitudinal direction it is considered as $68-168$\,m ($100-300$\,m)  \cite{Alpigiani:2020iam, Curtin:2018mvb}.}\label{mathusla}
	\end{center}
\end{figure*}


In this subsection, we study the distributions of the transverse and longitudinal decay lengths of  $N^\pm$ for three different centre of mass energies 14, 27 and 100 TeV, respectively for BP1 with three different Yukawa couplings. The displaced decays can be detected either in CMS, ATLAS or in a new proposed detector called MATHUSLA. CMS and ATLAS have transverse and longitudinal ranges  of  7.5, 12.5 and 22, 44 meter \cite{CMS:2007sch,ATLAS:design} respectively. However, the new proposed detector MATHUSLA is around 100 m from the CMS or ATLAS in the beam axis as well as in the transverse direction \cite{Curtin:2018mvb,MATHUSLA:2019qpy} as shown in \autoref{mathusla}. The length of the detector is 200 m with height of 25 m which give extra reach for the particle with late decay. According to a recent update, the MATHUSLA detector will be placed 68 metres away from the CMS/ATLAS interaction point and will have a volume of  $25\times 100\times 100\, \rm m^3$ \cite{Alpigiani:2020iam}. In the next few paragraphs, we show how different benchmark points pan out in different detectors. The choices of Yukawa here are motivated from the collider searches of the different displaced decays and any further constraint would restrict the parameter points  that we are interested in \cite{Ashanujjaman:2021jhi,CS:TypeI}.

\begin{figure*}[hbt]
	\begin{center}
		\hspace*{-0.5cm}
		\mbox{\subfigure[]{\includegraphics[width=0.3\linewidth,angle=-0]{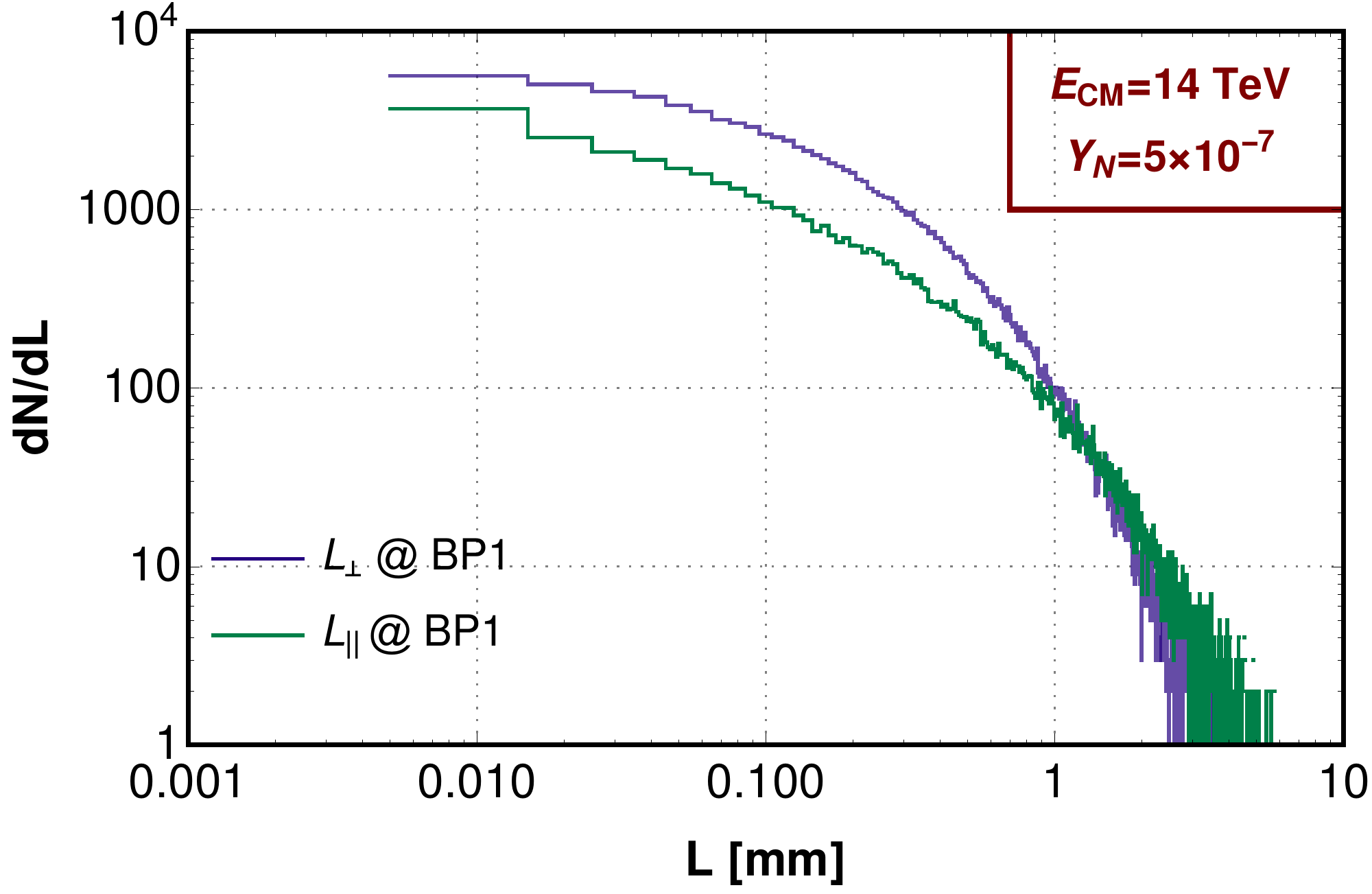}\label{}}\quad
			\subfigure[]{\includegraphics[width=0.3\linewidth,angle=-0]{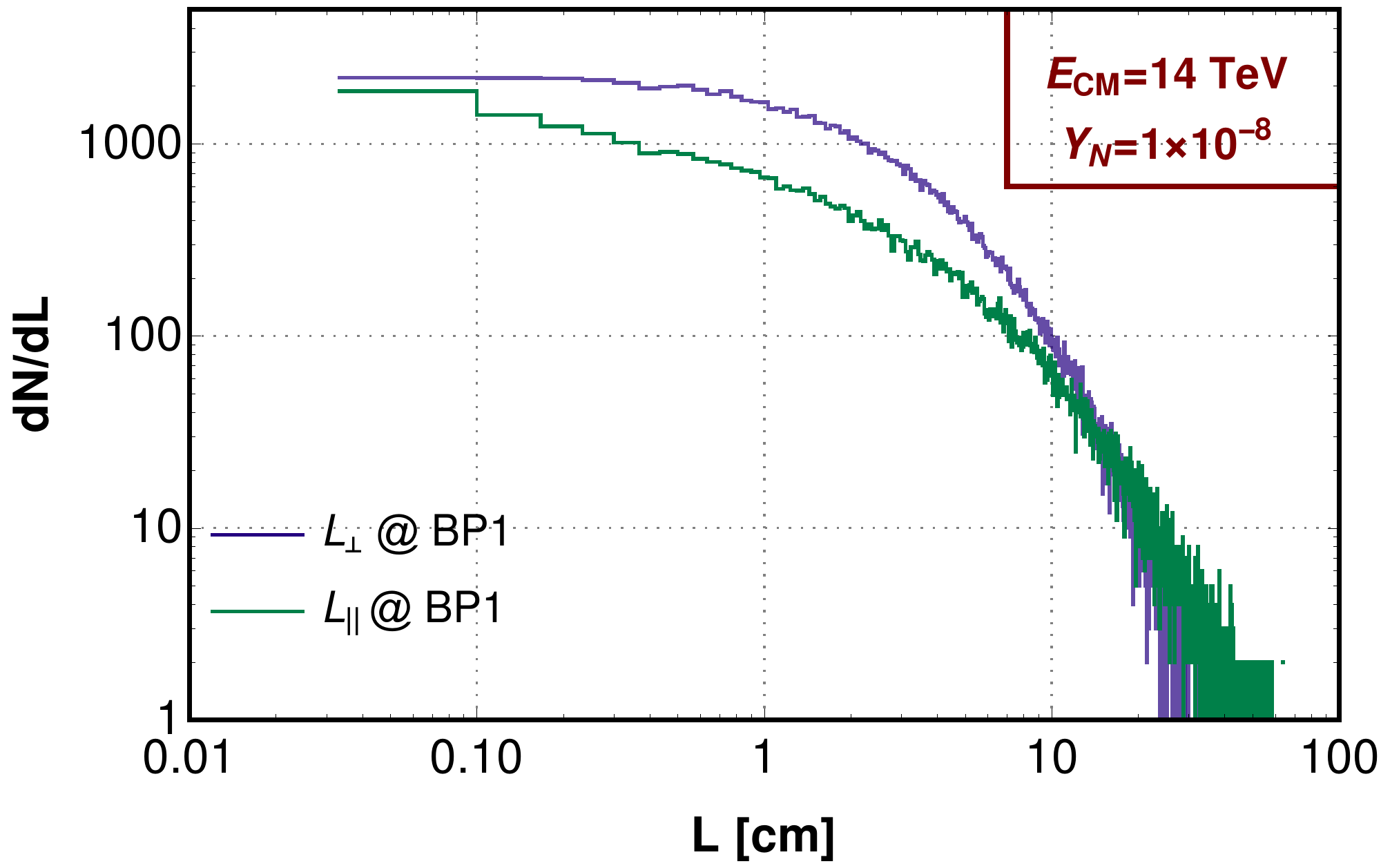}\label{}}\quad
			\subfigure[]{\includegraphics[width=0.3\linewidth,angle=-0]{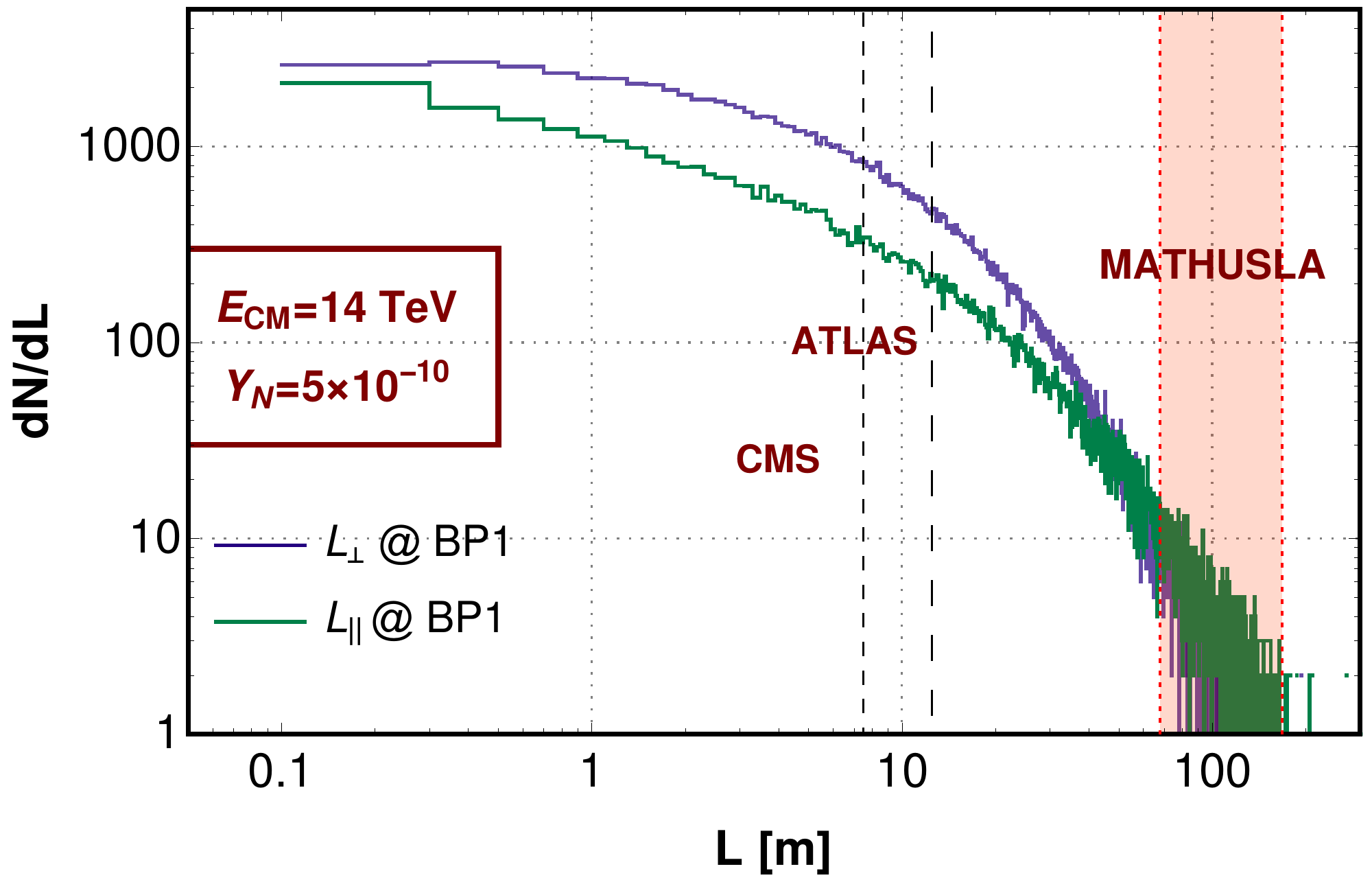}\label{}}}
		\hspace*{-0.5cm}
		\mbox{\subfigure[]{\includegraphics[width=0.3\linewidth,angle=-0]{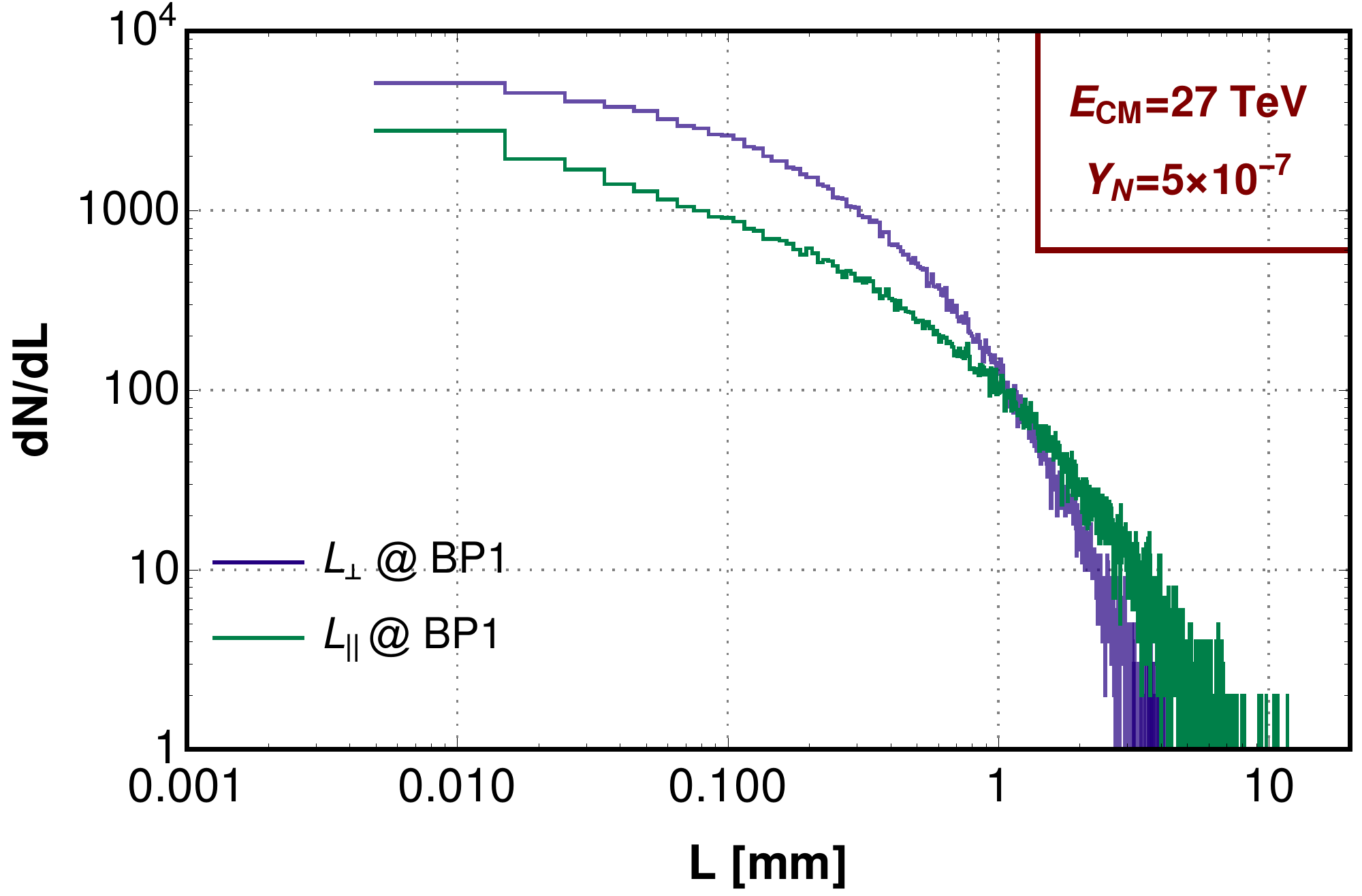}\label{}}\quad
			\subfigure[]{\includegraphics[width=0.3\linewidth,angle=-0]{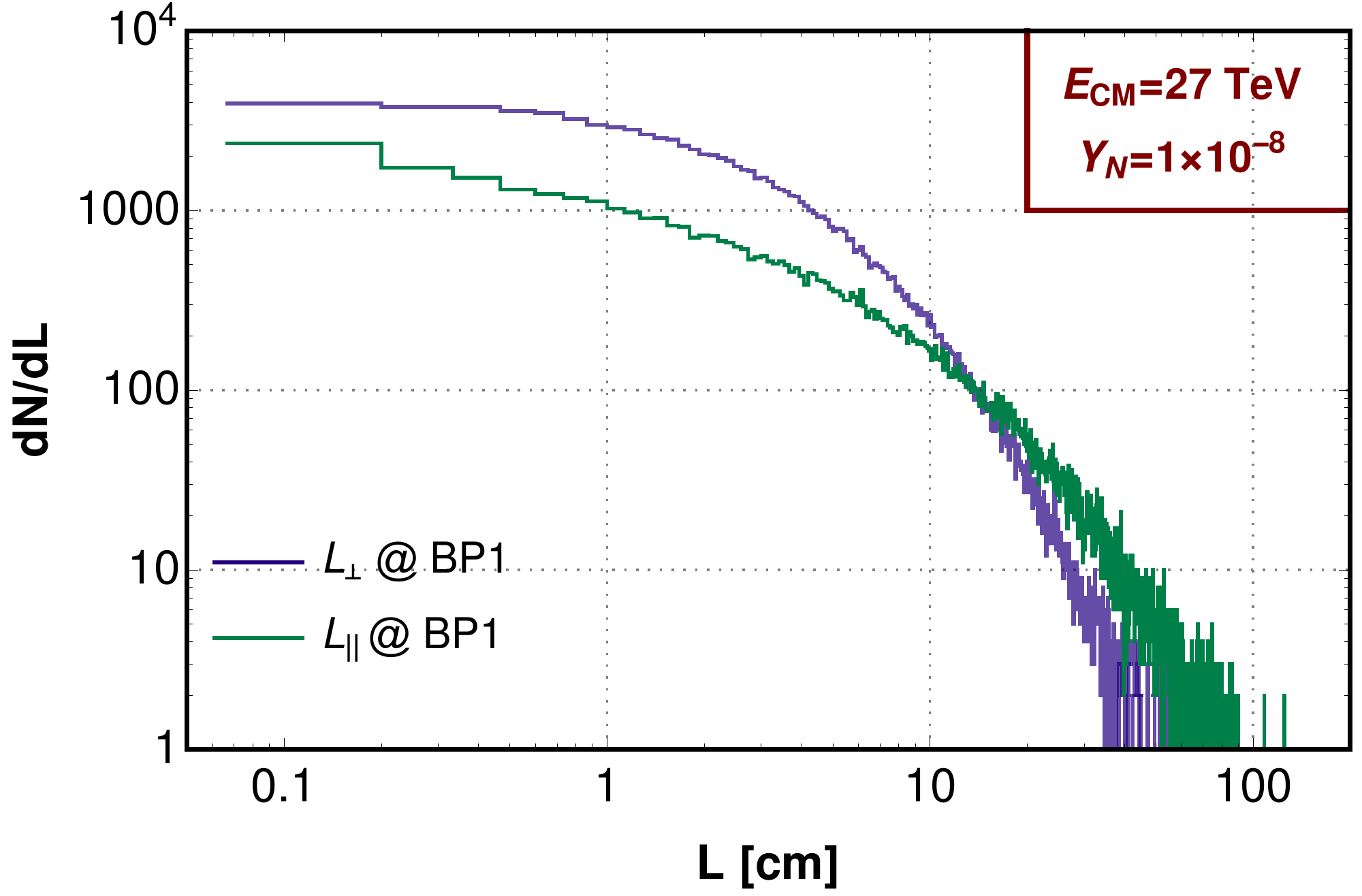}\label{}}\quad
			\subfigure[]{\includegraphics[width=0.3\linewidth,angle=-0]{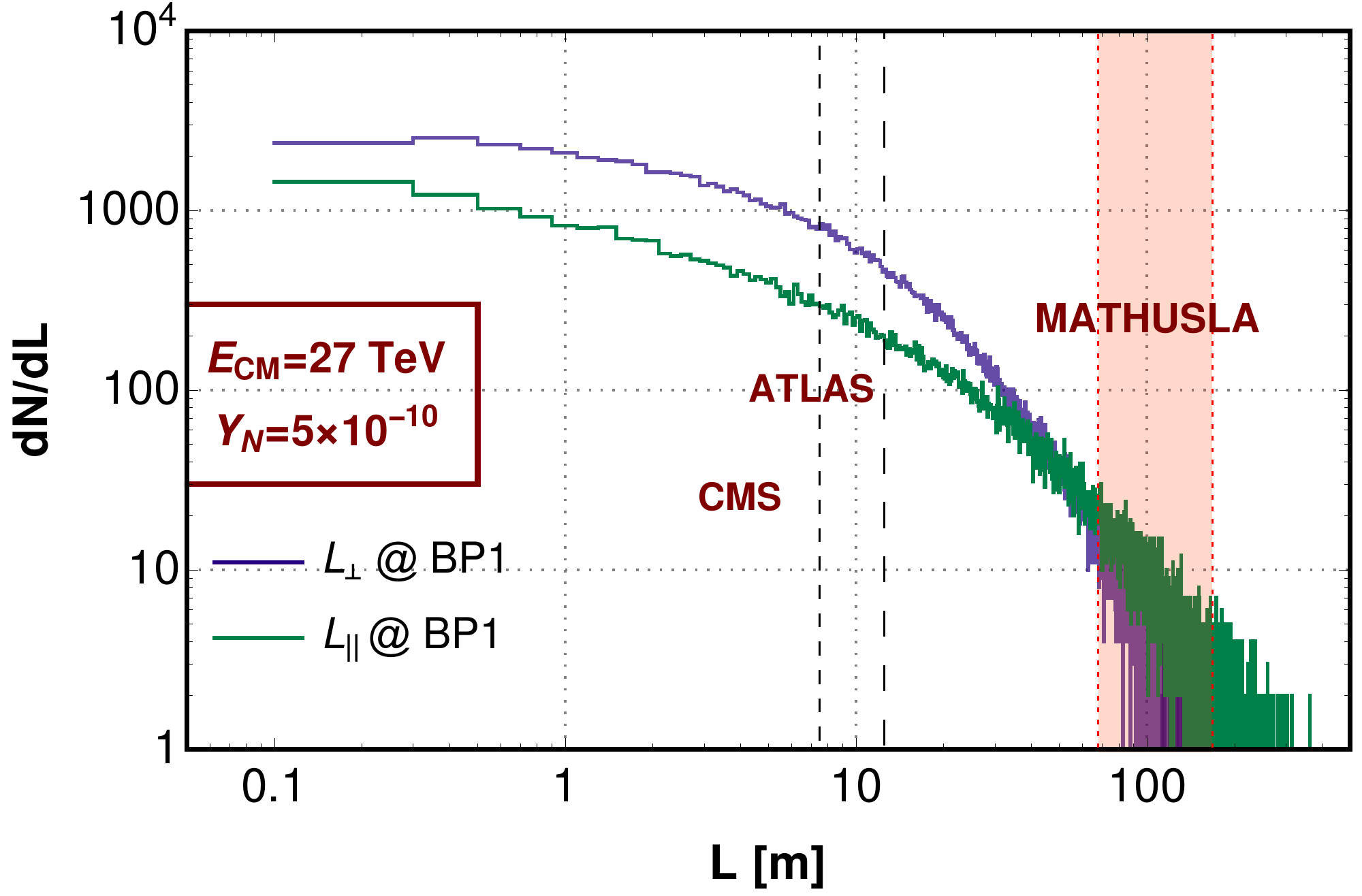}\label{}}}
		\hspace*{-0.5cm}
		\mbox{\subfigure[]{\includegraphics[width=0.3\linewidth,angle=-0]{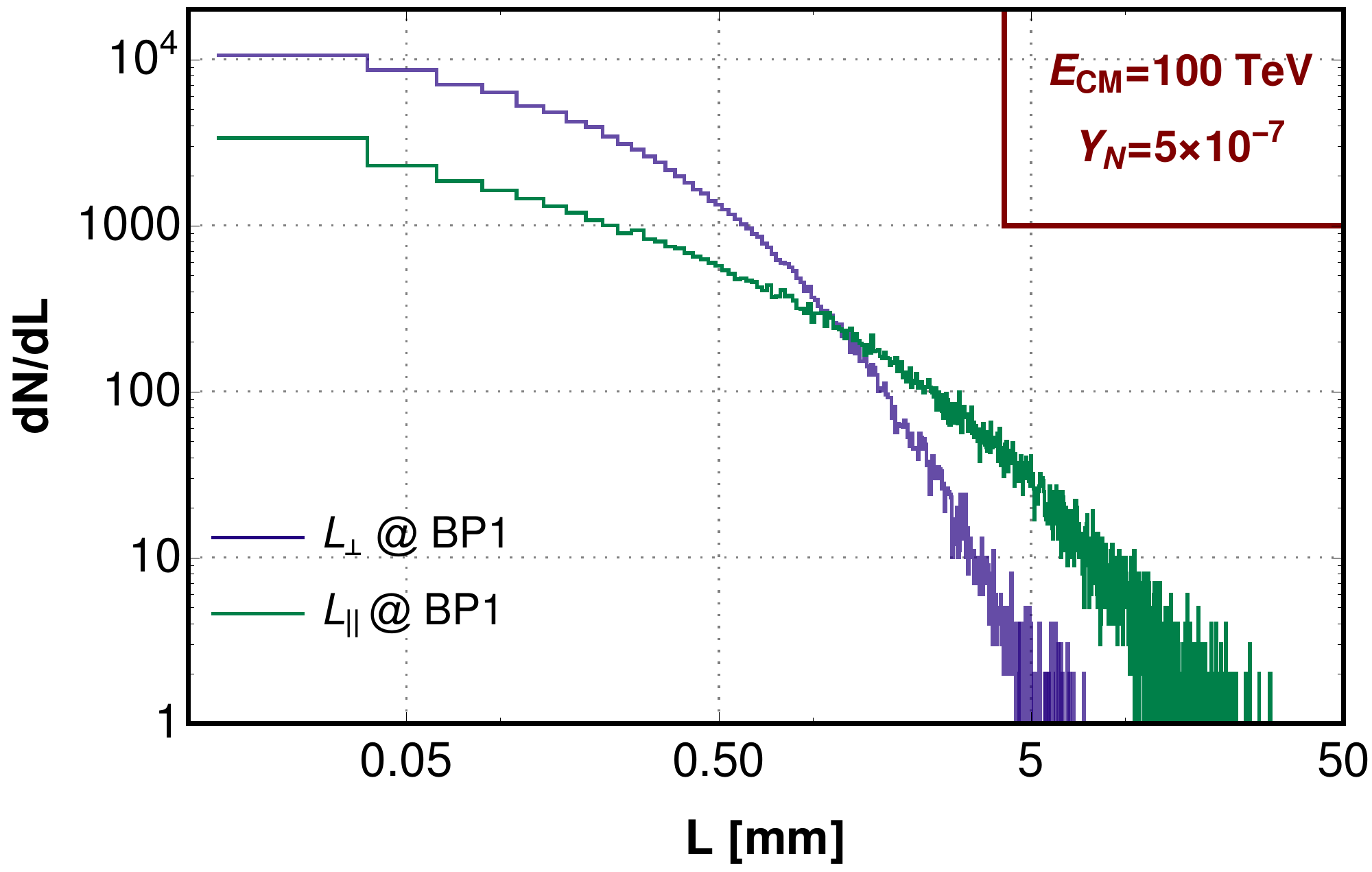}\label{}}\quad
			\subfigure[]{\includegraphics[width=0.3\linewidth,angle=-0]{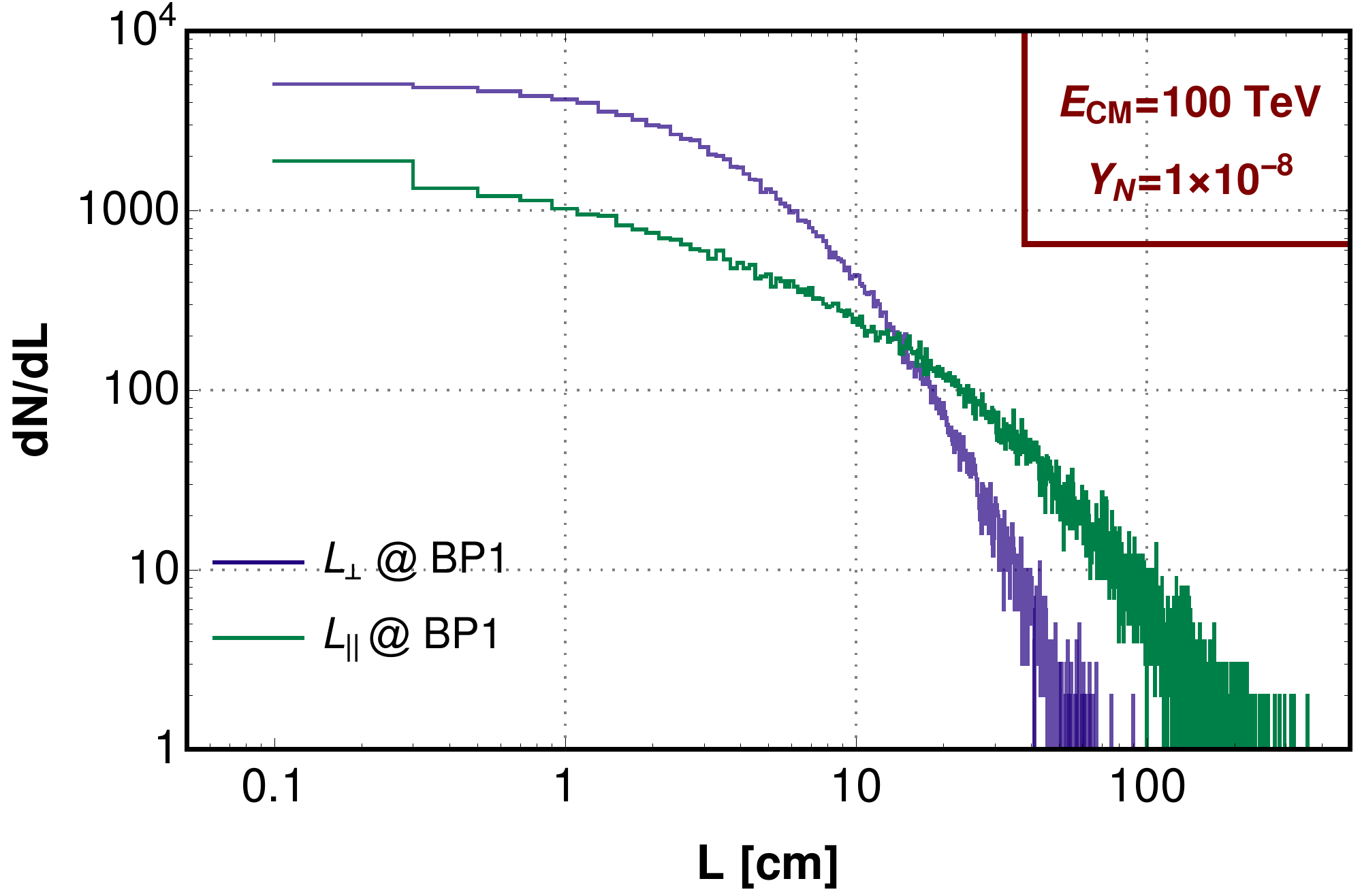}\label{}}\quad
			\subfigure[]{\includegraphics[width=0.3\linewidth,angle=-0]{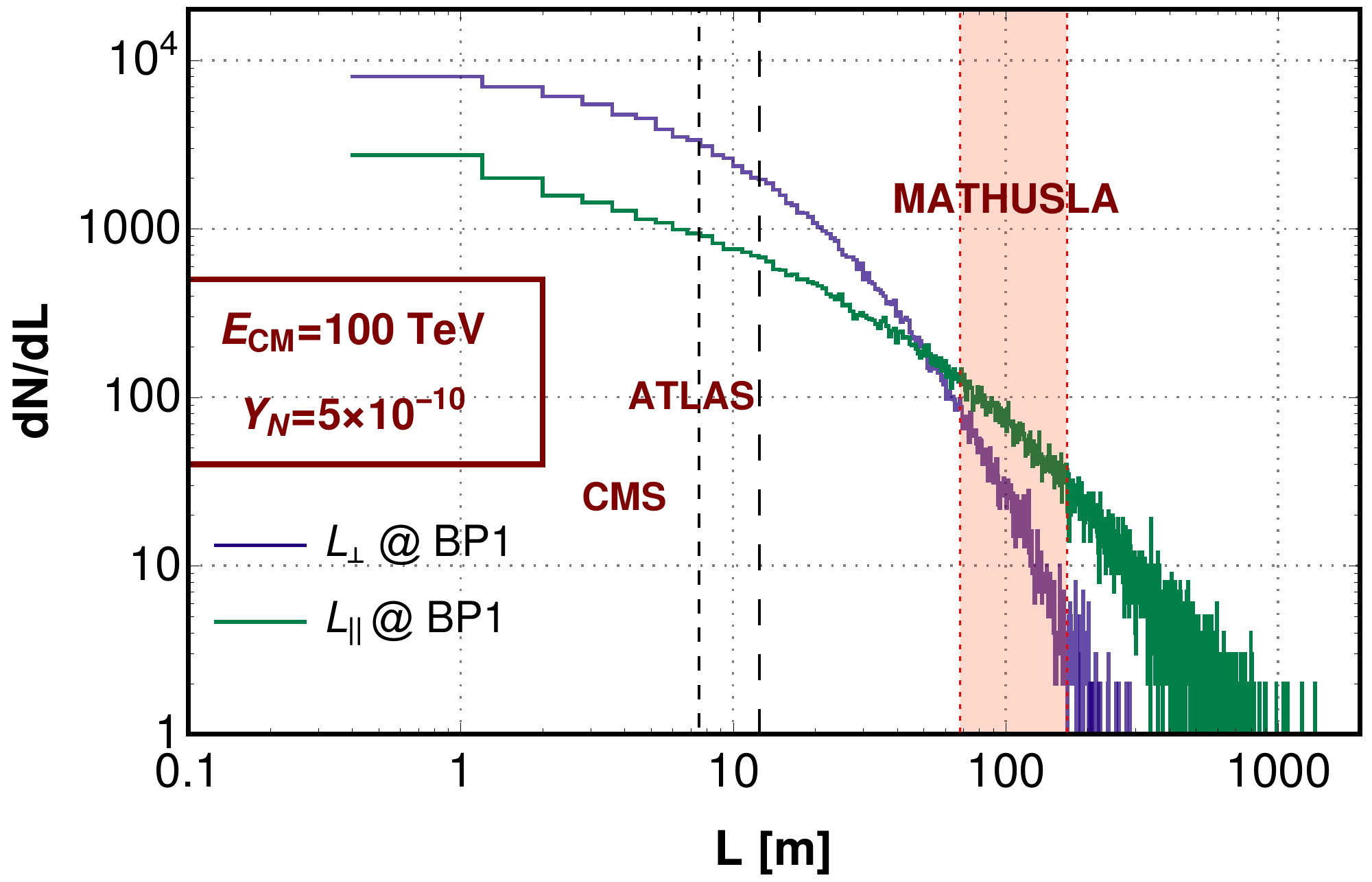}\label{}}}
		\caption{Displaced total transverse ($L_\perp$ in purple) and total longitudinal ($L_{||}$ in green) decay length distributions for the  $N^{\pm}$, coming from the pair productions at the LHC with the centre of mass energies 14\,TeV (a, b, c), 27\,TeV (d, e, f) and 100\,TeV (g, h, i)  for the benchmark points. Yukawa couplings $Y_N= 5\times 10^{-7},\,1\times 10^{-8} $, respectively are used for (a, d, g) and (b, e, h)  whereas the third column(c, f, i) depicts $Y_N=5\times 10^{-10}$. The dotted-dashed and dashed line indicates the upper limit of CMS and ATLAS, respectively. The light red band (68\,m$-$168\,m) denotes the MATHUSLA region. }\label{decay_pp}
	\end{center}
\end{figure*}

\autoref{decay_pp} depicts a comparative study of total transverse displaced decay and total longitudinal displaced decay, where $N^\pm$ are pair produced for BP1 ($M_N=1\,\rm TeV$). Here, we have used three panels for three different centre of mass energies i.e., 14\,TeV (a, b, c), 27\,TeV (d, e, f) and 100\,TeV (g, h, i), respectively.  The transverse $L_{\perp}$ and the longitudinal $L_{||}$   decay lengths are described via purple and green lines, respectively. The first column  (a, d, g) describes the case for $Y_N=5\times 10^{-7}$, where $N^\pm$ travels $\mathcal{O} (1-10)$ mm before dominantly decay to SM particles i.e. $h\ell^\pm,\, Z \ell^\pm,\, W^\pm \nu$, giving the first recoil. If we move further lower in Yukawa coupling i.e., $Y_N \leq 10^{-8}$,  the $N^\pm \to \pi^\pm N^0$ decay modes becomes dominant and the first displacement happens in around cm which remains almost unchanged  for further lower Yukawa couplings. $N^0$, thus produced will further give displaced decays depending on the Yukawa couplings, resulting total two recoils.  In \autoref{decay_pp} (b, e, h) we illustrate the total displaced decay lengths  including first and second recoils for i.e., $Y_N=1\times 10^{-8}$ for $M_N= \mathcal{O}(100)$ GeV.  The total decay lengths corresponding to the Yukawa coupling of  $5 \times 10^{-10}$ for the benchmark points are illustrated in (c, f, i), that give rise to decay lengths  in the MATHUSLA range. The dotted-dashed and dashed lines indicate the upper limits of CMS and ATLAS, respectively, while the light red band represents the MATHUSLA region ($68\,\rm{m}-168\,\rm{m}$) \cite{Alpigiani:2020iam}. It is apparent that, as the centre of mass energy goes from 14 TeV to 100 TeV,  the transverse boost effect being negligible fails to enhance the corresponding decay lengths $L_{\perp}$.  Only points with $Y_N=5\times 10^{-10}$ in (c, f, i) can reach MATHUSLA region and rest of other Yukawa couplings correspond to less than a meter range.

The situation however changes drastically as we see the longitudinal displaced decay lengths $L_{||}$, due to larger boost effects as discussed before.  However,  such effects are rather small for 14 TeV centre of mass energy compared to 100 TeV. As we approach higher centre of mass energies, the increase in longitudinal decay length $L_{||}$ becomes more pronounced. For example, if we consider the Yukawa coupling $Y_N=5\times 10^{-10}$ (c, f, i), the longitudinal decay length$L_{||}$ is 1.5 times that of the transverse one $L_{\perp}$ at 14 TeV, 2.5 times at 27 TeV and   7 times at 100 TeV centre of mass energies, respectively . It is worth mentioning that for a centre of mass energy of 100 TeV, we get events  $L_{||}\gsim1, \, 500$ m for  $Y_N =10^{-8}, \, 5 \times 10^{-10}$, respectively.

\begin{table}[hbt]	
	\begin{center}
		\hspace*{-1.8cm}
		\renewcommand{\arraystretch}{1.4}
		\begin{tabular}{|c|c|c|c|c|c|c|c|}
			\cline{2-8}
			\multicolumn{1}{c|}{} &\multirow{3}{*}{Modes} &\multicolumn{6}{c|}{Events inside MATHUSLA with $Y_N=5\times 10^{-10}$.}\\
			\cline{3-8}
			\multicolumn{1}{c|}{}&  &
			\multicolumn{2}{c|}{$E_{CM}=14$\,TeV}&\multicolumn{2}{c|}{$E_{CM}=27$\,TeV} &\multicolumn{2}{c|}{$E_{CM}=100$\,TeV}\\
			\cline{3-8}
			\multicolumn{1}{c|}{}& &\multicolumn{1}{c|}{BP1}&\multicolumn{1}{c|}{BP2}&\multicolumn{1}{c|}{BP1}&\multicolumn{1}{c|}{BP2}&\multicolumn{1}{c|}{BP1}&\multicolumn{1}{c|}{BP2}\\ 
			\hline
			\multirow{1.5}{*}{$60\,\rm{m}\leq L_{\perp} \leq 85\,\rm{m}$} & $  N^0\, N^\pm$ & 91.8\,(13.1) & 1.0\,(0.1) & 310.8\,(60.9) & 11.1\.(1.2)  & 1420.4\,(347.9) & 120.4\,(19.9) \\
			\cline{2-8}
			\multirow{2.0}{*}{($100\,\rm{m}\leq L_{\perp} \leq 125\,\rm{m}$)}& $  N^+\, N^-$ & 39.5\,(6.3) & 0.4\,(0.0) & 136.9\,(28.5) & 4.8\,(0.6) & 721.3\,(182.8) & 59.9\,(10.7) \\
			\cline{2-8}
			& Total & 131.3\,(19.4) & 1.4\,(0.1) & 447.7\,(89.4) & 15.9\,(1.8) & 2141.7\,(530.7) & 180.3\,(30.6)\\
			\hline \hline 	
			\multirow{1.5}{*}{$68\,\rm{m}\leq L_{||} \leq 168\,\rm{m}$} & $  N^0\, N^\pm$ & 151.5\,(66.2) & 1.7\,(0.4)  & 748.0\,(443.8) & 30.8\,(11.2) & 4603.3\,(4072.1) & 654.5\,(459.1) \\	
			\cline{2-8}
			\multirow{2.0}{*}{($100\,\rm{m}\leq L_{||} \leq 300\,\rm{m}$)}& $  N^+\, N^-$ & 64.0\,(28.0) & 0.7\,(0.2) & 328.6\,(197.7) & 13.8\,(5.5) & 2367.9\,(2051.6) & 335.3\,(228.2) \\
			\cline{2-8}
			& Total & 215.5\,(94.2) & 1.7\,(0.6) & 1076.6\,(641.5) & 44.6\,(16.7) & 6971.2\,(6123.7) & 989.8\,(687.3)\\
			\hline 
		\end{tabular}
		\caption{The observed number of events in MATHUSLA detector in transverse and longitudinal direction for all benchmark points, with $Y_N=5\times 10^{-10}$ at the centre of mass energies of 14, 27 and 100\,TeV  at an integrated luminosity of  3000\,fb$^{-1}$, 1000\,fb$^{-1}$ and 300\,fb$^{-1}$, respectively. In transverse direction ($L_{\perp}$) the range of MATHUSLA is taken as $60-85$\,m ($100-125$\,m), whereas in the longitudinal direction ($ L_{\parallel} $) it is considered as $68-168$\,m ($100-300$\,m)  \cite{Alpigiani:2020iam}.} \label{MathuslaTabpp}
	\end{center}	
\end{table}


In \autoref{MathuslaTabpp}, we present the number of events  corresponding  to transverse and longitudinal decay length inside MATHUSLA for the benchmark points conterminous with $Y_N =5\times 10^{-10}$ for the centre mass energies of 14, 27 and 100 TeV at an integrated luminosity of 3000\,fb$^{-1}$, 1000\,fb$^{-1}$ and 300\,fb$^{-1}$ respectively. The numbers in brackets refer to an earlier proposed detector dimension $25 \times 100  \times 200$ \cite{Curtin:2018mvb}, while the rest refer to a more recent one, $25 \times 100  \times 100$ \cite{Alpigiani:2020iam}. It is clear that the newly proposed detector dimension is better suited for probing the new physics via the Type-III seesaw model. The longitudinal mode $L_{||}$ clearly has the advantage in terms of event numbers, producing nearly one order of magnitude more events as compared to the transverse one i.e. $L_{\perp}$ .  We will see that the situation is quite opposite  in muon collider as the collision  happens in centre of mass energy.  In the next subsection, we summarise the event numbers for various finalstates as well as the invariant mass reconstructions of the Higgs boson and the Type III fermions.

\subsection{Results}\label{results_pp}

In this section, we focus on the decay channels $N^\pm \to h \ell^\pm$  and $N^0 \to h \nu$, where the Higgs boson further decays to $b\bar{b}$. We consider the possible finalstate with 2$b$-jets and more, which will  later reconstruct the Higgs boson via their invariant mass peak. We also demand one or more charged leptons in the finalstate. The most dominant decay modes of $N^{\pm(0)} \to W^\pm \nu(\ell^\mp)$ for $Y_N \sim 10^{-7}$,  only contributes to the finalstates with 2$b$-jets when the other one decays in $h\ell^{\pm}(\nu)$.  In \autoref{ppCol14}, we present the number of events for the finalstate topologies with 4$b$-jets and 2$b$-jets at the LHC, at the centre of mass energies of 14, 27 and 100 TeV with integrated luminosities of ($\mathcal{L}_{\text{int}}$=) 3000\,fb$^{-1}$, 1000\,fb$^{-1}$ and 300\,fb$^{-1}$, respectively, for all benchmark points and  $Y_N=5\times 10^{-7}$. As the cross-section of $N^\pm N^0$ is almost twice as the pair production of $N^\pm$, $N^0 N^\pm$ is the dominant contributor as compared with the pair production of $N^\pm$ for mono-leptonic finalstates. $N^+ N^-$  is the major contributor to the finalstates containing two charged leptons. We see that the integrated luminosity can be one order of magnitude lower as we go from 14 TeV to 100 TeV collisions even for larger number of events. 

\begin{table*}[hbt]	
	\begin{center}
		\renewcommand{\arraystretch}{1.2}
		\begin{tabular}{|c|c||c|c||c|c||c|c|}
			\cline{1-8}
			\multirow{2}{*}{Topologies} &\multirow{2}{*}{Modes}	& 
			\multicolumn{2}{c||}{$E_{CM}=14$\,TeV}&\multicolumn{2}{c||}{$E_{CM}=27$\,TeV} &\multicolumn{2}{c|}{$E_{CM}=100$\,TeV}\\
			\cline{3-8}
			&	&\multicolumn{1}{c|}{BP1}&\multicolumn{1}{c||}{BP2}&\multicolumn{1}{c|}{BP1}&\multicolumn{1}{c||}{BP2} &\multicolumn{1}{c|}{BP1}&\multicolumn{1}{c|}{BP2}\\ 
			\hline
			\multirow{3}{*}{$4b + 1\ell$ }&$ N^0 \, N^{\pm}$  & 11.0 & 0.3  & 21.7 & 1.4 & 45.4 & 5.1 \\	
			&$ N^+ \, N^{-}$  & 2.1 & 0.1 & 4.3 & 0.4 & 15.1 & 2.1 \\	
			\cline{2-8}
			&Total & 13.1  & 0.4 & 26.0 & 1.8 & 60.5 & 7.2  \\		
			\hline\hline
			\multirow{3}{*}{$4b+ \rm{OSD}$  }&	$  N^0\, N^\pm$	  & 0.3 & 0.0 & 0.6 & 0.0 & 1.3 & 0.1 \\	
			&	$  N^+\, N^-$	 	 & 4.0 & 0.1 & 7.6 & 0.5  & 17.2 & 2.0 \\	
			\cline{2-8}
			&Total  & 4.3 & 0.1 & 8.2 & 0.5 & 18.5 & 2.1  \\	
			\hline \hline
			\multirow{3}{*}{$2b + 4j + 1\ell $ }&	$  N^0\, N^\pm$	  & 66.8 & 4.0 & 150.4 & 17.6 & 382.3 & 73.3 \\		
			&	$  N^+\, N^-$	 & 34.1 & 1.9 & 76.1 & 8.4 & 211.6 & 37.1  \\		
			\cline{2-8}
			&Tota & 100.9 & 5.9 & 226.5 & 26.0 & 593.9 & 110.4 \\	
			\hline \hline
			\multirow{3}{*}{$2b + 4j + 2\ell $ }&	$  N^0\, N^\pm$	  & 26.3 & 1.2 & 52.4 & 4.9 & 101.4 & 16.7 \\	
			&	$  N^+\, N^-$  & 21.3 & 1.3 & 48.9 & 6.0 & 121.2 & 23.8 \\	
			\cline{2-8}
			&Total  & 47.6 & 2.5 & 101.3 & 10.9 & 222.6 & 40.5 \\	
			\hline \hline
			\multirow{3}{*}{$2b + 2j + 2\ell $ }&	$  N^0\, N^\pm$	   & 6.7 & 0.4 & 14.9 & 1.2 & 27.2 & 4.6 \\	
			&	$  N^+\, N^-$	  & 15.2 & 1.3 & 29.7 & 4.4 & 70.2 & 16.0\\	
			\cline{2-8}
			&Total & 21.9 & 1.7 & 44.6 & 5.6 & 97.4 & 20.6 \\	
			\hline \hline
			\multirow{3}{*}{$2b + 2j + 3\ell $ }&	$  N^0\, N^\pm$	 	 & 7.7 & 0.3 & 12.7 & 1.2 & 23.3 & 2.8 \\	
			&	$  N^+\, N^-$	 	 & 0.4 & 0.0 & 0.6 & 0.1 & 2.4 & 0.3 \\	
			\cline{2-8}
			&Total& 8.1 & 0.3 & 13.3 & 1.3 & 25.7 & 3.1  \\	
			\hline 
			
		\end{tabular}
		\caption{Number of events for finalstate topologies containing at least two $b$-jets for the benchmark points with the centre of mass energies of  14\,TeV, 27\,TeV and 100\,TeV at the LHC with integrated luminosities of ($\mathcal{L}_{\text{int}}$=) 3000\,fb$^{-1}$, 1000\,fb$^{-1}$ and 300\,fb$^{-1}$ respectively for $Y_N=5\times 10^{-7}$. }  \label{ppCol14}
	\end{center}	
\end{table*}



\begin{table*}[hbt]	
	\begin{center}
		\hspace*{-1cm}
		\renewcommand{\arraystretch}{1.2}
		\begin{tabular}{|c|c||c|c||c|c||c|c|}
			\cline{1-8}
			\multirow{2}{*}{Topologies} &\multirow{2}{*}{Modes}	& 
			\multicolumn{2}{c||}{$E_{CM}=14$\,TeV}&\multicolumn{2}{c||}{$E_{CM}=27$\,TeV} &\multicolumn{2}{c|}{$E_{CM}=100$\,TeV}\\
			\cline{3-8}
			&	&\multicolumn{1}{c|}{BP1}&\multicolumn{1}{c||}{BP2}&\multicolumn{1}{c|}{BP1}&\multicolumn{1}{c||}{BP2} &\multicolumn{1}{c|}{BP1}&\multicolumn{1}{c|}{BP2}\\ 
			\hline
			\multirow{3}{*}{$4b + 1\ell$ }&$ N^0 \, N^{\pm}$ & 7.1 & 0.2 & 12.7 & 0.9 & 27.6 & 3.7 \\	
			&$ N^+ \, N^{-}$  & 3.2 & 0.1  & 6.3 & 0.4 & 12.9 & 1.7 \\	
			\cline{2-8}
			&Total & 10.3 & 0.3  & 19.0 & 1.3 & 40.5 & 5.4 \\	
			\hline\hline
			\multirow{3}{*}{$4b+ \rm{OSD}$  }&	$  N^0\, N^\pm$	 	 & 0.1 & 0.0  & 0.2 & 0.1  & 1.6 & 0.2 \\
			&	$  N^+\, N^-$	 	 & 1.0 & 0.0 & 2.6  & 0.1  & 6.1 & 0.8  \\
			\cline{2-8}
			&Total& 1.1  & 0.0 & 2.8 & 0.1 & 7.7 & 1.0 \\
			\hline \hline
			\multirow{3}{*}{$2b + 4j + 1\ell $ }&	$  N^0\, N^\pm$	  & 65.5 & 3.4 & 140.0  & 15.0 & 311.7  & 58.7 \\	
			&	$  N^+\, N^-$	 	& 30.1 & 1.8 & 63.8 & 7.5 & 188.3 & 35.5 \\	
			\cline{2-8}
			&Total & 95.6 & 5.2 & 203.8 & 22.5 & 500.0 & 94.2 \\
			\hline \hline
			\multirow{3}{*}{$2b + 4j + 2\ell $ }&	$  N^0\, N^\pm$	 	 & 16.4 & 0.9 & 32.2 & 3.6 & 68.8 & 12.2 \\	
			&	$  N^+\, N^-$	  & 12.4 & 0.8 & 26.7 & 3.8 & 70.9 & 14.3 \\	
			\cline{2-8}
			&Total  & 28.8 & 1.7 & 58.9 & 7.4 & 139.7 & 26.5 \\
			\hline \hline
			\multirow{3}{*}{$2b + 2j + 2\ell $ }&	$  N^0\, N^\pm$	 	  & 11.7 & 0.5 & 20.0 & 1.5 & 37.0 & 4.9 \\	
			&	$  N^+\, N^-$	& 7.1 & 0.6 & 13.9 & 2.2 & 34.8 & 7.6 \\	
			\cline{2-8}
			&Total & 18.8 & 1.1 & 33.9 & 3.7 & 71.8 & 12.5 \\
			\hline \hline
			\multirow{3}{*}{$2b + 2j + 3\ell $ }&	$  N^0\, N^\pm$	 	 & 3.7 & 0.2 & 6.0  & 0.6 & 12.0 & 2.7 \\	
			&	$  N^+\, N^-$	 & 1.7 & 0.1 & 3.0 & 0.3 & 5.6 & 0.8 \\	
			\cline{2-8}
			&Total  & 5.4 & 0.3 & 9.0 & 0.9 & 17.6 & 3.5 \\
			\hline 
			
		\end{tabular}
		\caption{Number of events for finalstate topologies containing at least two $b$-jets for the benchmark points with the centre of mass energies of  14\,TeV, 27\,TeV and 100\,TeV at the LHC with integrated luminosities of ($\mathcal{L}_{\text{int}}$=) 3000\,fb$^{-1}$, 1000\,fb$^{-1}$ and 300\,fb$^{-1}$ respectively for $Y_N=1\times 10^{-8}$. }  \label{ppColYN2}
	\end{center}	
\end{table*}


The similar finalstate topologies are also investigated in \autoref{ppColYN2} for relatively lower Yukawa coupling i.e. $Y_N=1\times 10^{-8}$,  for the centre of mass energies of 14, 27 and 100 TeV at the integrated luminosities of ($\mathcal{L}_{\text{int}}$=) 3000\,fb$^{-1}$, 1000\,fb$^{-1}$ and 300\,fb$^{-1}$, respectively for the benchmark point.  Here,  $N^\pm $ dominantly decays to $ N^0 \pi^{\pm}$ with  $\mathcal{B}(N^{\pm}\to N^0 \pi^{\pm})\sim 47.3\%$ for $M_N = 1\,\rm{TeV}$, resulting a first recoil at around a few cm distance as discussed before. $N^0$ then follows the standard decay branching ratios to the SM modes. The presence of  a charged pion instead of a charged lepton reduces the number of charged lepton in the finalstate. Along with that, it contributes to the hadronic jets, and lepton number can further shrink due to jet-lepton isolation criteria. Thus, the multi-lepton finalstates are a rarity as compared to the case of  $Y_N=5\times 10^{-7}$. The main source of charged leptons in this case are from $N^0\to W^{\pm}\ell^{\mp}$ and the futher decays of the gauge bosons into leptons. This of course decreases the number of $b$-jets in the finalstate, which can be measured from the $4b+1\ell, \,4b+2\ell$ finalstates. This effect is even more pronounced for $Y_N=5\times 10^{-10}$, where  $\mathcal{B}(N^{\pm}\to N^0 \pi^{\pm})\sim 97\%$ for both the benchmark points. For example, in case of the most dominant finalstate $2b+4j+1\ell$, the event number reduced by $5\%, \, 16.5\%$  for $Y_N=1\times 10^{-8}, \, 5\times 10^{-10}$, respectively, compared to  $Y_N=5\times 10^{-7}$ for BP1 at 14 TeV centre of mass energy.

\begin{figure*}[h]
	\begin{center}
		\hspace*{-0.5cm}
		\mbox{\subfigure[]{\includegraphics[width=0.47\linewidth,angle=-0]{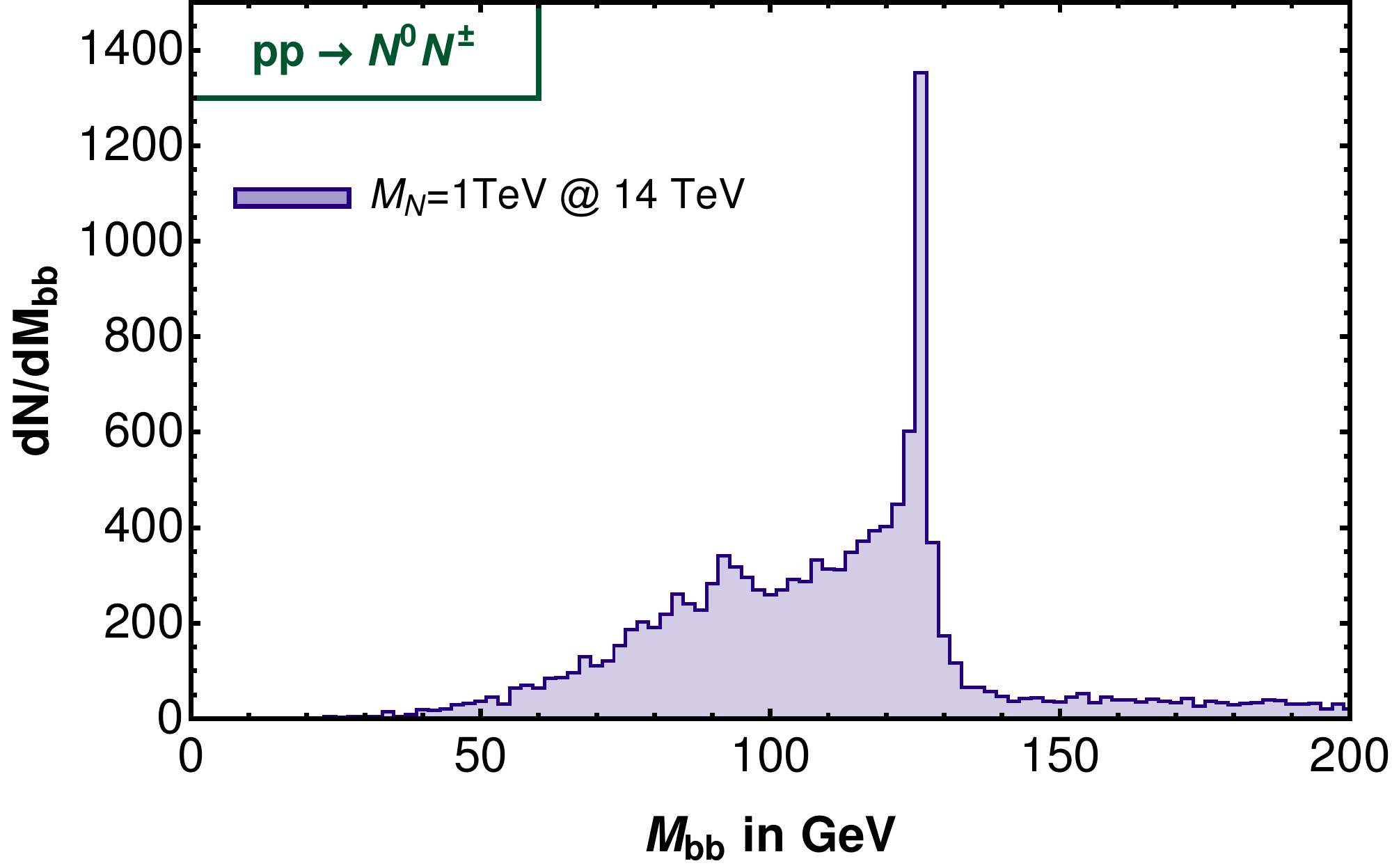}\label{}}\quad \quad
		\subfigure[]{\includegraphics[width=0.47\linewidth,angle=-0]{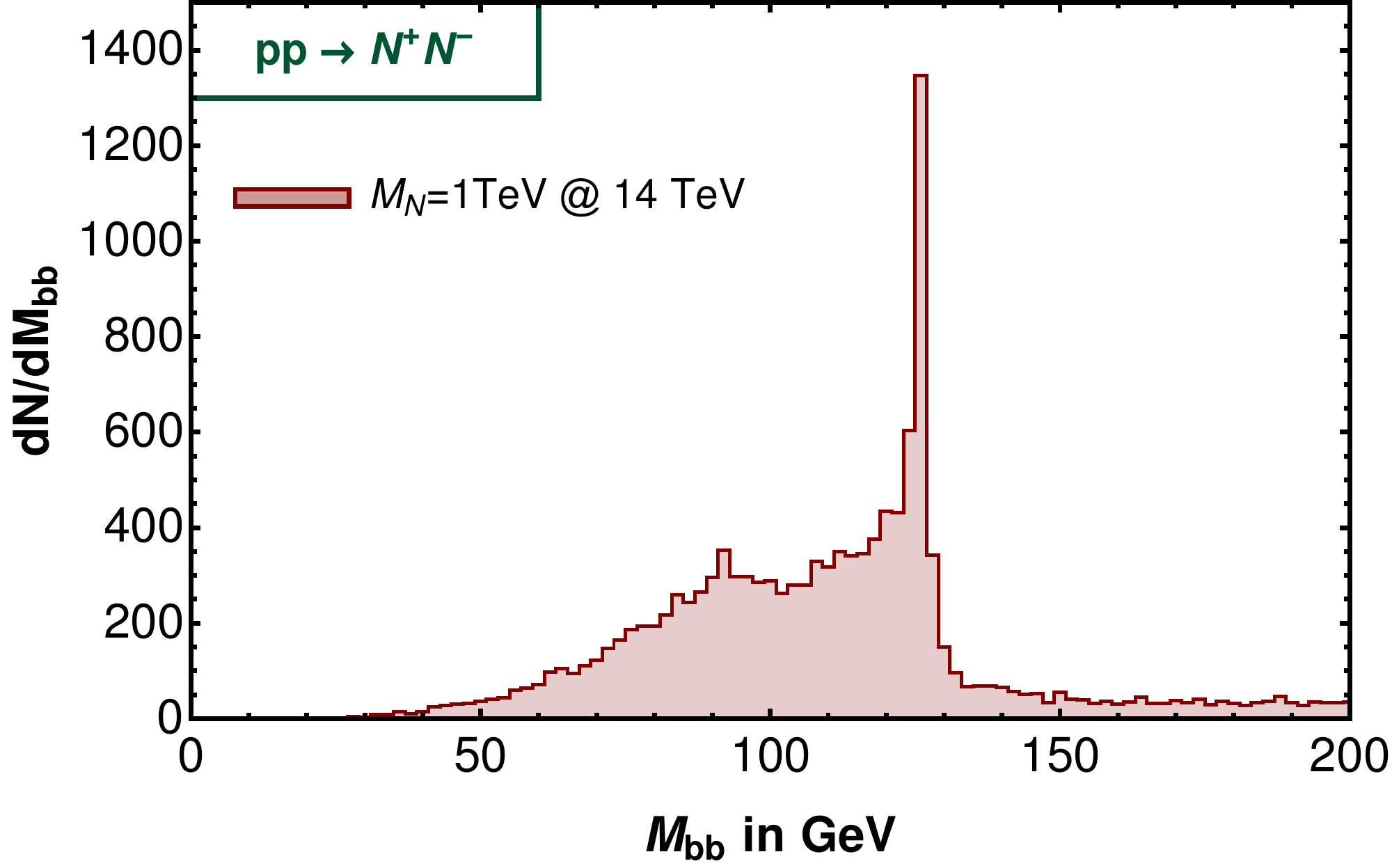}\label{}}}		
		\caption{Di-$b$-jet invariant mass ($M_{bb}$) distributions for the process $p\,p\to N^0\,N^{\pm}$ (a) and $p\,p\to N^+\,N^-$ (b) at the LHC with $M_N=1\,\rm{TeV}$ (BP1), $Y_N=5\times 10^{-7}$ and center of mass energy of 14\,TeV. The dominant peak around the SM Higgs mass (125.5\,GeV) along with a small peak around the $Z$ boson mass (91.1\,GeV) are visible.} \label{bbInvM_pp}
	\end{center}
\end{figure*}


\begin{table}[h]	
	\begin{center}
		\hspace*{-1cm}
		\renewcommand{\arraystretch}{1.2}
		\begin{tabular}{|c|c|c|c|c|c|c|c|}
			\cline{1-8}
			\multirow{2}{*}{Topologies} & \multirow{2}{*}{Modes}	& 
			\multicolumn{2}{c|}{$E_{CM}=14$\,TeV}&\multicolumn{2}{c|}{$E_{CM}=27$\,TeV} &\multicolumn{2}{c|}{$E_{CM}=100$\,TeV}\\
			\cline{3-8}
			&	&\multicolumn{1}{c|}{BP1}&\multicolumn{1}{c|}{BP2}&\multicolumn{1}{c|}{BP1}&\multicolumn{1}{c|}{BP2} &\multicolumn{1}{c|}{BP1}&\multicolumn{1}{c|}{BP2}\\ 
			\hline
			\multirow{1.5}{*}{$2b$ \&  }&$ N^0 \, N^{\pm}$  & 139.7 & 6.8  & 278.8  & 26.5  & 709.2 & 111.0 \\	
			$|M_{bb}-125.5|\leq 10$\,GeV &$ N^+ \, N^{-}$  & 58.2 & 3.0 & 125.8 & 12.5 & 365.0  & 54.3 \\
			\cline{2-8}
			&Total  & 197.9 & 9.8  & 404.6 & 39.0 & 1074.2 & 165.3 \\
			\hline\hline
			\multirow{1.5}{*}{$4b$ \& }&$ N^0 \, N^{\pm}$  & 2.6 & 0.1  & 5.8  &  0.2  & 30.1 & 1.2 \\
			$|M_{bb}-125.5|\leq 10$\,GeV &$ N^+ \, N^{-}$  & 1.4 &  0.0  & 2.2 & 0.2  & 4.3 & 0.5 \\
			\cline{2-8}
			&Total  & 4.0 & 0.1 & 8.0 & 0.4 & 34.4 & 1.7 \\	
			\hline
		\end{tabular}
		\caption{The number of events for single Higgs boson and di-Higgs boson mass peaks after the window cuts around the peak of the invariant mass  distributions at the LHC/FCC  at the centre of mass energies of  14\,TeV, 27\,TeV and 100\,TeV at the integrated luminosity ($\mathcal{L}_{\text{int}}$=) 3000\,fb$^{-1}$, 1000\,fb$^{-1}$ and 300\,fb$^{-1}$, respectively for the benchmark points with  $Y_N=5\times 10^{-7}$.}   \label{HiggsTabpp}
	\end{center}	
\end{table}


Thereafter we attempt to reconstruct the Higgs bosons  from  the $bb$ invariant mass distribution  coming from both $N^\pm \to h \ell^{\pm}$ and $N^0 \to h \nu$, as shown in \autoref{bbInvM_pp}. We can observe the sharp peaks of  Higgs boson around 125 GeV for both cases and a smaller peak around the $Z$ boson mass, which comes from  $N^\pm \to Z \ell^{\pm}$ and $N^0 \to Z \nu$ decays. The events around the Higgs mass certainly guarantees the finalstate with at least one Higgs boson and two Higgs bosons, which are described in \autoref{HiggsTabpp}.  A demand of 	$|M_{bb}-125.5|\leq 10$  GeV  is made for single Higgs boson reconstruction. The numbers for the finalstates of 2$b$- and 4$b$-jets  with the mass windows are shown in \autoref{HiggsTabpp} at the centre of mass energies of  14\,TeV, 27\,TeV and 100\,TeV at the integrated luminosity ($\mathcal{L}_{\text{int}}$=) 3000\,fb$^{-1}$, 1000\,fb$^{-1}$ and 300\,fb$^{-1}$  for  benchmark points with $Y_N=5\times 10^{-7}$. We see that though the number of events are quite healthy for single Higgs boson reconstruction, the same  cannot be said for two Higgs boson reconstructions coming from the two legs of  Type-III fermions.

\begin{figure}[h]
	\begin{center}
		\hspace*{-0.5cm}
		\mbox{\subfigure[]{\includegraphics[width=0.47\linewidth,angle=-0]{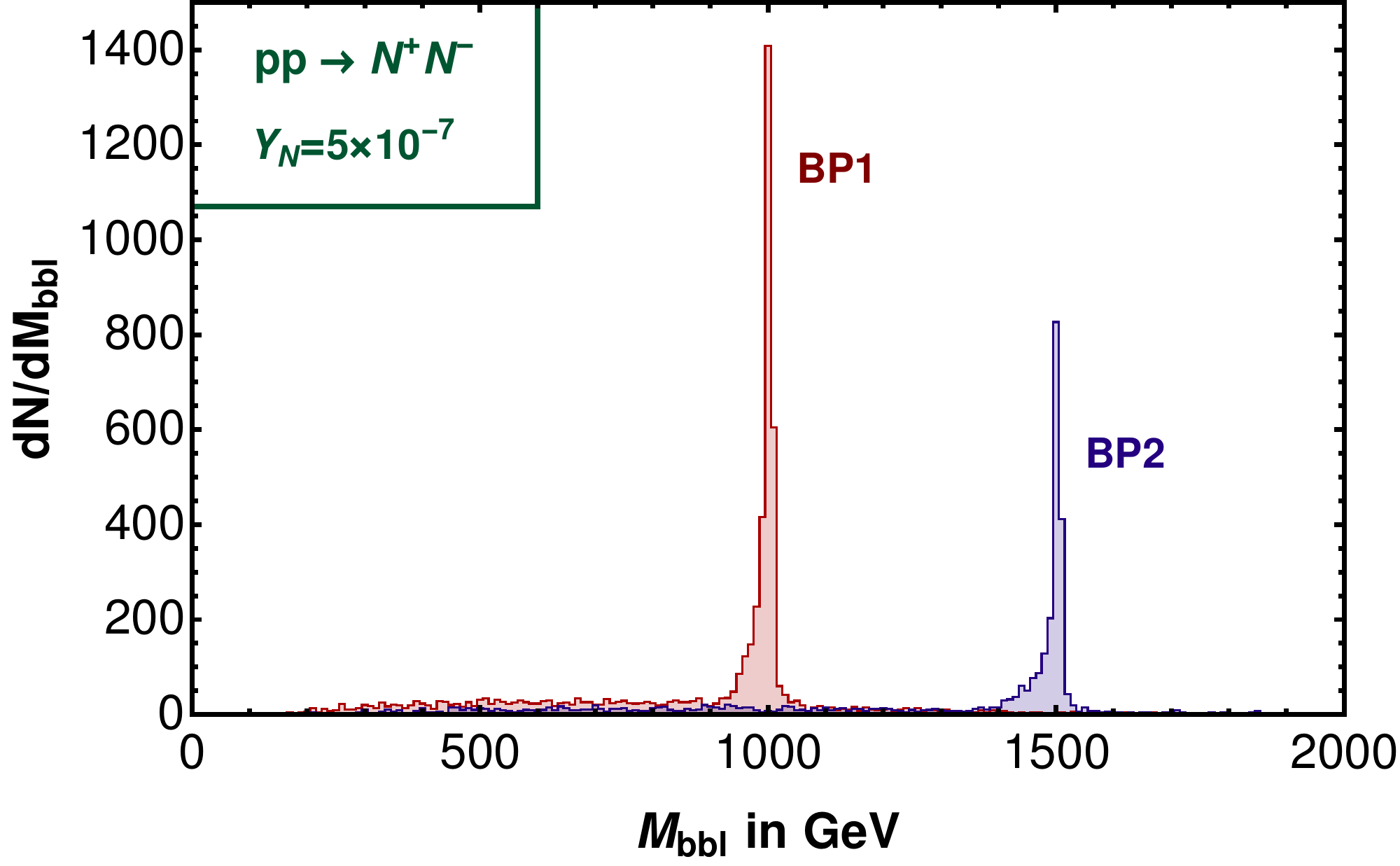}\label{}}\quad \quad
		\subfigure[]{\includegraphics[width=0.47\linewidth,angle=-0]{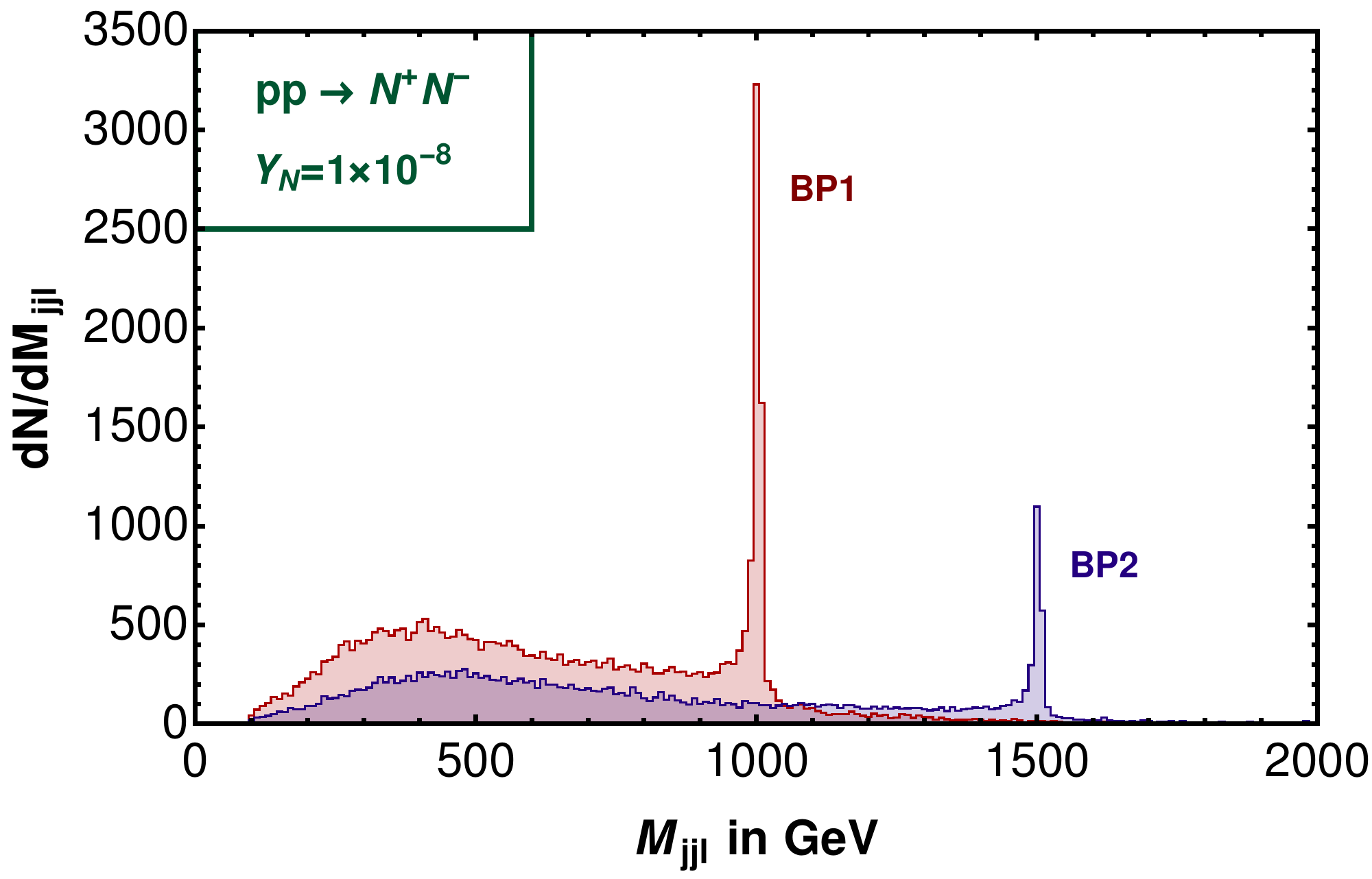}\label{}}}		
		\caption{(a) Di-bjej-mono-lepton invariant mass ($M_{bb\ell}$) distribution for the process $p\,p\to N^+\,N^-$  with $Y_N=5\times 10^{-7}$ (b) di-jet-mono-lepton ($M_{jj\ell}$) invariant mass distributions for the process $p\,p\to N^+\,N^-$ with $Y_N=1\times 10^{-8}$ 
			for the benchmark points at the LHC with centre of mass energy 14\,TeV. } \label{bblInvM_pp}
	\end{center}
\end{figure}


\begin{table*}[h]	
	\begin{center}
		\hspace*{-1cm}
		\renewcommand{\arraystretch}{1.2}
		\begin{tabular}{|c|c|c|c|c|}
			\hline
			Benchmark&\multirow{2}{*}{Topologies} & \multicolumn{3}{c|}{Centre of mass energy}\\
			\cline{3-5}
			Points & & 
			14\,TeV  & 27\,TeV  & 100\,TeV \\
			\hline
			\multirow{2}{*}{BP1} & $2b+ 1\ell$ \& & \multirow{2}{*}{38.0} & \multirow{2}{*}{79.5} & \multirow{2}{*}{194.6} \\	
			& $|M_{bb\ell}-1000.0|\leq 10$\,GeV & & & \\
			\hline\hline
			\multirow{2}{*}{BP2} & $2b+ 1\ell$ \& & \multirow{2}{*}{1.8} & \multirow{2}{*}{7.4} & \multirow{2}{*}{28.1} \\
			& $|M_{bb\ell}-1500.0|\leq 10$\,GeV & & & \\
			\hline
		\end{tabular}
		\caption{The number of events in $M_{bb\ell}$ distributions for $N^+ N^-$after the window cuts around the mass peak at the LHC/FCC for the benchmark points at the centre of mass energies of 14\,TeV, 27\,TeV and 100\,TeV at the integrated luminosities of  ($\mathcal{L}_{\text{int}}$=) 3000\,fb$^{-1}$, 1000\,fb$^{-1}$ and 300\,fb$^{-1}$ respectively with $Y_N=5\times 10^{-7}$. }  \label{NeutTabpp}
	\end{center}	
\end{table*}


\begin{table*}[h]	
	\begin{center}
		\hspace*{-1cm}
		\renewcommand{\arraystretch}{1.2}
		\begin{tabular}{|c|c|c|c|c|}
			\hline
			Benchmark&\multirow{2}{*}{Topologies} & \multicolumn{3}{c|}{Centre of mass energy}\\
			\cline{3-5}
			Points & & 
			14\,TeV  & 27\,TeV  & 100\,TeV \\
			\hline
			\multirow{2}{*}{BP1} & $2j+ 1\ell$ \& & \multirow{2}{*}{90.9} & \multirow{2}{*}{192.8} & \multirow{2}{*}{507.6} \\	
			& $|M_{jj\ell}-1000.0|\leq 10$\,GeV & & & \\
			\hline\hline
			\multirow{2}{*}{BP2} & $2j+ 1\ell$ \& & \multirow{2}{*}{2.5} & \multirow{2}{*}{11.0} & \multirow{2}{*}{45.6} \\
			& $|M_{jj\ell}-1500.0|\leq 10$\,GeV & & & \\
			\hline
		\end{tabular}
		\caption{The number of events in $M_{jj\ell}$ distributions after the window cuts around the mass peak at the LHC/FCC for the benchmark points at the centre of mass energies of 14\,TeV, 27\,TeV and 100\,TeV at the integrated luminosities of  ($\mathcal{L}_{\text{int}}$=) 3000\,fb$^{-1}$, 1000\,fb$^{-1}$ and 300\,fb$^{-1}$ respectively with $Y_N=1\times 10^{-8}$.}  \label{Neut0Tabpp}
	\end{center}	
\end{table*}


Finally, we plot the invariant mass distributions of $bb\ell$  in \autoref{bblInvM_pp}(a) in order to reconstruct the mass of Type-III fermion $N^\pm$. Here we ensure $|M_{bb}- 125.5|\leq 10$ GeV while reconstructing  $M_{bb\ell}$.  The two peaks visible at 1000 and 1500 GeV are reconstructed for BP1 and BP2, respectively for $Y_N=5\times 10^{-7}$, where, $\mathcal{B}(N^{\pm}\to h\ell^{\pm})\sim 25\%$. Similar to the previous case, here also we collect the events with a mass window of  $|M_{bb\ell}-M_N|\leq 10$ GeV, to predict the number of events in \autoref{NeutTabpp}. The table describes event numbers for the benchmark points for the  centre of mass energies of 14\,TeV, 27\,TeV and 100\,TeV at the integrated luminosities of ($\mathcal{L}_{\text{int}}$=) 3000\,fb$^{-1}$, 1000\,fb$^{-1}$ and 300\,fb$^{-1}$ respectively with $Y_N=5\times 10^{-7}$. As we move from BP1 to BP2, keeping the centre of mass energy same we see event number drops as the cross-section decreases simultaneously. Certainly, the analysis predicts LHC/FCC with higher energy and luminosity will have much higher reach,  which we discuss in \autoref{reach}.

In  \autoref{bblInvM_pp}(b) we present the invariant mass of di-jet-mono-lepton coming from $N^0$, which is produced from $ N^\pm \to \pi^\pm N^0$  decay  for lower Yukawa $Y_N= 10^{-8}$ for the benchmark points at the centre of mass energy of 14 TeV. The $N^0$ produced from such decays give rise to $W^\pm \ell^\mp$ dominantly  and we reconstruct such $W^\pm$ bosons from the di-jet invariant mass peak and select the events within $|M_{jj} -M_{W}| \leq 10$ GeV  for further reconstruction of  $M_{jj\ell}$. The corresponding events numbers are presented in \autoref{Neut0Tabpp} for the centre of mass energies of  14\, TeV, 27\,TeV and 100\,TeV at the integrated luminosities of  ($\mathcal{L}_{\text{int}}$=) 3000\,fb$^{-1}$, 1000\,fb$^{-1}$ and 300\,fb$^{-1}$ respectively.  Due to large boost effect the two-jets coming from $W^\pm$ decay combine as a single Fat-jet and we loose some events while reconstructing di-jet-mono-lepton invariant mass distribution.  The more we go for higher Type-III fermion mass the more prominent is such effect as the cross-sections still governs by on-shell productions.

\begin{figure}[hbt]
	\begin{center}
		\includegraphics[width=0.5\linewidth,angle=-0]{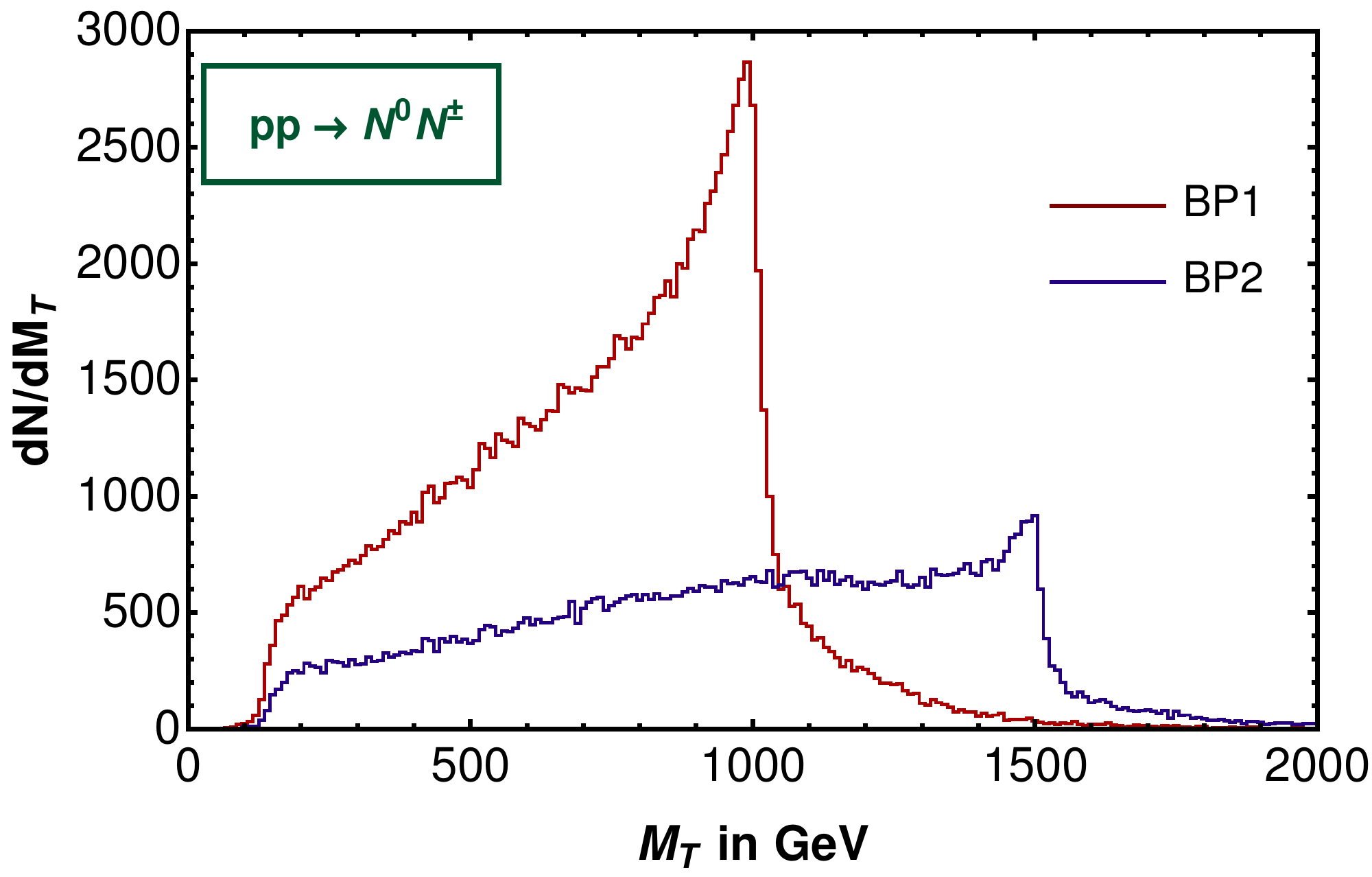}
		\caption{The transverse invariant mass ($M_T$) distribution for the process $p\,p\to N^0\,N^{\pm}$, for the benchmark points at the LHC with $Y_N=5\times 10^{-7}$ and center of mass energy 14\,TeV.  } \label{TransM_pp}
	\end{center}
\end{figure}


Next, we indulge in measuring the transverse invariant mass distribution, where we have only one neutrino in the finalstate. While we produce $N^\pm N^0$ at the LHC, $N^\pm \to Z \ell$ and $N^0\to h \nu$ decays give rise to finalstate with one neutrino, keeping other leg completely visible.  The transverse  invariant  mass \cite{Tovey:2010de} of $N^0$  in such finalstate is constructed via the transverse energy and momenta of Higgs boson reconstituted from the di-$b$-jet and missing neutrino as
\begin{equation}
M^2_T=m^2_h + 2\left( E^{h}_T \etmiss -\vec{p^{h}_T}.\vec{\ptmiss}\right),
\end{equation}
where $E^{h}_T ,\, p^{h}_T$ are the transverse energy and momenta, respectively of the Higgs boson reconstructed via di-$b$-jets.
In  \autoref{TransM_pp}, we depict the transverse invariant mass of  $bb\nu$ coming from $N^0$, for $Y_N=5\times 10^{-7}$, where the other leg is tagged via complete visible mode such  that, there is only one neutrino in the finalstate. In this case, missing $p_T$ is equivalent to neutrino $p_T$.  From \autoref{TransM_pp} we see the edge of the distributions depict the $N^0$ masses for the two benchmark points at the centre of mass energy of 14 TeV.

\begin{table}[hbt]	
	\begin{center}
		\hspace*{-1cm}
		\renewcommand{\arraystretch}{1.15}
		\begin{tabular}{|c|c|c|c|c|}
			\hline
			Benchmark&\multirow{2}{*}{Topologies} & \multicolumn{3}{c|}{Centre of mass energy}\\
			\cline{3-5}
			Points & & 
			14\,TeV  & 27\,TeV  & 100\,TeV \\
			\hline
			\multirow{2}{*}{BP1} & $2b$ \& & \multirow{2}{*}{1.4} & \multirow{2}{*}{2.5} & \multirow{2}{*}{5.3} \\	
			& $|M_{T}-1000.0|\leq 10$\,GeV & & & \\
			\hline\hline
			\multirow{2}{*}{BP2} & $2b$ \& & \multirow{2}{*}{0.0} & \multirow{2}{*}{0.1} & \multirow{2}{*}{0.4} \\
			& $|M_{T}-1500.0|\leq 10$\,GeV & & & \\
			\hline
		\end{tabular}
		\caption{The number of events in $M_{T}$ distributions after the window cuts around the mass peak at the LHC/FCC for the benchmark points at the centre of mass energies of 14\,TeV, 27\,TeV and 100\,TeV at the integrated luminosities of  ($\mathcal{L}_{\text{int}}$=) 3000\,fb$^{-1}$, 1000\,fb$^{-1}$ and 300\,fb$^{-1}$ respectively with $Y_N=5\times 10^{-7}$.}  \label{TransTabpp}
	\end{center}	
\end{table}


The number of events correlated with the transverse invariant mass distribution, i.e. $|M_{T}-M_N|\leq 10$ GeV is presented in \autoref{TransTabpp} for the benchmark points with the centre of mass energies of 14, 27 and 100\,TeV at the integrated luminosities of  ($\mathcal{L}_{\text{int}}$=) 3000\,fb$^{-1}$, 1000\,fb$^{-1}$ and 300\,fb$^{-1}$ respectively with $Y_N=5\times 10^{-7}$. We ensure that the di-$b$jet invariant mass is peaking at the Higgs mass by selecting the events under $|M_{bb}-125.5|\leq 10$\,GeV.

\section{Simulation at the muon collider}\label{muonc}

Muon collider  is proposed for the precision measurements  as well as sensitive to processes involving  lepton flavours. There is a recent buzz among the physicist where different BSM scenarios are explored \cite{AlAli:2021let,Costantini:2020stv,Buttazzo:2018qqp,Huang:2021nkl,Huang:2021biu,Asadi:2021gah,Capdevilla:2020qel,Long:2020wfp,Han:2020pif,Han:2020uak,Capdevilla:2021rwo,Han:2021udl,AK:LepQ}. The proposed optimistic reach  of muon collider is around 90 ab$^{-1}$  with  centre of mass energy of 30 TeV \cite{AlAli:2021let}. Lack of initial state QCD radiation and known centre of mass energy for the collision makes it a superior machine over LHC. In this section, we investigate the displaced Higgs production by observing charged tracks coming from the decays of Type-III fermions. Unlike at LHC, in a muon collider, the centre of mass energy is equal to the parton level collision energy; thus the kinematical distributions will differ from that of LHC.  Below we describe the kinematical distributions before presenting the results. We follow the same isolation criteria and minimum $p_T$ cut for the jets and the leptons as described in \autoref{setup}. 

\begin{figure*}[h]
	\begin{center}
		\mbox{\subfigure[]{\includegraphics[width=0.4\linewidth,angle=-0]{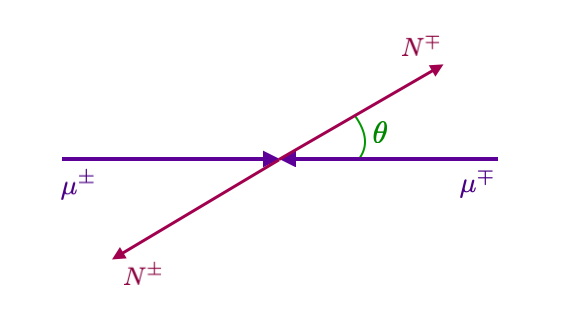}\label{}}\quad
			\subfigure[]{\includegraphics[width=0.4\linewidth,angle=-0]{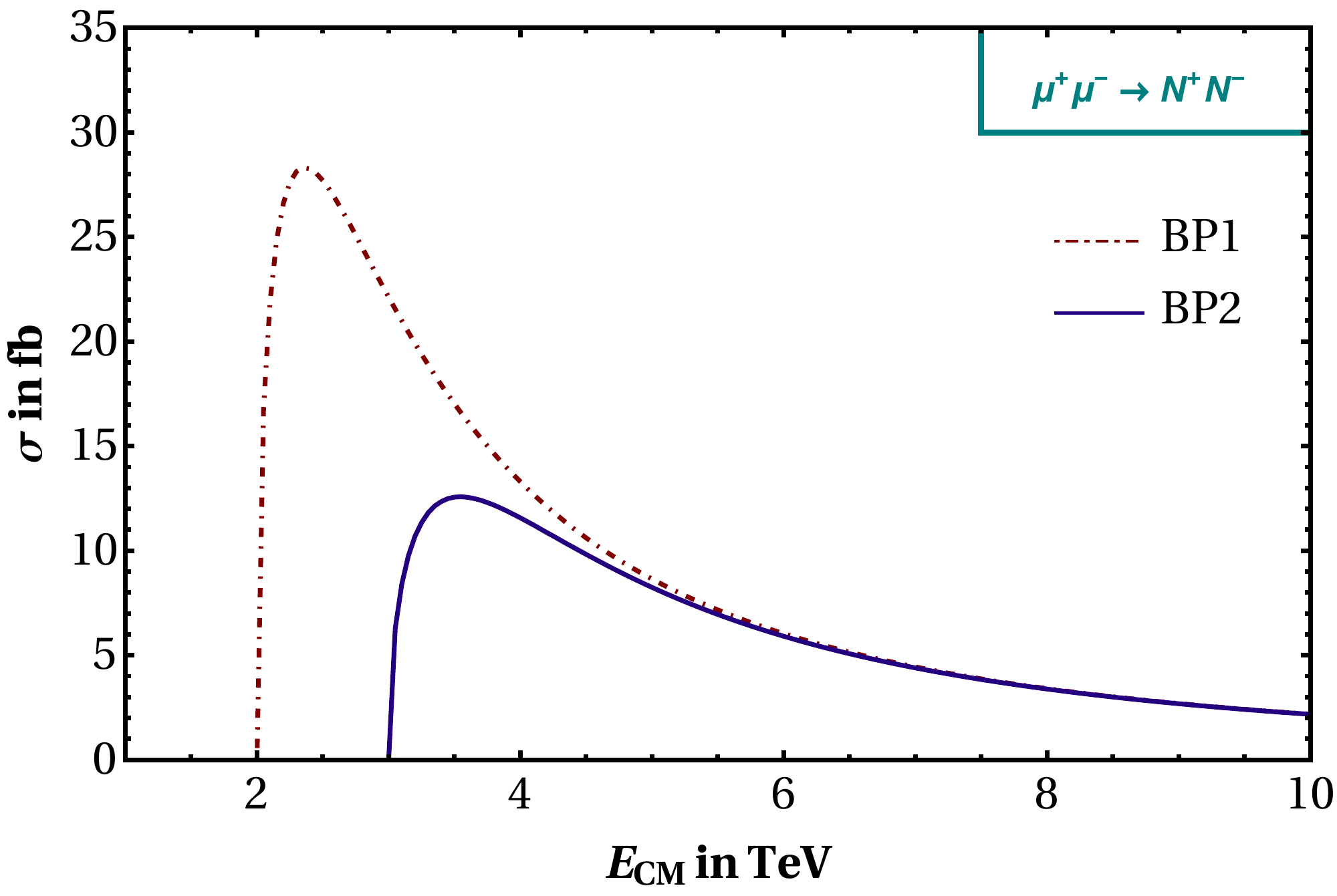}\label{}}}		
		\caption{(a) The Feynman diagram and (b) the cross-sections (in fb) as a function of centre of mass energy ($E_{\rm CM}$) for the benchmark points at muon collider for the process $\mu^+\mu^- \to N^+N^-$. Here $\theta$ is the polar angle with the beam axis.}\label{mucrosecfig}
	\end{center}
\end{figure*}


\begin{table*}[h]
	\begin{center}
		\renewcommand{\arraystretch}{1.2}
		\begin{tabular}{|c||c|c|c|}\hline 
			{\multirow{2}{*}{\diagbox[width=4.5cm]{ Benchmark points}{$E_{CM}$}}}&
			\multicolumn{3}{c|}{$\sigma_{\mu^+\mu^- \to N^+ N^-}$ (in fb)}\\
			\cline{2-4}
			&\multicolumn{1}{c|}{3.5\,TeV}&\multicolumn{1}{c|}{14\,TeV}&\multicolumn{1}{c|}{30\,TeV}\\ 
			\cline{2-4}
			\hline	
			BP1 &  17.02 & 1.11 & 0.24 \\ \hline
			BP2 &  12.56 & 1.11 & 0.24 \\ \hline 
		\end{tabular}
		\caption{ The production cross-sections (in fb) of $ N^+ N^-$ for the benchmark points at muon collider for the centre of mass energies of 3.5\,TeV, 14\,TeV and 30\,TeV.}\label{mucrosec}
	\end{center}
\end{table*}


At the muon collider, $N^\pm$ can be pair produced via $Z$ boson and photon while production of $N^\pm N^0$ is not possible, unlike at the LHC/FCC, but pair production of $N^0$ is possible via $N^{\pm}$ decay for relatively lower Yukawa couplings i.e. $Y_N \leq 10^{-8}$. \autoref{mucrosecfig}(a)  presents the Feynman diagram of production of $N^+ N^-$ and $\theta$ represents the polar angle with the beam axis.  \autoref{mucrosecfig}(b)  present the cross-section for $\mu^+ \mu^- \to N^+ N^-$ with respect to centre of mass energy for  the two benchmark points. It can be seen that  the off-shell  cross-sections drop very quickly, which implies  very high energy is not efficient to probe these heavy neutrinos, unlike at the LHC. The cross-section can always find the on-shell resonant mode due to the parton distribution function at the LHC.

The cross-sections for the benchmark points at three different energies of muon colliders are presented in \autoref{mucrosec}. We see that the  cross-sections drop  by an order for each as we increase the centre of mass energy. Certainly, we need to achieve 30 ab$^{-1}$ of integrated luminosity for the centre of mass energy of 30 TeV. In the following subsections, we study the kinematics of  the  hard scattering, i.e. of $N^\pm$. As opposed to LHC, here the system is in the centre of momentum frame keeping the three momenta constant along  with the energy, and the only variable is the polar angle with  the beam axis, i.e. $\theta$.

\subsection{Kinematical distributions}\label{kinm_mu}

\begin{figure}[hbt]
	\begin{center}
		\hspace*{-0.5cm}
		\mbox{\subfigure[]{\includegraphics[width=0.38\linewidth,angle=-0]{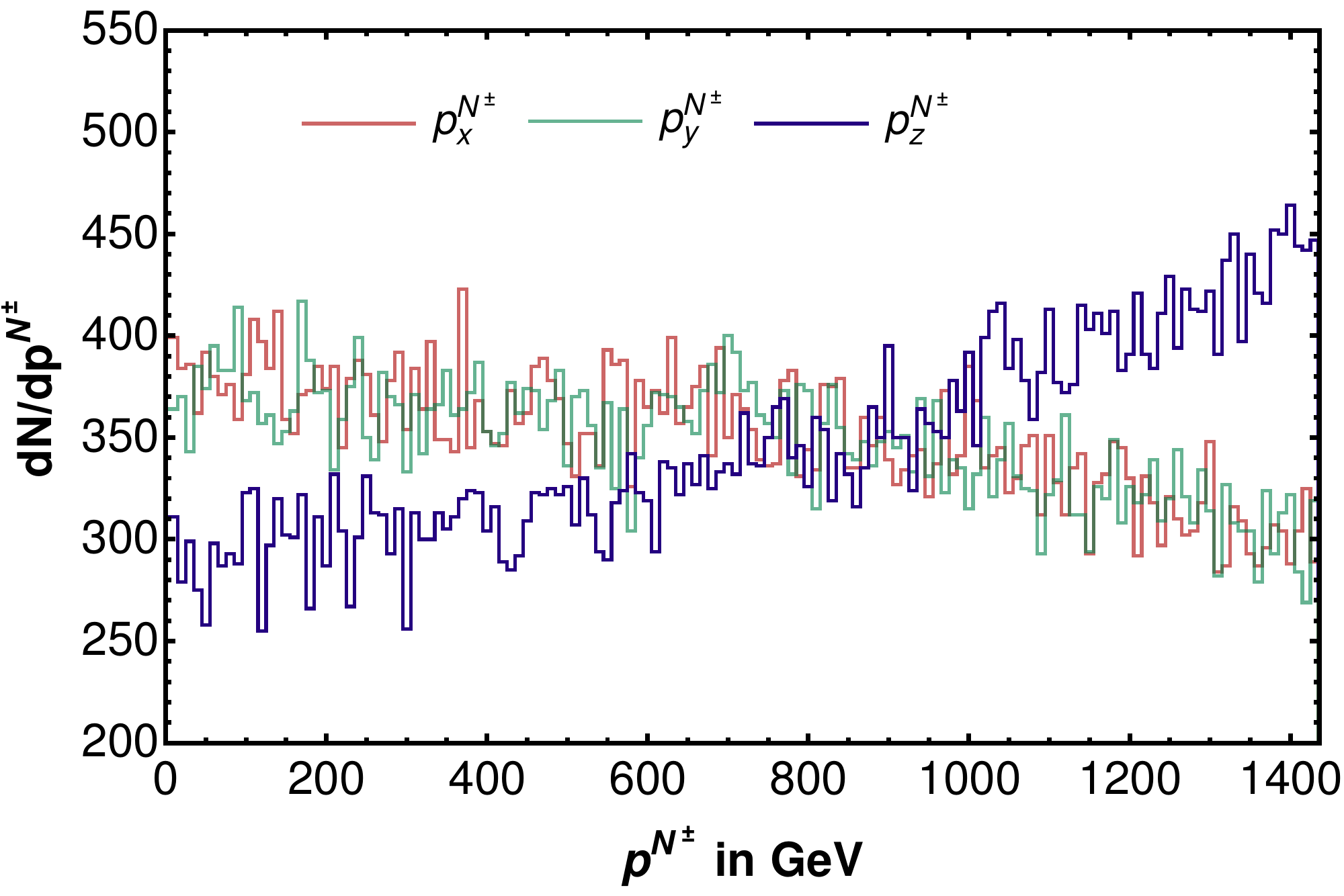}\label{}}\quad \quad
		\subfigure[]{\includegraphics[width=0.39\linewidth,angle=-0]{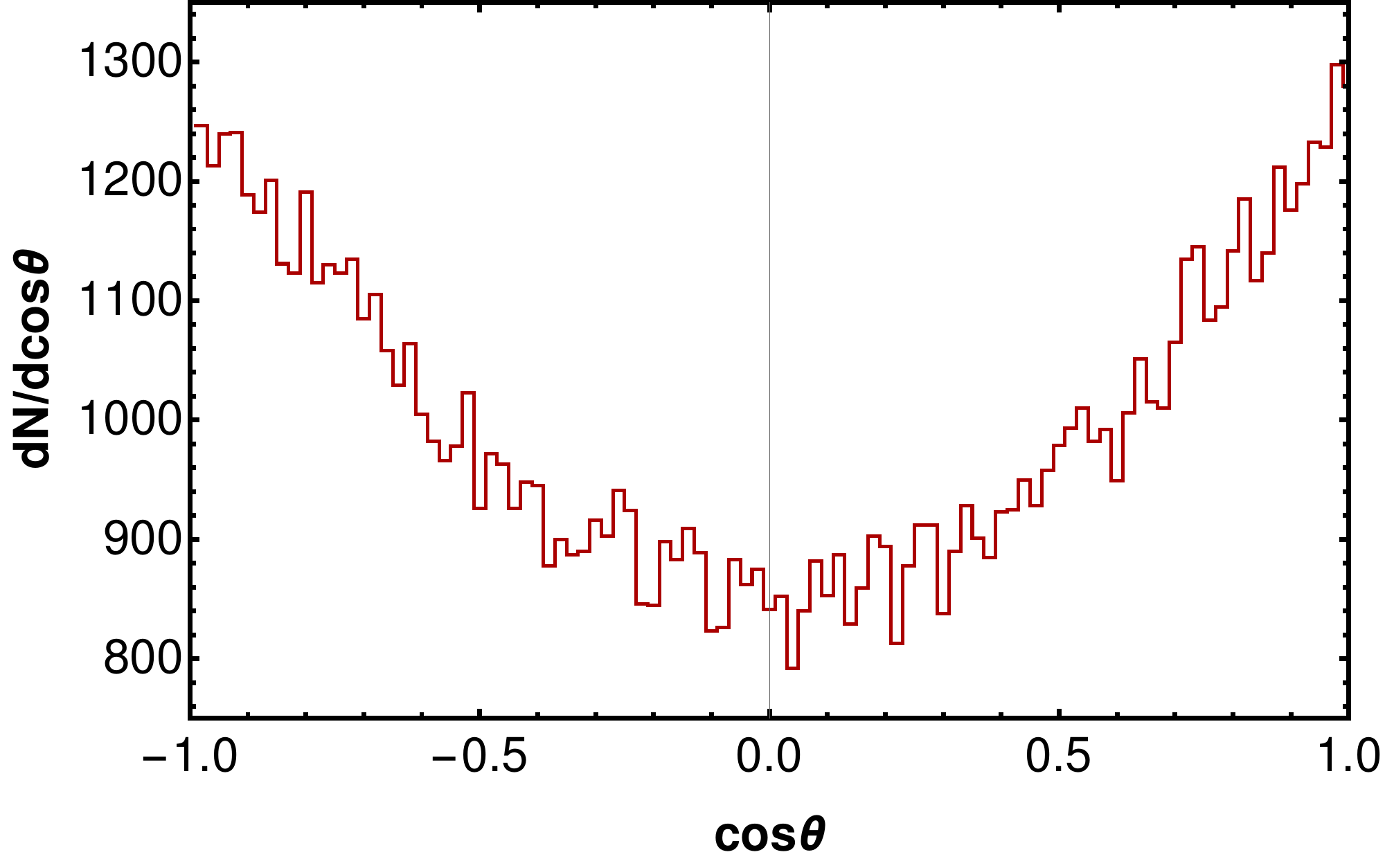}\label{}}}
		\hspace*{-0.5cm}
		\mbox{\subfigure[]{\includegraphics[width=0.38\linewidth,angle=-0]{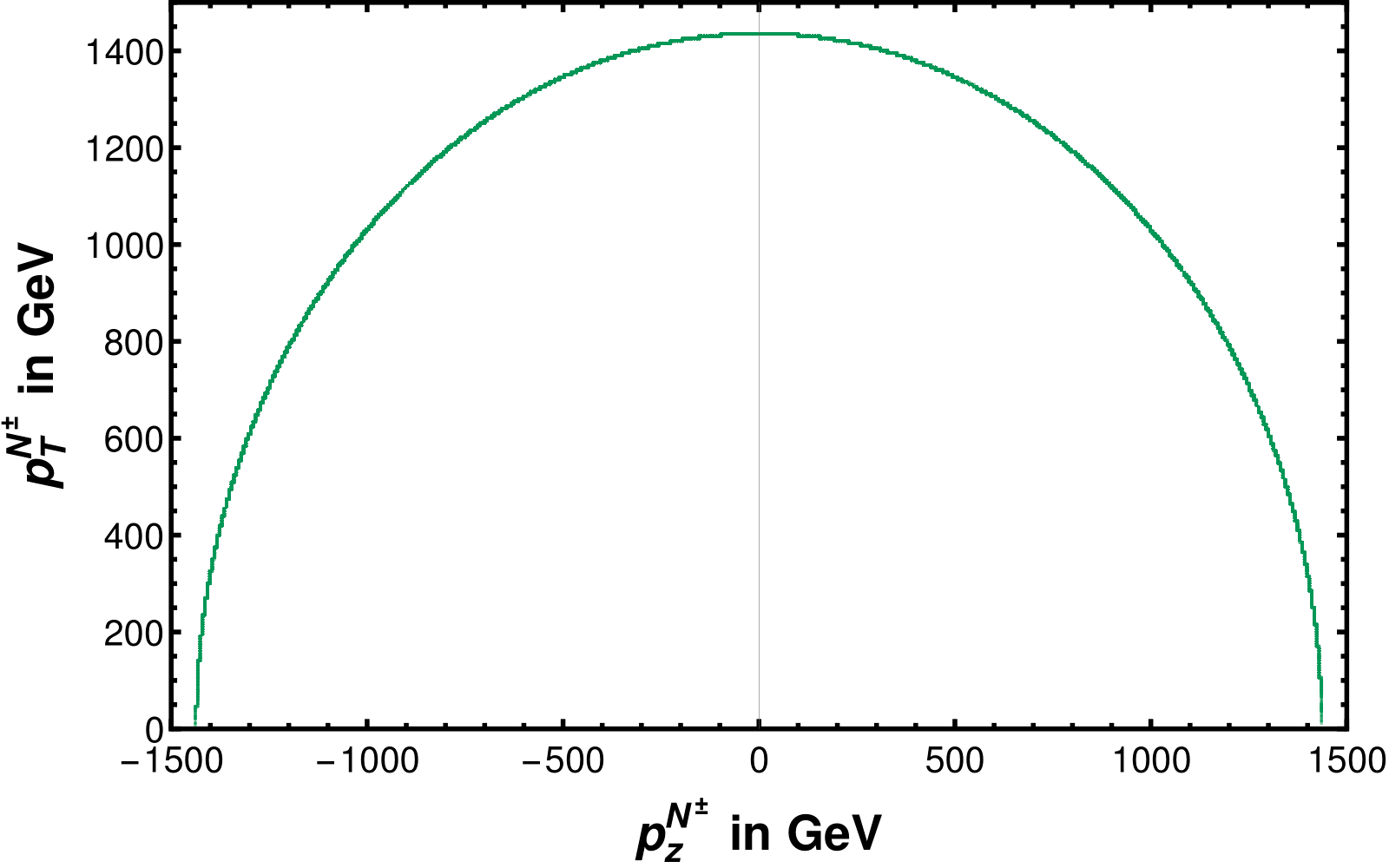}\label{}}\quad \quad
		\subfigure[]{\includegraphics[width=0.375\linewidth,angle=-0]{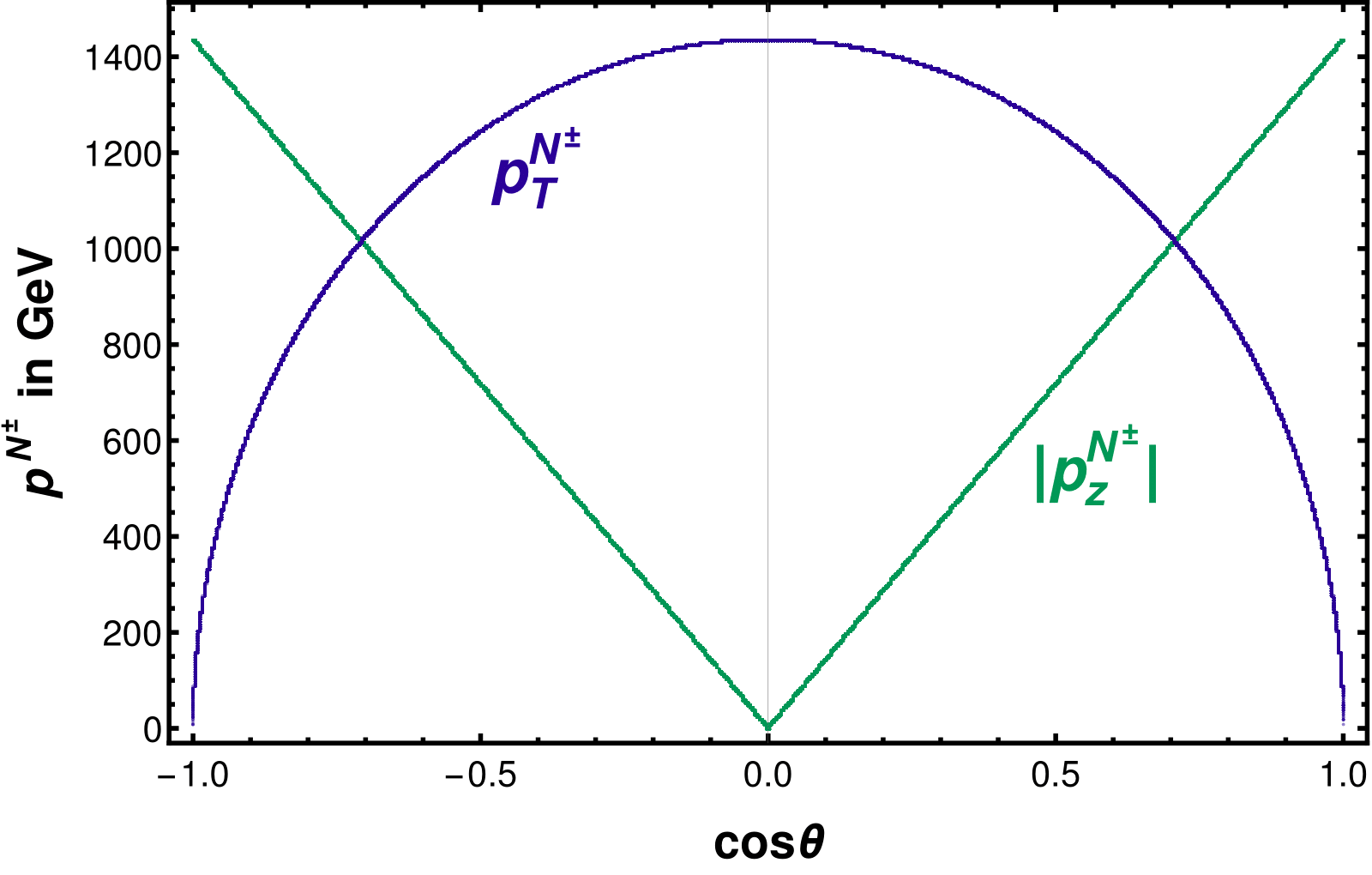}\label{}}}
		\caption{Various correlations of longitudinal ($p^{N^\pm}_z$) and transverse ($p^{N^\pm}_T$) momenta of $N^{\pm}$ at muon collider with centre of mass energy of  3.5 TeV for BP1 for the process of $\mu^+ \mu ^- \to N^+ N^-$. (a) Describes $p^{N^\pm}_x, p^{N^\pm}_y, p^{N^\pm}_z$, (b) presents the angular distribution of  $N^{\pm}$. (c) describes correlation between  $p^{N^\pm}_z$ and $p^{N^\pm}_T$  and (d) presents the correlation of $|p^{N^\pm}_z|$ and $p^{N^\pm}_T$  with  respect to  $\cos{\theta}$.}\label{corelation}
	\end{center}
\end{figure}

We begin by simulating the hard process, i.e. the kinematics of  the $N^\pm$, which is summarised in \autoref{corelation}. The components of  the three momentum of $N^\pm$, i.e. $p^{N^\pm}_x,\, p^{N^\pm}_y,\, p^{N^\pm}_z$ are shown in \autoref{corelation}(a), where it is evident that the events are populated at larger $p^{N^\pm}_z$ as compared to  $p^{N^\pm}_x,\, p^{N^\pm}_y$. This doesn't come as surprise since the angular distribution of $\mu^+ \mu ^- \to N^+ N^-$ in the centre of mass frame as given in \autoref{corelation} (b), i.e. $\frac{d\sigma}{d\cos{\theta}} \sim (1 + \cos^2{\theta})$ \cite{Bandyopadhyay:2020wfv}, thus the probability is more for $\cos{\theta}= \pm 1$, where  $p^{N^\pm}_z$ is large. This can also be interpreted as follows:  the longitudinal momentum $p^{N^\pm}_z=p^{N^\pm} \cos{\theta}$ and the transverse momentum $p^{N^\pm}_T=p^{N^\pm}\sin{\theta}$ are anti-correlated, which is apparent from  \autoref{corelation} (c). $p^{N^\pm}_z$ and $p^{N^\pm}_T$ peak at different angles , i.e. $\theta=0, \pi$ for the former and $\theta=\pi/2,\, 3\pi/2$ for the latter. As the total momentum is conserved and finite, the distribution of  $p^{N^\pm}_T$ forms a circular pattern with radius $p$, and $p^{N^\pm}_z$  project the gradient as $|\vec{p}|$ which can be seen from \autoref{corelation}(d). 

However, if we plot the differential distributions with respect to $p^{N^\pm}_z$, it boils down to
\begin{equation} \frac{d\sigma}{dp^{N^\pm}_z}=\frac{d\sigma}{d\cos{\theta}}\frac{d\cos{\theta}}{dp^{N^\pm}_z} =\frac{1}{p^{N^\pm}}\frac{d\sigma}{d\cos{\theta}}\simeq \frac{1}{p^{N^\pm}}(1+\cos^2{\theta}).
\end{equation}
Thus given a constant momentum for the $N^\pm$, i.e. $p^{N^\pm}= \rm{constant}\neq 0$,  $\frac{d\sigma}{dp^{N^\pm}_z}$ never diverges. On the contrary, the situation is quite different for transverse momentum distributions as
\begin{eqnarray}
\frac{d\sigma}{dp^{N^\pm}_T}=\frac{d\sigma}{d\cos{\theta}}\frac{d\cos{\theta}}{dp^{N^\pm}_T}=\frac{1}{p^{N^\pm}}\frac{d\sigma}{d\cos{\theta}}\frac{d\cos{\theta}}{d\sin{\theta}}\nonumber\\
\simeq \frac{\tan{\theta}}{-p^{N^\pm}}(1+\cos^2{\theta}).
\end{eqnarray}

\begin{figure}[hbt]
	\begin{center}
		\mbox{\subfigure{\includegraphics[width=0.46\linewidth,angle=-0]{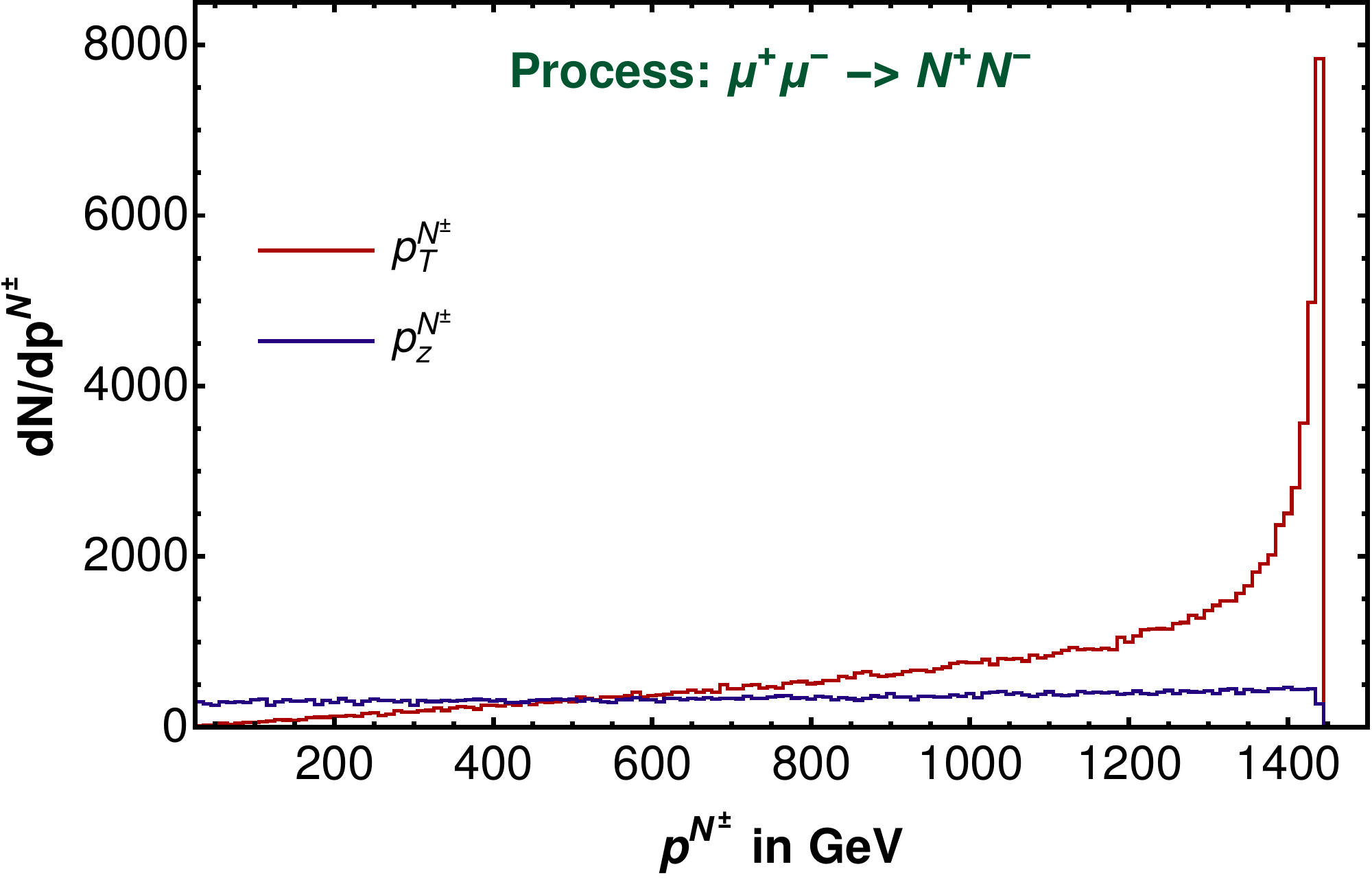}\label{}}}		
		\caption{The transverse ($p^{N^\pm}_T$) and longitudinal ($p^{N^\pm}_z$) momenta distribution of the $N^{\pm}$ at the muon collider with $M_N=1\,\rm{TeV}$ (BP1), $Y_N=5\times 10^{-7}$ and center of mass energy 3.5\,TeV. }\label{Hard_pzpT}
	\end{center}
\end{figure}


\begin{figure}[hbt]
	\begin{center}
		\hspace*{-0.5cm}
		\mbox{\subfigure[]{\includegraphics[width=0.44\linewidth,angle=-0]{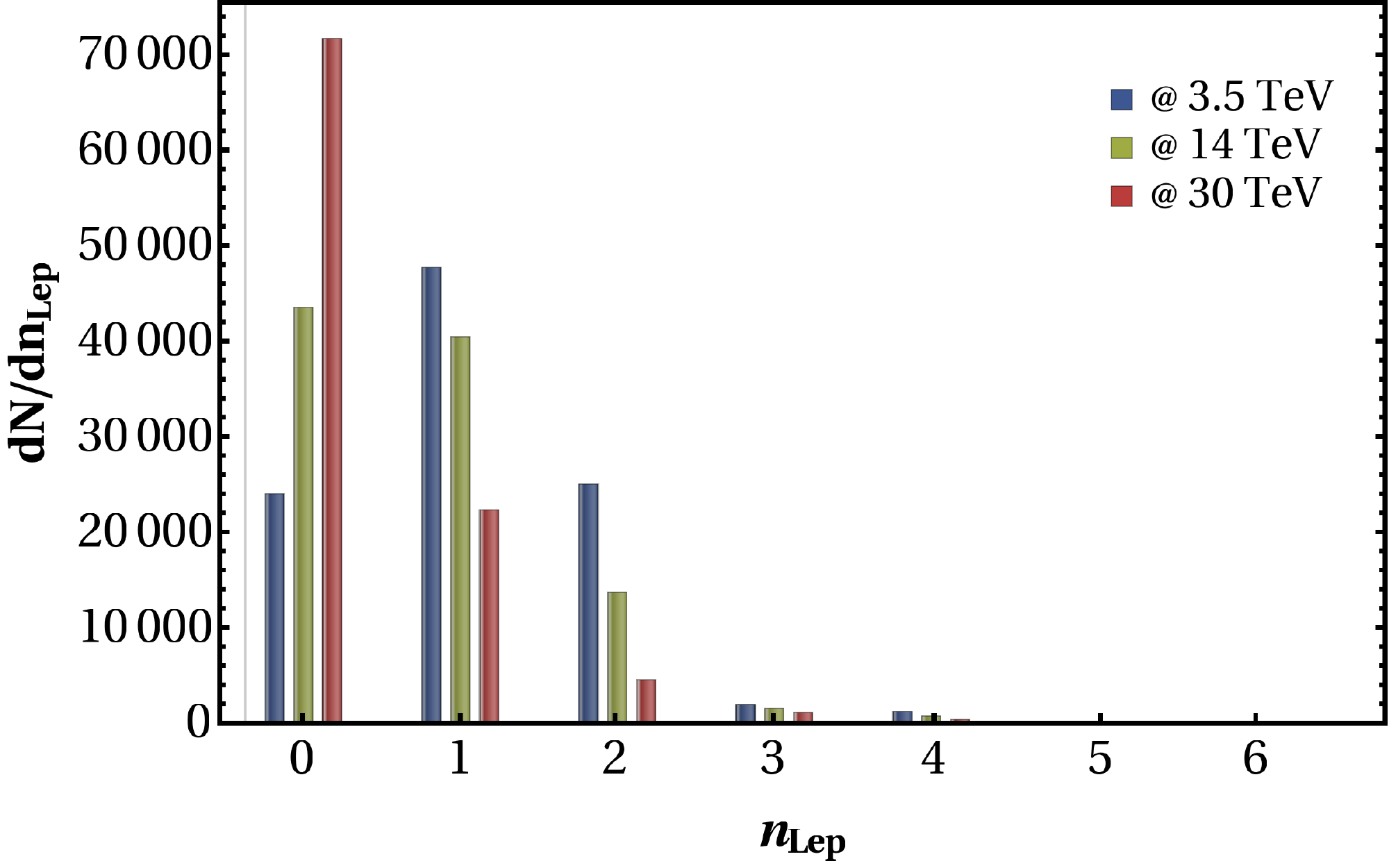}\label{}}\quad \quad \quad
		\subfigure[]{\includegraphics[width=0.44\linewidth,angle=-0]{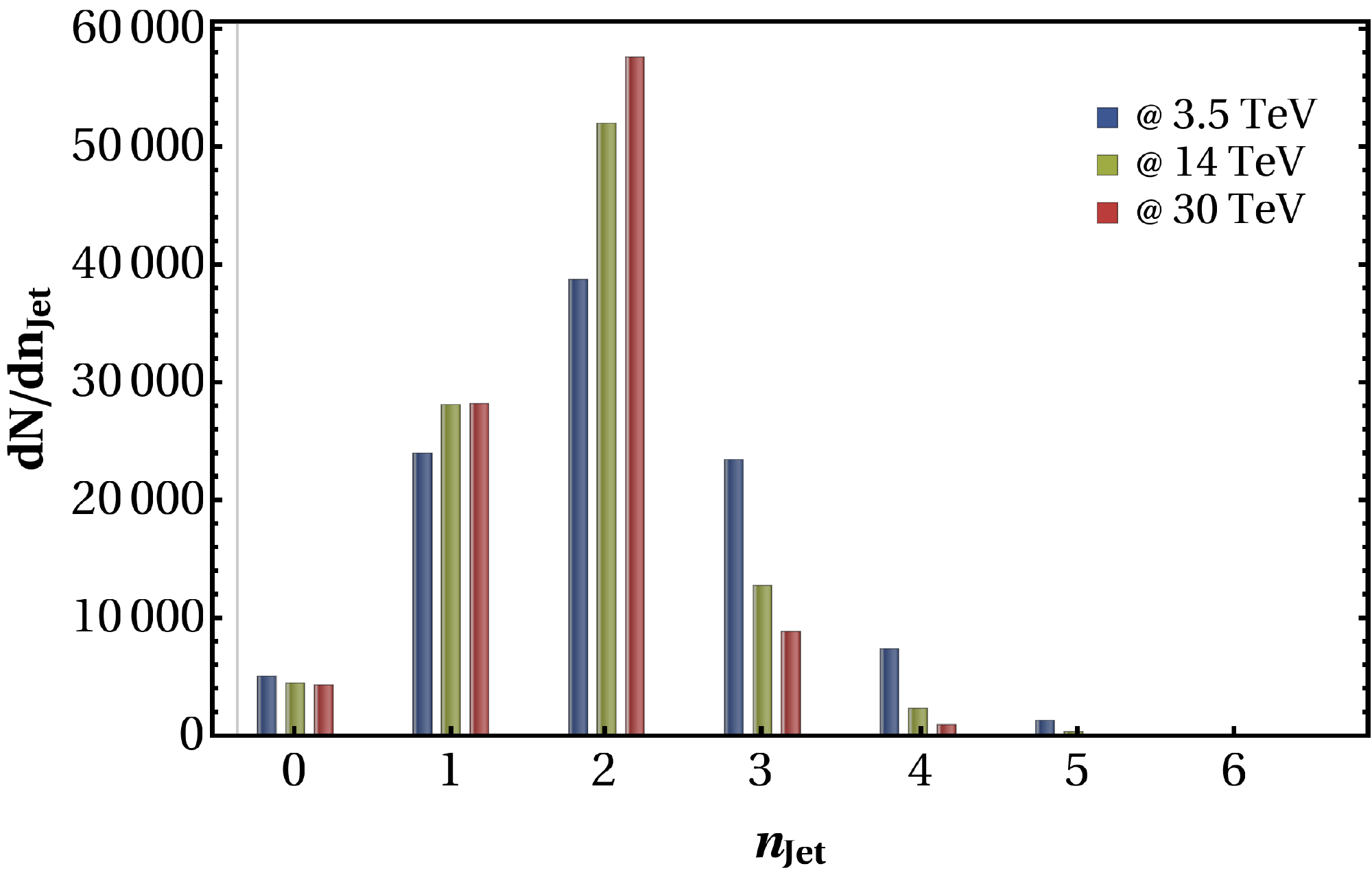}\label{}}}
		\caption{Multiplicity distributions (a) for the charged leptons ($n_{\rm Lep}$) and (b) for the jets ($n_{\rm Jet}$) at the centre of mass energies of 3.5\,TeV, 14\,TeV and 30\,TeV for $M_N=1\,\rm{TeV}$ (BP1) and $Y_N=5\times 10^{-7}$ at the muon collider.}\label{mu_mul}
	\end{center}
\end{figure}

$p^{N^\pm}_T$   diverges at $\theta=\pi/2, 3\pi/2$, which is evident from \autoref{Hard_pzpT}. Thus unlike at the LHC, here at the  muon  collider, the transverse momentum dominates, and so transverse decay length can expected to be slightly larger compared to the LHC/FCC. These effects are then transferred to the decayed leptons and jets from $N^\pm$. We should remember that this divergence also depends on the Lorentz structure of the matrix element, governed by the spin of the initial, finalstates and the propagator \cite{Bandyopadhyay:2020wfv}. 

The kinematical features at the hard scattering  level can decline when it comes to the finalstate decay products. Here we are focusing on $N^\pm/N^0$ giving rise to $h \ell^\pm/\nu$, and thus  2$b$-jets and one lepton ($e,\, \mu$) or neutrino will be present in the finalstate. The jets and leptons are tagged via similar cuts as described in \autoref{setup}. \autoref{mu_mul} depicts the lepton and jet multiplicity distributions for BP1 with $Y_N=5 \times 10^{-7}$ at three different energies. \autoref{mu_mul}(a) presents the lepton multiplicity distributions which remains similar to that at the LHC. However, in \autoref{mu_mul}(b) we have the jet multiplicity distributions, and the higher multiplicity is much lesser than at the LHC as muon collider is devoid of any initial state QCD radiations.

\begin{figure}[h]
	\begin{center}
		\hspace*{-1.1cm}
		\mbox{\subfigure[]{\includegraphics[width=0.35\linewidth,angle=-0]{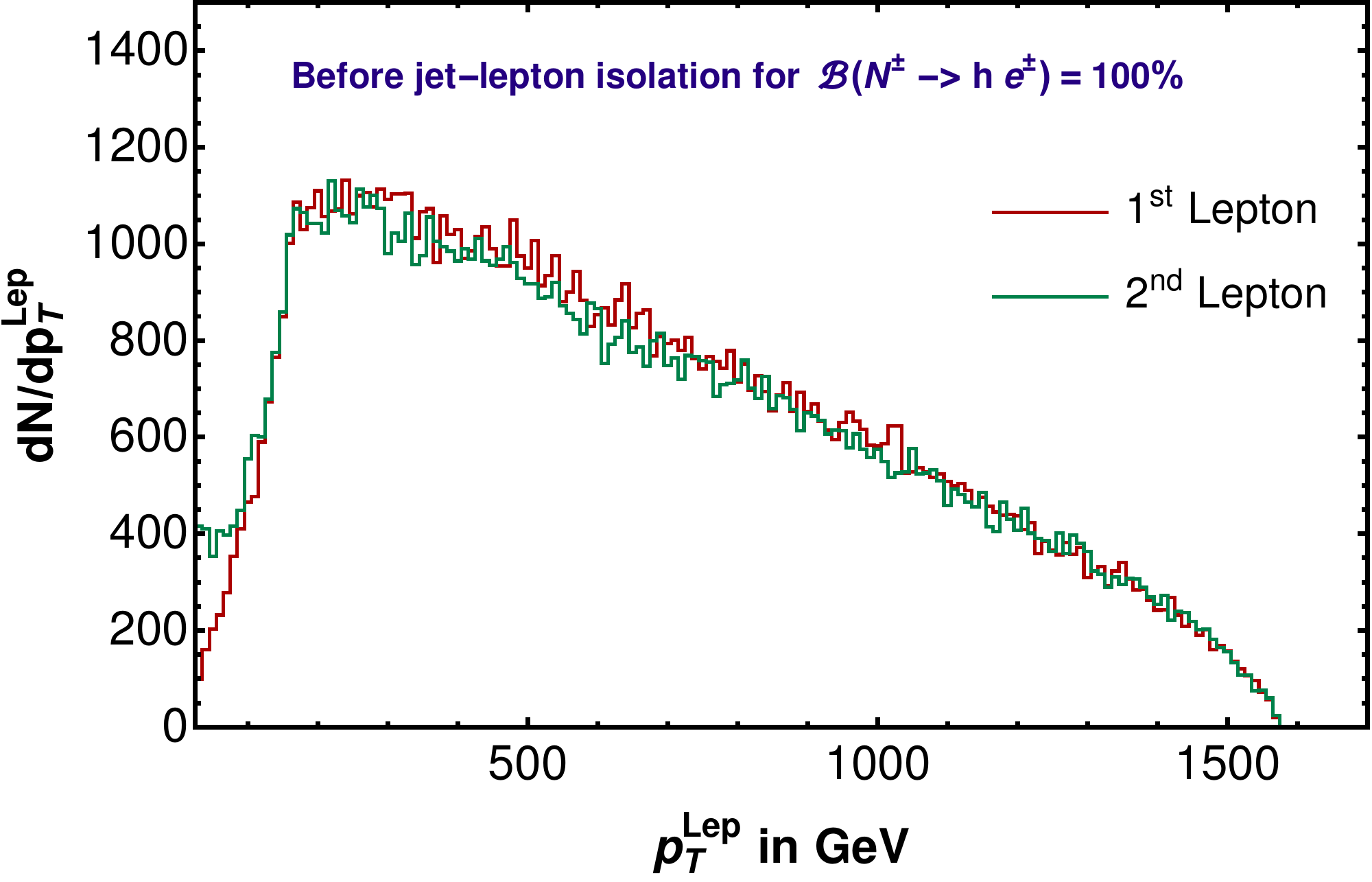}\label{}}\quad
			\subfigure[]{\includegraphics[width=0.35\linewidth,angle=-0]{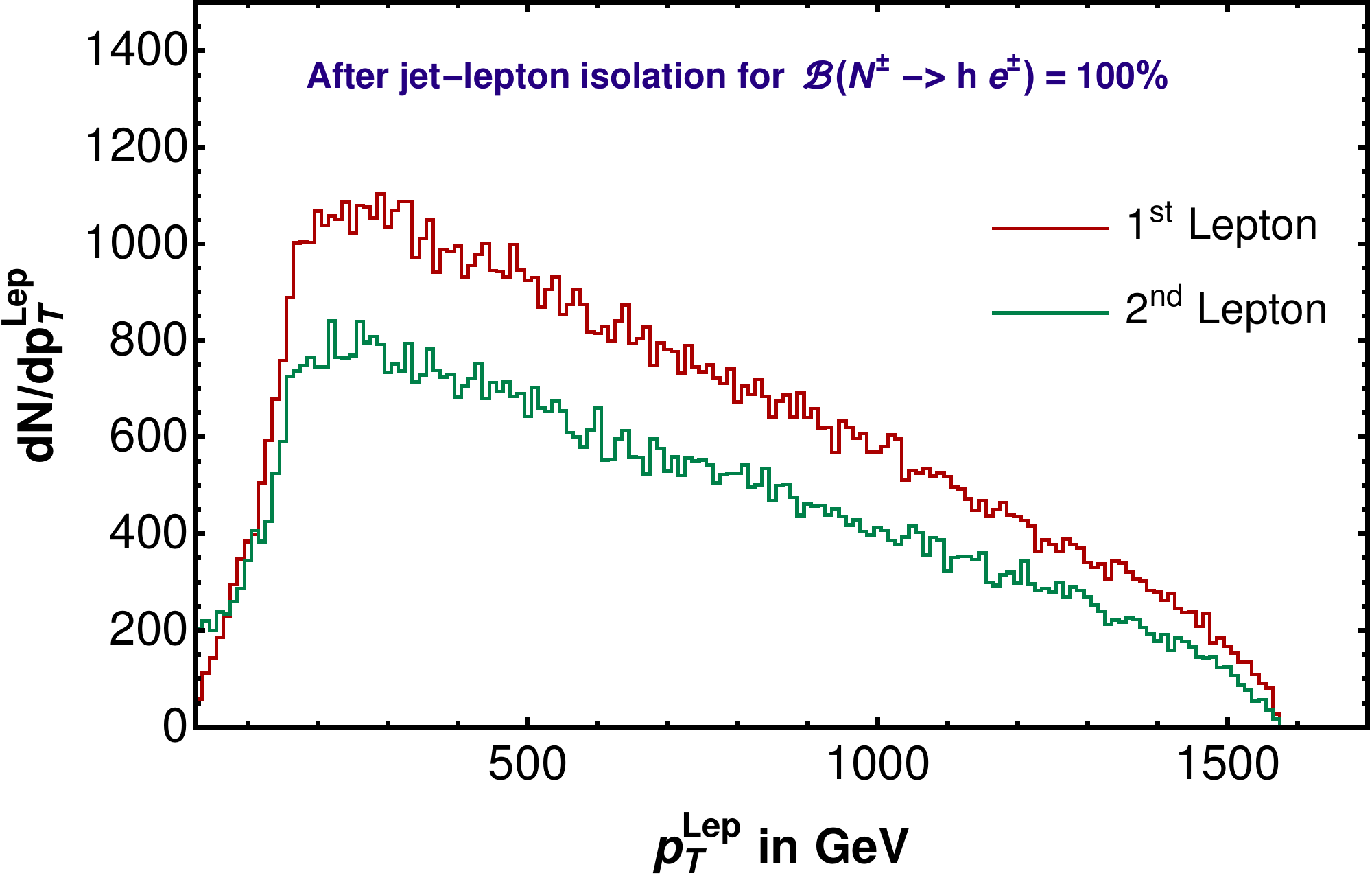}\label{}}\quad
			\subfigure[]{\includegraphics[width=0.35\linewidth,angle=-0]{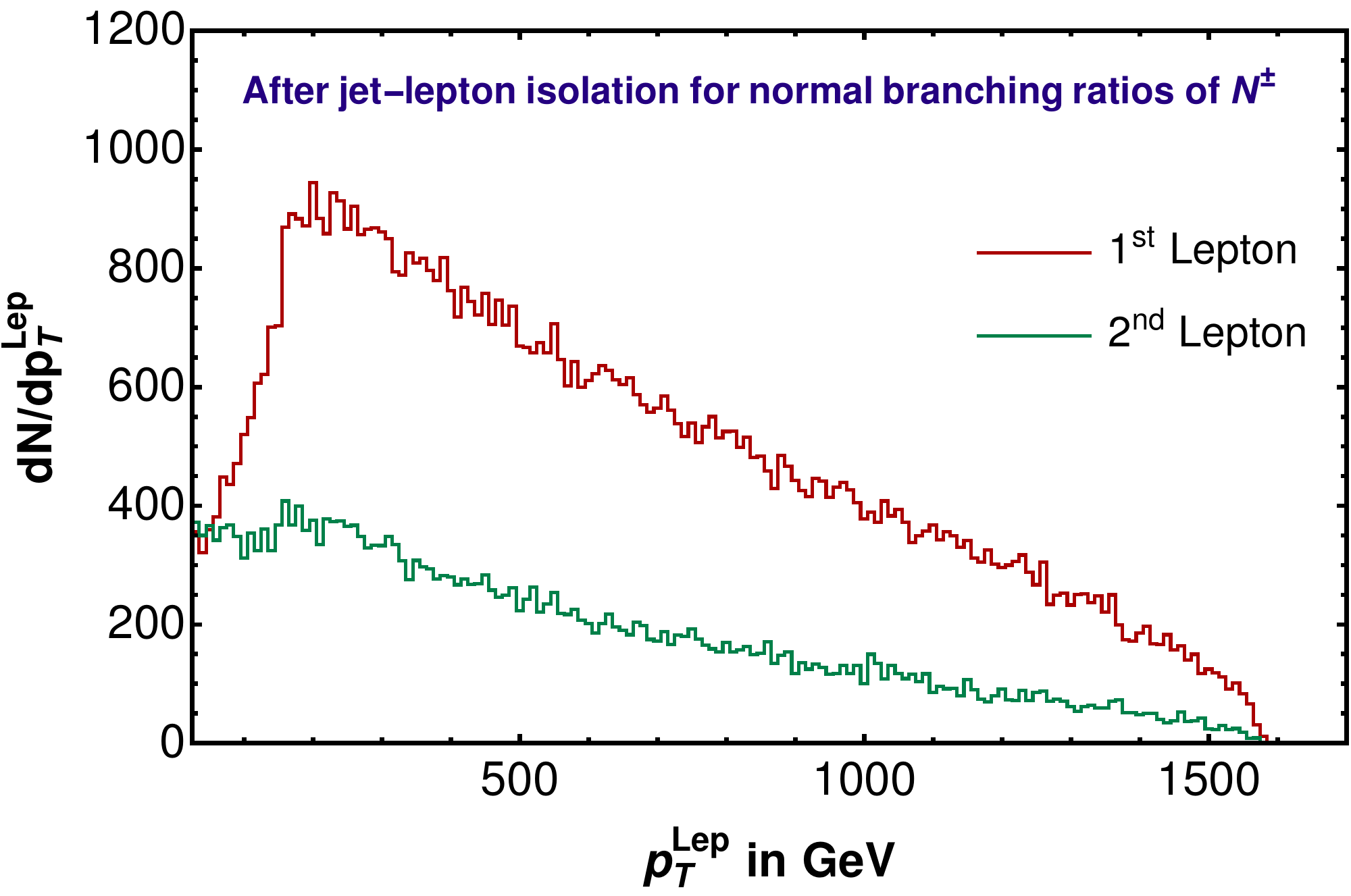}\label{}}}		
		\caption{The transverse momentum ($p_T$) distributions of charged leptons ($1^{st}$ and $2^{nd}$) at the muon collider for $M_N=1\,\rm{TeV}$, $Y_N=5\times 10^{-7}$ and center of mass energy 3.5\,TeV. (a) and (b) represent the distributions of two leptons for pre and post jet-lepton isolation, respectively for $\mathcal{B}(N^{\pm} \to h e^{\pm})= 100\%$. (c) describes the scenario for the isolated leptons obeying the branching ratio of BP1.}\label{lepPT_mu}
	\end{center}
\end{figure}


\begin{figure}[h]
	\begin{center}
		\mbox{\subfigure[]{\includegraphics[width=0.44\linewidth,angle=-0]{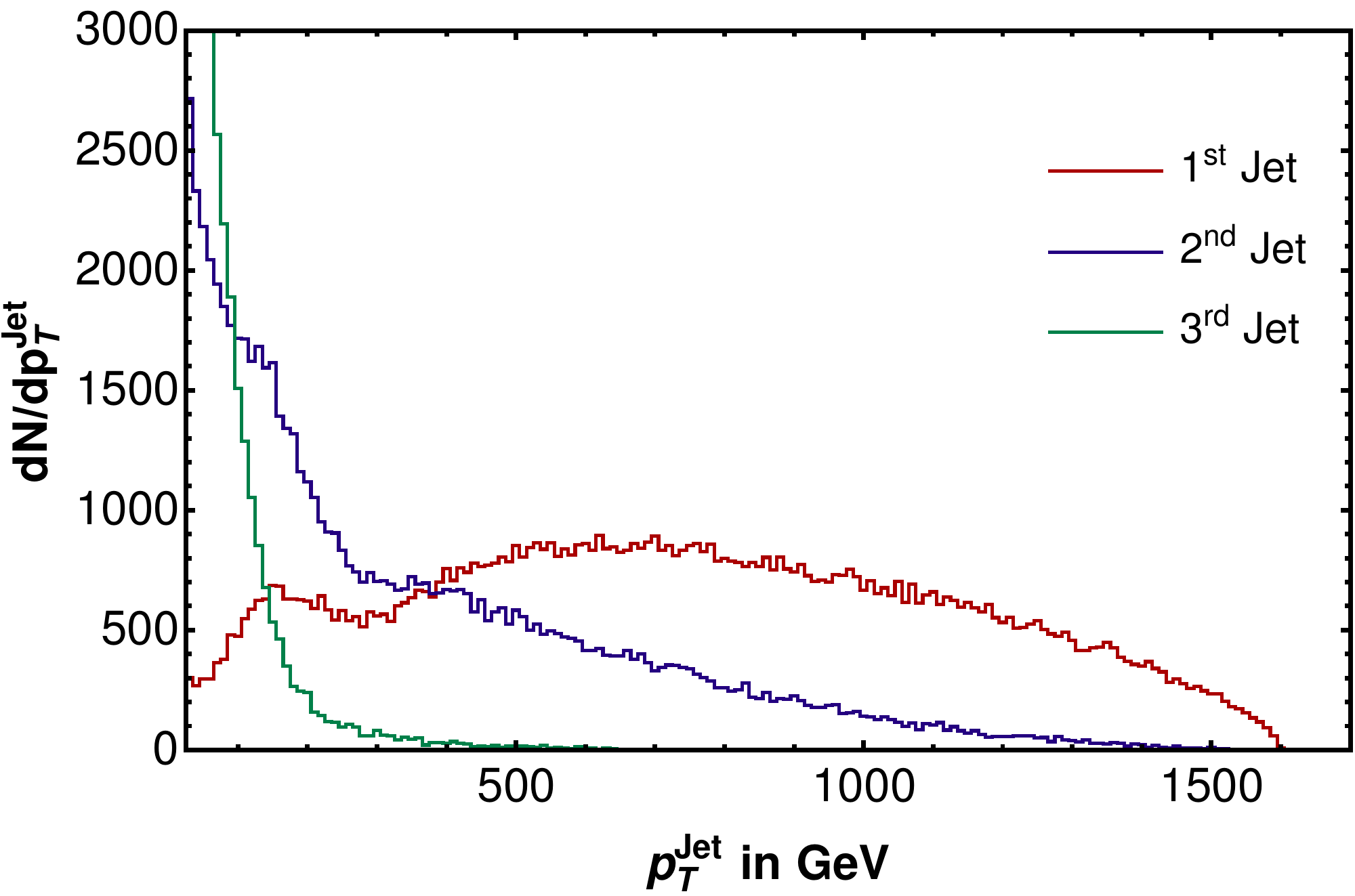}\label{}}\quad \quad 
		\subfigure[]{\includegraphics[width=0.44\linewidth,angle=-0]{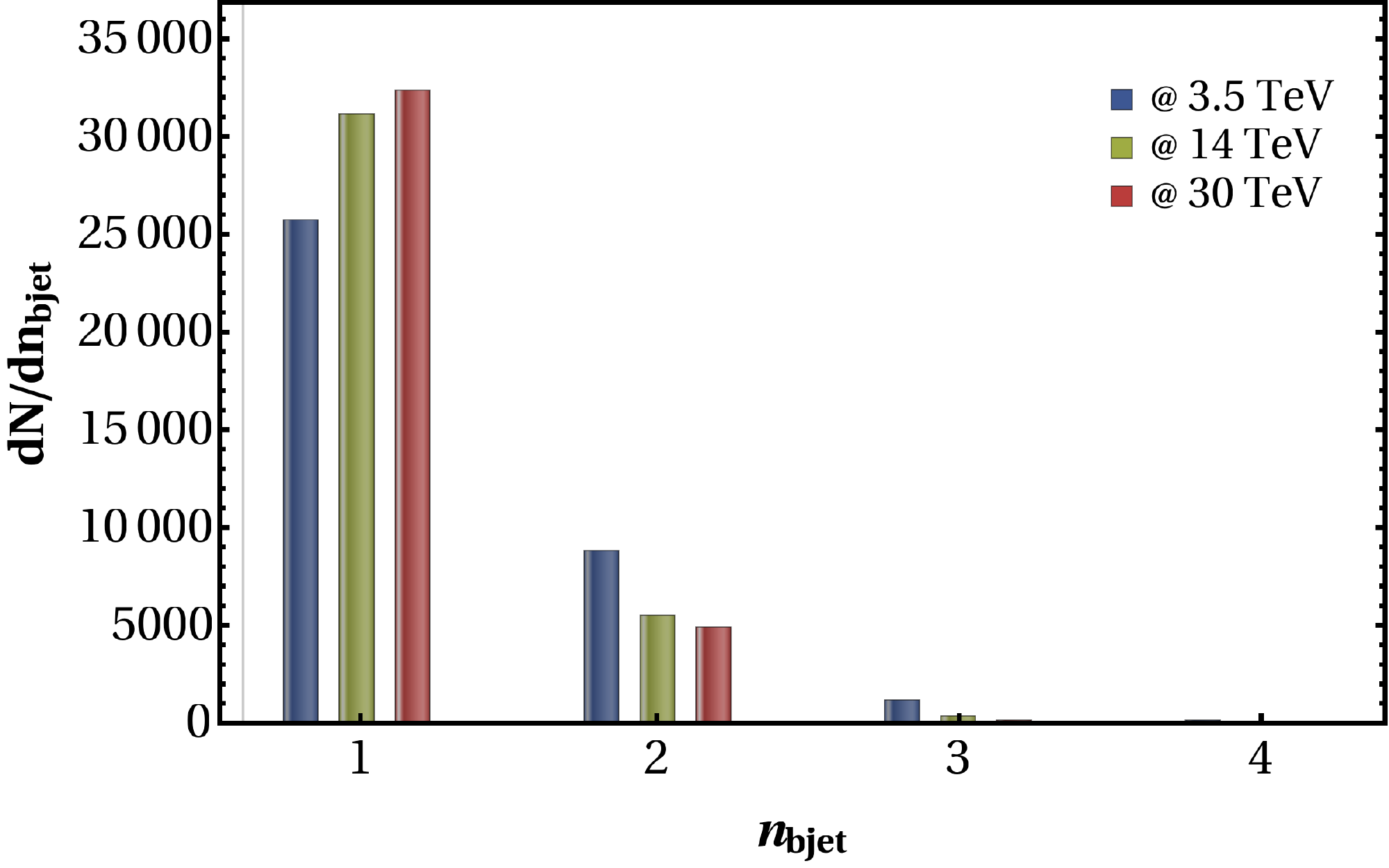}\label{}}}
		\caption{(a)The transverse momentum ($p_T$) distribution of the first three $p_T$ ordered jets with the centre of mass energy of 3.5\,TeV and (b) $b$-jet multiplicity distribution ($n_{\rm bjet}$) at the muon collider for the centre of mass energies of 3.5\,TeV, 14\,TeV and 30\,TeV, with $M_N=1\,\rm{TeV}$ (BP1), $Y_N=5\times 10^{-7}$. }\label{jetPT_mu}
	\end{center}
\end{figure}


\autoref{lepPT_mu} shows the lepton $p_T$ distributions  coming from $\mu^+ \mu^- \to  N^+ N^-$ at the centre of mass energy of 3.5 TeV for the first benchmark point (BP1), where  $M_N=1$\,TeV and $Y_N=5\times 10^{-7}$.  \autoref{lepPT_mu}(a) presents the case, where the $N^\pm$ is forced decayed to $e^\pm h$ and both the electrons have similar $p_T$ distributions. \autoref{lepPT_mu}(b) depicts the case after the jet-lepton isolation where the yield of the second lepton is reduced. Finally, \autoref{lepPT_mu}(c) shows the distributions with the branching ratios of BP1. Here it can be noticed that the second lepton can either come from $N^\pm$ or gauge bosons, while the first lepton mostly comes from $N^\pm$ decays. The interesting point is the end point of the $p_T$ distribution governed by the momentum conservation is limited to $\frac{E_{\rm CM}}{2}$, unlike at the LHC/FCC.

\autoref{jetPT_mu} depicts the $p_T$ distributions of the first three $p_T$ ordered jets for BP1 at the centre of mass energy of  3.5 TeV. The red curve representing the leading jet has two smooth peaks. The earlier one can be from the jets coming from the gauge bosons  as well as from the Higgs boson decay. However, the higher peaks occur when two of these jets are merged due to the sufficient boost \cite{CS:TypeI}. The second and third jets can come from gauge boson decays. Along with these jets, there could be additional jets due to finalstate radiation (FSR) as evident from the jet multiplicity distributions in \autoref{mu_mul}(b).  Contrary to the LHC/FCC, there cannot be any QCD radiation from the initial state in this case, so ISR  effects  are absent.

\autoref{jetPT_mu}(b) presents the $b$-jet multiplicity distributions. In spite of having four $b$-jets coming from the two Higgs bosons for BP1 at the centre of mass energy of 3.5 TeV, not all are tagged as $b$-jets due to the $b$-tagging efficiency. There can be $b$-jets coming from the $Z$ boson decays as well. For the Higgs mass reconstructions we require at least two $b$-jets to be tagged which we discuss in the \autoref{resulsm}.

\subsection{Boost effects at the muon collider}

\begin{figure}[hbt]
	\begin{center}
		\hspace*{-0.5cm}
		\mbox{\subfigure[]{\includegraphics[width=0.4\linewidth,angle=-0]{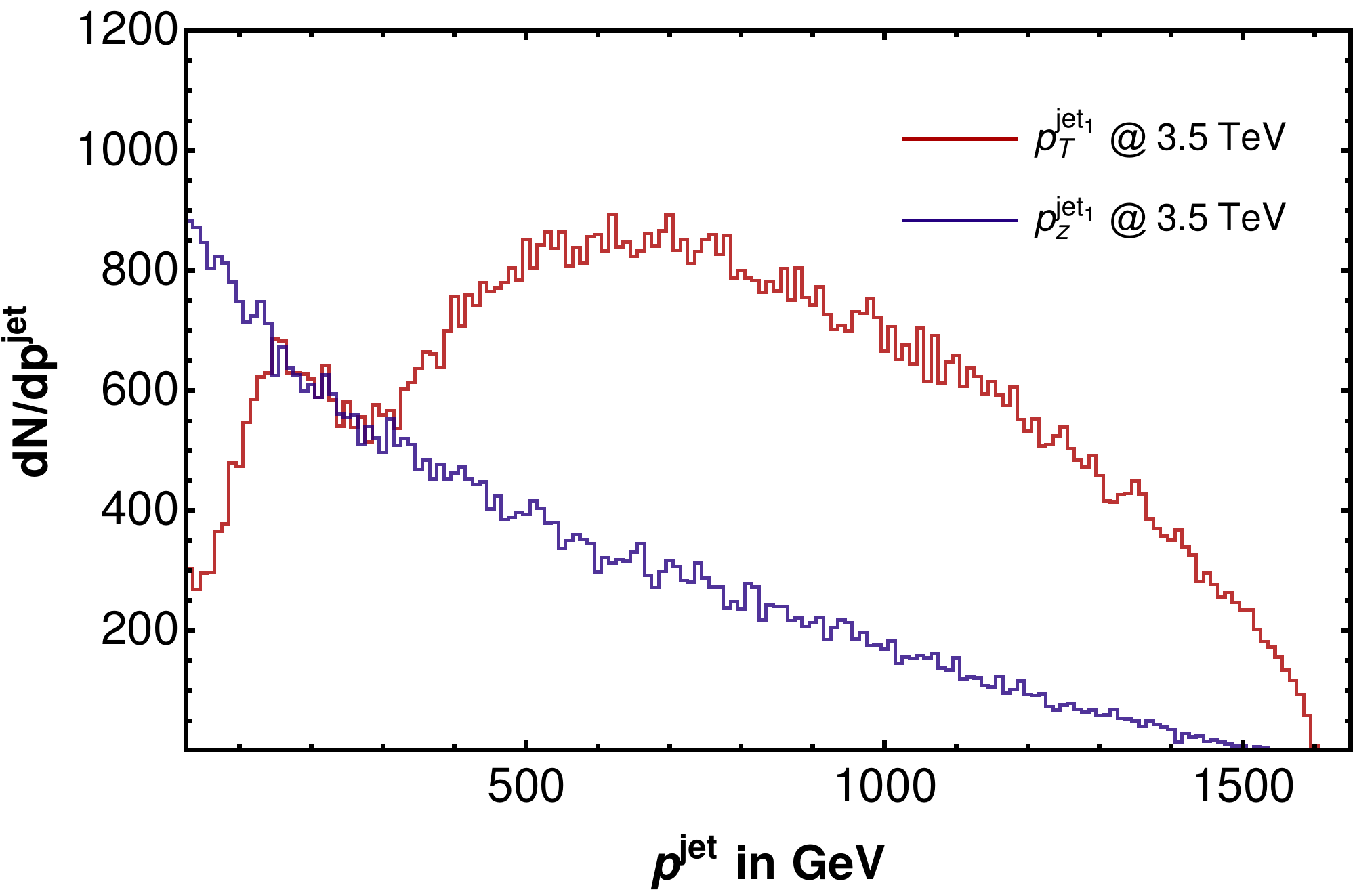}\label{}}\quad \quad 
		\subfigure[]{\includegraphics[width=0.4\linewidth,angle=-0]{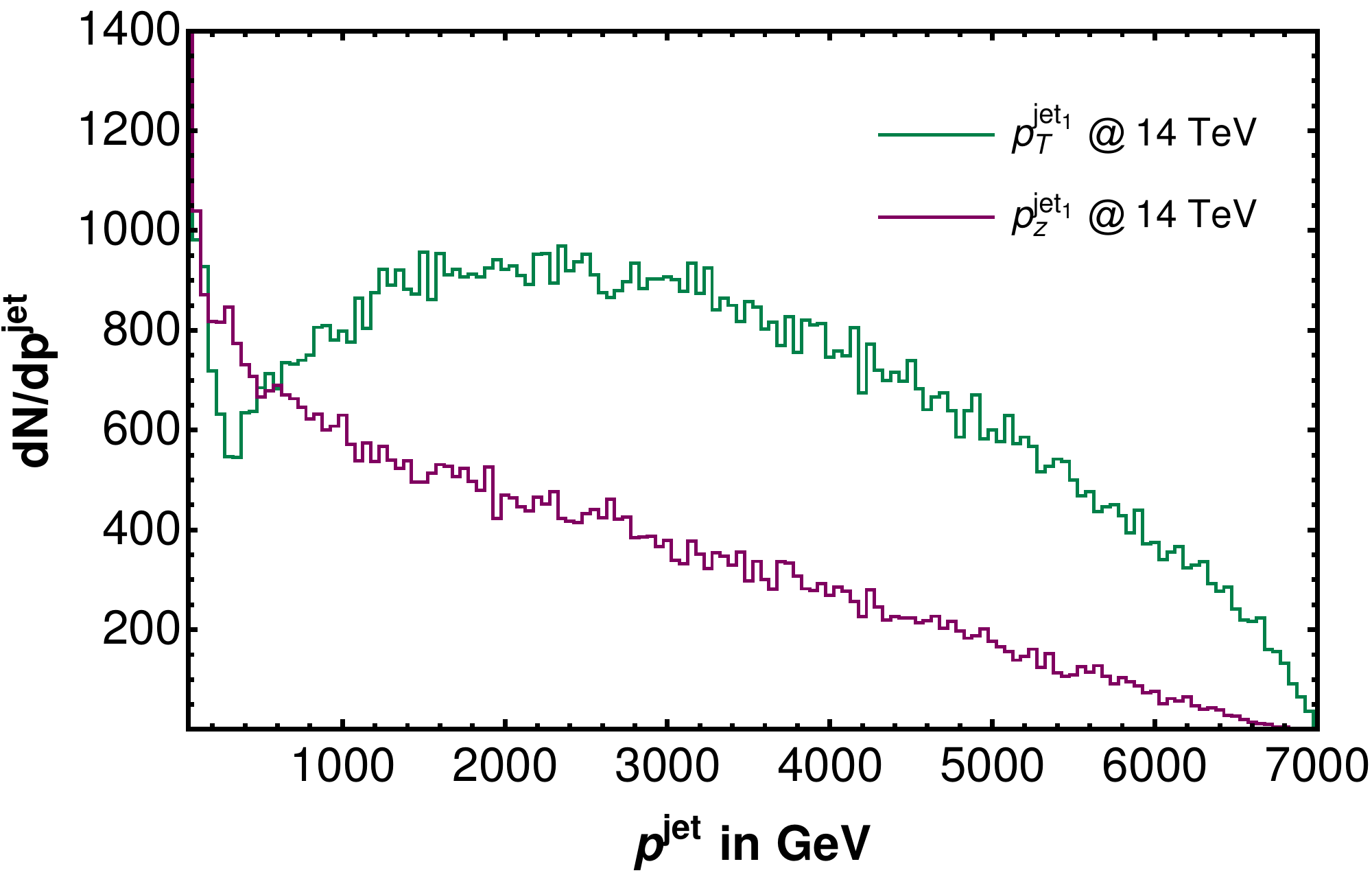}\label{}}}
		\hspace*{-0.5cm}
		\mbox{\subfigure[]{\includegraphics[width=0.4\linewidth,angle=-0]{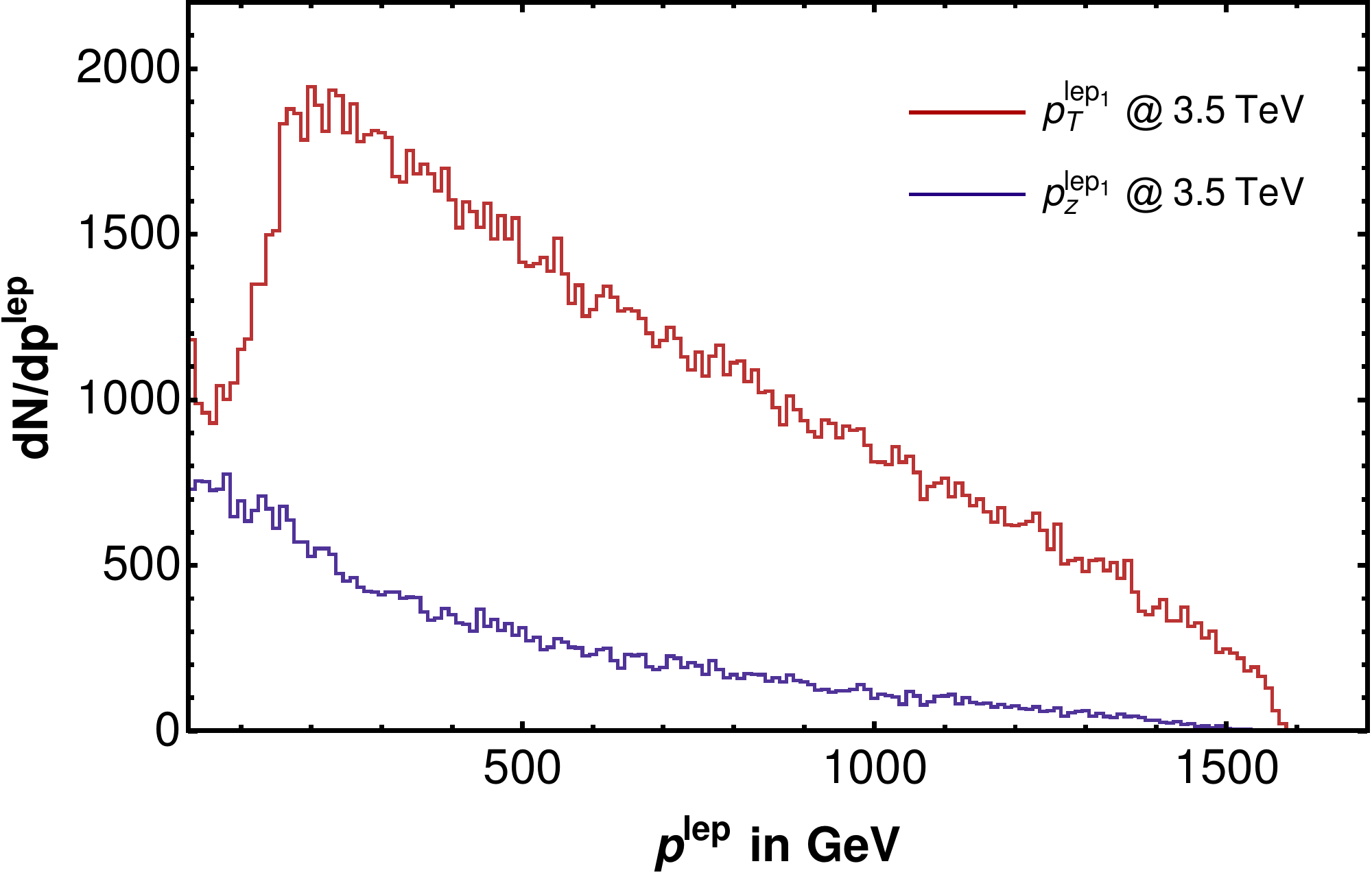}\label{}}\quad \quad 
		\subfigure[]{\includegraphics[width=0.4\linewidth,angle=-0]{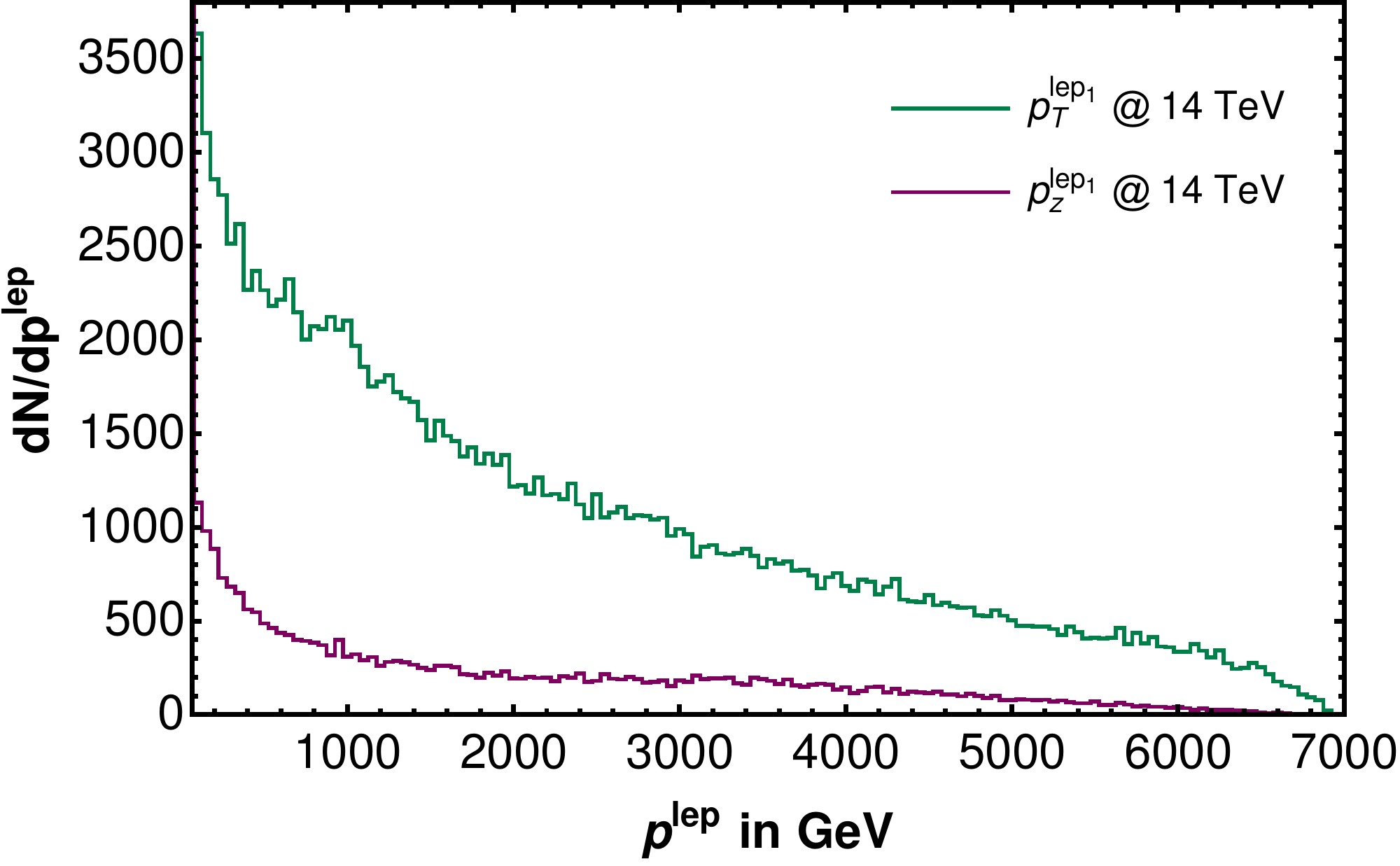}\label{}}}		
		\caption{The transverse ($p_T$) and the longitudinal ($p_z$) momenta distributions at the muon collider for the process $\mu^+\,\mu^-\to N^+\,N^{-}$,  with $M_N=1\,\rm{TeV}$ (BP1), $Y_N=5\times 10^{-7}$. (a) and (b) represent the distributions for leading jet with the centre of mass energies of 3.5\,TeV and 14\,TeV, respectively. (c) and (d) depict the similar distributions for isolated charged leptons.} \label{pzpT_mu}
	\end{center}
\end{figure}


Unlike at the LHC/FCC (as shown in \autoref{kindis}) in muon collider, the momentum conservation restricts  the boost. Again, the angular distributions govern the behaviour of $p_T$ and $p_z$, with the former diverging around $|\eta|\sim 0$. The resultant of these two, leads to $p_T$ being more dominant over $p_z$ for both jets and leptons, as described in \autoref{pzpT_mu}. \autoref{pzpT_mu}(a) and  \autoref{pzpT_mu}(b)  show the $p_T$ and  $p_z$ distributions of the leading jet for BP1 at  the centre of mass  energy of 3.5 and 14 TeV, respectively. We see the two-hump behaviour as before, which declines as we move  from \autoref{pzpT_mu}(a) to \autoref{pzpT_mu}(b) due to larger boost which tends to create fat-jet like signatures \cite{Chakraborty:2018khw,Bhardwaj:2018lma,Bandyopadhyay:2010ms,CS:TypeI} at higher $p_T$ values. Similar plots can be seen for the leptons in \autoref{pzpT_mu}(c) to \autoref{pzpT_mu}(d) where $p_T$ remains dominant over $p_z$. Thus the boost effect on the decay length will be more on the transverse decay  length than the  longitudinal ones. This is contrary to what we observe at the LHC/FCC, where the partonic system is often boosted along $z-$ direction governed by the parton distribution function, resulting in longer displaced decay length for the longitudinal ones as compared to the transverse one. This is an artifact that at the LHC/FCC the partonic momentum is  not fixed, unlike at the muon collider, where $\theta $ is the only parameter that varies.

\subsection{Displaced decays at the muon collider}

\begin{figure}[hbt]
	\begin{center}
		\hspace*{-0.5cm}
		\mbox{\subfigure[]{\includegraphics[width=0.33\linewidth,angle=-0]{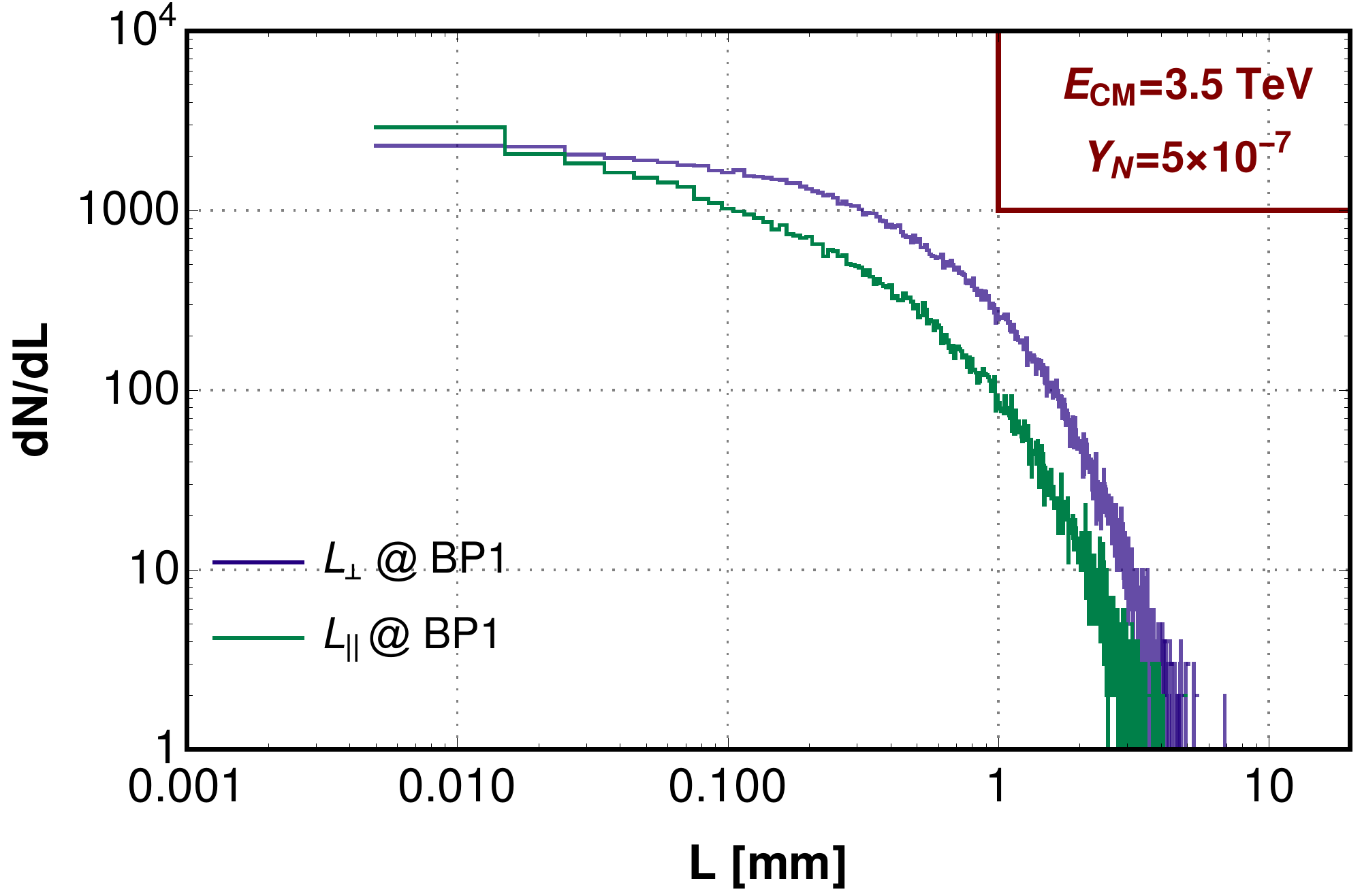}\label{}}\quad
			\subfigure[]{\includegraphics[width=0.33\linewidth,angle=-0]{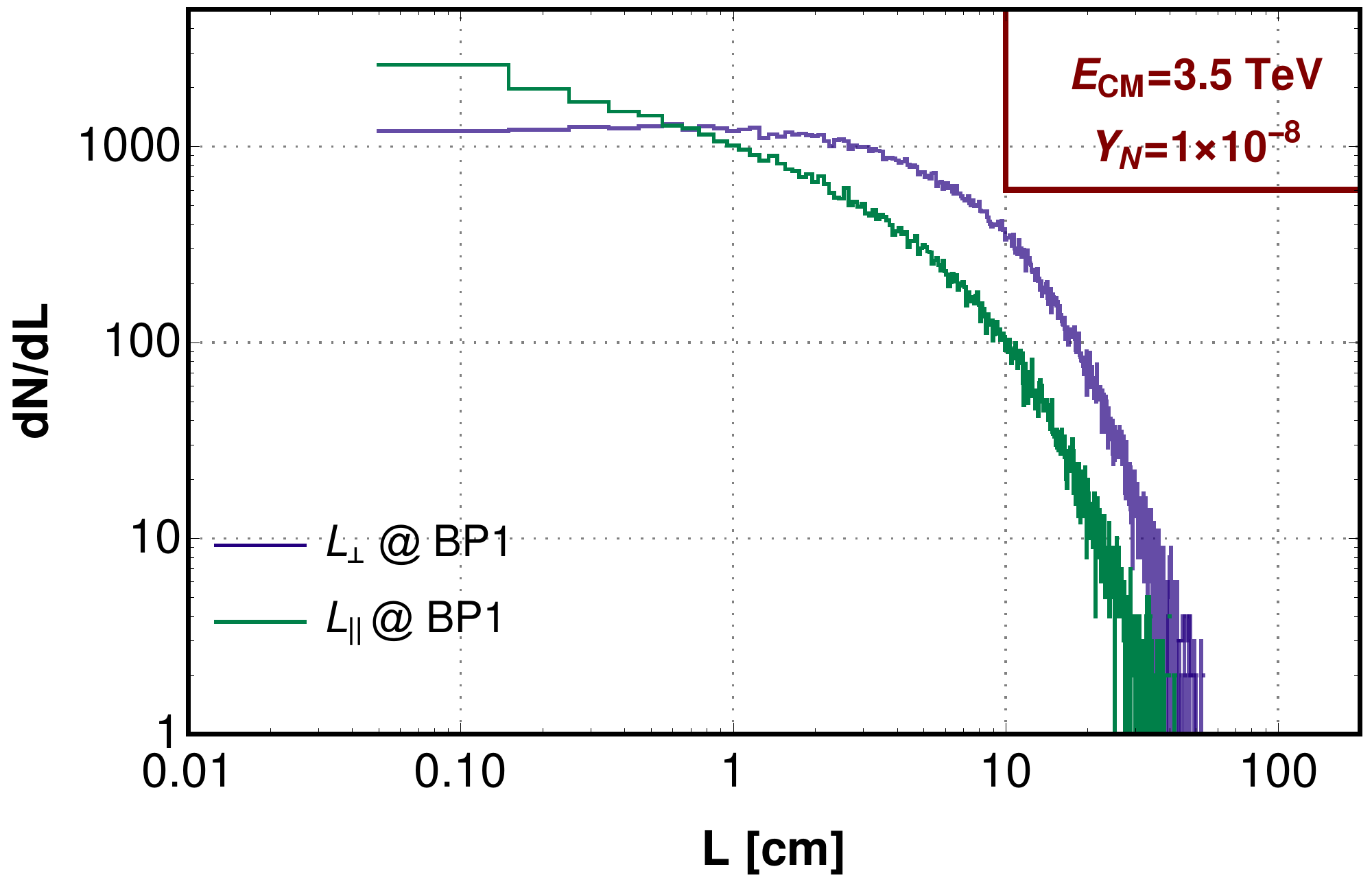}\label{}}\quad
			\subfigure[]{\includegraphics[width=0.33\linewidth,angle=-0]{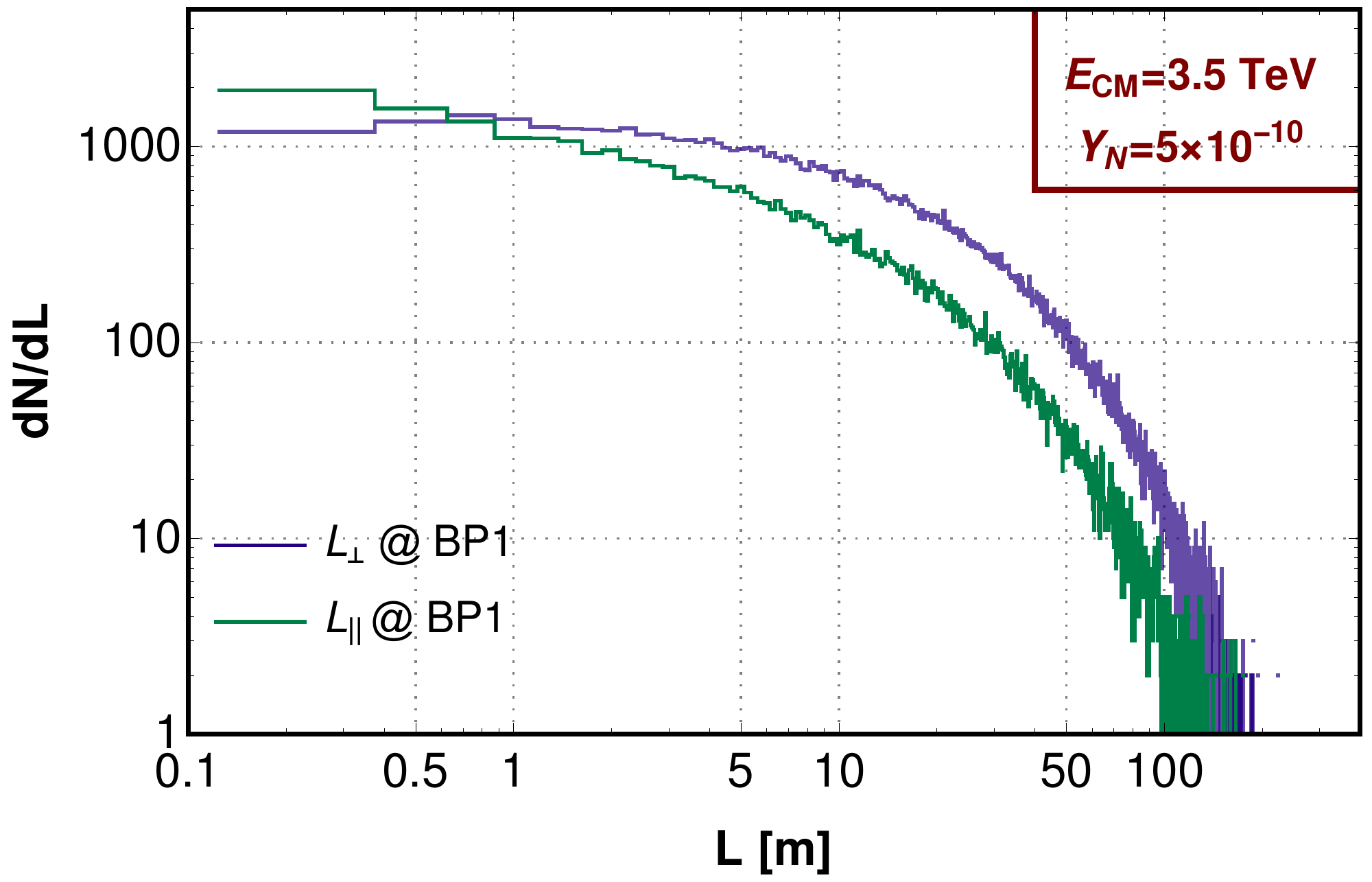}\label{}}}
		\hspace*{-0.5cm}
		\mbox{\subfigure[]{\includegraphics[width=0.33\linewidth,angle=-0]{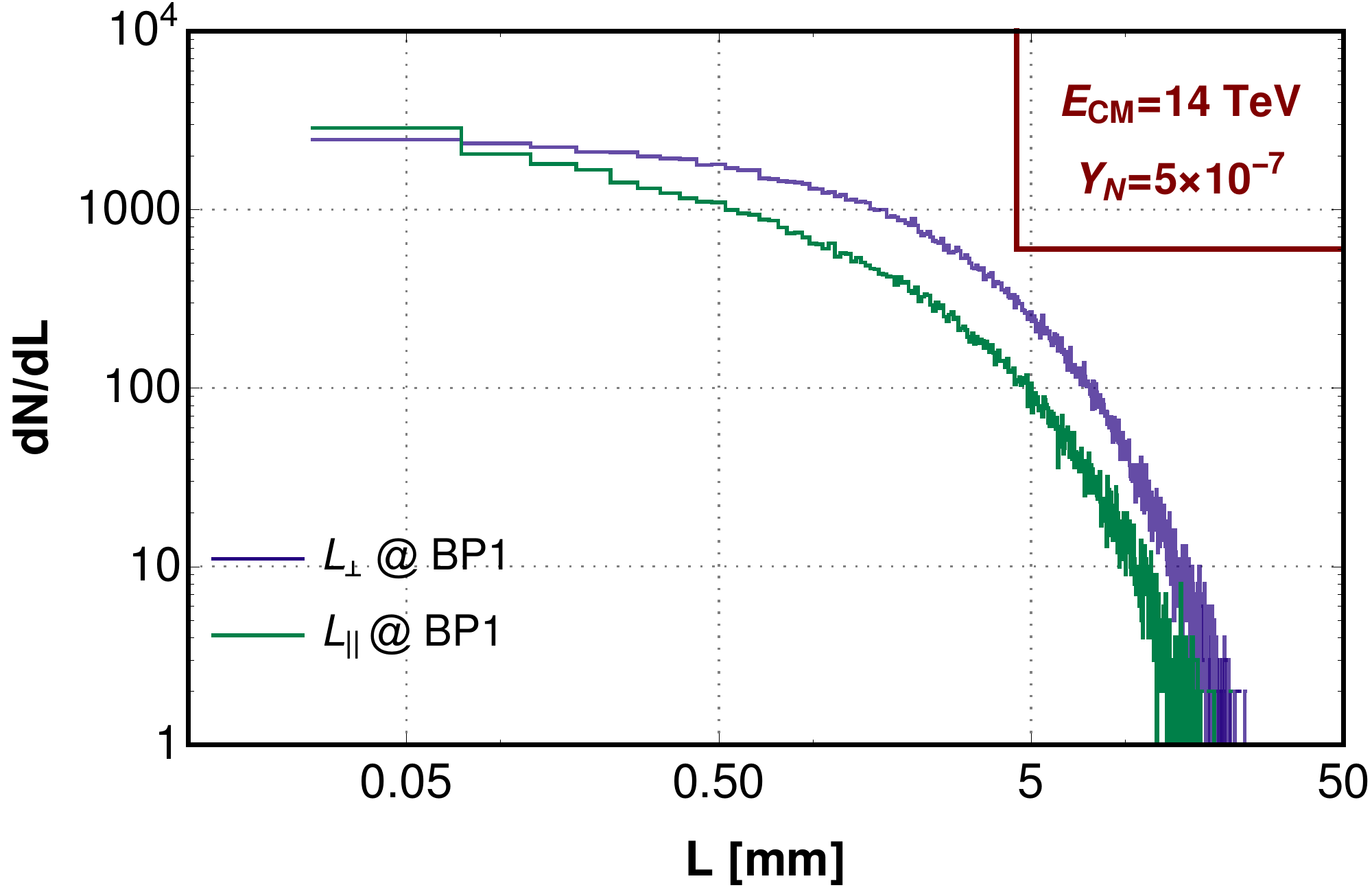}\label{}}\quad
			\subfigure[]{\includegraphics[width=0.33\linewidth,angle=-0]{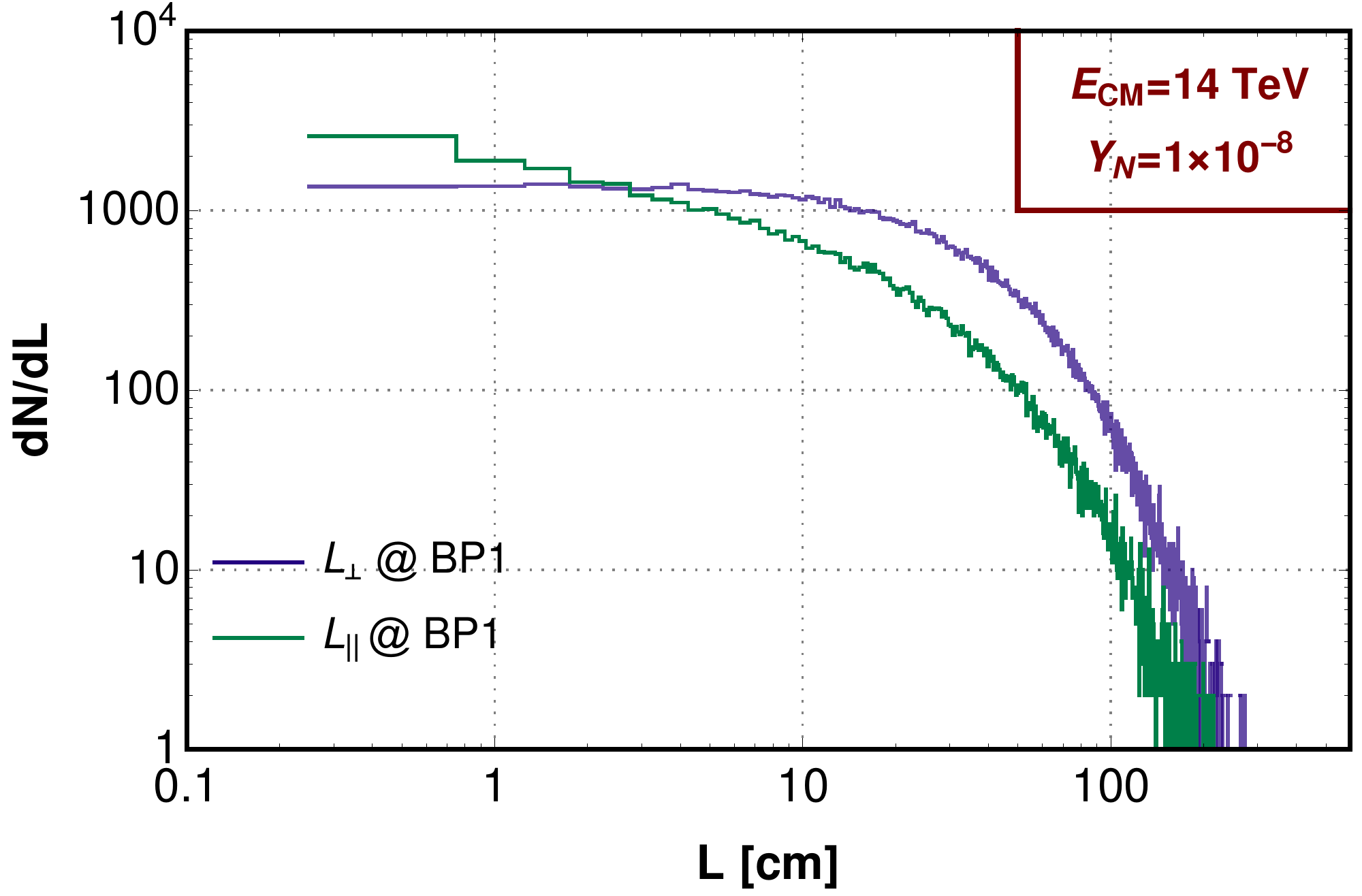}\label{}}\quad
			\subfigure[]{\includegraphics[width=0.33\linewidth,angle=-0]{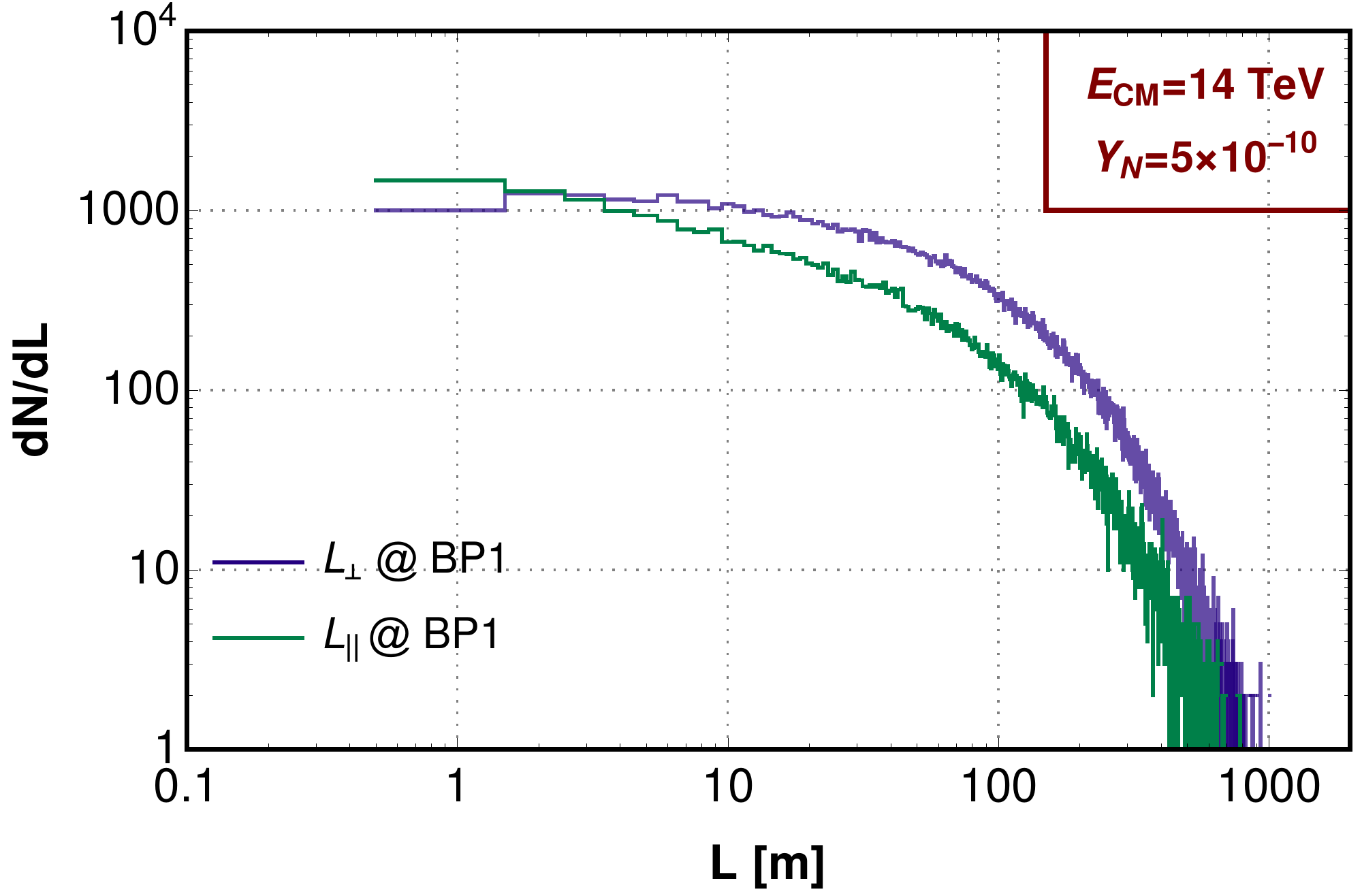}\label{}}}
		\hspace*{-0.5cm}
		\mbox{\subfigure[]{\includegraphics[width=0.33\linewidth,angle=-0]{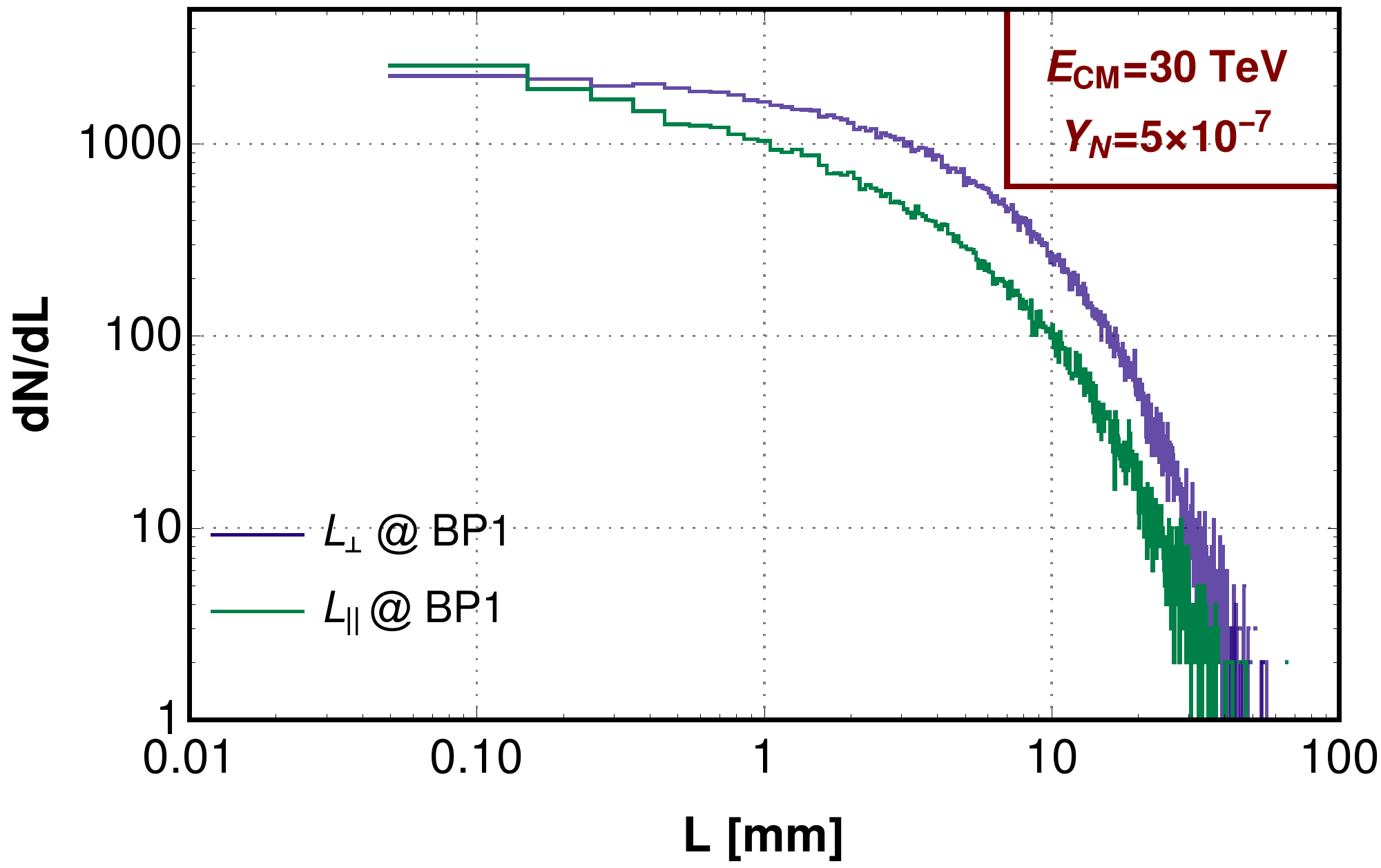}\label{}}\quad
			\subfigure[]{\includegraphics[width=0.33\linewidth,angle=-0]{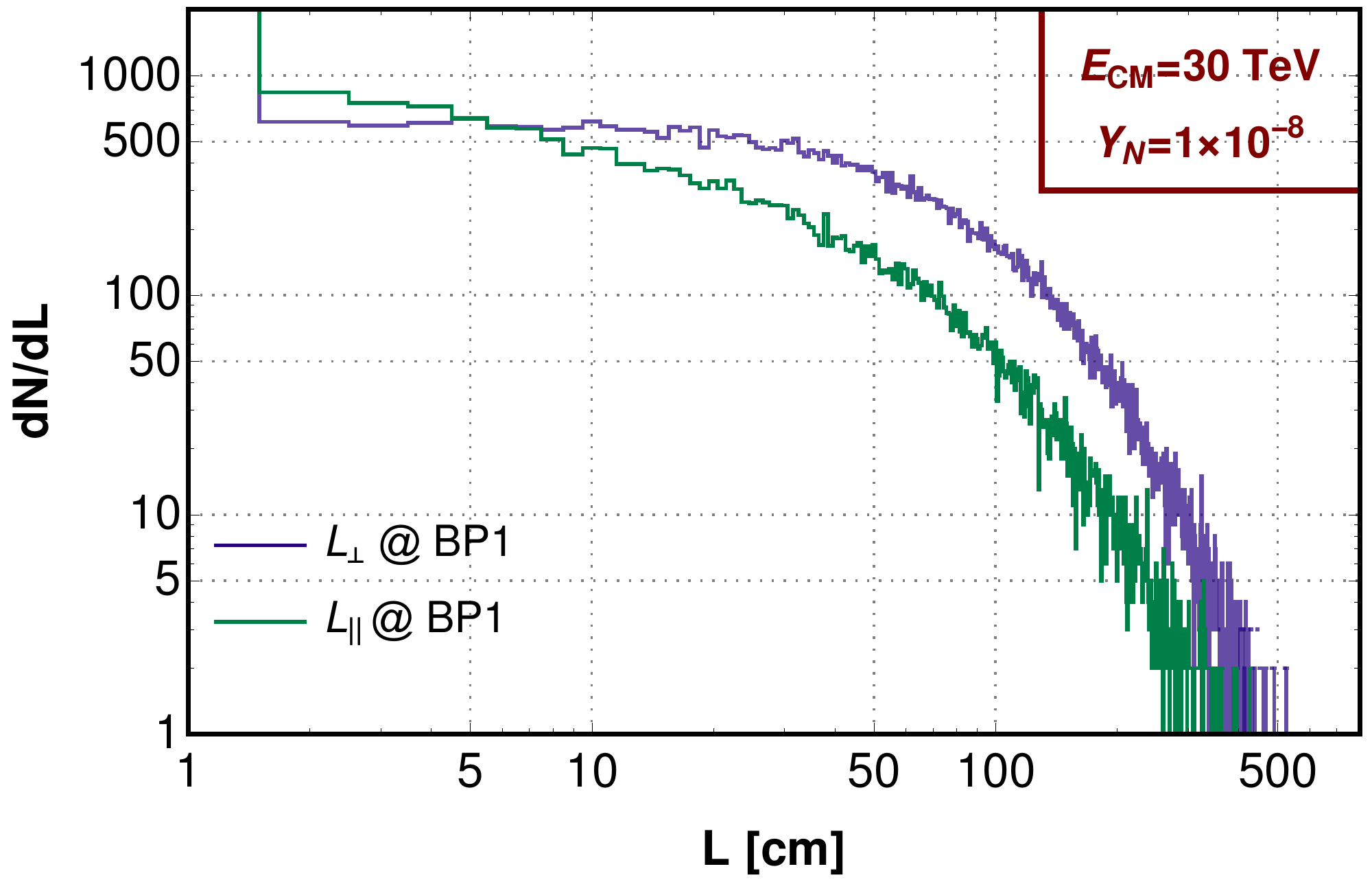}\label{}}\quad
			\subfigure[]{\includegraphics[width=0.33\linewidth,angle=-0]{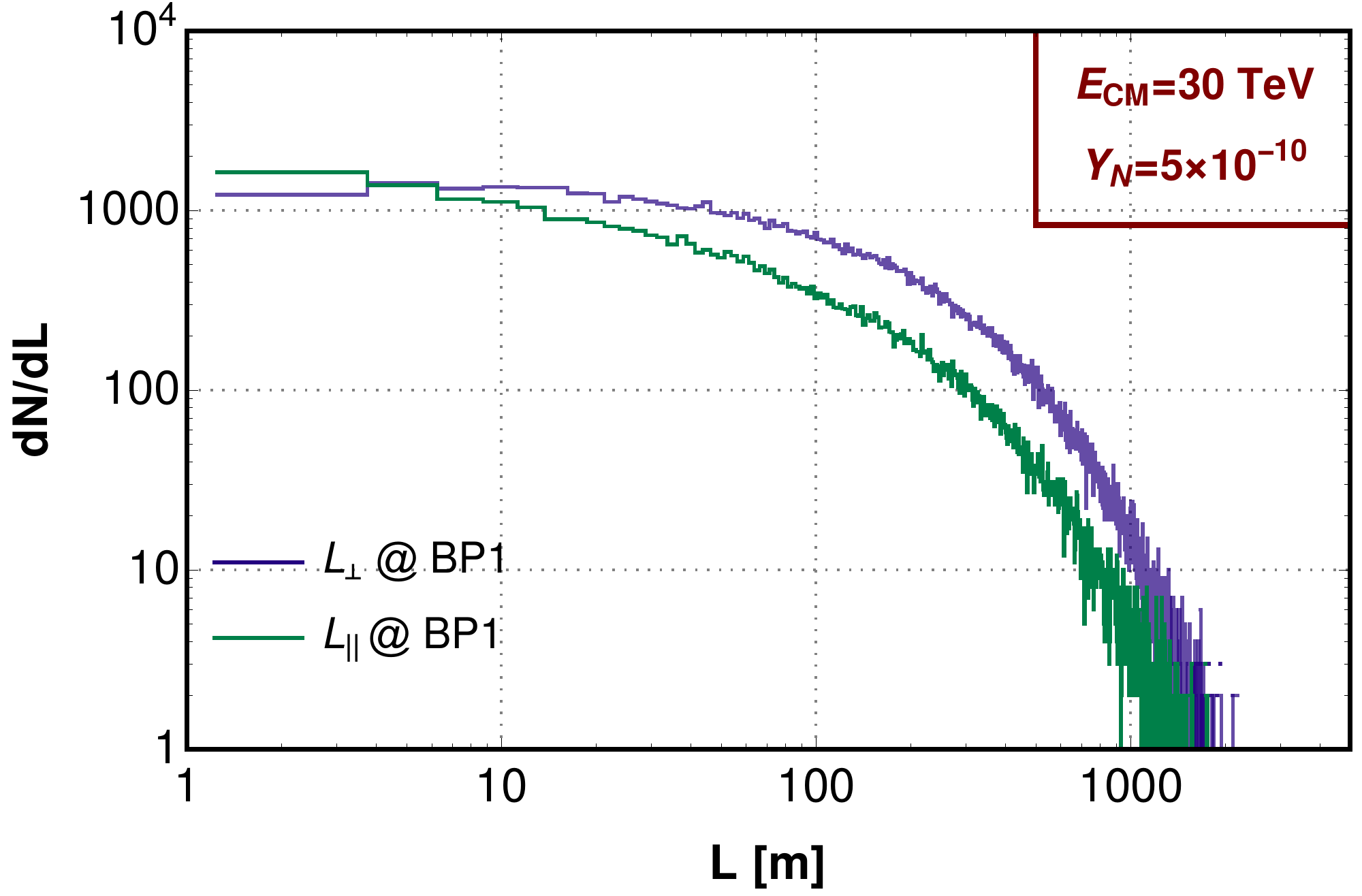}\label{}}}
		\caption{Displaced total transverse ($L_\perp$ in purple) and longitudinal ($L_\parallel$ in green) decay length distributions for the $SU(2)_L$ triplet heavy fermions $N^{\pm}$, at the muon collider with the centre of mass energies of 3.5\,TeV (a, b, c), 14\,TeV (d, e, f) and 30\,TeV (g, h, i) for BP1 ($M_N=1$\,TeV). Yukawa coupling $Y_N= 5\times 10^{-7},\, 1\times 10^{-8}$  are used in (a, d, g) and in (b, e, h), respectively, whereas the third column (c, f, i)  depicts  $Y_N=5\times 10^{-10}$. }\label{CompDecay_mu}
	\end{center}
\end{figure}

Displaced decays can also be observed at muon collider by observing charged tracks from $N^\pm$ as they fly a certain distance before its decay.  In \autoref{CompDecay_mu}, we describe the longitudinal $L_{||}$ (in green) and transverse $L_{\perp}$ (in purple) displaced decay length distributions for the charged $SU(2)_L$ triplet heavy fermions $N^{\pm}$ coming   from $\mu^+ \mu^- \to N^+N^-$  for the centre of  mass energies of  3.5, 14 and 30 TeV for BP1 ($M_N=1\, \rm TeV$). Due to enhanced transverse boost contrary to LHC as mentioned before, the total transverse displaced decay lengths are relatively inflated. Column wise they describe the cases for three   different Yukawa couplings  $5\times 10^{-7},\, 1\times 10^{-8}, \, 5\times 10^{-10}$.  In the first column corresponds to $Y_N= 5\times 10^{-7}$, where the maximum transverse decay length  is around 7 mm (purple curve) at the centre of mass energy 3.5 TeV, 
which  is enhanced due to the boost effect to  25 mm and 70 mm at the centre of mass energies of 14 and 30 TeV, respectively. $Y_N=1\times 10^{-8}, \, 5\times 10^{-10}$ are used in the second and third columns to represent the larger reaches, where we expect two recoils, the first one due to the dominant decay mode of $N^\pm \to N^0 \pi^\pm$, while the second one comes from the decay of $N^0$ as discussed before. The total displaced decay lengths coming from both recoils are depicted in \autoref{CompDecay_mu}. The first one  is for the  $N^\pm \to N^0 \pi^\pm$, which is $\mathcal{O}(1)$ cm, whereas the second one is $\mathcal{O}(0.1-1000)$ m depending on the Yukawa couplings, coming from  $N^0$ decay. It is evident that only for $Y_N=5\times 10^{-10}$ (third column), the total transverse decay length falls in the range of  $\gtrsim 100$ m for all three energies, while for the rest, we observe the signatures within $10$ m range.

\begin{table*}[h]	
	\begin{center}
		\hspace*{-1cm}
		\renewcommand{\arraystretch}{1.3}
		\begin{tabular}{|c|c|c|c|c|c|c|}
			\cline{2-7}
			\multicolumn{1}{c|}{}&\multicolumn{6}{c|}{Number of events with $Y_N=5\times 10^{-10}$.}\\
			\cline{2-7}
			\multicolumn{1}{c|}{}&  
			\multicolumn{2}{c|}{$E_{CM}=3.5$\,TeV}&\multicolumn{2}{c|}{$E_{CM}=14$\,TeV} &\multicolumn{2}{c|}{$E_{CM}=30$\,TeV}\\
			\cline{2-7}
			\multicolumn{1}{c|}{}&\multicolumn{1}{c|}{BP1}&\multicolumn{1}{c|}{BP2}&\multicolumn{1}{c|}{BP1}&\multicolumn{1}{c|}{BP2}&\multicolumn{1}{c|}{BP1}&\multicolumn{1}{c|}{BP2}\\ 
			\hline
			\multirow{1}{*}{$1\,\rm{mm}\leq L_{\perp} \leq 10\,\rm{m}$} & 10522.6 & 11966.2 & 2308.6 & 4369.3 & 784.4  & 1551.6 \\
			\hline 	
			$10\,\rm{m} < L_{\perp} \leq 100\,\rm{m}$  & 13804.6 & 4424.0 & 8594.8 & 9709.1 & 4033.6 & 5662.9 \\	
			\hline \hline
			\multirow{1}{*}{$1\,\rm{mm}\leq L_{||} \leq 10\,\rm{m}$} & 9314.5 & 11214.3 & 2154.0  & 3786.1 & 765.6 & 1442.0 \\
			\hline 	
			$10\,\rm{m}\leq L_{||} \leq 100\,\rm{m}$ & 7570.7  & 1330.2 & 6058.3 & 6503.1 & 2964.9 & 3978.9  \\	
			\hline 
		\end{tabular}
		\caption{The number of events within the range 1\,mm - 10\,m and 10\,m - 100\,m in the transverse ($L_\perp$) and longitudinal ($L_\parallel$) direction for the benchmark points at the centre of mass energies of 3.5, 14 and 30\,TeV with the integrated luminosities of  ($\mathcal{L}_{\text{int}}$=) 1000\,fb$^{-1}$, 10000\,fb$^{-1}$ and 30000\,fb$^{-1}$, respectively for $Y_N=5\times 10^{-10}$. }  \label{LengthTabmuon}
	\end{center}	
\end{table*}


In \autoref{LengthTabmuon}, we present the number of events for the longitudinal and transverse decay lengths for the benchmark points for the centre of mass energies of 3.5, 14 and 30\,TeV at the integrated luminosity ($\mathcal{L}_{\text{int}}$=) 1000\,fb$^{-1}$, 10000\,fb$^{-1}$ and 30000\,fb$^{-1}$, respectively with $Y_N=5\times 10^{-10}$. We separate two different regions, 1mm-10 m and 10-100 m, for those event numbers. For other Yukawa couplings chosen earlier, the events lie within the 10 m range. One  very interesting feature to observe is that  due to larger transverse  boost the event number for the transverse decay length are more unlike at the LHC/FCC (\autoref{MathuslaTabpp}). 

\subsection{Results}\label{resulsm}

\begin{table*}[h]	
	\begin{center}
		\renewcommand{\arraystretch}{1.3}
		\begin{tabular}{||c||c|c|c||c|c|}
			\cline{1-6}
			Center of mass & \multirow{2}{*}{Topologies} & 
			\multicolumn{2}{c||}{$Y_N=5\times 10^{-7}$}  & 
			\multicolumn{2}{c|}{$Y_N=1\times 10^{-8}$}\\
			\cline{3-6}
			energy&	&\multicolumn{1}{c|}{BP1}&\multicolumn{1}{c||}{BP2}&\multicolumn{1}{c|}{BP1}&\multicolumn{1}{c|}{BP2}\\ 
			\hline
			\multirow{6}{*}{3.5\,TeV}
			&	$4b + 1\ell$ & 7.3  & 4.0  & 13.0 & 6.6 \\	
			\cline{2-6}
			&	$4b+\rm{OSD}$  & 16.7 & 7.4  & 3.5 & 4.8 \\
			\cline{2-6}
			&	$2b + 4j + 1\ell $  & 168.5 & 89.0  & 126.1 & 76.9 \\	
			\cline{2-6}
			&	$2b + 4j + 2\ell $  & 69.8 &  35.3 & 40.6 & 18.3 \\
			\cline{2-6}
			&	$2b + 2j + 2\ell $  & 354.6 & 269.6  & 137.3 & 137.2 \\
			\cline{2-6}
			&	$2b + 2j + 3\ell $   & 4.6  & 3.3  & 27.1 & 18.4 \\	
			\hline \hline
			\multirow{6}{*}{14\,TeV}
			&	$4b + 1\ell$  & 1.0 & 0.6  & 0.2 & 0.8 \\	
			\cline{2-6}
			&	$4b+\rm{OSD}$  & 0.2 & 0.4  & 0.0 & 0.4 \\
			\cline{2-6}
			&	$2b + 4j + 1\ell $  & 22.3 &  35.9 & 14.0 & 23.6\\	
			\cline{2-6}
			&	$2b + 4j + 2\ell $  & 5.3 & 6.7  & 2.3  & 4.7 \\
			\cline{2-6}
			&	$2b + 2j + 2\ell $  & 121.3 & 256.8 & 39.2 & 110.0 \\
			\cline{2-6}
			&	$2b + 2j + 3\ell $  & 0.8 & 0.7 & 3.2 & 3.8\\			
			\hline \hline
			\multirow{6}{*}{30\,TeV}
			&	$4b + 1\ell$  & 0.0 & 0.1  & 0.0 & 0.0 \\	
			\cline{2-6}
			&	$4b+\rm{OSD}$  & 0.0 & 0.0  & 0.0 & 0.0 \\
			\cline{2-6}
			&	$2b + 4j + 1\ell $  & 3.1 & 7.1 & 1.8 & 4.4 \\	
			\cline{2-6}
			&	$2b + 4j + 2\ell $  & 0.8  & 0.9 & 0.2 & 0.7 \\
			\cline{2-6}
			&	$2b + 2j + 2\ell $  & 13.2 & 42.1 & 5.7 & 15.6 \\
			\cline{2-6}
			&	$2b + 2j + 3\ell $  & 0.1 & 0.7  & 0.8 & 1.1 \\		
			\hline			
		\end{tabular}
		\caption{Number of events for finalstate topologies containing at least two $b$-jets for the benchmark points with the centre of mass energies of 3.5\,TeV, 14\,TeV and 30\,TeV at the muon collider with integrated luminosities of ($\mathcal{L}_{\text{int}}=$) 1000\,fb$^{-1}$, 10000\,fb$^{-1}$, 30000\,fb$^{-1}$, respectively with $Y_N=5\times 10^{-7}$ (columns 3, 4) and  with  $Y_N=1\times 10^{-8}$ (columns 5, 6). }  \label{muonCol}
	\end{center}	
\end{table*}


In this section, we describe the event numbers for the finalstates coming from the $N^\pm$ decay and at least one Higgs bosons which is tagged via $b$-jets. Thus in \autoref{muonCol} (columns 3 and 4) we describe the finalstates with two and four $b$-jets for the benchmark points at the  centre of mass energies 3.5\,TeV, 14\,TeV and 30\,TeV at the muon collider with integrated luminosity ($\mathcal{L}_{\text{int}}=$) 1000\,fb$^{-1}$, 10000\,fb$^{-1}$, 30000\,fb$^{-1}$, respectively for $Y_N=5\times 10^{-7}$.  The $b$-tagging efficiency is taken to be maximum of 85\%  and dynamically set while simulating the events via secondary vertex reconstruction as explained earlier \cite{CMS:2012jki,CMS:2017wtu}.  The number of 4$b$-jets events are one order of magnitude less than that of 2$b$-jets events, where they are ideally coming via $N^\pm \to h\ell^\pm$, from both the legs. We see that the number of events for various finalstates reduce from 3.5 TeV to 30 TeV as we go more and more off-shell for all the benchmark points. One interesting point to note is that, in case of muon collider, the most dominant final state is $2b+2j+2\ell$,  which is $2b+4j+1\ell$ for the LHC. This is due to the fact that  at  muon collider, unlike LHC, there is no initial state QCD radiation, inflating  the chance of more isolated leptons.  All the finalstates described here are displaced ones and the event number $\geq 3$ can be probed at 95\% confidence level for the background-less signals \cite{pdg,pdg1}.

Similarly, we present the event numbers for $Y_N=10^{-8} $ in \autoref{muonCol} (columns 4 and 5) for the above mentioned finnalstates for the benchmark points with the centre of mass energies of 3.5\,TeV, 14\,TeV and 30\,TeV at an integrated luminosities of ($\mathcal{L}_{\text{int}}=$) 1000\,fb$^{-1}$, 10000\,fb$^{-1}$, 30000\,fb$^{-1}$, respectively.  The multi-lepton finalstates suffer  compared to the case of $Y_N=5 \times 10^{-7}$ due to lost charged lepton by the new decay mode of $N^\pm \to N^0 \pi^\pm$, which is very similar to that of the LHC results in \autoref{ppColYN2}.

\begin{figure}[hbt]
	\begin{center}
		\hspace*{-0.5cm}
		\includegraphics[width=0.46\linewidth,angle=-0]{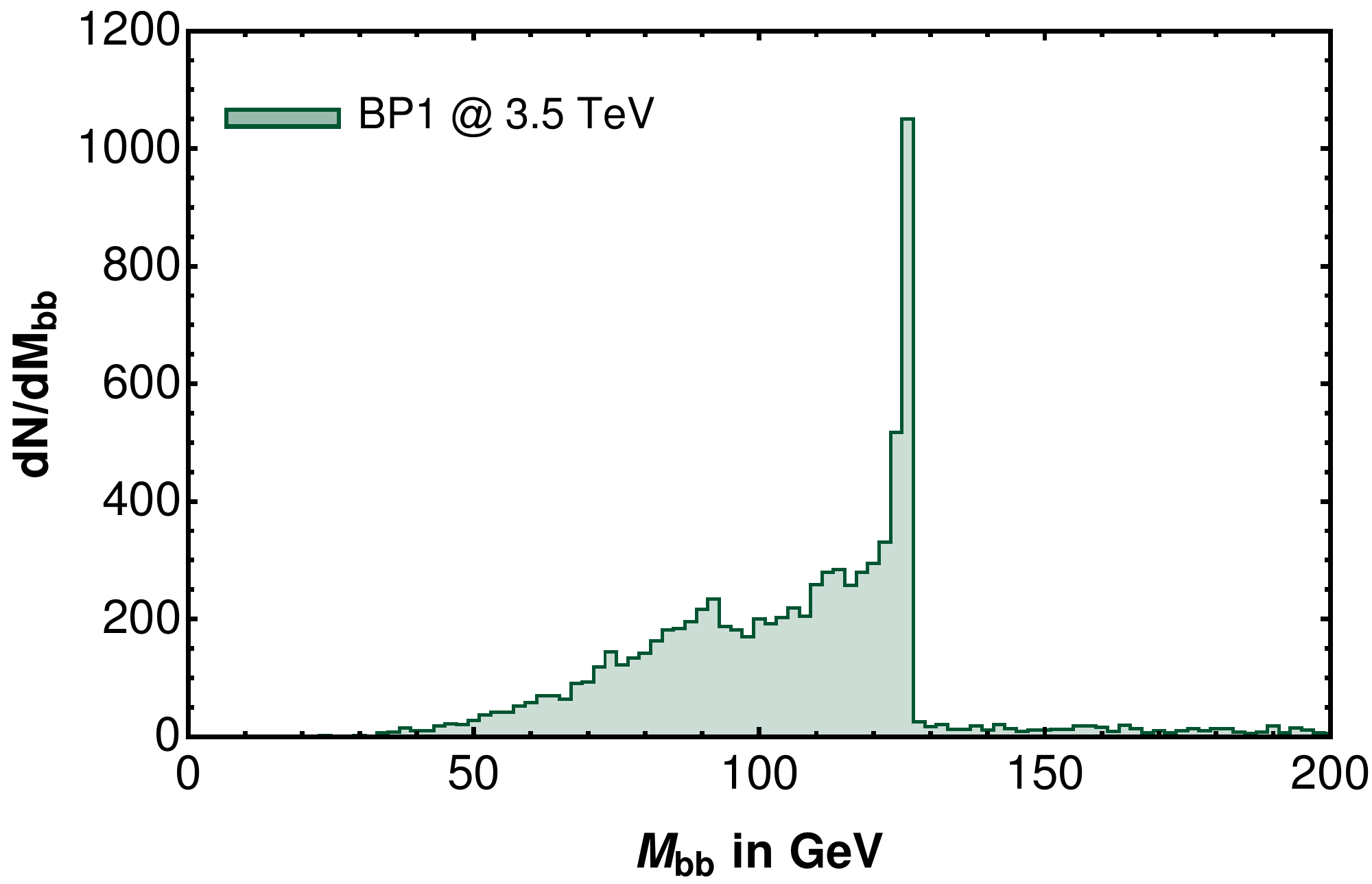}\label{}\caption{Di-$b$-jet ($M_{bb}$) invariant mass distributions for BP1 ($M_N=1\,$TeV) at the  muon collider  with $Y_N=5\times 10^{-7}$ and centre of mass energy 3.5\,TeV. The dominant peak around the SM Higgs boson mass (125.5\,GeV) along with a small peak around the $Z$ boson mass (91.1\,GeV) are visible.} \label{bblInvM_mu}
	\end{center}
\end{figure}


\begin{table}[hbt]	
	\begin{center}
		\hspace*{-1cm}
		\renewcommand{\arraystretch}{1.3}
		\begin{tabular}{|c||c|c||c|c||c|c|}
			\cline{1-7}
			\multirow{2}{*}{Topologies} & 
			\multicolumn{2}{c||}{$E_{CM}=3.5$\,TeV}&\multicolumn{2}{c||}{$E_{CM}=14$\,TeV} &\multicolumn{2}{c|}{$E_{CM}=30$\,TeV}\\
			\cline{2-7}
			&\multicolumn{1}{c|}{BP1}&\multicolumn{1}{c||}{BP2}&\multicolumn{1}{c|}{BP1}&\multicolumn{1}{c||}{BP2} &\multicolumn{1}{c|}{BP1}&\multicolumn{1}{c|}{BP2}\\ 
			\hline
			$2b$ \& $|M_{bb}-125.5|\leq 10$\,GeV & 354.2 & 254.4 & 50.6 & 57.5 & 13.7 & 14.0 \\	
			\hline\hline
			$4b$ \& $|M_{bb}-125.5|\leq 10$\,GeV  & 3.2 & 3.1 & 0.2 & 0.0 & 0.0 & 0.1 \\	
			\hline
		\end{tabular}
		\caption{Number of events for single Higgs boson and di-Higgs boson mass peaks after the window cuts around the peak of the invariant mass  distributions for the benchmark points at the muon collider with the centre of mass energies of 3.5\,TeV, 14\,TeV and 30\,TeV at the integrated luminosities of ($\mathcal{L}_{\text{int}}$=) 1000\,fb$^{-1}$, 10000\,fb$^{-1}$ and 30000\,fb$^{-1}$, respectively with $Y_N=5\times 10^{-7}$. }  \label{HiggsTabMuon}
	\end{center}	
\end{table}

Subsequently, we focus on the di-$b$-jet invariant mass reconstruction as shown in \autoref{bblInvM_mu} for BP1. We can see the two peaks coming from the Higgs and $Z$ bosons distinctively. We impose the constraint of $|M_{bb}-125.5|\leq 10$ GeV, essential to reconstruct one Higgs boson mass peak. Following this approach, we tag at least one Higgs boson for the finalstate of $2b$ and two Higgs bosons for $4b$, which are listed in \autoref{HiggsTabMuon} for the centre of mass energies of 3.5\,TeV, 14\,TeV and 30\,TeV at the integrated luminosity ($\mathcal{L}_{\text{int}}$=) 1000\,fb$^{-1}$, 10000\,fb$^{-1}$ and 30000\,fb$^{-1}$, respectively for all benchmark points with $Y_N=5\times 10^{-7}$. From the event numbers we realise that only one Higgs boson reconstruction is feasible.
\begin{figure}[hbt]
	\begin{center}
		\hspace*{-0.5cm}
		\mbox{
			\subfigure[]{\includegraphics[width=0.42\linewidth,angle=-0]{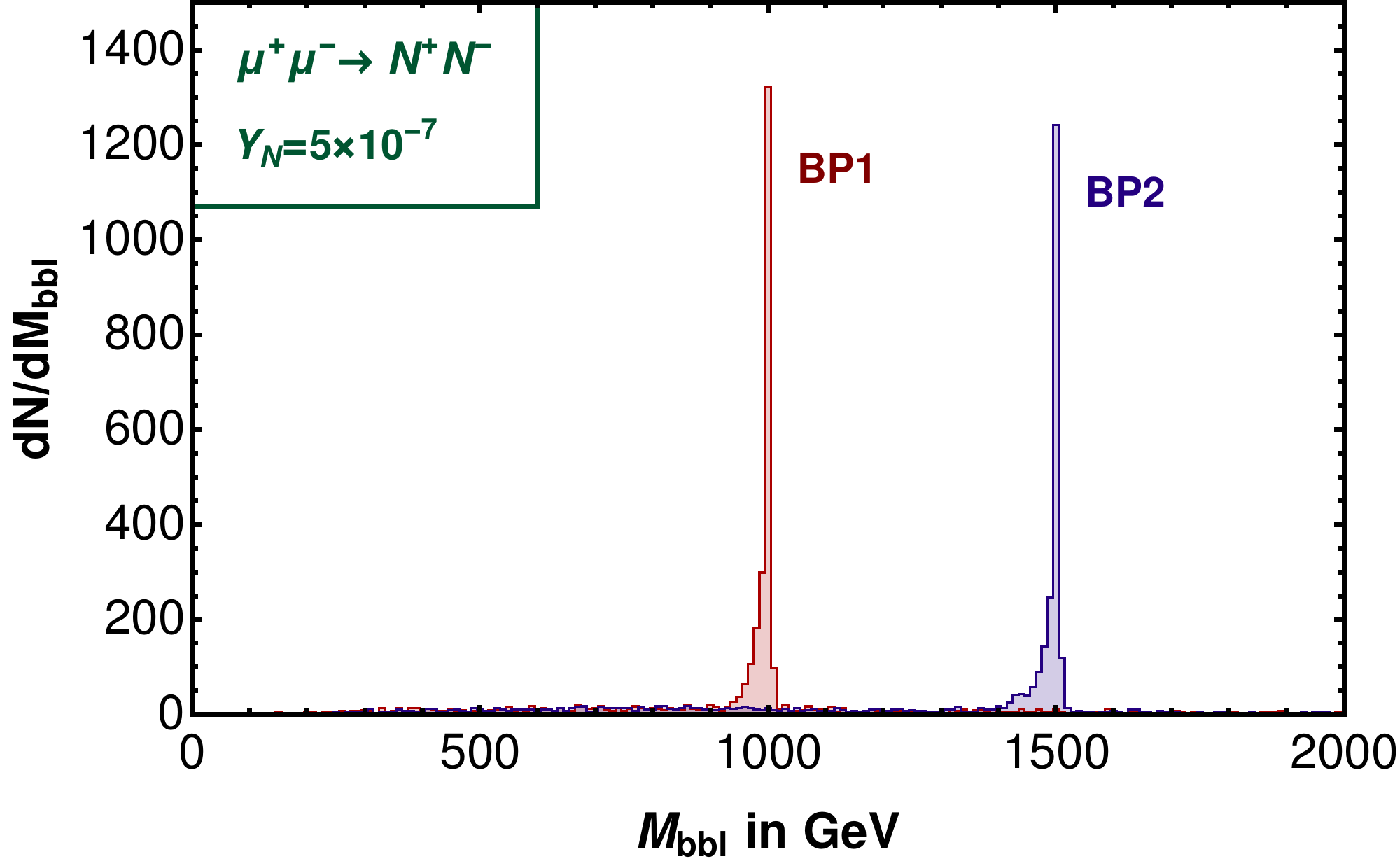}\label{}}\quad
			\subfigure[]{	\includegraphics[width=0.42\linewidth,angle=-0]{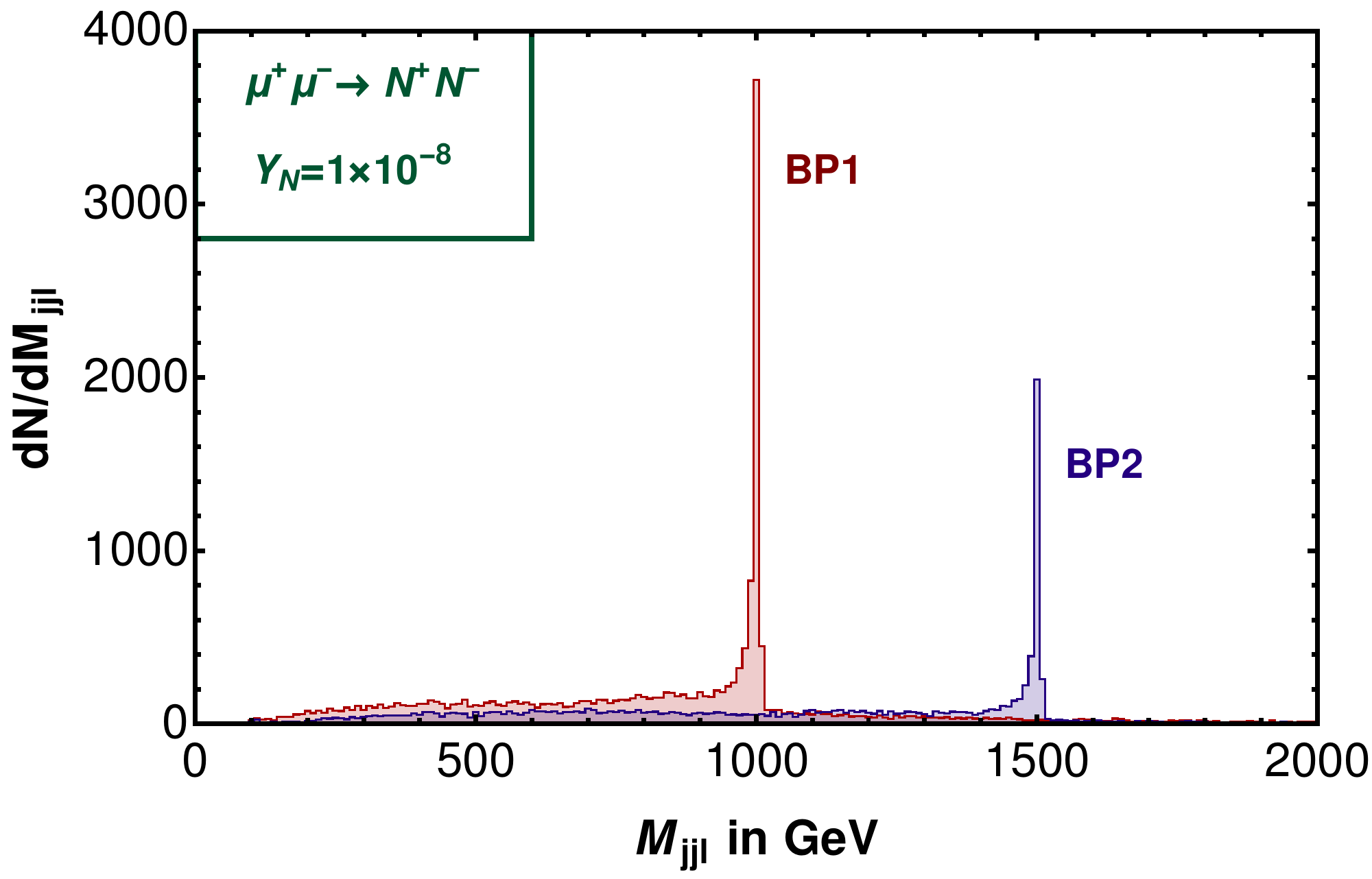}}}
		\caption{(a) Di-bjej-mono-lepton ($M_{bb\ell}$) invariant mass  distribution with $Y_N=5\times 10^{-7}$, (b) di-jet-mono-lepton ($M_{jj\ell}$) invariant mass distributions with $Y_N=1\times 10^{-8}$ for the benchmark points at the  muon collider  with the centre of mass energy 3.5\,TeV. } \label{jjlInvM_mu}
	\end{center}
\end{figure}


In \autoref{jjlInvM_mu}(a), we plot the $bb\ell$ invariant mass to reconstruct the $N^\pm$ mass for the benchmark points. We can clearly see the two peaks for the two benchmark points. The corresponding number of events are projected in \autoref{NeutTabmuon} for the  centre of mass energies of 3.5\,TeV, 14\,TeV and 30\,TeV at the integrated luminosity ($\mathcal{L}_{\text{int}}$=) 1000\,fb$^{-1}$, 10000\,fb$^{-1}$ and 30000\,fb$^{-1}$,  respectively with $Y_N=5\times 10^{-7}$.  Thus the visible mode leading to $bb\ell$ can successfully reconstruct the Type-III fermion masses. 
\begin{table*}[h]	
	\begin{center}
		\hspace*{-1cm}
		\renewcommand{\arraystretch}{1.2}
		\begin{tabular}{|c|c|c|c|c|}
			\hline
			Benchmark& \multirow{2}{*}{Topologies} & \multicolumn{3}{c|}{Centre of mass energy} \\
			\cline{3-5}
			points &  & 
			3.5\,TeV  & 14\,TeV  & 30\,TeV \\
			\hline
			\multirow{2}{*}{BP1} & $2b+ 1\ell$ \& & \multirow{2}{*}{238.3} & \multirow{2}{*}{28.2} & \multirow{2}{*}{1.8} \\	
			& $|M_{bb\ell}-1000.0|\leq 10$\,GeV & & & \\
			\hline\hline
			\multirow{2}{*}{BP2} & $2b+ 1\ell$ \& & \multirow{2}{*}{169.7} & \multirow{2}{*}{27.8} & \multirow{2}{*}{6.3} \\
			& $|M_{bb\ell}-1500.0|\leq 10$\,GeV & & & \\
			\hline
		\end{tabular}
		\caption{The number of events in $M_{bb\ell}$ distribution after the window cut around the mass peak for the benchmark points at the muon collider with the  centre of mass energies  of 3.5\,TeV, 14\,TeV and 30\,TeV at the integrated luminosities of ($\mathcal{L}_{\text{int}}$=) 1000\,fb$^{-1}$, 10000\,fb$^{-1}$ and 30000\,fb$^{-1}$, respectively with $Y_N=5\times 10^{-7}$. }  \label{NeutTabmuon}
	\end{center}	
\end{table*}


In \autoref{jjlInvM_mu}(b), we reconstruct the $N^0$ invariant mass for the case of  $Y_N=1\times 10^{-8}$, where $N^\pm \to N^0 \pi^\pm$ is the dominant decay mode. The $N^0 \to W^\pm \ell^\mp$ is first reconstructed via the di-jet invariant mass peak around $W^\pm$ mass by demanding $|M_{jj}-M_W|\leq 10$ GeV. Later events are collected around that mass window to constitute $M_{jj\ell}$.   It is evident from  \autoref{jjlInvM_mu}(b)  that $M_{jj\ell}$ peaks around $M_N$ for the benchmark points at the centre of mass energy of 14 TeV.  The corresponding event numbers are presented in \autoref{Neu0Tabmuon} for the benchmark points with the  centre of mass energies  of 3.5\,TeV, 14\,TeV and 30\,TeV at the integrated luminosities of ($\mathcal{L}_{\text{int}}$=) 1000\,fb$^{-1}$, 10000\,fb$^{-1}$ and 30000\,fb$^{-1}$, respectively. We see similar effects of boost reducing the jet multiplicity for BP2 (see \autoref{Neut0Tabpp}) and thus reducing the number of events for $M_{jj\ell}$. For higher energies of 14 and 30 TeV such effects are negligible as the off-shell cross-section remain very low.
\begin{table*}[h]	
	\begin{center}
		\hspace*{-1cm}
		\renewcommand{\arraystretch}{1.2}
		\begin{tabular}{|c|c|c|c|c|}
			\hline
			Benchmark& \multirow{2}{*}{Topologies} & \multicolumn{3}{c|}{Centre of mass energy} \\
			\cline{3-5}
			points &  & 
			3.5\,TeV  & 14\,TeV  & 30\,TeV \\
			\hline
			\multirow{2}{*}{BP1} & $2j+ 1\ell$ \& & \multirow{2}{*}{698.3} & \multirow{2}{*}{81.4} & \multirow{2}{*}{11.5} \\	
			& $|M_{jj\ell}-1000.0|\leq 10$\,GeV & & & \\
			\hline\hline
			\multirow{2}{*}{BP2} & $2j+ 1\ell$ \& & \multirow{2}{*}{279.5} & \multirow{2}{*}{79.7} & \multirow{2}{*}{15.7} \\
			& $|M_{jj\ell}-1500.0|\leq 10$\,GeV & & & \\
			\hline
		\end{tabular}
		\caption{The number of events in $M_{jj\ell}$ distribution after the window cut around the mass peak for the benchmark points at the muon collider with the  centre of mass energies  of 3.5\,TeV, 14\,TeV and 30\,TeV at the integrated luminosities of ($\mathcal{L}_{\text{int}}$=) 1000\,fb$^{-1}$, 10000\,fb$^{-1}$ and 30000\,fb$^{-1}$, respectively with $Y_N=1\times 10^{-8}$. }  \label{Neu0Tabmuon}
	\end{center}	
\end{table*}

\section{Reach plots in Yukawa versus mass plane}\label{reach}

\begin{figure*}[h]
	\begin{center}
		\hspace*{-0.6cm}
		\mbox{\subfigure[]{\includegraphics[width=0.45\linewidth,angle=-0]{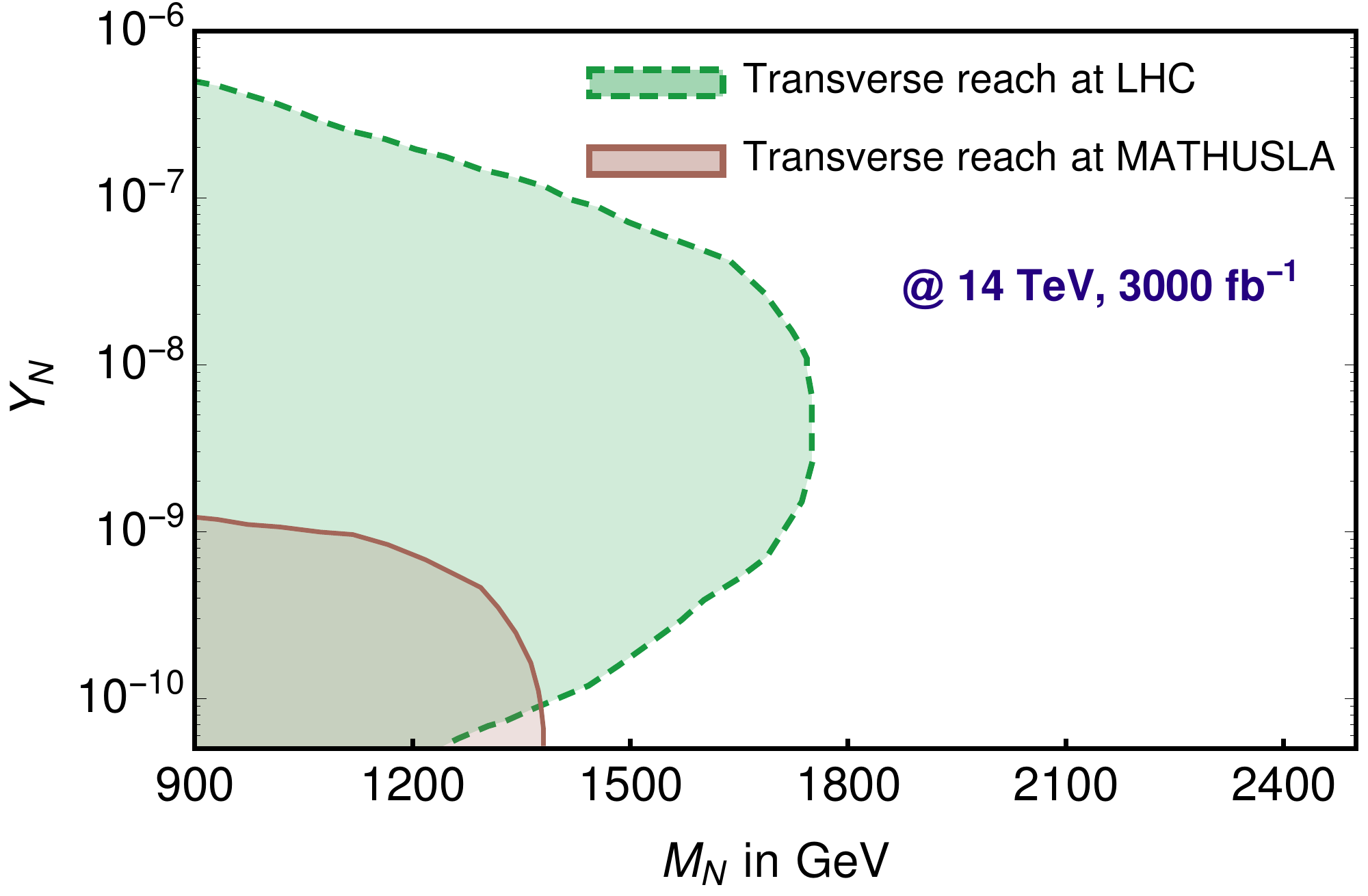}\label{}} \quad \quad
		\subfigure[]{\includegraphics[width=0.45\linewidth,angle=-0]{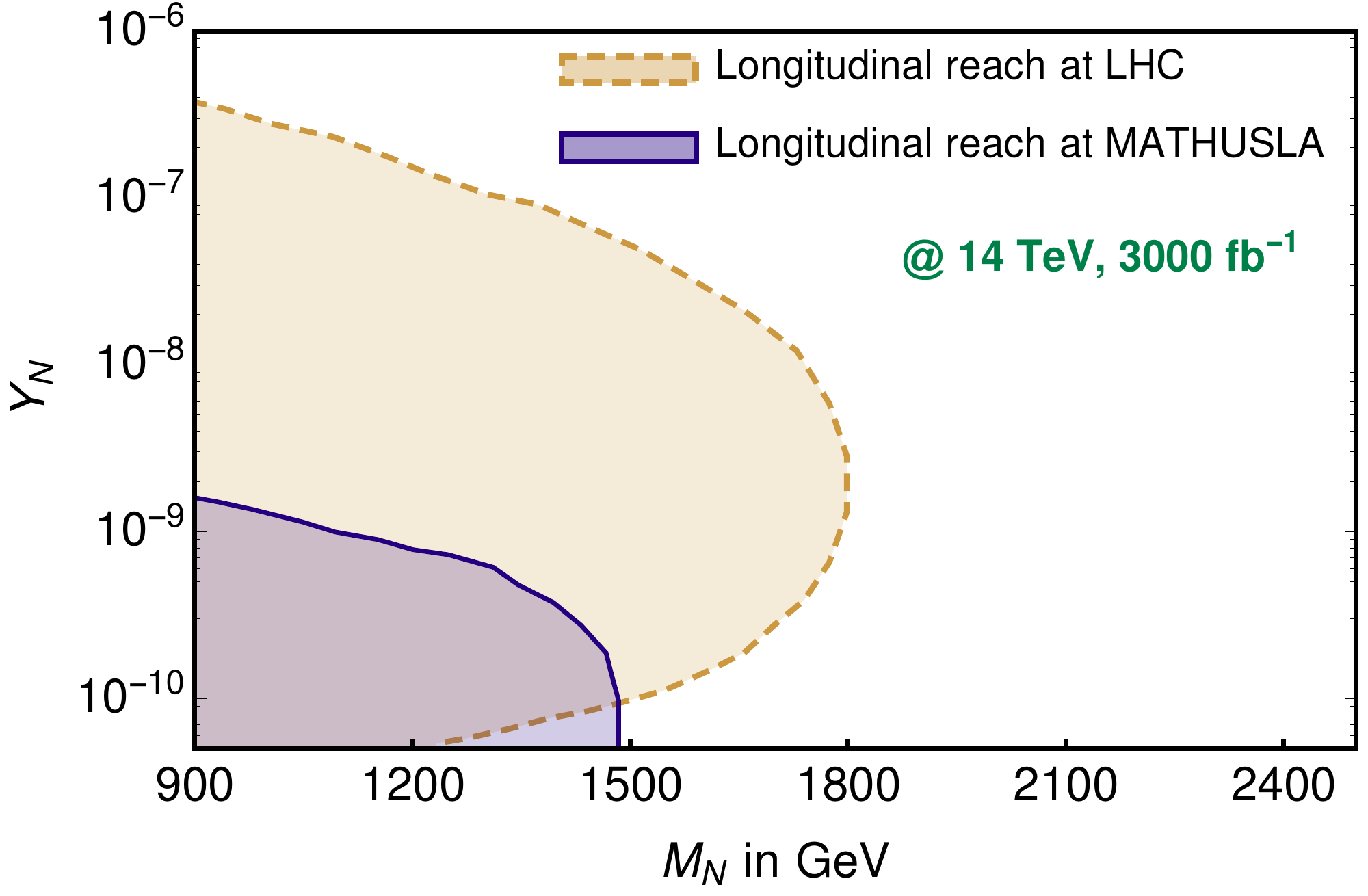}\label{}}}
		\hspace*{-0.1cm}
		\mbox{\subfigure[]{\includegraphics[width=0.47\linewidth,angle=-0]{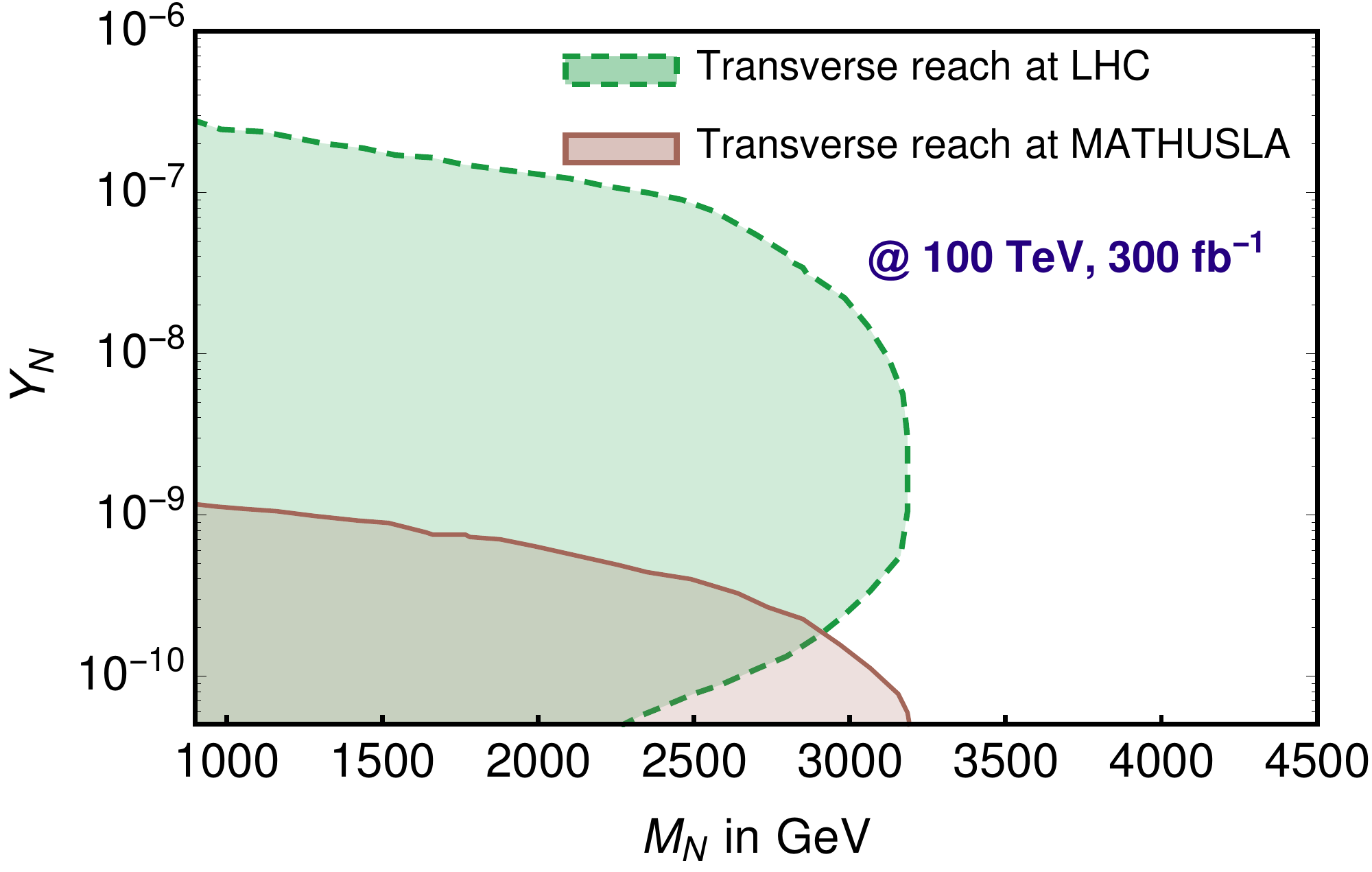}\label{}}\quad \quad 
		\subfigure[]{\includegraphics[width=0.47\linewidth,angle=-0]{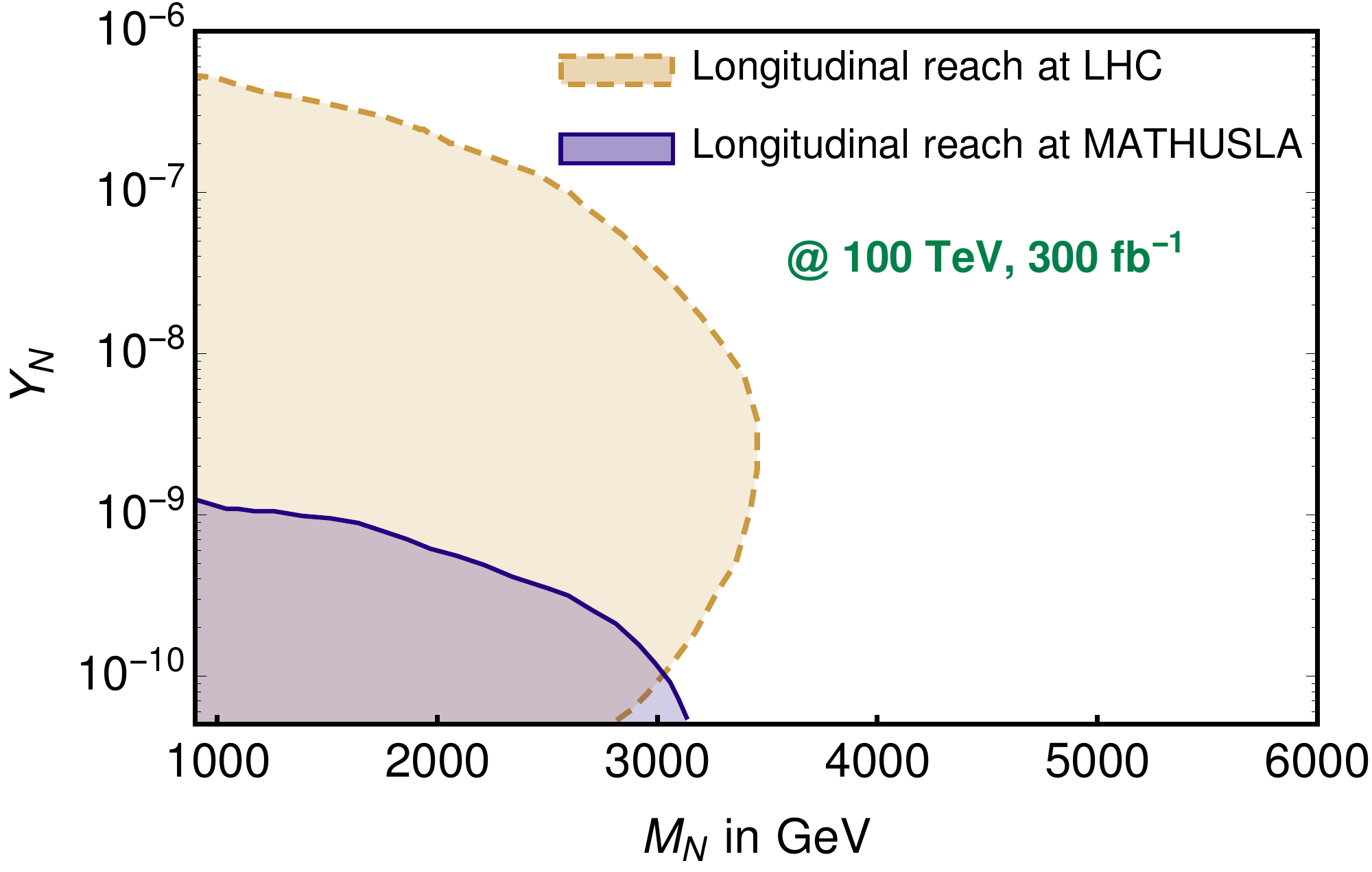}\label{}}}	
		\caption{Limits obtained via the inclusive measurements from $N^0N^\pm,\, N^+N^-$ productions for the Yukawa coupling versus Type-III fermion mass for the transverse (a, c) and the longitudinal (b, d) decay lengths for the finalstates containing at least one displaced Higgs boson with 14 (a, b), 100 (c, d) TeV centre of mass energies at the LHC/FCC with integrated luminosities of 3000 and 300 fb$^{-1}$ respectively.}  \label{pp_reach}
	\end{center}
\end{figure*}

In this section, we plot the regions with at least one displaced Higgs boson reconstructed from di-$b$-jet invariant mass using the window cut $|M_{bb}-125.5|\leq 10\,\rm GeV$  in Yukawa versus mass plane that can be probed at the LHC/FCC and at the muon collider. The regions are obtained from the zero background analysis  with a confidence level of 95\% at a given luminosity \cite{pdg,pdg1}.  \autoref{pp_reach} presents the regions in a plane of Type-III Yukawa coupling versus mass at the LHC with 14 TeV and 100 TeV centre of mass energies that can be probed at an integrated luminosities of 3000 and 300 fb$^{-1}$, respectively where QCD corrections to the cross-sections\cite{Ruiz:2015zca} are taken into account.  The bounds obtained here are exclusively based on the displaced decay signature as we present them in \autoref{pp_reach} in $Y_N - M_N$ plane.


In \autoref{pp_reach}(a), we see the bounds obtained from the transverse decay lengths at the CMS  and ATLAS (in light green) and MATHUSLA (in brown) at the centre of mass energy of 14 TeV with an integrated luminosity of  3000 fb$^{-1}$.  The corresponding longitudinal decay length bounds are presented in \autoref{pp_reach}(b).  The maximum  Type-III fermion mass  of 1.75 TeV, 1.8 TeV can be probed via the transverse and the longitudinal  displaced decay length at the LHC with centre of mass energy of 14 TeV with an integrated luminosity of 3000 fb$^{-1}$, respectively for the reconstruction of atleast one Higgs boson mass. The bounds at 100 TeV centre of mass energy are  3.2 TeV and 3.6 TeV,  respectively for the transverse and the longitudinal decay lengths at an integrated luminosity of 300 fb$^{-1}$. At larger luminosity of 3000 fb$^{-1}$  we can probe even higher mass range around 4.25\,TeV. Though it is very small, we observe slight enhancement on the mass bounds in the longitudinal mode as compared to the transverse one. Yukawa couplings in the  range of  $5\times 10^{-11}- 5 \times 10^{-7}$ can be probed if we consider the inclusive measurements of  LHC/FCC and MATHUSLA. The transverse and longitudinal reaches at the MATHUSLA are quite low as compared to CMS and ATLAS. Considering both the transverse and the  longitudinal decay lengths, we see that  Yukawa coupling $> 10^{-9}$  is out of the reach of MATHUSLA, however can be addressed at CMS and ATLAS.

\begin{figure}[hbt]
	\begin{center}
		\hspace*{-0.5cm}
		\mbox{\subfigure[]{\includegraphics[width=0.45\linewidth,angle=-0]{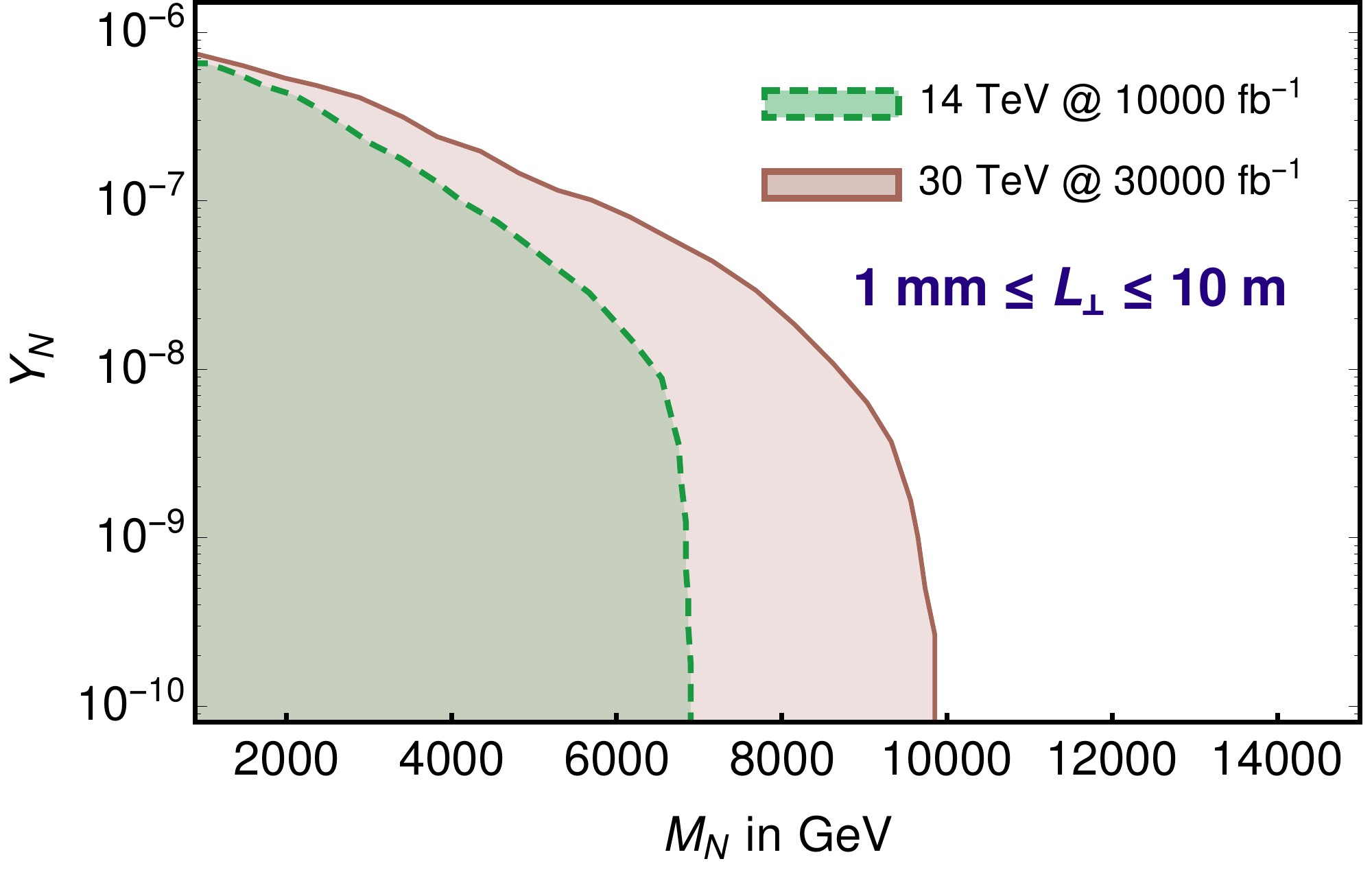}\label{}}\quad \quad
		\subfigure[]{\includegraphics[width=0.45\linewidth,angle=-0]{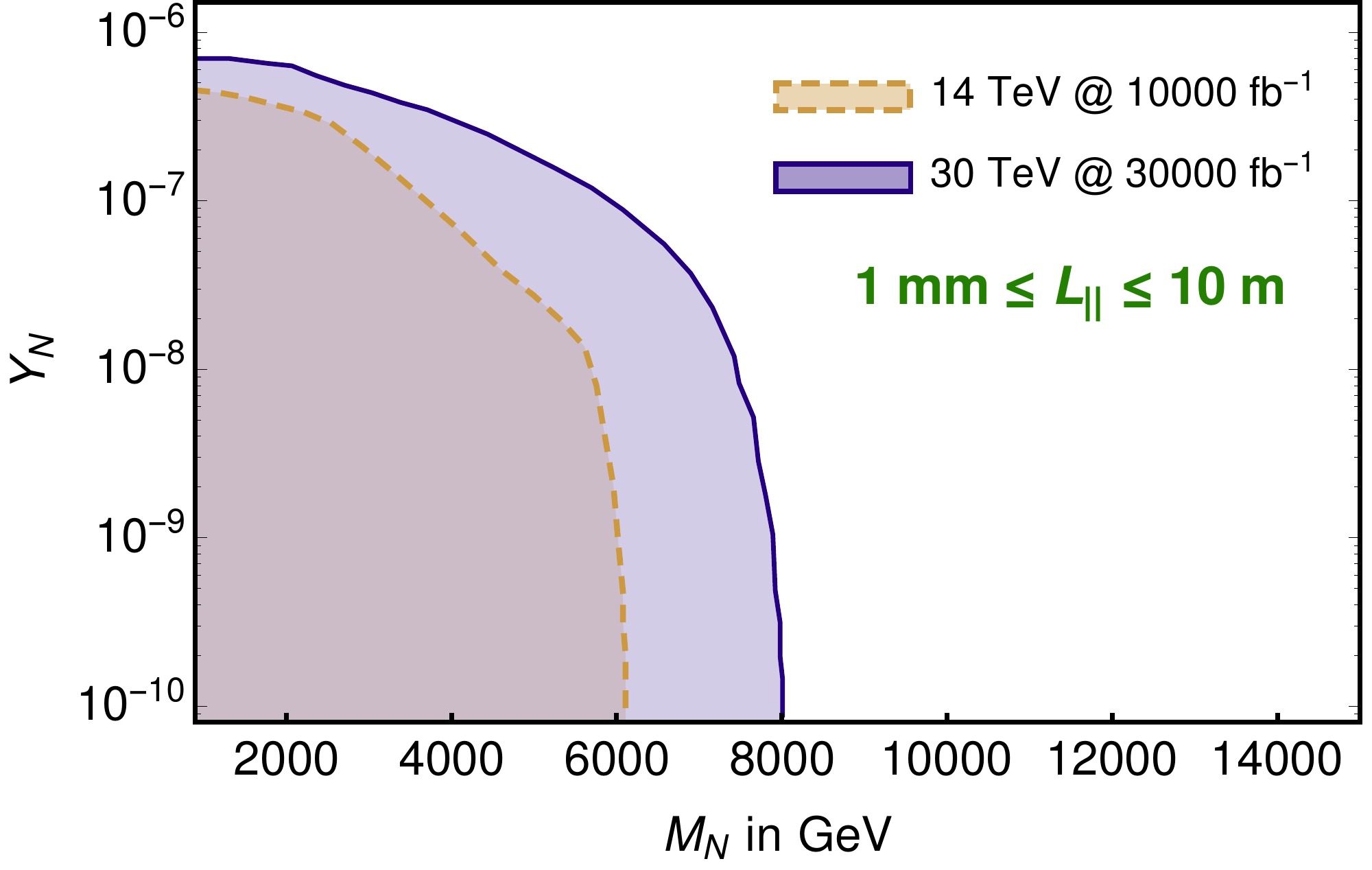}\label{}}}
		\hspace*{-0.5cm}
		\mbox{\subfigure[]{\includegraphics[width=0.45\linewidth,angle=-0]{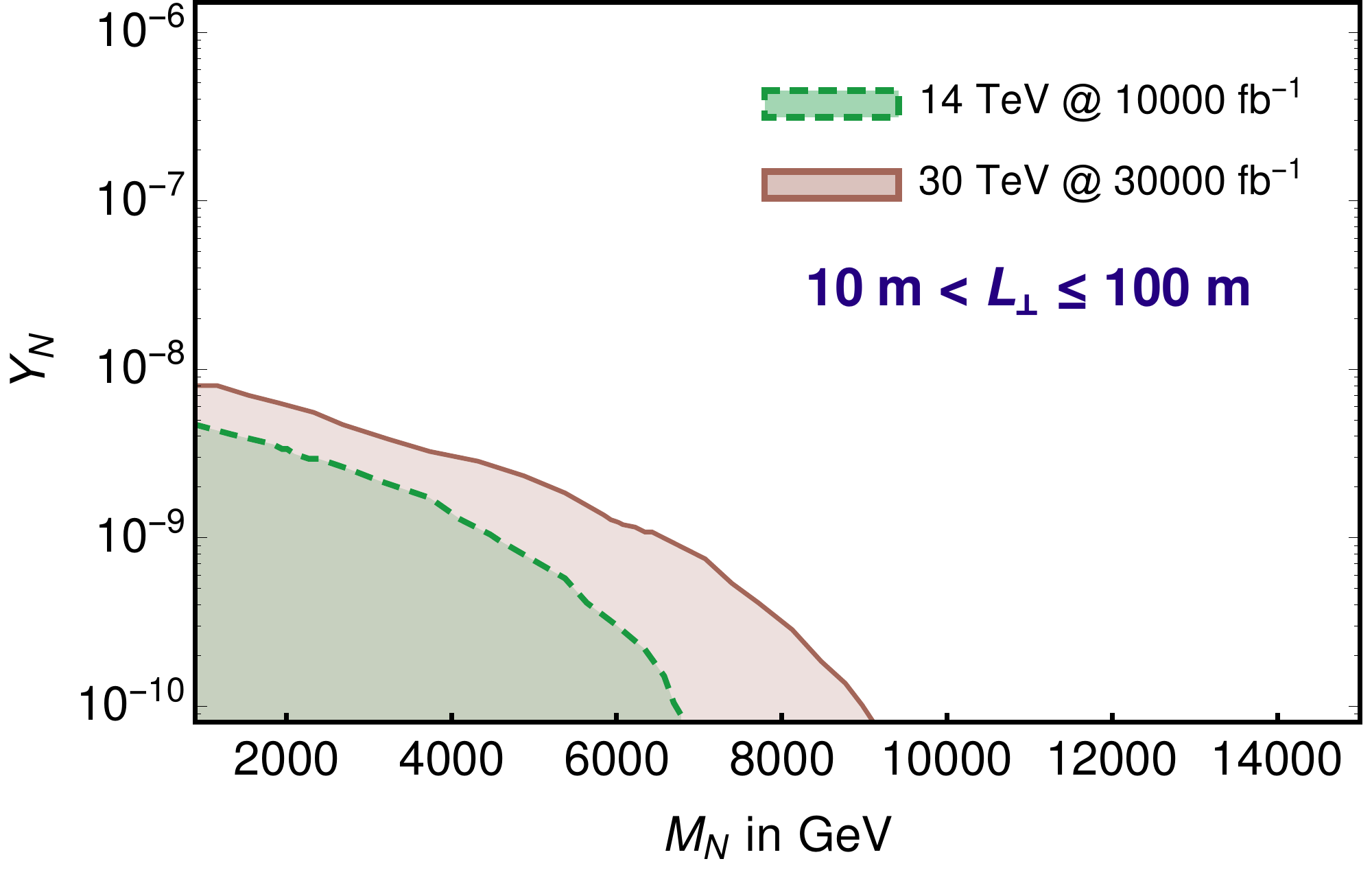}\label{}}\quad \quad
		\subfigure[]{\includegraphics[width=0.45\linewidth,angle=-0]{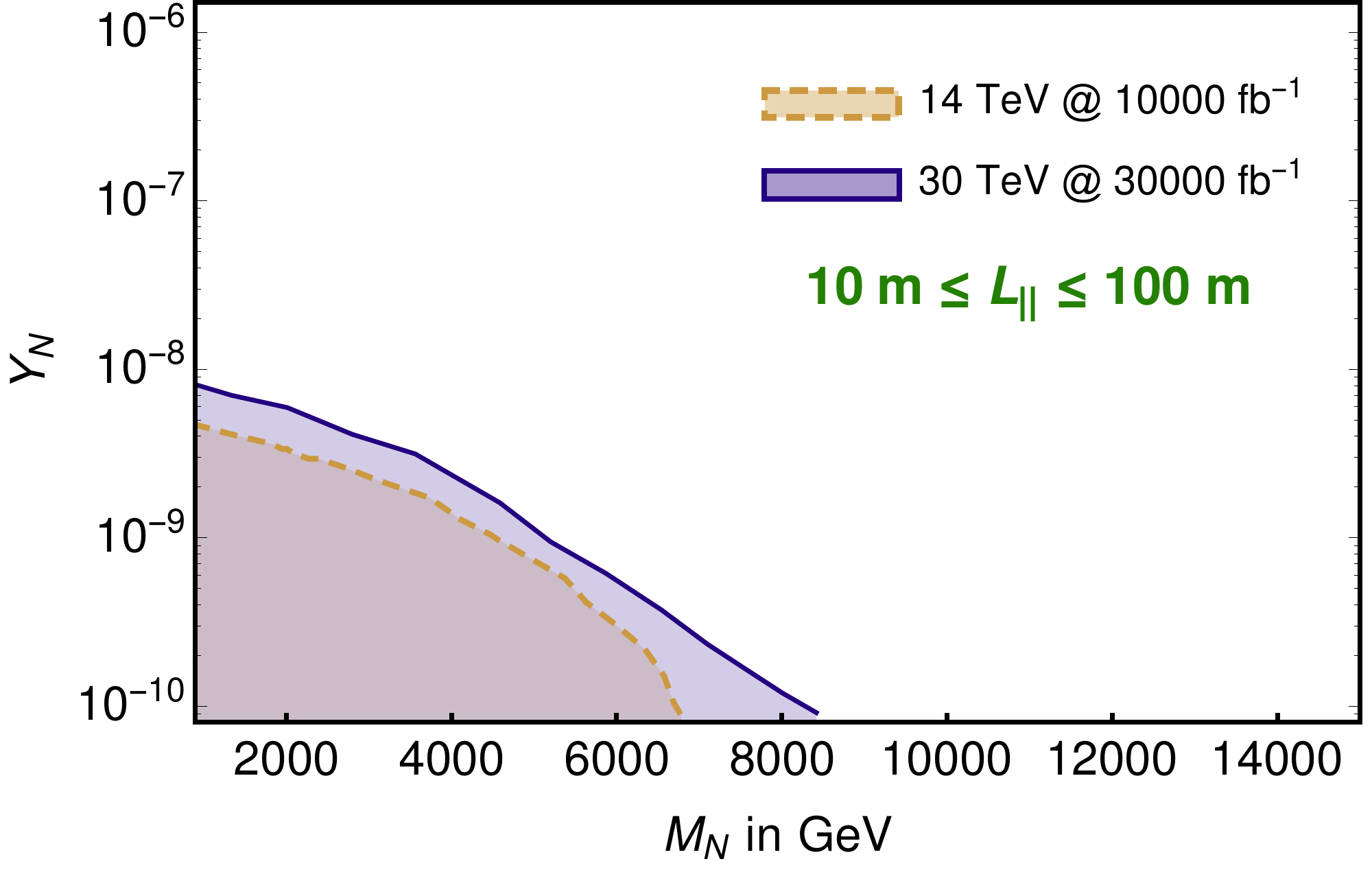}\label{}}}	
		\caption{Limits obtained for Yukawa coupling versus Type-III fermion mass via transverse (a, c) and longitudinal (b, d) decay lengths for the finalstates containing at least one displaced Higgs boson with the centre of mass energies of 14 and 30 TeV  at the muon collider with integrated luminosities of 10000 and 30000 fb$^{-1}$, respectively. (a, b) represent 1\,mm $\leq L_{\perp, \, \parallel}\leq $ 10\,m and (c, d) depict 10\,m $\leq L_{\perp, \, \parallel}\leq $ 100\,m regions.  }  \label{muon_reach}
	\end{center}
\end{figure}


\begin{figure*}[h]
	\begin{center}		
		\hspace*{-0.5cm}
		\mbox{\subfigure[]{\includegraphics[width=0.45\linewidth,angle=-0]{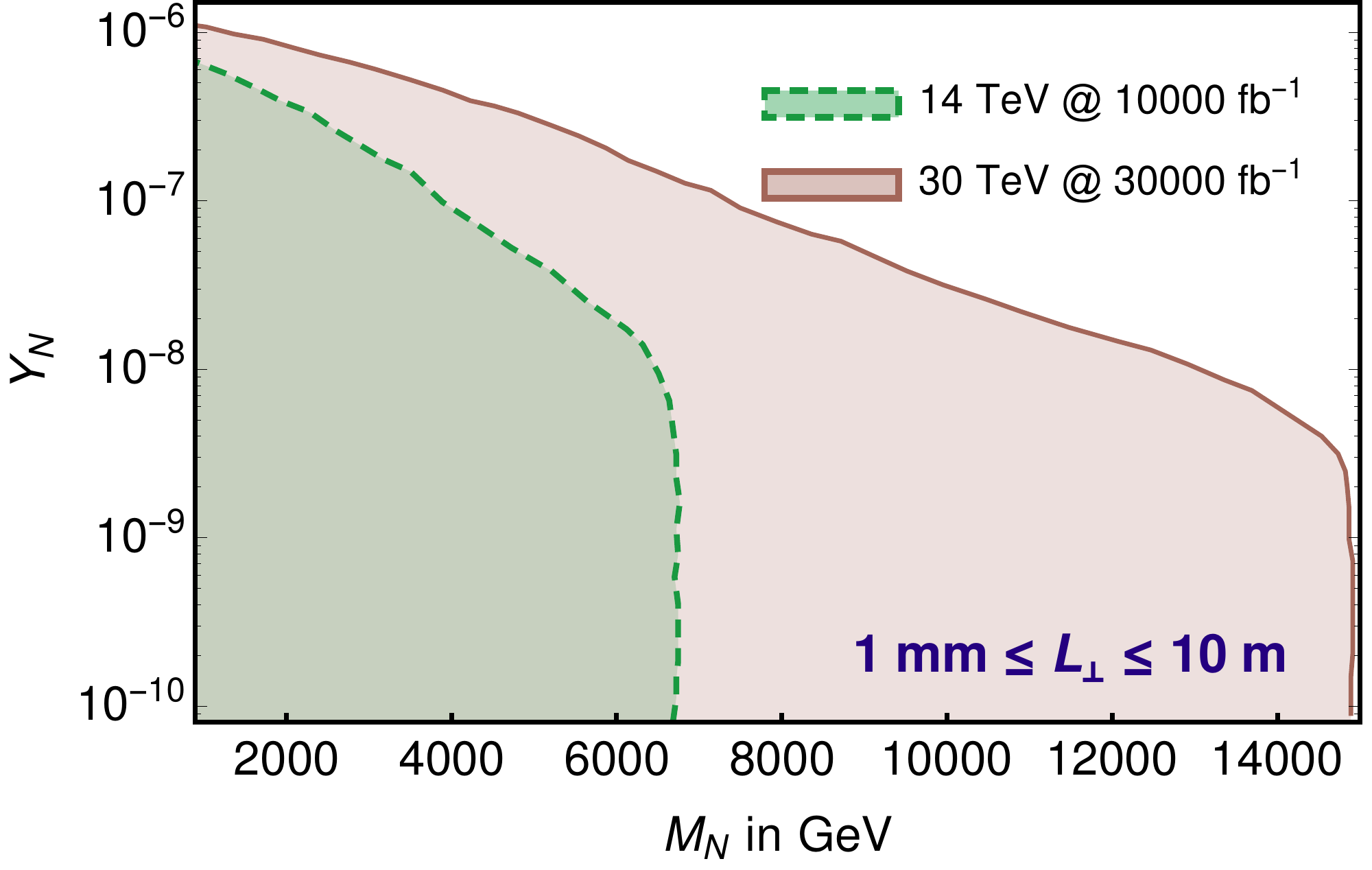}\label{}}\quad \quad 
		\subfigure[]{\includegraphics[width=0.45\linewidth,angle=-0]{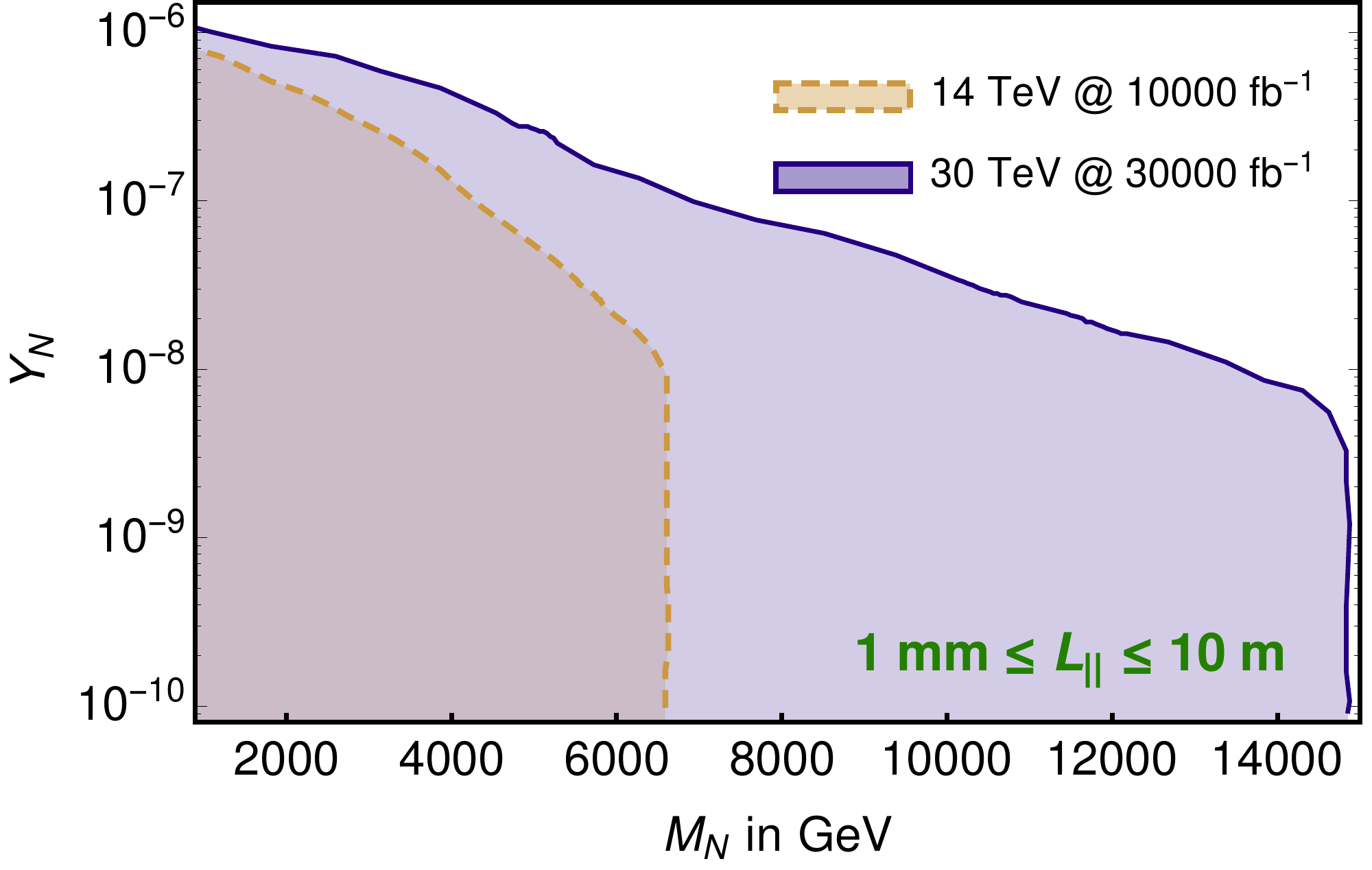}\label{}}}
		\hspace*{-0.5cm}
		\mbox{\subfigure[]{\includegraphics[width=0.45\linewidth,angle=-0]{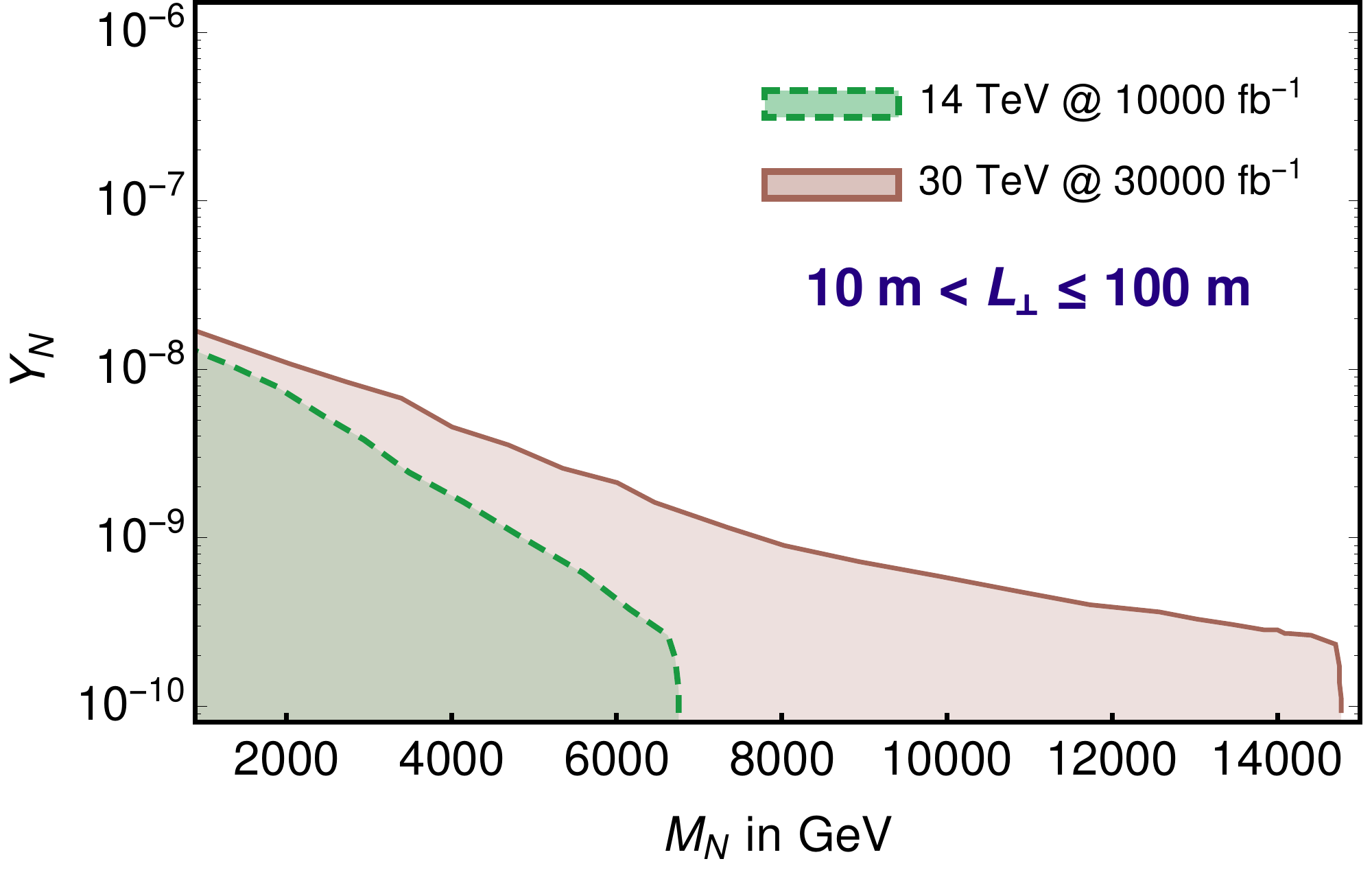}\label{}}\quad \quad 
		\subfigure[]{\includegraphics[width=0.45\linewidth,angle=-0]{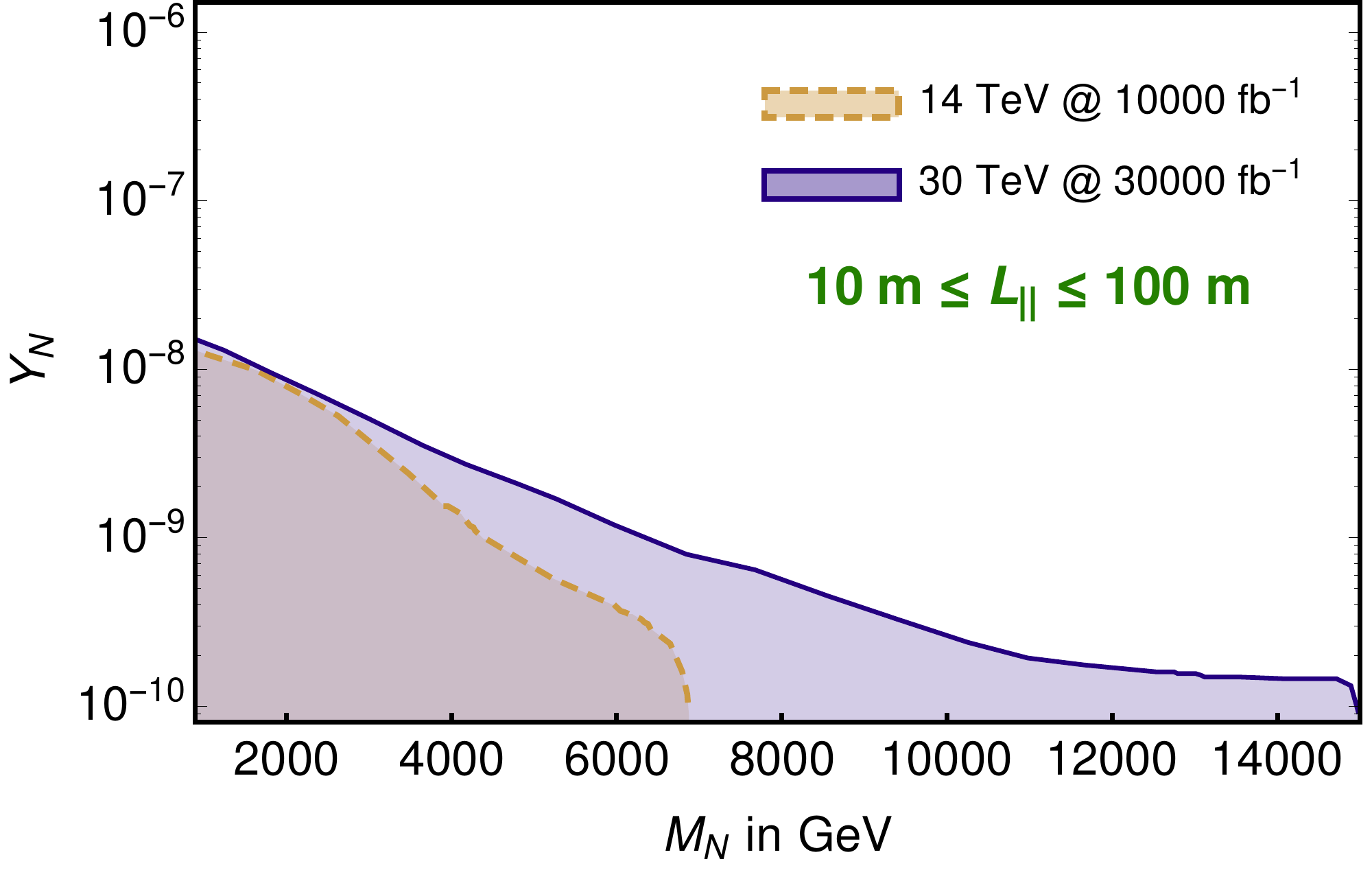}\label{}}}	
		\caption{Limits obtained for Yukawa coupling versus Type-III fermion mass via transverse (a, c) and longitudinal (b, d) decay lengths for the finalstates containing atleast 2$b$-jets with the centre of mass energies of 14 and 30 TeV  at the muon collider with integrated luminosities of 10000 and 30000 fb$^{-1}$, respectively. (a, b) represent 1\,mm $\leq L_{\perp, \, \parallel}\leq $ 10\,m and (c, d) depict 10\,m $\leq L_{\perp, \, \parallel}\leq $ 100\,m regions.  }  \label{muon_reach_noh}
	\end{center}
\end{figure*}


In \autoref{muon_reach}, we depict the region plots for obtaining at least one displaced Higgs boson at the muon collider for 14 TeV and 30 TeV centre of mass energies. In \autoref{muon_reach}(a) and (b), we present the regions that can be probed with transverse ($L_{\perp}$) and longitudinal ($L_{||}$) decay lengths of  1\,mm to 10\,m at the centre of mass energies of 14 and 30 TeV with integrated luminosities of 10000 and 30000 fb$^{-1}$, respectively. From \autoref{muon_reach} (a) we see that one can probe $M_N$ up to 6.8 and 9.9 TeV in the transverse direction for the centre of mass energies of 14, 30 TeV at the integrated luminosities of 10000 and 30000 fb$^{-1}$, respectively. The corresponding bounds from longitudinal decays can be estimated from  \autoref{muon_reach} (b)  as 6.0 and 8.0 TeV, respectively for the centre of mass energies of 14 and 30 TeV. In case of muon collider, the reach is a little inflated in the transverse direction for the similar displacement length reach for longitudinal and transverse direction. \autoref{muon_reach} (c), (d) present the limits obtained from the decay lengths of $10-100$ meter probing very low Yukawa couplings $\lesssim 10^{-8}$.

In \autoref{muon_reach_noh}, we also have presented our results in the $Y_N$ verses $M_N$ plane for the  $2b$-jet finalstate, however the  reconstruction of Higgs mass is not demanded. \autoref{muon_reach_noh} (a) and (b) describes the probable region in the transverse and longitudinal direction, respectively for the displaced range within 1\,mm to 10\,m with the centre of mass energies of 14 and 30 TeV and integrated luminosities of 10000 and 30000 fb$^{-1}$, respectively. \autoref{muon_reach_noh} (c) and (d) represents the same for the decay length within 10\,m to 100\, m at muon collider. Clearly, here the reach is higher than \autoref{muon_reach} as the  Higgs boson mass reconstruction efficiency  reduces the number for the latter case,  which can be seen from \autoref{HiggsTabMuon}. Unlike at the  LHC, in muon collider the reach is restricted to the centre of mass energy, thus higher  centre of mass energy would probe higher values of  Type-III fermion mass. 


\section{Conclusion}\label{concl}

Type-III seesaw model is motivated to explain the tiny neutrino mass scale which predicts the heavy charged and neutral leptons. They mix with SM charged and neutral leptons via the Yukawa couplings and the vev of the SM Higgs doublet at the electroweak symmetry breaking scale. Current lower limit of  this SU(2) triplet fermion obtained using the prompt charged lepton finalstates is around 740 GeV \cite{CMS:2017ybg,ATLAS:2020wop,Biggio:2019eeo,CMS:2019lwf} at $2\sigma$ with possible QCD corrections \cite{Ruiz:2015zca} but LHC with higher energy and luminosity can probe even higher mass range. The same Yukawa coupling also gives direct coupling to SM Higgs boson with the charged and neutral heavy fermions. Thus Higgs bosons can be produced from the decays of such heavy fermions proportional to the square of the Yukawa couplings. In particular, we focus on the parameter space with low Yukawa, where the displaced decays of $N^\pm$ and $N^0$ occur.

In this article, we explored the displaced Higgs production from such decays both at the LHC/FCC and at the muon collider. The Higgs boson comes from these decays of  $N^\pm \to h \ell^\pm$ and $N^0 \to h \nu$, where the former leads to the complete visible displaced finalstate for $Y_N\simeq 5\times 10^{-7}$. Different Yukawa coupling ranges are probed, which are compatible with the atmospheric and the solar neutrino mass scale as well as very light ones which can be explored by the proposed detector, MATHUSLA. However, for lower Yukawa couplings i.e. $Y_N \leq 10^{-8}$, we observe two decay recoils: first from the decay of $N^\pm \to N^0 \pi^\pm$ and the second one is from $N^0$ decays to the SM particles. We also notice that the longitudinal boost at the LHC/FCC can lead to the enhancement of the displaced decay lengths ($L_{||}$), and thus has a better reach compared to the transverse ($L_{\perp}$) one. The detailed analyses at the LHC with centre of mass energies of 14, 27 and 100 TeV are presented along with the bounds on Yukawa couplings and Type-III fermion masses. A very early data of 300fb$^{-1}$ can probe Type-III fermion mass of $\sim $ 3.6 TeV at the centre of mass energy of 100 TeV (FCC) can probe mass range of 4.25 TeV with the integrated luminosity of 3000\,fb$^{-1}$. 

Displaced Higgs production in a supersymmetric Type-III seesaw mechanism is already looked into \cite{Bandyopadhyay:2010wp}. In general, the seesaw models with lower couplings predict displaced decays of right-handed neutrinos in case of Type-I seesaw \cite{Bandyopadhyay:2011qm,Bandyopadhyay:2015iij} but we do not expect any flying charged track in those scenarios. In case of Type-III seesaw at the LHC, $N^\pm$ gives charged track before it decays, whereas $N^0$ has invisible track before the decay. Models with inert SU(2) triplet scalars also predicts displaced charged track but mostly leads to very soft leptons and jets due to phase space suppression induced by $Z_2$ symmetry \cite{Jangid:2020qgo, Bandyopadhyay:2020otm,Snehasish1}. In a supersymmetric extension of such triplet scalar, the fermionic partners stay nearly degenerate at the tree-level as well as couple to SM fermions via the mixing with doublet like Higgsinos and gauginos, resulting displaced charged track  along with large missing energy due to $R$-parity\cite{SabanciKeceli:2018fsd}. 
Thus, a demand of one visible and another invisible charged tracks, i.e. first and second recoils along with one(two) hard lepton(s) but not so large missing energy can segregate the scenario from others. The reconstruction of the Higgs bosons in this scenario, can distinguish it from other possible BSM scenarios having long-lived particles. In the context of supersymmetry and UED, where we see long lived particles often have larger missing energy in the finalstates, thus can also be easily segregated. In $R$-parity violating framework the displaced decays of the Higgs boson is studied in \cite{Bandyopadhyay:2010cu} though produced promptly.

Unlike at the LHC/FCC, at the muon collider, we can only produce $N^+ N^-$, and thus two such displaced charged tracks will be visible before the recoil. At the muon collider, the mass ranges that can be probed is identical for both the transverse and the longitudinal modes due to fixed total momentum in each collision. However, one interesting fact we observe is that the transverse momentum of the $N^\pm$ diverges perpendicular to the beam axis, as opposed to the LHC. This leads to higher momentum for transverse compared to longitudinal one, and predicts larger number of events for the transverse decay lengths. We plot the sensitivity regions in the Yukawa versus mass plane for two different decay length regions, i.e. 1\,mm - 10\,m and 10-100\,m for the centre of mass energies of  14 and 30 TeV and it is evident that a detector with length around 10 m is sufficient in probing the larger ranges of the Yukawa coupling. Certainly, muon collider has good prospects in probing such scenarios having them in centre of mass frame and without initial state radiation. 

\section*{Acknowledgements}
PB acknowledges SERB CORE Grant CRG/2018/004971 and MATRICS Grant MTR/2020/000668 for the financial and computational support towards the work. CS would like to thank MOE Government of India for an JRF fellowship. PB and CS also thank Anirban Karan for some useful discussions. CS wishes to thank Antonio Costantini and  Snehashis Parashar for some useful discussions. PB wants to thank Kirtiman Ghosh for some crucial information regarding the simulation.

\bibliography{Reference_TypeIII}

\providecommand{\href}[2]{#2}\begingroup\raggedright\begin{thebibliography}{10}

\bibitem{nuOscl1}
P.F.~de~Salas, D.V.~Forero, S.~Gariazzo, P.~Mart\'\i{}nez-Mirav\'e, O.~Mena,
  C.A.~Ternes et~al., \emph{{2020 global reassessment of the neutrino
  oscillation picture}},
  \href{https://doi.org/10.1007/JHEP02(2021)071}{\emph{JHEP} {\bfseries 02}
  (2021) 071} [\href{https://arxiv.org/abs/2006.11237}{{\ttfamily
  2006.11237}}].

\bibitem{nuOscl2}
D.S.~Hajdukovic, \emph{{On the absolute value of the neutrino mass}},
  \href{https://doi.org/10.1142/S0217732311035948}{\emph{Mod. Phys. Lett. A}
  {\bfseries 26} (2011) 1555}
  [\href{https://arxiv.org/abs/1106.5810}{{\ttfamily 1106.5810}}].

\bibitem{Foot:1988aq}
R.~Foot, H.~Lew, X.G.~He and G.C.~Joshi, \emph{{Seesaw Neutrino Masses Induced
  by a Triplet of Leptons}}, \href{https://doi.org/10.1007/BF01415558}{\emph{Z.
  Phys. C} {\bfseries 44} (1989) 441}.

\bibitem{Bajc:2006ia}
B.~Bajc and G.~Senjanovic, \emph{{Seesaw at LHC}},
  \href{https://doi.org/10.1088/1126-6708/2007/08/014}{\emph{JHEP} {\bfseries
  08} (2007) 014} [\href{https://arxiv.org/abs/hep-ph/0612029}{{\ttfamily
  hep-ph/0612029}}].

\bibitem{Franceschini:2008pz}
R.~Franceschini, T.~Hambye and A.~Strumia, \emph{{Type-III see-saw at LHC}},
  \href{https://doi.org/10.1103/PhysRevD.78.033002}{\emph{Phys. Rev. D}
  {\bfseries 78} (2008) 033002}
  [\href{https://arxiv.org/abs/0805.1613}{{\ttfamily 0805.1613}}].

\bibitem{Bajc:2007zf}
B.~Bajc, M.~Nemevsek and G.~Senjanovic, \emph{{Probing seesaw at LHC}},
  \href{https://doi.org/10.1103/PhysRevD.76.055011}{\emph{Phys. Rev. D}
  {\bfseries 76} (2007) 055011}
  [\href{https://arxiv.org/abs/hep-ph/0703080}{{\ttfamily hep-ph/0703080}}].

\bibitem{Ma:1998dn}
E.~Ma, \emph{{Pathways to naturally small neutrino masses}},
  \href{https://doi.org/10.1103/PhysRevLett.81.1171}{\emph{Phys. Rev. Lett.}
  {\bfseries 81} (1998) 1171}
  [\href{https://arxiv.org/abs/hep-ph/9805219}{{\ttfamily hep-ph/9805219}}].

\bibitem{Arhrib:2009mz}
A.~Arhrib, B.~Bajc, D.K.~Ghosh, T.~Han, G.-Y.~Huang, I.~Puljak et~al.,
  \emph{{Collider Signatures for Heavy Lepton Triplet in Type I+III Seesaw}},
  \href{https://doi.org/10.1103/PhysRevD.82.053004}{\emph{Phys. Rev. D}
  {\bfseries 82} (2010) 053004}
  [\href{https://arxiv.org/abs/0904.2390}{{\ttfamily 0904.2390}}].

\bibitem{Bandyopadhyay:2009xa}
P.~Bandyopadhyay, S.~Choubey and M.~Mitra, \emph{{Two Higgs Doublet Type III
  Seesaw with mu-tau symmetry at LHC}},
  \href{https://doi.org/10.1088/1126-6708/2009/10/012}{\emph{JHEP} {\bfseries
  10} (2009) 012} [\href{https://arxiv.org/abs/0906.5330}{{\ttfamily
  0906.5330}}].

\bibitem{Bandyopadhyay:2011aa}
P.~Bandyopadhyay, S.~Choi, E.J.~Chun and K.~Min, \emph{{Probing Higgs bosons
  via the type III seesaw mechanism at the LHC}},
  \href{https://doi.org/10.1103/PhysRevD.85.073013}{\emph{Phys. Rev. D}
  {\bfseries 85} (2012) 073013}
  [\href{https://arxiv.org/abs/1112.3080}{{\ttfamily 1112.3080}}].

\bibitem{Eboli:2011ia}
O.J.P.~Eboli, J.~Gonzalez-Fraile and M.C.~Gonzalez-Garcia, \emph{{Neutrino
  Masses at LHC: Minimal Lepton Flavour Violation in Type-III See-saw}},
  \href{https://doi.org/10.1007/JHEP12(2011)009}{\emph{JHEP} {\bfseries 12}
  (2011) 009} [\href{https://arxiv.org/abs/1108.0661}{{\ttfamily 1108.0661}}].

\bibitem{Cai:2017mow}
Y.~Cai, T.~Han, T.~Li and R.~Ruiz, \emph{{Lepton Number Violation: Seesaw
  Models and Their Collider Tests}},
  \href{https://doi.org/10.3389/fphy.2018.00040}{\emph{Front. in Phys.}
  {\bfseries 6} (2018) 40} [\href{https://arxiv.org/abs/1711.02180}{{\ttfamily
  1711.02180}}].

\bibitem{Goswami:2017jqs}
D.~Goswami and P.~Poulose, \emph{{Direct searches of Type III seesaw triplet
  fermions at high energy $e^+e^-$ collider}},
  \href{https://doi.org/10.1140/epjc/s10052-017-5478-1}{\emph{Eur. Phys. J. C}
  {\bfseries 78} (2018) 42} [\href{https://arxiv.org/abs/1702.07215}{{\ttfamily
  1702.07215}}].

\bibitem{delAguila:2008hw}
F.~del Aguila and J.A.~Aguilar-Saavedra, \emph{{Electroweak scale seesaw and
  heavy Dirac neutrino signals at LHC}},
  \href{https://doi.org/10.1016/j.physletb.2009.01.010}{\emph{Phys. Lett. B}
  {\bfseries 672} (2009) 158}
  [\href{https://arxiv.org/abs/0809.2096}{{\ttfamily 0809.2096}}].

\bibitem{delAguila:2008cj}
F.~del Aguila and J.A.~Aguilar-Saavedra, \emph{{Distinguishing seesaw models at
  LHC with multi-lepton signals}},
  \href{https://doi.org/10.1016/j.nuclphysb.2008.12.029}{\emph{Nucl. Phys. B}
  {\bfseries 813} (2009) 22} [\href{https://arxiv.org/abs/0808.2468}{{\ttfamily
  0808.2468}}].

\bibitem{Bandyopadhyay:2020mnp}
P.~Bandyopadhyay, A.~Karan and C.~Sen, \emph{{Discerning Signatures of Seesaw
  Models and Complementarity of Leptonic Colliders}},
  \href{https://arxiv.org/abs/2011.04191}{{\ttfamily 2011.04191}}.

\bibitem{Agostinho:2017biv}
N.R.~Agostinho, O.J.P.~Eboli and M.C.~Gonzalez-Garcia, \emph{{LHC Run I Bounds
  on Minimal Lepton Flavour Violation in Type-III See-saw: A Case Study}},
  \href{https://doi.org/10.1007/JHEP11(2017)118}{\emph{JHEP} {\bfseries 11}
  (2017) 118} [\href{https://arxiv.org/abs/1708.08456}{{\ttfamily
  1708.08456}}].

\bibitem{Ibanez:2009du}
D.~Ibanez, S.~Morisi and J.W.F.~Valle, \emph{{Inverse tri-bimaximal type-III
  seesaw and lepton flavor violation}},
  \href{https://doi.org/10.1103/PhysRevD.80.053015}{\emph{Phys. Rev. D}
  {\bfseries 80} (2009) 053015}
  [\href{https://arxiv.org/abs/0907.3109}{{\ttfamily 0907.3109}}].

\bibitem{Das:2020uer}
A.~Das and S.~Mandal, \emph{{Bounds on the triplet fermions in type-III seesaw
  and implications for collider searches}},
  \href{https://doi.org/10.1016/j.nuclphysb.2021.115374}{\emph{Nucl. Phys. B}
  {\bfseries 966} (2021) 115374}
  [\href{https://arxiv.org/abs/2006.04123}{{\ttfamily 2006.04123}}].

\bibitem{Ashanujjaman:2021jhi}
S.~Ashanujjaman and K.~Ghosh, \emph{{Type-III see-saw: Phenomenological
  implications of the information lost in decoupling from high-energy to
  low-energy}},
  \href{https://doi.org/10.1016/j.physletb.2021.136403}{\emph{Phys. Lett. B}
  {\bfseries 819} (2021) 136403}
  [\href{https://arxiv.org/abs/2102.09536}{{\ttfamily 2102.09536}}].

\bibitem{Das:2020gnt}
A.~Das, S.~Mandal and T.~Modak, \emph{{Testing triplet fermions at the
  electron-positron and electron-proton colliders using fat jet signatures}},
  \href{https://doi.org/10.1103/PhysRevD.102.033001}{\emph{Phys. Rev. D}
  {\bfseries 102} (2020) 033001}
  [\href{https://arxiv.org/abs/2005.02267}{{\ttfamily 2005.02267}}].

\bibitem{Abada:2008ea}
A.~Abada, C.~Biggio, F.~Bonnet, M.B.~Gavela and T.~Hambye, \emph{{mu
  ---\ensuremath{>} e gamma and tau ---\ensuremath{>} l gamma decays in the
  fermion triplet seesaw model}},
  \href{https://doi.org/10.1103/PhysRevD.78.033007}{\emph{Phys. Rev. D}
  {\bfseries 78} (2008) 033007}
  [\href{https://arxiv.org/abs/0803.0481}{{\ttfamily 0803.0481}}].

\bibitem{Escribano:2021css}
P.~Escribano, J.~Terol-Calvo and A.~Vicente,
  \emph{{$\boldsymbol{(g-2)_{e,\mu}}$ in an extended inverse type-III seesaw
  model}}, \href{https://doi.org/10.1103/PhysRevD.103.115018}{\emph{Phys. Rev.
  D} {\bfseries 103} (2021) 115018}
  [\href{https://arxiv.org/abs/2104.03705}{{\ttfamily 2104.03705}}].

\bibitem{Jana:2020qzn}
S.~Jana, N.~Okada and D.~Raut, \emph{{Displaced Vertex and Disappearing Track
  Signatures in type-III Seesaw}},
  \href{https://arxiv.org/abs/1911.09037}{{\ttfamily 1911.09037}}.

\bibitem{Goswami:2018jar}
S.~Goswami, K.N.~Vishnudath and N.~Khan, \emph{{Constraining the minimal
  type-III seesaw model with naturalness, lepton flavor violation, and
  electroweak vacuum stability}},
  \href{https://doi.org/10.1103/PhysRevD.99.075012}{\emph{Phys. Rev. D}
  {\bfseries 99} (2019) 075012}
  [\href{https://arxiv.org/abs/1810.11687}{{\ttfamily 1810.11687}}].

\bibitem{Bandyopadhyay:2020djh}
P.~Bandyopadhyay, S.~Jangid and M.~Mitra, \emph{{Scrutinizing Vacuum Stability
  in IDM with Type-III Inverse seesaw}},
  \href{https://doi.org/10.1007/JHEP02(2021)075}{\emph{JHEP} {\bfseries 02}
  (2021) 075} [\href{https://arxiv.org/abs/2008.11956}{{\ttfamily
  2008.11956}}].

\bibitem{CMS:2017ybg}
{\scshape CMS} collaboration, \emph{{Search for Evidence of the Type-III Seesaw
  Mechanism in Multilepton Final States in Proton-Proton Collisions at
  $\sqrt{s}=13\text{ }\text{ }\mathrm{TeV}$}},
  \href{https://doi.org/10.1103/PhysRevLett.119.221802}{\emph{Phys. Rev. Lett.}
  {\bfseries 119} (2017) 221802}
  [\href{https://arxiv.org/abs/1708.07962}{{\ttfamily 1708.07962}}].

\bibitem{ATLAS:2020wop}
{\scshape ATLAS} collaboration, \emph{{Search for type-III seesaw heavy leptons
  in dilepton final states in $pp$ collisions at $\sqrt{s}$ = 13 TeV with the
  ATLAS detector}},
  \href{https://doi.org/10.1140/epjc/s10052-021-08929-9}{\emph{Eur. Phys. J. C}
  {\bfseries 81} (2021) 218}
  [\href{https://arxiv.org/abs/2008.07949}{{\ttfamily 2008.07949}}].

\bibitem{CMS:2019lwf}
{\scshape CMS} collaboration, \emph{{Search for physics beyond the standard
  model in multilepton final states in proton-proton collisions at $\sqrt{s} =$
  13 TeV}}, \href{https://doi.org/10.1007/JHEP03(2020)051}{\emph{JHEP}
  {\bfseries 03} (2020) 051}
  [\href{https://arxiv.org/abs/1911.04968}{{\ttfamily 1911.04968}}].

\bibitem{Biggio:2019eeo}
C.~Biggio, E.~Fernandez-Martinez, M.~Filaci, J.~Hernandez-Garcia and
  J.~Lopez-Pavon, \emph{{Global Bounds on the Type-III Seesaw}},
  \href{https://doi.org/10.1007/JHEP05(2020)022}{\emph{JHEP} {\bfseries 05}
  (2020) 022} [\href{https://arxiv.org/abs/1911.11790}{{\ttfamily
  1911.11790}}].

\bibitem{Ibarra:2003up}
A.~Ibarra and G.G.~Ross, \emph{{Neutrino phenomenology: The Case of two
  right-handed neutrinos}},
  \href{https://doi.org/10.1016/j.physletb.2004.04.037}{\emph{Phys. Lett. B}
  {\bfseries 591} (2004) 285}
  [\href{https://arxiv.org/abs/hep-ph/0312138}{{\ttfamily hep-ph/0312138}}].

\bibitem{Ibarra:2005qi}
A.~Ibarra, \emph{{Reconstructing the two right-handed neutrino model}},
  \href{https://doi.org/10.1088/1126-6708/2006/01/064}{\emph{JHEP} {\bfseries
  01} (2006) 064} [\href{https://arxiv.org/abs/hep-ph/0511136}{{\ttfamily
  hep-ph/0511136}}].

\bibitem{Bandyopadhyay:2010wp}
P.~Bandyopadhyay and E.J.~Chun, \emph{{Displaced Higgs production in type III
  seesaw}}, \href{https://doi.org/10.1007/JHEP11(2010)006}{\emph{JHEP}
  {\bfseries 11} (2010) 006} [\href{https://arxiv.org/abs/1007.2281}{{\ttfamily
  1007.2281}}].

\bibitem{Cepeda:2019klc}
M.~Cepeda et~al., \emph{{Report from Working Group 2}: {Higgs Physics at the
  HL-LHC and HE-LHC}},
  \href{https://doi.org/10.23731/CYRM-2019-007.221}{\emph{CERN Yellow Rep.
  Monogr.} {\bfseries 7} (2019) 221}
  [\href{https://arxiv.org/abs/1902.00134}{{\ttfamily 1902.00134}}].

\bibitem{MATHUSLA:2019qpy}
{\scshape MATHUSLA} collaboration, \emph{{Explore the lifetime frontier with
  MATHUSLA}},
  \href{https://doi.org/10.1088/1748-0221/15/06/C06026}{\emph{JINST} {\bfseries
  15} (2020) C06026} [\href{https://arxiv.org/abs/1901.04040}{{\ttfamily
  1901.04040}}].

\bibitem{Chou:2016lxi}
J.P.~Chou, D.~Curtin and H.J.~Lubatti, \emph{{New Detectors to Explore the
  Lifetime Frontier}},
  \href{https://doi.org/10.1016/j.physletb.2017.01.043}{\emph{Phys. Lett. B}
  {\bfseries 767} (2017) 29}
  [\href{https://arxiv.org/abs/1606.06298}{{\ttfamily 1606.06298}}].

\bibitem{Curtin:2018mvb}
D.~Curtin et~al., \emph{{Long-Lived Particles at the Energy Frontier: The
  MATHUSLA Physics Case}},
  \href{https://doi.org/10.1088/1361-6633/ab28d6}{\emph{Rept. Prog. Phys.}
  {\bfseries 82} (2019) 116201}
  [\href{https://arxiv.org/abs/1806.07396}{{\ttfamily 1806.07396}}].

\bibitem{Coccaro:2016lnz}
A.~Coccaro, D.~Curtin, H.J.~Lubatti, H.~Russell and J.~Shelton,
  \emph{{Data-driven Model-independent Searches for Long-lived Particles at the
  LHC}}, \href{https://doi.org/10.1103/PhysRevD.94.113003}{\emph{Phys. Rev. D}
  {\bfseries 94} (2016) 113003}
  [\href{https://arxiv.org/abs/1605.02742}{{\ttfamily 1605.02742}}].

\bibitem{Alpigiani:2020iam}
C.~Alpigiani, \emph{{Exploring the lifetime and cosmic frontier with the
  MATHUSLA detector}},
  \href{https://doi.org/10.1088/1748-0221/15/09/C09048}{\emph{JINST} {\bfseries
  15} (2020) C09048} [\href{https://arxiv.org/abs/2006.00788}{{\ttfamily
  2006.00788}}].

\bibitem{Palmer:1996gs}
R.~Palmer et~al., \emph{{Muon collider design}},
  \href{https://doi.org/10.1016/0920-5632(96)00417-3}{\emph{Nucl. Phys. B Proc.
  Suppl.} {\bfseries 51} (1996) 61}
  [\href{https://arxiv.org/abs/acc-phys/9604001}{{\ttfamily
  acc-phys/9604001}}].

\bibitem{Ankenbrandt:1999cta}
C.M.~Ankenbrandt et~al., \emph{{Status of muon collider research and
  development and future plans}},
  \href{https://doi.org/10.1103/PhysRevSTAB.2.081001}{\emph{Phys. Rev. ST
  Accel. Beams} {\bfseries 2} (1999) 081001}
  [\href{https://arxiv.org/abs/physics/9901022}{{\ttfamily physics/9901022}}].

\bibitem{Delahaye:2019omf}
J.P.~Delahaye, M.~Diemoz, K.~Long, B.~Mansouli\'e, N.~Pastrone, L.~Rivkin
  et~al., \emph{{Muon Colliders}},
  \href{https://arxiv.org/abs/1901.06150}{{\ttfamily 1901.06150}}.

\bibitem{Bartosik:2019dzq}
N.~Bartosik et~al., \emph{{Preliminary Report on the Study of Beam-Induced
  Background Effects at a Muon Collider}},
  \href{https://arxiv.org/abs/1905.03725}{{\ttfamily 1905.03725}}.

\bibitem{Bartosik:2020xwr}
N.~Bartosik et~al., \emph{{Detector and Physics Performance at a Muon
  Collider}},
  \href{https://doi.org/10.1088/1748-0221/15/05/P05001}{\emph{JINST} {\bfseries
  15} (2020) P05001} [\href{https://arxiv.org/abs/2001.04431}{{\ttfamily
  2001.04431}}].

\bibitem{AlAli:2021let}
H.~Al~Ali et~al., \emph{{The Muon Smasher's Guide}},
  \href{https://arxiv.org/abs/2103.14043}{{\ttfamily 2103.14043}}.

\bibitem{Costantini:2020stv}
A.~Costantini, F.~De~Lillo, F.~Maltoni, L.~Mantani, O.~Mattelaer, R.~Ruiz
  et~al., \emph{{Vector boson fusion at multi-TeV muon colliders}},
  \href{https://doi.org/10.1007/JHEP09(2020)080}{\emph{JHEP} {\bfseries 09}
  (2020) 080} [\href{https://arxiv.org/abs/2005.10289}{{\ttfamily
  2005.10289}}].

\bibitem{Buttazzo:2018qqp}
D.~Buttazzo, D.~Redigolo, F.~Sala and A.~Tesi, \emph{{Fusing Vectors into
  Scalars at High Energy Lepton Colliders}},
  \href{https://doi.org/10.1007/JHEP11(2018)144}{\emph{JHEP} {\bfseries 11}
  (2018) 144} [\href{https://arxiv.org/abs/1807.04743}{{\ttfamily
  1807.04743}}].

\bibitem{Huang:2021nkl}
G.-y.~Huang, F.S.~Queiroz and W.~Rodejohann, \emph{{Gauged
  $L^{}_{\mu}{-}L^{}_{\tau}$ at a muon collider}},
  \href{https://doi.org/10.1103/PhysRevD.103.095005}{\emph{Phys. Rev. D}
  {\bfseries 103} (2021) 095005}
  [\href{https://arxiv.org/abs/2101.04956}{{\ttfamily 2101.04956}}].

\bibitem{Huang:2021biu}
G.-y.~Huang, S.~Jana, F.S.~Queiroz and W.~Rodejohann, \emph{{Probing the RK(*)
  anomaly at a muon collider}},
  \href{https://doi.org/10.1103/PhysRevD.105.015013}{\emph{Phys. Rev. D}
  {\bfseries 105} (2022) 015013}
  [\href{https://arxiv.org/abs/2103.01617}{{\ttfamily 2103.01617}}].

\bibitem{Asadi:2021gah}
P.~Asadi, R.~Capdevilla, C.~Cesarotti and S.~Homiller, \emph{{Searching for
  leptoquarks at future muon colliders}},
  \href{https://doi.org/10.1007/JHEP10(2021)182}{\emph{JHEP} {\bfseries 10}
  (2021) 182} [\href{https://arxiv.org/abs/2104.05720}{{\ttfamily
  2104.05720}}].

\bibitem{Capdevilla:2020qel}
R.~Capdevilla, D.~Curtin, Y.~Kahn and G.~Krnjaic, \emph{{Discovering the
  physics of $(g-2)_\mu$ at future muon colliders}},
  \href{https://doi.org/10.1103/PhysRevD.103.075028}{\emph{Phys. Rev. D}
  {\bfseries 103} (2021) 075028}
  [\href{https://arxiv.org/abs/2006.16277}{{\ttfamily 2006.16277}}].

\bibitem{Long:2020wfp}
K.~Long, D.~Lucchesi, M.~Palmer, N.~Pastrone, D.~Schulte and V.~Shiltsev,
  \emph{{Muon colliders to expand frontiers of particle physics}},
  \href{https://doi.org/10.1038/s41567-020-01130-x}{\emph{Nature Phys.}
  {\bfseries 17} (2021) 289}
  [\href{https://arxiv.org/abs/2007.15684}{{\ttfamily 2007.15684}}].

\bibitem{Han:2020pif}
T.~Han, D.~Liu, I.~Low and X.~Wang, \emph{{Electroweak couplings of the Higgs
  boson at a multi-TeV muon collider}},
  \href{https://doi.org/10.1103/PhysRevD.103.013002}{\emph{Phys. Rev. D}
  {\bfseries 103} (2021) 013002}
  [\href{https://arxiv.org/abs/2008.12204}{{\ttfamily 2008.12204}}].

\bibitem{Han:2020uak}
T.~Han, Z.~Liu, L.-T.~Wang and X.~Wang, \emph{{WIMPs at High Energy Muon
  Colliders}}, \href{https://doi.org/10.1103/PhysRevD.103.075004}{\emph{Phys.
  Rev. D} {\bfseries 103} (2021) 075004}
  [\href{https://arxiv.org/abs/2009.11287}{{\ttfamily 2009.11287}}].

\bibitem{Capdevilla:2021rwo}
R.~Capdevilla, D.~Curtin, Y.~Kahn and G.~Krnjaic, \emph{{No-lose theorem for
  discovering the new physics of (g-2)\ensuremath{\mu} at muon colliders}},
  \href{https://doi.org/10.1103/PhysRevD.105.015028}{\emph{Phys. Rev. D}
  {\bfseries 105} (2022) 015028}
  [\href{https://arxiv.org/abs/2101.10334}{{\ttfamily 2101.10334}}].

\bibitem{Han:2021udl}
T.~Han, S.~Li, S.~Su, W.~Su and Y.~Wu, \emph{{Heavy Higgs bosons in 2HDM at a
  muon collider}},
  \href{https://doi.org/10.1103/PhysRevD.104.055029}{\emph{Phys. Rev. D}
  {\bfseries 104} (2021) 055029}
  [\href{https://arxiv.org/abs/2102.08386}{{\ttfamily 2102.08386}}].

\bibitem{Bandyopadhyay:2021pld}
P.~Bandyopadhyay, A.~Karan and R.~Mandal, \emph{{Distinguishing signatures of
  scalar leptoquarks at hadron and muon colliders}},
  \href{https://arxiv.org/abs/2108.06506}{{\ttfamily 2108.06506}}.

\bibitem{Liu:2021jyc}
W.~Liu and K.-P.~Xie, \emph{{Probing electroweak phase transition with
  multi-TeV muon colliders and gravitational waves}},
  \href{https://doi.org/10.1007/JHEP04(2021)015}{\emph{JHEP} {\bfseries 04}
  (2021) 015} [\href{https://arxiv.org/abs/2101.10469}{{\ttfamily
  2101.10469}}].

\bibitem{NNPDF:2014otw}
{\scshape NNPDF} collaboration, \emph{{Parton distributions for the LHC Run
  II}}, \href{https://doi.org/10.1007/JHEP04(2015)040}{\emph{JHEP} {\bfseries
  04} (2015) 040} [\href{https://arxiv.org/abs/1410.8849}{{\ttfamily
  1410.8849}}].

\bibitem{Cirelli:2005uq}
M.~Cirelli, N.~Fornengo and A.~Strumia, \emph{{Minimal dark matter}},
  \href{https://doi.org/10.1016/j.nuclphysb.2006.07.012}{\emph{Nucl. Phys. B}
  {\bfseries 753} (2006) 178}
  [\href{https://arxiv.org/abs/hep-ph/0512090}{{\ttfamily hep-ph/0512090}}].

\bibitem{Li:2009mw}
T.~Li and X.-G.~He, \emph{{Neutrino Masses and Heavy Triplet Leptons at the
  LHC: Testability of Type III Seesaw}},
  \href{https://doi.org/10.1103/PhysRevD.80.093003}{\emph{Phys. Rev. D}
  {\bfseries 80} (2009) 093003}
  [\href{https://arxiv.org/abs/0907.4193}{{\ttfamily 0907.4193}}].

\bibitem{Casas:2001sr}
J.A.~Casas and A.~Ibarra, \emph{{Oscillating neutrinos and $\mu \to e,
  \gamma$}}, \href{https://doi.org/10.1016/S0550-3213(01)00475-8}{\emph{Nucl.
  Phys. B} {\bfseries 618} (2001) 171}
  [\href{https://arxiv.org/abs/hep-ph/0103065}{{\ttfamily hep-ph/0103065}}].

\bibitem{Ruiz:2015zca}
R.~Ruiz, \emph{{QCD Corrections to Pair Production of Type III Seesaw Leptons
  at Hadron Colliders}},
  \href{https://doi.org/10.1007/JHEP12(2015)165}{\emph{JHEP} {\bfseries 12}
  (2015) 165} [\href{https://arxiv.org/abs/1509.05416}{{\ttfamily
  1509.05416}}].

\bibitem{Sjostrand:2014zea}
T.~Sj\"ostrand, S.~Ask, J.R.~Christiansen, R.~Corke, N.~Desai, P.~Ilten et~al.,
  \emph{{An introduction to PYTHIA 8.2}},
  \href{https://doi.org/10.1016/j.cpc.2015.01.024}{\emph{Comput. Phys. Commun.}
  {\bfseries 191} (2015) 159}
  [\href{https://arxiv.org/abs/1410.3012}{{\ttfamily 1410.3012}}].

\bibitem{Staub:2013tta}
F.~Staub, \emph{{SARAH 4 : A tool for (not only SUSY) model builders}},
  \href{https://doi.org/10.1016/j.cpc.2014.02.018}{\emph{Comput. Phys. Commun.}
  {\bfseries 185} (2014) 1773}
  [\href{https://arxiv.org/abs/1309.7223}{{\ttfamily 1309.7223}}].

\bibitem{Belyaev:2012qa}
A.~Belyaev, N.D.~Christensen and A.~Pukhov, \emph{{CalcHEP 3.4 for collider
  physics within and beyond the Standard Model}},
  \href{https://doi.org/10.1016/j.cpc.2013.01.014}{\emph{Comput. Phys. Commun.}
  {\bfseries 184} (2013) 1729}
  [\href{https://arxiv.org/abs/1207.6082}{{\ttfamily 1207.6082}}].

\bibitem{CMS:2012jki}
{\scshape CMS} collaboration, \emph{{Search for heavy bottom-like quarks in 4.9
  inverse femtobarns of $pp$ collisions at $\sqrt{s}=7$ TeV}},
  \href{https://doi.org/10.1007/JHEP05(2012)123}{\emph{JHEP} {\bfseries 05}
  (2012) 123} [\href{https://arxiv.org/abs/1204.1088}{{\ttfamily 1204.1088}}].

\bibitem{CMS:2017wtu}
{\scshape CMS} collaboration, \emph{{Identification of heavy-flavour jets with
  the CMS detector in pp collisions at 13 TeV}},
  \href{https://doi.org/10.1088/1748-0221/13/05/P05011}{\emph{JINST} {\bfseries
  13} (2018) P05011} [\href{https://arxiv.org/abs/1712.07158}{{\ttfamily
  1712.07158}}].

\bibitem{CMS:2007sch}
{\scshape CMS} collaboration, \emph{{CMS technical design report, volume II:
  Physics performance}},
  \href{https://doi.org/10.1088/0954-3899/34/6/S01}{\emph{J. Phys. G}
  {\bfseries 34} (2007) 995}.

\bibitem{ATLAS:design}
{\scshape ATLAS} collaboration, C.G.L.E.C..~LHCC, ``{ATLAS detector and physics
  performance: Technical Design Report, 1}.''
  \url{https://cds.cern.ch/record/391176?ln=en}.

\bibitem{CS:TypeI}
P.~Bandyopadhyay, E.J.~Chun and C.~Sen, \emph{{Displaced flavour violating
  leptonic jet signature at LHC and MATHUSLA}},  \href{https://arxiv.org/abs/To
  be appeared soon}{{\ttfamily To be appeared soon}}.

\bibitem{Tovey:2010de}
D.R.~Tovey, \emph{{Transverse mass and invariant mass observables for measuring
  the mass of a semi-invisibly decaying heavy particle}},
  \href{https://doi.org/10.1007/JHEP11(2010)148}{\emph{JHEP} {\bfseries 11}
  (2010) 148} [\href{https://arxiv.org/abs/1008.3837}{{\ttfamily 1008.3837}}].

\bibitem{Bandyopadhyay:2020wfv}
P.~Bandyopadhyay, S.~Dutta, M.~Jakkapu and A.~Karan, \emph{{Distinguishing
  Leptoquarks at the LHC/FCC}},
  \href{https://doi.org/10.1016/j.nuclphysb.2021.115524}{\emph{Nucl. Phys. B}
  {\bfseries 971} (2021) 115524}
  [\href{https://arxiv.org/abs/2007.12997}{{\ttfamily 2007.12997}}].

\bibitem{Chakraborty:2018khw}
S.~Chakraborty, M.~Mitra and S.~Shil, \emph{{Fat Jet Signature of a Heavy
  Neutrino at Lepton Collider}},
  \href{https://doi.org/10.1103/PhysRevD.100.015012}{\emph{Phys. Rev. D}
  {\bfseries 100} (2019) 015012}
  [\href{https://arxiv.org/abs/1810.08970}{{\ttfamily 1810.08970}}].

\bibitem{Bhardwaj:2018lma}
A.~Bhardwaj, A.~Das, P.~Konar and A.~Thalapillil, \emph{{Looking for Minimal
  Inverse Seesaw scenarios at the LHC with Jet Substructure Techniques}},
  \href{https://doi.org/10.1088/1361-6471/ab7769}{\emph{J. Phys. G} {\bfseries
  47} (2020) 075002} [\href{https://arxiv.org/abs/1801.00797}{{\ttfamily
  1801.00797}}].

\bibitem{Bandyopadhyay:2010ms}
P.~Bandyopadhyay and B.~Bhattacherjee, \emph{{Boosted top quarks in
  supersymmetric cascade decays at the LHC}},
  \href{https://doi.org/10.1103/PhysRevD.84.035020}{\emph{Phys. Rev. D}
  {\bfseries 84} (2011) 035020}
  [\href{https://arxiv.org/abs/1012.5289}{{\ttfamily 1012.5289}}].

\bibitem{pdg}
{\scshape Particle Data Group} collaboration, \emph{{Review of Particle
  Physics}}, \href{https://doi.org/10.1093/ptep/ptaa104}{\emph{PTEP} {\bfseries
  2020} (2020) 083C01}.

\bibitem{pdg1}
G.~Cowan, ``Statistics (40).''
  \url{https://pdg.lbl.gov/2017/reviews/rpp2017-rev-statistics.pdf}, September,
  2017.

\bibitem{Bandyopadhyay:2011qm}
P.~Bandyopadhyay, E.J.~Chun and J.-C.~Park, \emph{{Right-handed sneutrino dark
  matter in $\mathbf{U(1)'}$ seesaw models and its signatures at the LHC}},
  \href{https://doi.org/10.1007/JHEP06(2011)129}{\emph{JHEP} {\bfseries 06}
  (2011) 129} [\href{https://arxiv.org/abs/1105.1652}{{\ttfamily 1105.1652}}].

\bibitem{Bandyopadhyay:2015iij}
P.~Bandyopadhyay, \emph{{Displaced lepton flavour violating signatures of
  right-handed sneutrinos in $U(1)'$ supersymmetric models}},
  \href{https://doi.org/10.1007/JHEP09(2017)052}{\emph{JHEP} {\bfseries 09}
  (2017) 052} [\href{https://arxiv.org/abs/1511.03842}{{\ttfamily
  1511.03842}}].

\bibitem{Jangid:2020qgo}
S.~Jangid and P.~Bandyopadhyay, \emph{{Distinguishing Inert Higgs Doublet and
  Inert Triplet Scenarios}},
  \href{https://doi.org/10.1140/epjc/s10052-020-8271-5}{\emph{Eur. Phys. J. C}
  {\bfseries 80} (2020) 715}
  [\href{https://arxiv.org/abs/2003.11821}{{\ttfamily 2003.11821}}].

\bibitem{Bandyopadhyay:2020otm}
P.~Bandyopadhyay and A.~Costantini, \emph{{Obscure Higgs boson at Colliders}},
  \href{https://doi.org/10.1103/PhysRevD.103.015025}{\emph{Phys. Rev. D}
  {\bfseries 103} (2021) 015025}
  [\href{https://arxiv.org/abs/2010.02597}{{\ttfamily 2010.02597}}].

\bibitem{Snehasish1}
P.~Bandyopadhyay, S.~Jangid, A.~KT and S.~Parashar, \emph{{Discerning the
  Triplet charged Higgs bosons in BSM scenarios at the LHC and MATHUSLA}},
  \href{https://arxiv.org/abs/To be appeared soon}{{\ttfamily To be appeared
  soon}}.

\bibitem{SabanciKeceli:2018fsd}
A.~Sabanc\i{}~Keceli, P.~Bandyopadhyay and K.~Huitu, \emph{{Long-lived
  triplinos and displaced lepton signals at the LHC}},
  \href{https://doi.org/10.1140/epjc/s10052-019-6818-0}{\emph{Eur. Phys. J. C}
  {\bfseries 79} (2019) 345}
  [\href{https://arxiv.org/abs/1810.09172}{{\ttfamily 1810.09172}}].

\bibitem{Bandyopadhyay:2010cu}
P.~Bandyopadhyay, P.~Ghosh and S.~Roy, \emph{{Unusual Higgs boson signal in
  R-parity violating nonminimal supersymmetric models at the LHC}},
  \href{https://doi.org/10.1103/PhysRevD.84.115022}{\emph{Phys. Rev. D}
  {\bfseries 84} (2011) 115022}
  [\href{https://arxiv.org/abs/1012.5762}{{\ttfamily 1012.5762}}].

\end{thebibliography}\endgroup

\end{document}